# An investigation of cooling flows and general cluster properties from an X-ray image deprojection analysis of 207 clusters of galaxies

D.A. White[1], C. Jones[2] and W. Forman[2]
[1]Institute of Astronomy, Madingley Road, Cambridge CB3 OHA. (E-mail: daw@mail.ast.cam.ac.uk)
[2]Smithsonian Astrophysical Observatory, 60 Garden Street, Cambridge, MA02138, U.S.A

**ABSTRACT**

In this paper we present an X-ray image deprojection analysis of EINSTEIN OBSERVATORY imaging data on 207 clusters of galaxies. The resulting radial profiles for luminosity, temperature, and electron density variations are determined from the cluster surface-brightness profiles according to gravitational potential constraints from average X-ray temperatures and optical velocity dispersions. This enables us to determine cooling-flow and other cluster properties, such as baryon fractions, Sunyaev-Zeldovich microwave decrements, and Thomson depths. From the results, we have compiled a catalogue of the detected cooling flows, and investigated their effects on general cluster properties. To assist in the analysis, we have constructed self-consistent correlations between the cluster X-ray luminosity, temperature, and optical velocity-dispersion, using 'orthogonal distance' regression to account for errors in both dimensions of the data. These fits indicate that, in general, the temperatures of clusters are isothermal and that they have spectral $\beta$-values consistent with unity.

We find that the X-ray luminosity, temperature, and optical velocity dispersion relations depend significantly on the cooling flow mass-deposition rate, through characteristic differences in the density profiles. Clusters of similar cooling flow mass-deposition rate exhibit self-similar density profiles, with larger cooling flows showing higher central densities. This leads to scatter in the luminosity related correlations within the X-ray luminosity, temperature and optical velocity dispersion plane. The segregation in density also leads to dispersion in other related properties such as 'half-light radii' and baryon fractions. The baryon fraction in the cores of cooling flow clusters appears to be higher, but as the density profiles tend to a similar value at larger radii, irrespective of cooling flow property, so too do the baryon fraction profiles appear to rise to a concordant value of greater than 10 percent at 1 Mpc. Thus, this sample indicates that clusters, as a whole, are inconsistent with primordial nucleosynthesis baryon fraction prediction, for a flat Universe, of 6 percent.

## 1 INTRODUCTION

The technique of 'X-ray image deprojection' was first used in the study of the Cassiopeia-A supernova remnant (Fabian et al. 1980), but its subsequent application to the Perseus cluster (Fabian et al. 1981) showed that it was particularly successful in the analysis of clusters of galaxies, especially those with cooling flows because of their near spherically-symmetric appearance and highly peaked surface-brightness profiles. A benefit of the deprojection method (especially over detailed spectral analysis) is that it can easily be applied to fainter systems, allowing analysis for a large sample of clusters. Although it requires a-priori knowledge of a cluster's gravitational potential, sufficient constraints can be obtained from the optical velocity dispersion and global cluster X-ray temperature to allow the properties of the intracluster gas to be determined.

The first deprojection analysis of a significant number of clusters (36) was from EINSTEIN OBSERVATORY data by Stewart et al. (1984). A larger sample of approximately 100 clusters was analysed by Arnaud (1988), again using EINSTEIN OBSERVATORY imaging data, with particular emphasis on the study of cooling flows. From the analysis by Arnaud and the most recent analysis of the 50 brightest clusters observed by EXOSAT (Edge 1989; Edge & Stewart 1991a; Edge, Stewart, & Fabian 1992) the prevalence of cooling flows in clusters is estimated to be approximately 50 to 60 percent. However, Edge et al. (1992) have noted that this proportion may be a lower limit, as the spatial resolution of detectors can lead to a bias against the detection of more distant cooling flows at large redshift, and the true fraction of cooling flows in clusters could be up to 90 percent.

This paper describes the deprojection analysis of EINSTEIN OBSERVATORY IPC (Imaging Proportional Counter) and HRI (High Resolution Imager) X-ray images of 207 clusters – the largest sample of cluster deprojections yet published. The deprojected temperature and electron density radial profiles then enable the bolometric X-ray luminosity, gas mass, baryon fraction, reprojected temperature, Thomson depth, microwave decrement, and cooling flow properties to be derived. Integrated mass-deposition rates, which are accumulated out to the cooling radius (where the cooling time of the hot gas is less than the age of the Universe, $t_0$), are compiled into a catalogue of detected cooling flows. The aim of this paper is to investigate the relationship between cooling flows and general properties of clusters.

The paper is organised as follows. First, there is a brief description of the deprojection method and the derived results, followed by a discussion of the effect of uncertainties in the assumptions and input data. Correlations between (bolometric) X-ray luminosity ($L_{\rm X}$), X-ray temperature ($T_{\rm X}$) and optical velocity dispersion ($\sigma_{\rm opt}$) are then presented as they are required in the analysis to estimate unknown input data. The deprojection results are then summarised, and the spatial-resolution bias is determined to give a better estimate of the number of cooling flows in the sample. Subsequently, the $L_{\rm X}$, $T_{\rm X}$ and $\sigma_{\rm opt}$ correlations are re-examined with cooling flow mass-deposition rates as an additional parameter. Finally, the relationship between cooling flow mass-deposition rates and other cluster properties, especially the baryon fraction, are discussed.

Note, $H_0 = 50$ km s$^{-1}$ Mpc$^{-1}$ and $q_0 = 0.5$ are assumed throughout. This implies an age for the Universe of $t_0 = 1.3 \times 10^{10}$ yr.

## 2 DEPROJECTION METHOD

This section briefly describes the method used to deproject an observed cluster's X-ray surface-brightness profile (a more rigorous treatment is given by Fabian et al. 1980; Fabian et al. 1981; Kriss, Cioffi, & Canizares 1983). The procedure is best considered in two parts. In the first step the number of counts emitted per unit volume is determined as a function of cluster radius (*i.e.* the surface brightness profile is 'deprojected'), assuming spherical symmetry. Thus, the volume of each shell⋆ (numbered $i$ with radius) projected into any surface-brightness annulus (numbered $j$) can be determined analytically from simple geometrical considerations (Kriss, Cioffi, & Canizares 1983). Subsequently, the observed volume count-emissivity, $F(i)$, can be compared with the theoretically expected values, $\Im(i)$, for a thermal plasma of a specific metal abundance (we use a metallicity of 0.4 times Solar throughout). The theoretical emissivity is corrected for source distance, attenuation due to line-of-sight absorption, and also the response of the detector:

$$\Im(i) = \frac{n_{\rm p}(i) n_{\rm e}(i)}{4\pi D_L^2} \int Q(E) \exp[-\sigma(E) N_{\rm H}]\, \Lambda[Z, T(i), E]\ dE\ , \quad (1)$$

where $\Im(i)$ is the flux registered over the energy band of the detector; $\Lambda[Z, T(i), E]$ is the cooling function (Mewe, Gronenschild, & van den Oord 1985; Mewe, Lemen, & van den Oord 1986) which in turn depends on the metal-abundance $Z$ and the temperature $T(i)$ of the hot-gas in the $i$th shell; $\sigma$ is the absorption cross-section (Morrison & McCammon 1983) normalised by the equivalent absorbing column density of neutral hydrogen, $N_{\rm H}$; $n_{\rm e}(i)$ and $n_{\rm p}(i)$ are the electron and ion densities respectively [note $n_{\rm e} = 1.21\, n_{\rm p}$ is used for a fully ionised gas]; $Q$ is the response of the detector at energy $E$; and $D_L$ is the luminosity distance of the source.

The important quantities required from the calculation are the temperature and the electron density, thus for purposes of illustration, eq. (1) can be written as:

$$\Im(i) \mathrel{\substack{\propto\\ \sim}} \frac{n_{\rm e}(i)^2 T(i)^{1/2}}{D_L^2}\ , \quad (2)$$

assuming the cooling function follows $\Lambda \propto T^{1/2}$ (which is valid for $T > 3 \times 10^7$ K where bremsstrählung dominates; note this approximation is not used in the actual calculations). If we use the Perfect Gas Law, $P = n_{\rm T}\, \mu m_{\rm p}\, kT$ (where $n_{\rm T} = n_{\rm e} + n_{\rm p}$, $\mu \approx 0.6$ is the mean molecular weight at $kT > 0.1$ keV, and $m_{\rm p}$ is the mass of a proton), in eq. (2) this gives:

$$\Im(i) \quad \mathrel{\substack{\propto\\ \sim}} \quad \left[\frac{P(i)}{T(i)}\right]^2 \frac{T(i)^{1/2}}{D_L^2}\ . \quad (3)$$

Because $F(i)$ is known and $F(i) = \Im(i)$ is a requirement, we can then specify the pressure $P(i)$ and obtain $T(i)$ and thereby $n_{\rm e}(i)$. The pressure need only be given at one particular radius, say in the outer shell $P_{\rm out}$, as the equation of hydrostatic equilibrium can be used to define the pressure at all other radii:

$$\frac{dP}{dr} = -n_{\rm T}\, \mu m_{\rm p} \frac{d\phi}{dr}\ , \quad (4)$$

where $d\phi/dr$ represents the gradient of the cluster's gravitational potential at radius $r$ corresponding to any shell $i$.

Thus, if we specify $P_{\rm out}$ and $d\phi/dr$ we can determine $kT$, and then $n_{\rm e}$, at all radii.

### 2.1 Derived results

From the resulting $n_{\rm e}$ and $kT$ radial profiles, many other parameters can be derived. Of particular interest are the cooling flow properties of each cluster, but the deprojected profiles also can be used to determine baryon fraction profiles, or reprojected to give emission-weighted temperatures, Thomson depths, and microwave decrements (Sunyaev-Zeldovich effect; Sunyaev & Zel'dovich 1980).

• *Cooling flow properties*: The mass-deposition rate, $\dot{M}$, is defined by the mass of hot gas which loses all its thermal energy within a specified cooling timescale, where the cooling time, $t_{\rm cool}$, of hot gas at constant pressure is given by:

$$t_{\rm cool}(i) = \frac{5}{2} \frac{k}{n_{\rm T}(i) T(i)} \int_0^{T(i)} \frac{T}{\int \Lambda[Z, T(i), E]\ dE}\ dT\ , \quad (5)$$

for gas in a volume-shell $i$ to cool from a temperature of $T(i)$ to zero. The critical timescale is usually chosen to be the age of the Universe ($t_0 = 1.3 \times 10^{10}$ yr – used throughout this paper – for $H_0 = 50$ km s$^{-1}$ Mpc$^{-1}$ and $q_0 = 0.5$), although ideally it should be the time since the last major disruptive merger event in the history of the cluster. However, this is generally indeterminate and the chosen definition in this analysis provides a consistent timescale for all clusters in the sample. The mass of gas involved in cooling to zero-temperature can then be estimated, to first-order, from the X-ray luminosity of the cluster, $\dot{M} \propto L_{\rm X}/T_{\rm X}$. However, the deprojection analysis also allows the gravitational-work done on the gas to be calculated, if we assume that the gas flows inwards to maintain pressure as it cools. The luminosity in any shell is then due to: (i) the mass of gas that cools completely after crossing a fraction of the shell's radius; (ii) a contribution from gas passing all the way across the shell with the associated change in temperature due to the gravitational work. This can be written as:

$$L_{\rm X}(i) = \dot{M}(i)\left[h(i) + f(i)\Delta\phi(i)\right] + \sum_{i'=1}^{i'=i-1} \dot{M}(i')\left[\Delta h(i) + \Delta\phi(i)\right] , \quad (6)$$

where $\dot{M}(i)$ is the mass-deposited in shell $i$; $\sum_{i'=1}^{i'=i-1}\dot{M}(i')$ is the mass of gas that needs to pass through shell $i$ to give rise to the radiation and mass-deposition in interior shells ($i$ increases outwards from the centre to bin $n$); $\Delta\phi(i)$ is the change in the gravitational potential; $h(i) = \frac{5}{2}\frac{kT(i)}{\mu m_{\rm p}}$ is the temperature in units of energy per particle-mass of the hot gas. In eq. (6) the mass that drops out in shell $i$ is given by the first two terms, and the mass that flows through to the next interior shell is described by the second two terms. Note that the gas which drops out in a shell also has gravitational work associated with crossing a fraction $f(i)$ of the overall change in the cluster potential, $\Delta\phi(i)$. This factor $f(i)$ can be calculated to represent the volume-averaged radius at which the mass is expected to drop out (see Arnaud 1988), however in this analysis $f(i) = 1$ is used. We assume that mass deposition takes place at the inner edge of each bin, which maximises the work done and therefore results in conservative mass-deposition rates. It also eliminates a bin-size dependency which arises if the $f$ is set to be the volume-averaged radius. The mass-deposition rate profile of a cluster may then be determined from the radial profiles of luminosity, temperature and gravitational potential by re-arrangement of eq. (6):

$$\dot{M}(i) = \frac{L_{\rm X}(i) - \left[\Delta\phi(i) + \Delta h(i)\right]\sum_{i'=1}^{i'=i-1}\dot{M}(i')}{h(i) + f(i)\Delta\phi(i)} . \quad (7)$$

⋆ Note the term 'annulus' or 'bin' shall refer to projected regions while the term 'shell' will refer to the corresponding volume within two bounding radii.

Integrated 'mass-deposition rates' that we quote later correspond to $\sum_{i=1}^{i=i(r_{\rm cool})}\dot{M}(i)$, but will hereafter be referred to simply as $\dot{M}$ [or $d\dot{M}/dt(R < r_{\rm cool})$ in figures]. The summation is performed out to the 'cooling radius', $r_{\rm cool}$, which is defined to be the radius within which the cooling time of the hot gas is less than the critical timescale, *i.e.* $t_{\rm cool} \le t_0$, as noted above.

• *Baryon fractions*: As we neglect the stellar contribution, this quantity is given by $f_{\rm b} = M_{\rm gas}/M_{\rm grav}$, where $M_{\rm gas}$ is the integrated gas mass obtained from the density solution and $M_{\rm grav}$ is the integrated gravitational mass which is defined by the equation of hydrostatic equilibrium and the gravitational-potential parameters used in the analysis. (Note the only uncertainty in $f_{\rm b}$ arises from the uncertainty in $M_{\rm gas}$ determined from the Monte-Carlo deprojection results, as $M_{\rm grav}$ is fixed according to the $\sigma_{\rm opt}$ and $R_{\rm core}$ parameters used in each cluster's deprojection.)

• *Emission-weighted temperatures*: These are calculated for the comparison of deprojected temperatures with the spatial-average temperature constraints from observations. The projected average seen in a particular annulus is calculated from the temperature and emissivities in volume shells that contribute to each annulus:

$$k\bar{T}(j) = \frac{\sum_{i=n}^{i=j} kT(i)\, w(i,j)}{w(i,j)} , \quad (8)$$

where $w(i,j) = L_{\rm X}(i)\frac{V(i,j)}{V(i)}$ is the weighting factor (which does not account for the absorption or response of the detector in the weighting function as it uses the luminosity not flux), $V(i,j)$ is the volume of the $i$th shell projected into the $j$th annular bin, and $V(i)$ is the volume of shell $i$ (note that the summation proceeds inwards from the outermost annulus $n$ to the selected annulus $j$, because only exterior shells contribute in projection). A spatial-average over the whole deprojected region of the cluster is then given by the median statistic (with 10th and 90th percentile limits) of all ($j = 1$ to $n$) annular bins. This is done to give a general cluster temperature which can be compared with broad-beam observations.

• *Microwave decrements and Thomson depths*: The projected microwave decrement can be determined using:

$$\frac{\Delta T_{\rm mw}}{T_{\rm cmb}}(j) = -A\sum_{i=1}^{i=j}\frac{2{\rm k}T(i)}{m_{\rm e}c^2}\,\sigma_{\rm T} n_{\rm e}(i)\ \Delta l(i,j) , \quad (9)$$

where $\Delta T_{\rm mw}/T_{\rm cmb}$ is the microwave decrement relative to the cosmic microwave background temperature of $T_{\rm cmb} = 2.735$ K (Smoot et al. 1991), and $\sigma_{\rm T} = 6.652 \times 10^{-25}$ cm$^2$ is the Thomson cross-section. The summation is performed for the path-length, $\Delta l(i,j)$, through the centre of any particular annular bin (*i.e.* the projected contribution from a shell $i$ seen in annular bin $j$). The annular contribution is normalised to the total decrement seen in the aperture by $A = [r(j)^2 - r(j-1)^2]/R_{\rm aper}^2$ (where $R_{\rm aper} = 6$ arcmin for this analysis). Note that the Thomson depth is simply the radial integral $d\tau = \sum_{i=1}^{i=j}\sigma_{\rm T}\Delta l(i,j)/2$, and so represents the probability of a photon being scattered while traversing half way through the projected depth of the cluster at any particular radius.

### 2.2 Statistical uncertainties

The complexity of the deprojection calculation and interdependence of the various results requires that the statistical uncertainties must be estimated using the Monte-Carlo technique. The observed surface-brightness profile for each cluster is regenerated 100 times, perturbed according to the statistical errors on the original data, and deprojected using the same input parameters. In this paper most of the results are then quoted as mean values with standard deviation errors. However, because the results for temperature, cooling time, and mass-deposition rate are not always symmetrically distributed, these are quoted as the median values (50th percentile) with 10th and 90th percentile uncertainty limits (which are converted to pseudo $1\sigma$ errors *i.e.* 16th and 84th percentiles for plots and regression analysis).

It should be noted that the Monte-Carlo technique can sometimes lead to numerical problems if an annulus in the surface-brightness profile has a significantly greater number of counts than the next interior bin, either due to statistical noise or background/foreground sources which may not have been properly subtracted. Too many counts will be subtracted from the interior annulus and the volume count-emissivity in the interior shell may then be so low, or even negative, that a physically realistic temperature solution cannot be found. This situation is compounded by the Monte-Carlo regenerations which can exaggerate the statistical fluctuations from bin-to-bin. The problem can be rectified by effectively smoothing the data in the problematic region, by sharing the counts evenly between the bins. This procedure only usually affects the outer regions of fainter clusters where the signal-to-noise ratio is low, but allows deprojection of these clusters to greater radii.

## 3 DEPROJECTION ANALYSIS

The deprojection of each cluster surface-brightness profile requires a redshift ($z$), line-of-sight Galactic column density ($N_{\rm H}$), optical velocity dispersion ($\sigma_{\rm opt,ref}$) and a spatially-averaged cluster X-ray temperature ($T_{\rm X,ref}$) (the 'ref' subscript is used to indicate the values are reference data available from various sources which are collated in Table 1). Although uncertainties in these parameters will manifest in the deprojection results, the observational uncertainties are not available for the whole sample. However, in individual cases the effect of observational uncertainties can be quantified by varying the parameter to observe its effect. As this has previously been done in some detail (*e.g.* White et al. 1994; White & Fabian 1995) only the general effects of parameter uncertainties and assumptions will be described below.

### 3.1 Deprojection input data

#### *3.1.1 Cluster surface-brightness profiles*

The surface-brightness profiles used in this analysis were extracted from the EINSTEIN OBSERVATORY data archive at the Harvard-Smithsonian Centre for Astrophysics (IPC data by C. Stern, C. Jones & W. Forman; HRI data by K. Arnaud) in the form of azimuthally-summed profiles centred on the peak of each cluster's X-ray emission. Obvious contaminating sources were excluded, and the profiles were corrected for the effect of vignetting. The general background-emission is accounted for either during the extraction process or the deprojection analysis. The actual counts that were registered by the HRI depended only on the bandpass of the X-ray telescope (*i.e.* $<$ 4.5 keV), and the low-energy cutoff of the detector (0.4 keV), while the energy range of counts detected by the IPC also depended on the detector gain. The information on each cluster observation, including the exposure duration, is presented in Table 2. This table also indicates the energy range selected for use in each IPC deprojection, and other input parameters which are discussed below.

The majority of the surface-brightness profiles (200 of the 228) in this sample are from the IPC. Its greater efficiency, lower internal background-noise, and larger field of view (effectively 60 arcmin, as the region from $60 - 70$ arcmin is unusable due to the high particle background) compared to the HRI ( 25 arcmin) enables cluster deprojections to larger radii. However, for some brighter sources HRI data is available and gives superior spatial resolution [the HRI's point spread function (PSF) is quoted to be 2 arcsec $1\sigma$ Gaussian width compared to $30 - 120$ arcsec for the IPC; Giacconi et al. 1979]. Although each surface-brightness profile was extracted with a certain fixed bin-size, the data can be binned-up during the deprojection analysis to improve statistics. The ideal annular bin-size for a deprojection analysis is one that is larger than the PSF of the detector, has sufficient number of counts per bin to minimize numerical problems (see Section 2.2), but also is small enough to maintain spatial resolution. It should be noted that occasionally an IPC deprojection requires a bin-size which is smaller than the IPC's PSF to obtain a reasonable number of deprojected radial bins (*i.e.* where possible at least three); these cases are identified in Table 3 as IPC entries where the bin-size is less than 60 arcsec. One problem with the IPC is that its spatial resolution biases against the detection of cooling flows – essentially it cannot resolve a cooling flow of size $r_{\rm cool} = 200$ kpc beyond a redshift of $z \sim 0.14$ (assuming a PSF of 60 arcsec, $H_0 = 50$ km s$^{-1}$ Mpc$^{-1}$ and $q_0 = 0.5$). This bias is investigated later, in Section 4.1.

The main uncertainties in the cluster surface-brightness profiles are the background contribution and the departure of the cluster emission from the assumed spherical symmetry. In principle the background may be overestimated if the cluster fills the field of the detector (and the background is taken from outside the maximum extent of the deprojection). Thus, background contributions for the whole sample will, if anything, err on the side of being overestimated, and then the luminosities and derived parameters will be underestimated. In terms of the assumption of spherical symmetry, the morphology of X-ray emission from clusters can be very complex. Of course, highly disturbed clusters should be avoided, but even relaxed clusters can have elliptical isophotes. Analytical expressions for an ellipsoidal volume could be used, but this would require the true spatial orientation of the cluster's emission to be known and specified for each cluster. As White et al. (1994) showed, the deprojection of the surface-brightness distribution of A478 in individual quadrants gives results which scatter around those obtained by deprojecting a complete azimuthal profile. Thus, the deprojection of the complete azimuthal profile of most regular clusters, even those with larger ellipticities that A478, should give results which are representative of the average radial properties, as long as the centroid of the emission does not change significantly with radius. After we have excluded clusters which are clearly disturbed and/or irregular, the main potential problem remaining is that of significant, but unknown, elongation along the line of sight.

#### *3.1.2 Redshift and cluster distances*

Although some clusters have very good redshift estimates, others may be very uncertain (*e.g.* estimates from the magnitude of the 10th brightest galaxy). The main problem arises from the choice of cosmological parameters: $H_0$ and $q_0$. The Hubble constant comes into mass-deposition rate determinations approximately as $\dot{M} \propto h_{50}^{-3}$ (because, to first approximation, $\dot{M} \propto L_{\rm X}/T_{\rm X} \propto h_{50}^{-}2$ and $r_{\rm cool} \propto h_{50}^{-1}$), and X-ray baryon fraction determinations vary as $f_{\rm b} = h_{50}^{-5/2}$ †. As the calculated nucleosynthesis values vary as $\Omega_{\rm b} = (0.05 \pm 0.01)\Omega_0 h_{50}^{-2}$, no reasonable value of $H_0$ can reconcile observed baryon fractions with the nucleosynthesis calculations (Steigman 1989; White et al. 1993; White & Fabian 1995; David, Jones, & Forman 1995). The uncertainties in $q_0$ and redshift, $z$, do not have a significant affect on the gas or gravitational masses, but $q_0$ affects the detected proportion of cooling flows in the sample because the critical timescale, $t_0$, equals $2 \times 10^{10}$ yr for $q_0 = 0$, whereas it drops to $1.3 \times 10^{10}$ yr for $q_0 = 0.5$.

† White & Fabian 1995 argued, erroneously, that the deprojection results did not conform to this trend. This was due to an incorrectly hardwired Hubble constant in a subroutine. This only affected the $H_0 = 100$ km s$^{-1}$ Mpc$^{-1}$ test in their paper and none of the other results.

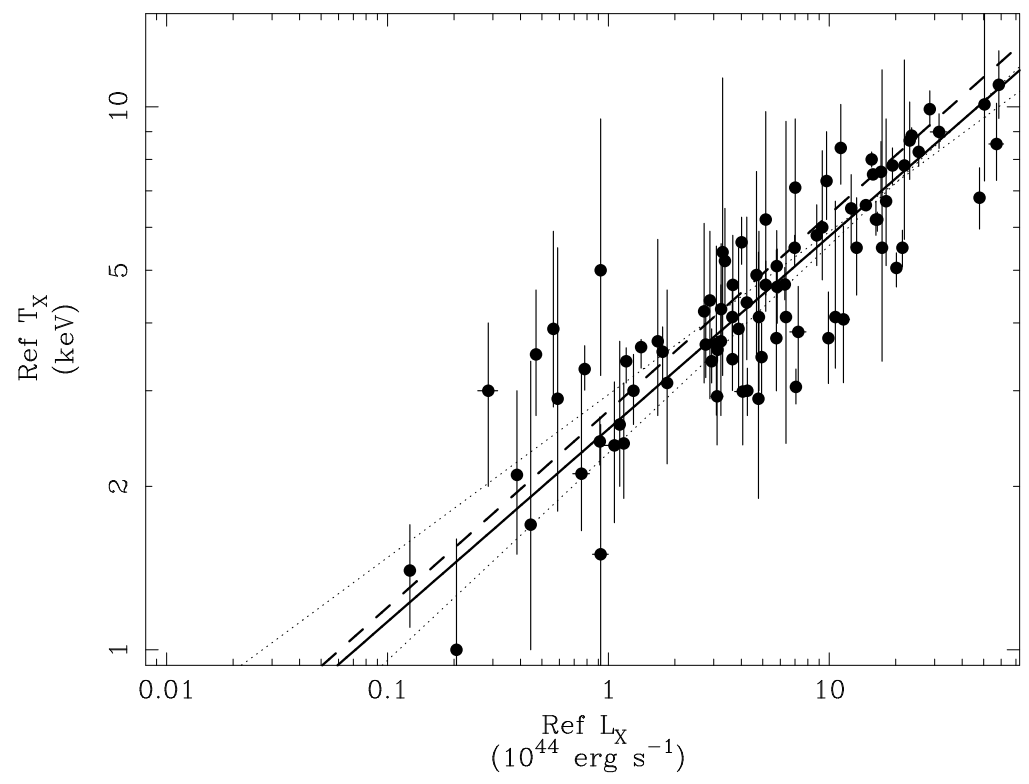

Double-sided weighting

$T_{\rm X,ref} = 2.55 \pm 0.09\, L_{\rm X,ref}{}^{0.35\pm0.01}$ (86)

$L_{\rm X,ref} = (6.08 \pm 5.92) \times 10^{-2}\, T_{\rm X,ref}{}^{2.77\pm1.84}$ (86)

Single-sided weighting

$T_{\rm X,ref} = 2.76 \pm 0.08\, L_{\rm X,ref}{}^{0.33\pm0.01}$ (86)

$L_{\rm X,ref} = (4.291 \pm 4.113) \times 10^{-2}\, T_{\rm X,ref}{}^{3.002\pm2.940}$ (86)

(a)

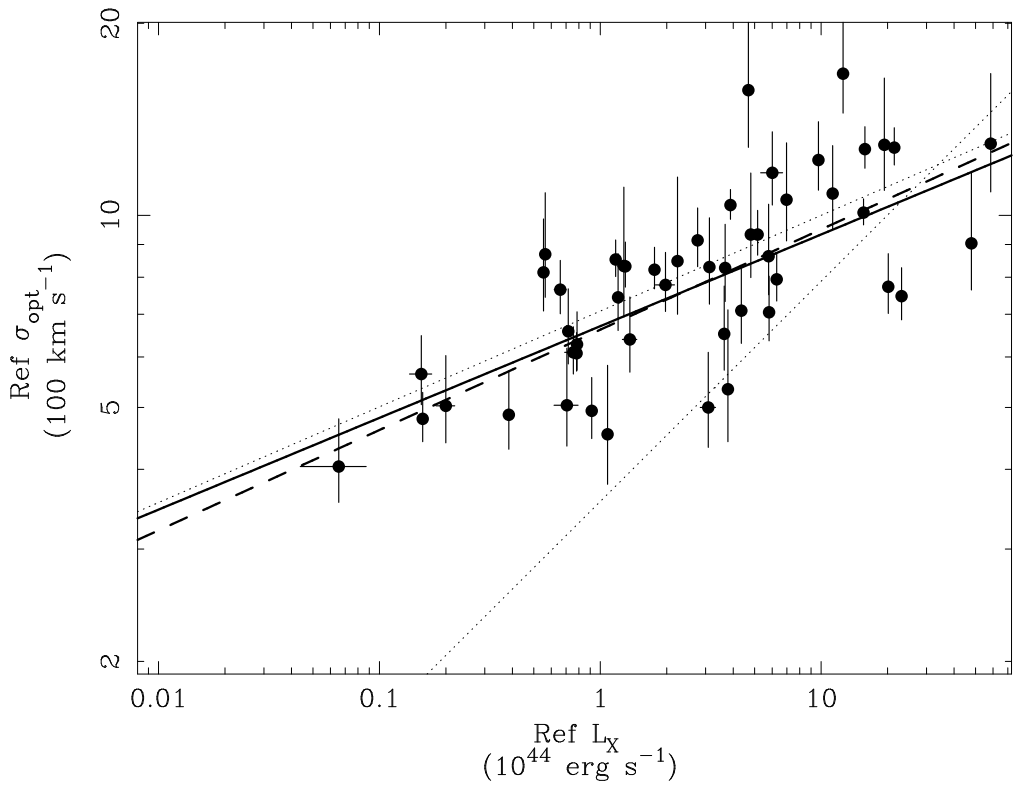

Double-sided weighting

$\sigma_{\rm opt,ref} = 6.70 \pm 0.19\, L_{\rm X,ref}{}^{0.14\pm0.01}$ (50)

$L_{\rm X,ref} = (6.00 \pm 3.48) \times 10^{-6}\, \sigma_{\rm opt,ref}{}^{6.36\pm0.27}$ (50)

Single-sided weighting

$\sigma_{\rm opt,ref} = 6.74 \pm 0.22\, L_{\rm X,ref}{}^{0.14\pm0.02}$ (50)

$L_{\rm X,ref} = (9.16 \pm 6.73) \times 10^{-6}\, \sigma_{\rm opt,ref}{}^{6.15\pm0.45}$ (50)

(b)

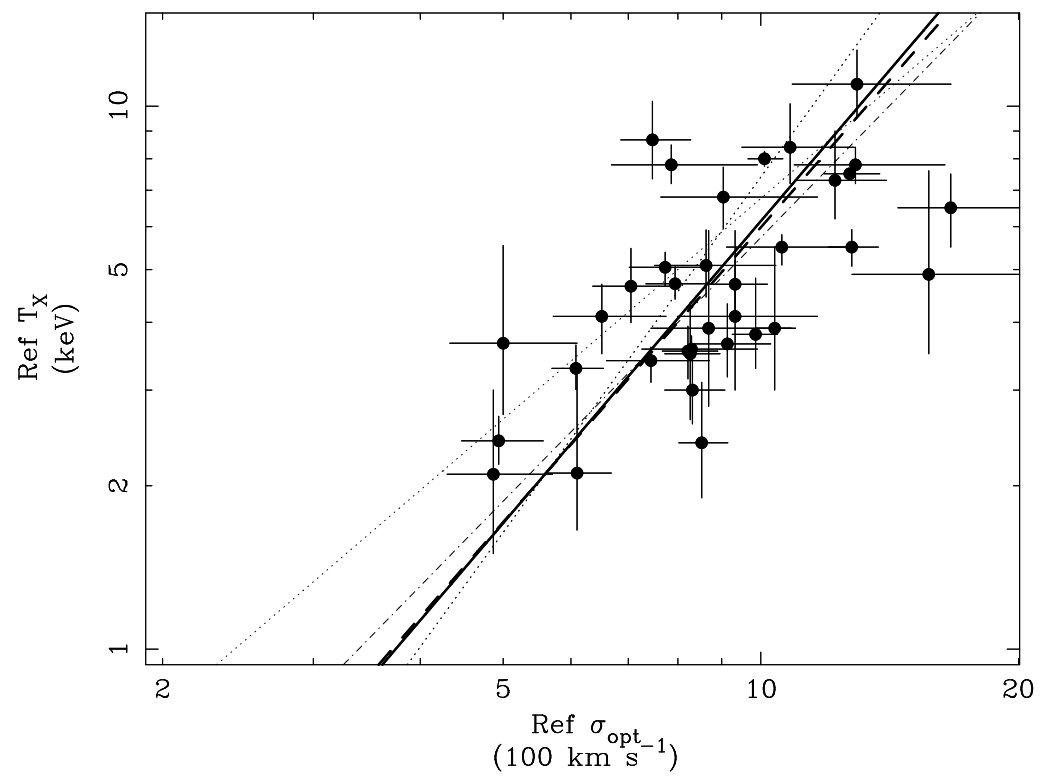

Double-sided weighting

$T_{\rm X,ref} = (8.74 \pm 4.90) \times 10^{-2}\, \sigma_{\rm opt,ref}{}^{1.85\pm0.24}$ (35)

$\sigma_{\rm opt,ref} = 3.71 \pm 0.52\, T_{\rm X,ref}{}^{0.55\pm0.08}$ (35) $\beta = 0.90$

($\sigma_{\rm opt,ref} = 4.03 \pm 0.10\, T_{\rm X,ref}{}^{[0.50\pm0.00]}$ (35) $\beta = 1.06$)

Single-sided weighting

$T_{\rm X,ref} = (7.93 \pm 5.17) \times 10^{-2}\, \sigma_{\rm opt,ref}{}^{1.88\pm0.27}$ (35)

$\sigma_{\rm opt,ref} = 3.75 \pm 0.58\, T_{\rm X,ref}{}^{0.54\pm0.10}$ (35) $\beta = 0.92$

($\sigma_{\rm opt,ref} = 4.01 \pm 0.11\, T_{\rm X,ref}{}^{[0.50\pm0.00]}$ (35) $\beta = 1.05$)

(c)

**Figure 1.**

These plots show the correlations between cluster X-ray luminosity ($L_{\rm X,ref}$; in $10^{44}$ erg s$^{-1}$), temperature ($T_{\rm X,ref}$; in keV) and optical velocity dispersion ($\sigma_{\rm opt,ref}$; 100 km s$^{-1}$). The data reference used are those with quoted uncertainties in Table 1. The thicker lines show the ODRPACK best fitting power-law functions, which are parameterised in the encapsulated tables for fits of the form $y = ax^b$ (solid line) and the inverse function $x = (y/a)^{(1/b)}$ (dashed line). The method produces results which are more stable than the standard least-squares fitting, *e.g.* see the comparable results from Edge & Stewart (1991a) shown by the dotted lines. The result from Lubin & Bahcall (1993) are also shown in panel (c) by the dashed-dotted line. Also shown in plot (c), in the tables of fit results between $T_{\rm X,ref}$ and $\sigma_{\rm opt,ref}$, are spectral $\beta$ values. [Note, the ODRPACK regression results, which are shown below each plot, indicate the best fits obtained from either weights calculated from the average of the absolute value of both positive and negative errors (these are the ones plotted), or those using only the positive errors. The parameters enclosed by square-brackets were fixed during the fits. The number of data points used in each fit is shown in parentheses. The errors quoted and used in the plots are $1\sigma$ standard deviations.]

### *3.1.3 Column density*

Neutral-hydrogen column densities, which are required to correct for the absorption of soft X-rays in our Galaxy, are generally estimated from radio 21 cm measurements (*e.g.* Stark et al. 1992). The discovery of intrinsic excess-absorption in the X-ray spectra of some clusters (White et al. 1991; Johnstone et al. 1992), shows that $N_{\rm H}$ may exceed the radio inferred values. However, excess column densities are not known for most clusters and so the use of the Stark et al. values produces conservative results with a consistent assumption. It also should be noted that the measured excess column densities of up to $10^{21}$ cm$^{-2}$ are probably only associated with the cooling flow region of clusters (White et al. 1991; Allen et al. 1993), and so possible underestimates in cluster luminosities and derived parameters ($t_{\rm cool}$, $\dot{M}$, and $M_{\rm gas}$) could be potentially larger in higher mass-deposition rate clusters.

### *3.1.4 Gravitational potential – temperature constraints*

To constrain the gravitational potentials used in the analysis requires an X-ray temperature and velocity dispersion for each cluster. If accurate X-ray temperature profiles were known then they could be used to constrain cluster gravitational potentials directly. However, it is only recently that detectors have gained sufficient spectral and spatial resolution to do so ( *e.g.* ROSAT Allen et al. 1993, and ASCA ) – the IPC data certainly do not provide the required constraints. Therefore, spatial-average cluster temperatures, from broad-beam detectors such as EXOSAT and GINGA , are used to constrain the deprojected temperature profiles, as they are known for a reasonable number of clusters. Most of these observed temperatures were taken from the compilations by Edge & Stewart (1991b), David et al. (1993), but also include some values from Jones & Forman (1996).

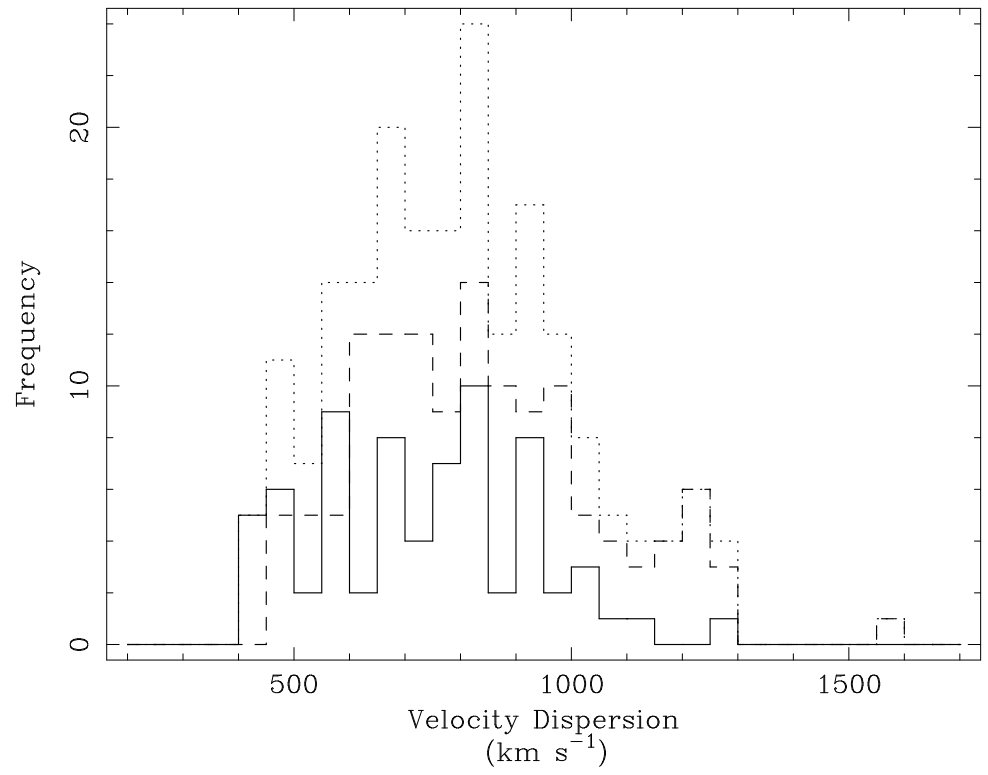

(a)

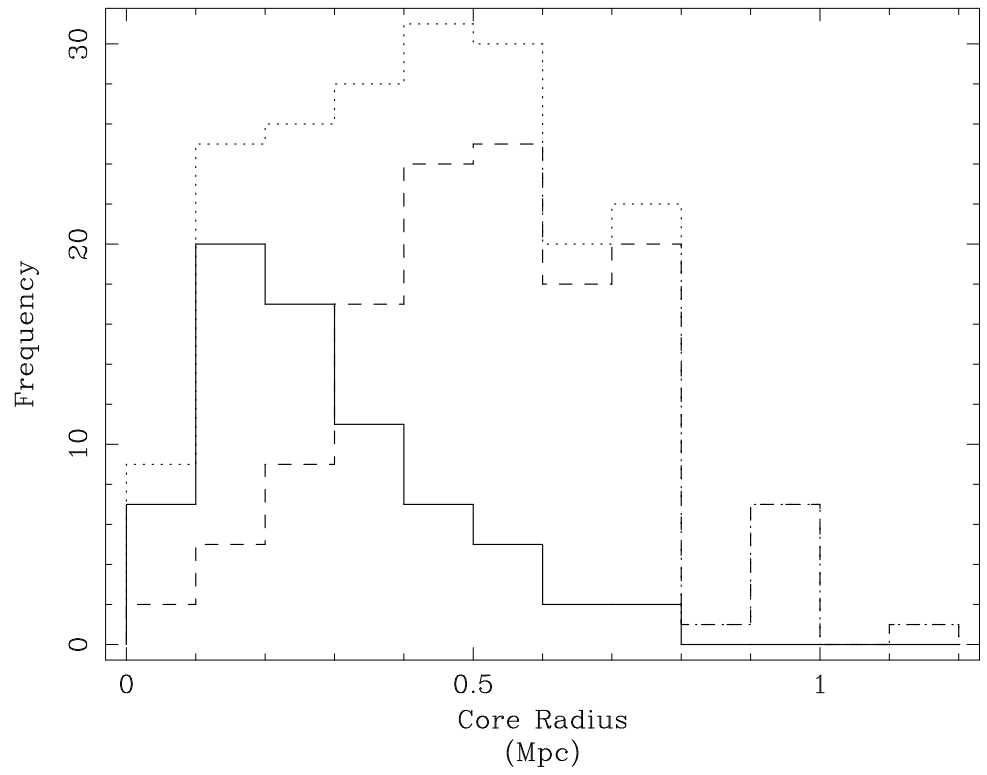

(b)

**Figure 2.**

These histograms show the distributions of (a) velocity dispersions, and (b) core radii used in the IPC deprojections. The distributions of cooling flow (CF) results are shown by the solid lines, the non cooling flow (NCF) results by the dashed lines, and the combined sample by the dotted lines. The median velocity dispersions are 782 km s$^{-1}$ for the combined sample, 752 km s$^{-1}$ for the CF, and 806 km s$^{-1}$ for the NCF sample. The median core radii are 0.5 Mpc, 0.3 Mpc and 0.6 Mpc, respectively.

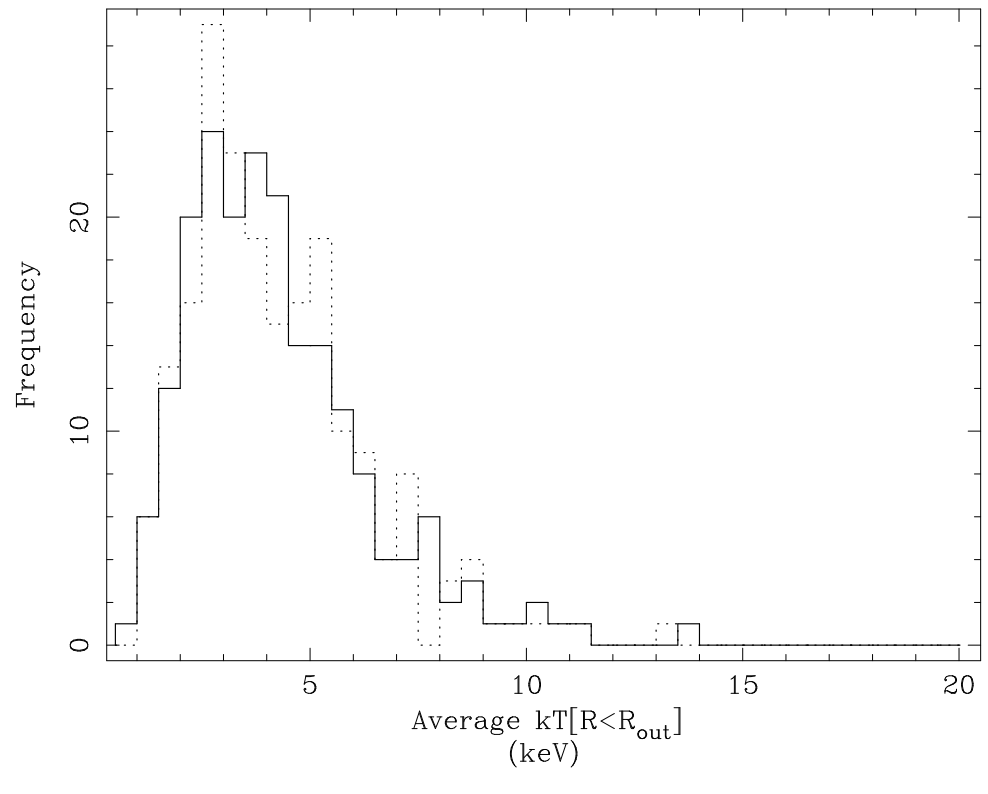

(a)

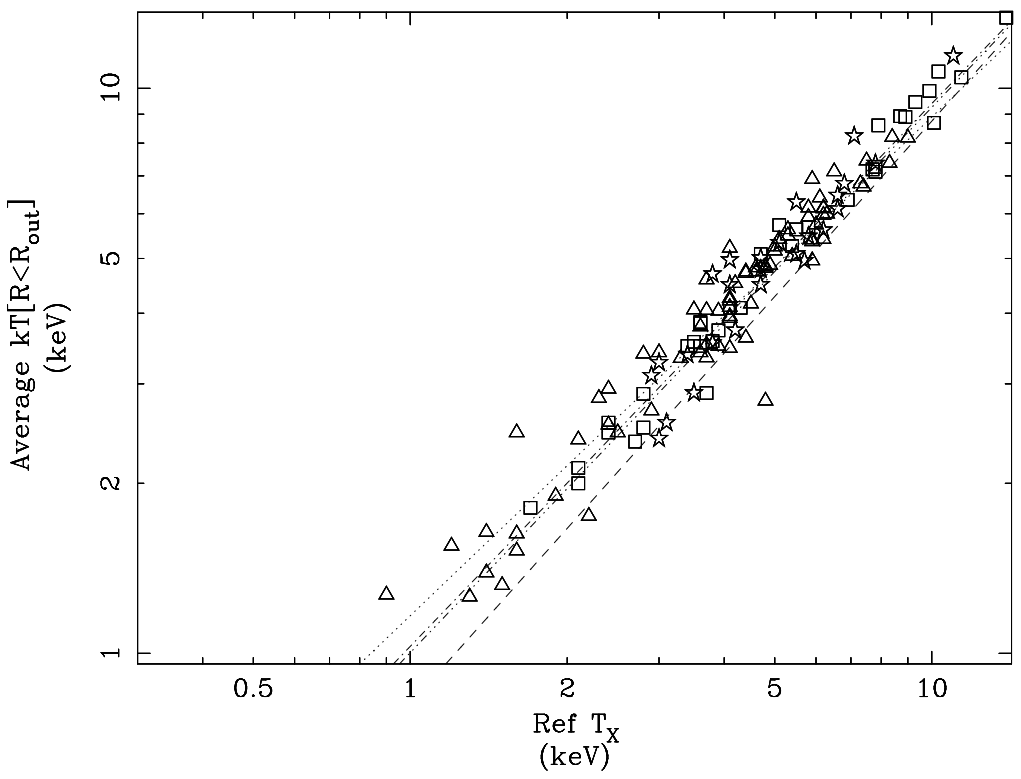

LCF: $\mathrm{k}T = 0.81 \pm 0.32\, T_{\mathrm{X,ref}}{}^{1.03\pm0.23}$ (22)
MCF: $\mathrm{k}T = 1.03 \pm 0.16\, T_{\mathrm{X,ref}}{}^{0.96\pm0.10}$ (30)
SCF: $\mathrm{k}T = 1.17 \pm 0.12\, T_{\mathrm{X,ref}}{}^{0.88\pm0.07}$ (57)
XCF: $\mathrm{k}T = 1.00 \pm 0.13\, T_{\mathrm{X,ref}}{}^{0.97\pm0.09}$ (46)

(b)

**Figure 3.**

These two plots compare the emission-weighted average deprojected temperatures against the reference values. The histogram in (a) shows the IPC deprojection results as solid lines against the reference values shown by the dotted line; these distributions are similar at the 96 per cent confidence level. In (b) the deprojected temperatures (both IPC and HRI) are plotted against the reference data as a function of mass-deposition rate using different symbol definitions (see Section 4.2). (Note the error bars are not plotted for clarity.)

All the remaining temperatures were interpolated from the $L_{\mathrm{X,ref}}$, $T_{\mathrm{X,ref}}$ and $\sigma_{\mathrm{opt,ref}}$ correlations, as will be described in Section 3.2.

For consistency in this analysis, all the cluster temperature profiles are assumed to be isothermal. It is known that the average cluster temperature in the core of large cooling flow clusters may decrease (*e.g.* Allen et al. 1993). However, new spectral results from ASCA (*e.g.* Allen, Fabian, & Kneib 1996 which also has gravitational lens constraints) indicate that in the cooling flow region the ICM (intracluster medium) is multiphase, and that there is probably an isothermal component as well as a cooler phase from the cooling flow. Thus the assumption of an isothermal potential and temperature profile may be physically realistic even in the core of a cooling flow cluster.

Reference temperatures, either from broad-beam observations of clusters, or interpolated values from $\sigma_{\mathrm{opt,ref}}$ and $L_{\mathrm{X,ref}}$, have uncertainties which primarily affect the normalisation of the gravitational mass, and thereby the baryon fraction and also mass-deposition rate profiles. As the effect on a gas mass determination is relatively small (because $L_X \lesssim n_e{}^2 T_X{}^{1/2}$), the uncertainties in temperature measurements can be translated to uncertainties in baryon fractions. However, these are unable to alleviate the baryon overdensity problem for all clusters (White & Fabian 1995).

#### *3.1.5 Gravitational potential – radial profile constraints*

The functional form of the cluster chosen for this analysis is a combination of two true-isothermal spheres (Binney & Tremaine 1987). These represent the central galaxy and the general cluster potentials; each is parameterised by a velocity dispersion $\sigma_{\rm opt}$ and core radius $R_{\rm core}$. The galaxy potential is fixed and assumed to have a velocity dispersion of 350 km s$^{-1}$ and core radius of 2 kpc (except for Fornax-A in which only the central galaxy potential is required). Velocity dispersions for the general cluster potential are taken from the literature when available, otherwise they are interpolated from the X-ray temperature or luminosity‡. Core radii constraints on the underlying mass profile are difficult to obtain observationally (*e.g.* a cooling flow can affect the X-ray determination of a core radius from the surface-brightness profile), and so this is treated as a free parameter which, in conjunction with the outer pressure ($P_{\rm out}$), produces a flat deprojected§ temperature profile (this assumption was justified in Section 3.1.5). As the effect of reducing $R_{\rm core}$ is to increase the deprojected gas temperature in the core, while increasing $P_{\rm out}$ raises the temperature at larger radii, a suitable combination of these parameters will usually produce a flat deprojected temperature profile which is consistent with the observational or interpolation constraint over most radii. For HRI data with complementary IPC observation the $\sigma_{\rm opt}$ and $R_{\rm core}$ parameters are fixed to that used in the IPC analysis, while $P_{\rm out}$ is set to the pressure at the equivalent radius in the IPC deprojection (as the IPC surface-brightness profiles usually extend to larger radii than the HRI data).

It should be noted that sometimes $R_{\rm core}$ is rather large (see Fig 2 and Table 2), but these values should not be taken too literally as they may be biased by: (i) our requirement that the deprojected temperature profile be as flat as possible, even in the core; (ii) the use of a galaxy potential in addition to the cluster potential; (iii) the PSF of the detector; (iv) possibly unresolved sources/structure in the cluster emission; and/or (v) uncertainties in $T_{\rm X,ref}$ (increasing this value enables a deprojection with a smaller $R_{\rm core}$). In relation to the second point, we note that a similar temperature profile results from the use of $R_{\rm core} = 0.6$ Mpc (instead of 0.8 Mpc) in the deprojection of the IPC data on A1736, if the galaxy potential is removed. The removal of the galaxy potential from all the deprojections would yield smaller core radii, however the magnitude of the effect would depend on the relative contribution of the galaxy to the original potential. We also note, in relation to the third point, that there does appear to be a correlation, although with large scatter, between $R_{\rm core}$ and redshift, which would be expected if the PSF had some effect on the choice of $R_{\rm core}$. Thus, there are probably several effects which can lead to core radii larger than might be expected. However, as this analysis is not attempting to say anything about cluster core radii (because this is highly dependant on the assumption about the isothermality of the temperature profiles), we do not consider the occasionally large core radii to be a problem.

The parameterisation of the gravitational potential is the most uncertain aspect of the deprojection analysis. The assumption of isothermality in the cluster core appears to be supported by the reference correlation which is approximately $\sigma_{\rm opt,ref} \propto T_{\rm X,ref}{}^{1/2}$ (see Section 3.2.1), while the use of $R_{\rm core}$ as a 'fitting parameter' means the only observational uncertainties are in $\sigma_{\rm opt,ref}$ because the galaxy generally has a negligible effect in the total gravitational potential. Uncertainties in $\sigma_{\rm opt,ref}$ can reduce baryon fractions and mass-deposition rate estimates (by increasing $\sigma_{\rm opt,ref}$) but the observational uncertainties on $\sigma_{\rm opt,ref}$ usually exclude gravitational masses which are sufficient to reduce baryon fractions to the expected primordial nucleosynthesis value (see White & Fabian 1995).

‡ In some deprojection results it was necessary to abandon the optical determination of $\sigma_{\rm opt,ref}$ as this produced a deprojected temperature profile which was far too hot compared to $T_{\rm X,ref}$, *e.g.* A426. In these cases $\sigma_{\rm opt,ref}$ was then interpolated.

§ In detail, the deprojected temperature profile is reprojected and an emission-weighted spatial average calculated to give the global average temperature for the cluster deprojection.

### 3.2 X-ray luminosity, temperature and optical velocity-dispersion correlations

As noted above, the deprojection analysis requires some constraints on the gravitational potential from X-ray temperatures and optical velocity dispersions. Correlations between X-ray luminosity, temperature and optical velocity dispersion are well known, and we use them to obtain missing temperatures or velocity dispersions when one of the three parameters is known. The correlations are re-evaluated using subsamples of $L_{\rm X,ref}$, $T_{\rm X,ref}$ and $\sigma_{\rm opt,ref}$ data on nearly 400 clusters (shown in Table 1). Before the procedure for interpolating unknown values is described, these results will be compared with the expected correlations, as they will be re-examined later in terms of the deprojection cooling flow results.

#### *3.2.1 Fit results and the expected scaling relations*

There have been many previous attempts at relating trends in $L_X$, $T_X$ and $\sigma_{\rm opt}$ observational data. Early investigation into the $T_X$ - $\sigma_{\rm opt}$ correlation were presented by Smith, Mushotzky, & Serlemitsos (1979), that of the $L_X$ - $T_X$ relation by Mitchell et al. (1979) and Henry & Tucker (1979), and $L_X$ - $\sigma_{\rm opt}$ by Quintana & Melnick (1982). Many subsequent investigations have been presented, *e.g.* Edge & Stewart (1991a) and Stewart (1994) (from the 50 brightest X-ray clusters), Lubin & Bahcall (1993), and recently Bird, Mushotzky, & Metzler (1995). However, in some of the previous analyses there have been problems obtaining consistent fits in some of these analyses because conventional least-squares regression analysis assumes the abscissae data have zero error. $T_X$ and $\sigma_{\rm opt}$ can have sufficient uncertainties to produce results which differ according to which data set is assigned to which axis. This is overcome by our use of an algorithm that takes into account errors in both dimensions of the data: 'orthogonal-distance regression' (ODRPACK; Boggs et al. 1990). (Note, in cases where there are no errors in the abscissae ODRPACK is equivalent to conventional least-squares fitting.) The resulting power-law fits are compared with those of Edge & Stewart (1991a) in Fig. 1. For $L_{\rm X,ref}$ versus $\sigma_{\rm opt,ref}$ we also show the results from Lubin & Bahcall (1993); they used a bootstrap method to determine consistent fit results. The ODRPACK fit and inverse-fit results show much better consistency than, for example, the Edge & Stewart (1991a) results, but the precise fit results depend on how the weighting factors are calculated

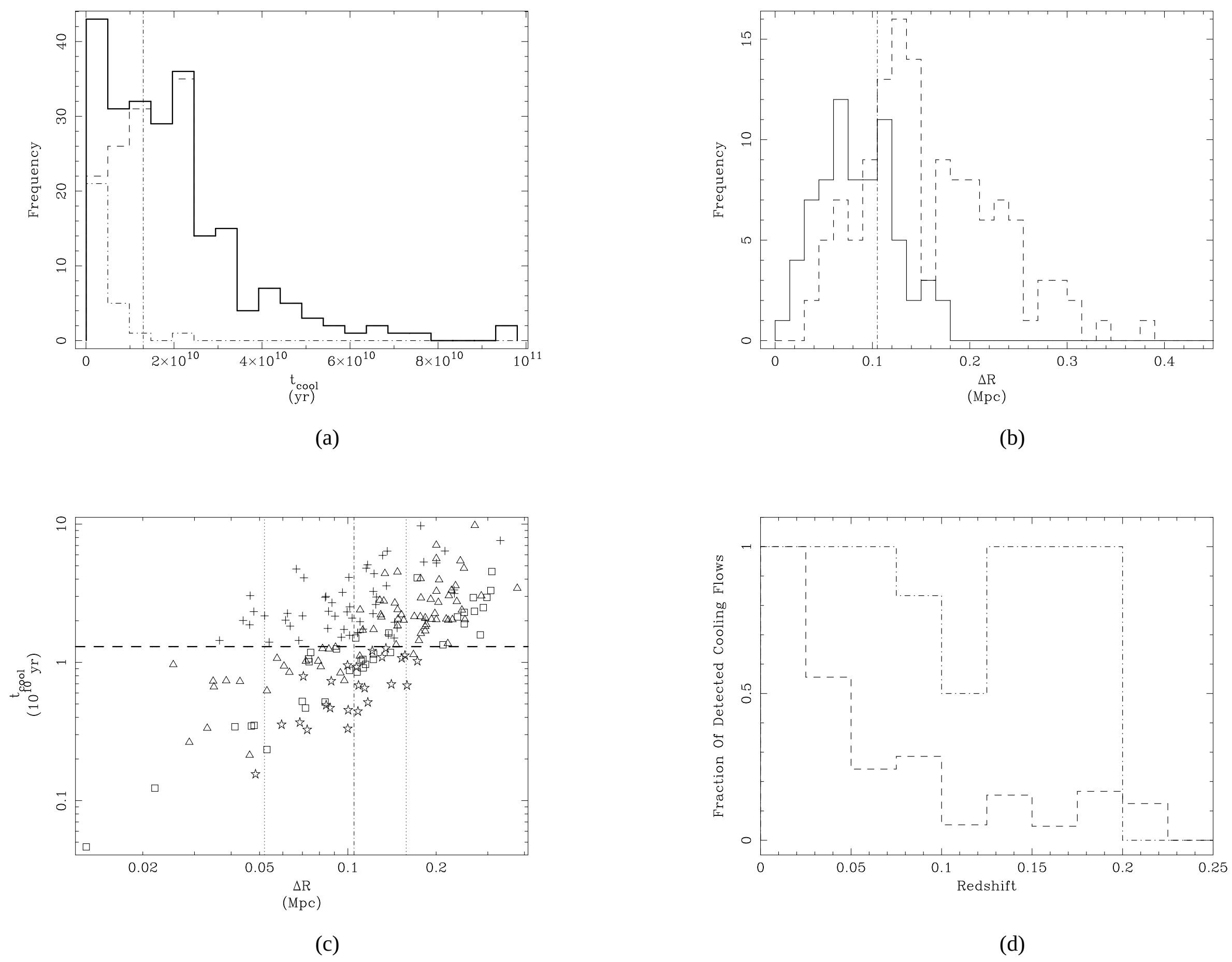

**Figure 4.**

The distribution of cooling time ($t_{\rm cool}$) values from the central bin of each deprojection of IPC data (dash line), HRI (dot-dash) data, and the combined sample (solid line) are shown in panel (a). Note that for a cluster to be classed as a cooling flow requires $t_{\rm cool} < 1.3 \times 10^{10}$ yr (delimited by the vertical dot-dash line). The distribution of the bin-sizes used in the IPC deprojection analysis, for cooling flow (solid line) and non cooling flow (dash line) results are shown in (b). There is a clear deficit of cooling flow results for bin-sizes $r > \overline{r}_{\rm cool}$ (shown by the vertical dot-dash line; derived from the HRI data which are well-resolved spatially). The correlation between the deprojection bin-size and central cooling time is presented in (c), for both the IPC and HRI data (error bars are omitted for clarity and the symbols indicate different cooling flow classes, as defined in Section 4.2). Note that there are very few cooling flows detected when the angular bin-size exceeds approximately 160 kpc. The horizontal line shows the critical cooling timescale, and the vertical lines indicate the mean cooling radius, $\overline{r}_{\rm cool}$, and uncertainties, determined from the HRI data. In (d) we plot the histogram of the fraction of cooling flows detected in the sample as a function of redshift [both IPC (dashed) and HRI (dash-dot) data are included here]. This decrease in the number of cooling flows detected by the IPC does not represent the evolution of cooling flows with redshift, but rather the bias in cooling flow detections due to spatial resolution.

from the errors in the data. These differences are shown, in encapsulated tables in Fig. 1, where the weighting is calculated from the average of the absolute values of the positive and negative errors, and alternatively from only the positive errors. The changes are relatively small, and in this analysis the positive and negative error weight fits are used throughout.

The ODRPACK power-law fits to observational data (*i.e.* excluding previously interpolated data) indicate that $L_{\rm X,ref} \propto \sigma_{\rm opt,ref}{}^{6}$, $L_{\rm X,ref} \propto T_{\rm X,ref}{}^{3}$, and $\sigma_{\rm opt,ref} \propto T_{\rm X,ref}{}^{1/2}$. (Note, different functions may provide a better description of the physical relationship being plotted, but this assumes that the actual physical relationship is known. Therefore, for consistency in this paper, we use power-law fits throughout.) The relationship between temperature and velocity dispersion is close to that expected for isothermal clusters, *i.e.* $\sigma_{\rm opt} \propto T_{\rm X}{}^{1/2}$, and the $\beta$-values (which also can be determined from the normalisation of this relation, as $\beta = \sigma_{\rm opt}{}^{2}/[{\rm k}T/(\mu m_{\rm p})]$) are consistent with unity, in agreement with Lubin & Bahcall (1993). However, the scaling relation between luminosity and temperature is expected to be $L_{\rm X} \propto T_{\rm X}{}^{2.5}$, if the X-ray gas is distributed similarly to the dark matter¶, and the observed correlation is sig-nificantly steeper than expected. This difference is amplified in the correlation between luminosity and velocity dispersion. Combining the above relations indicates that $L_{\rm X} \propto \sigma_{\rm opt}{}^5$, whereas a slope nearer $\sigma_{\rm opt,ref}{}^6$ is observed. The ODRPACK results shown in Fig. 1 by the thick lines show two fit results in each plot – one for fitting $y = ax^b$ and the other from fitting $x = (y/a)^{(1/b)}$. These results typically agree much better than the conventional results. Using these different weighting factors appears gives slightly different results, although the are still stable and well determined irrespective of which data set is assigned to be the dependant variable. It should also be noted that there is some discrepancy between the various correlations in the tables presented in Fig. 1, for example $L_{\rm X,ref} \propto T_{\rm X,ref}{}^{2.8\pm1.8}$ and $T_{\rm X,ref} \propto \sigma_{\rm opt,ref}{}^{1.8\pm0.2}$ implies that we should find $L_{\rm X,ref} \propto \sigma_{\rm opt,ref}{}^{5.0\pm3.3}$ whereas $L_{\rm X,ref} \propto \sigma_{\rm opt,ref}{}^{6.4\pm0.3}$ is actually the result (although the fits results are consistent with each other given the propagated uncertainties). It should also be noted that the same data are not used in each fit as only valid pairs of data are used. For example, a certain cluster may have $L_{\rm X,ref}$ and $T_{\rm X,ref}$ values used in the $L_{\rm X}\,T_{\rm X}$ correlation, but if the $\sigma_{\rm opt,ref}$ observed value is missing then this cluster does not come into correlations requiring $\sigma_{\rm opt,ref}$ data. In summary, the apparent internal inconsistency between the best fits of $L_{\rm X,ref} - T_{\rm X,ref} - \sigma_{\rm opt,ref}$ is not considered significant.

¶ If mass traces light then $M_{\rm gas} \propto M_{\rm grav}$ at all radii which is equivalent to $n_{\rm e} \propto M_{\rm gas} \propto M_{\rm grav} \propto \sigma_{\rm opt}{}^{2} \propto T_{\rm X}$, then as $L_{\rm X} \propto \int n_{\rm e}{}^{2} T^{1/2} dV$, $L_{\rm X} \propto T_{\rm X}{}^{2.5}$. (Note, the individual deprojected baryon fraction profiles are not constant, but appear to increase with radius, however the above argument holds if clusters are compared at similar and large enough radii.)

Despite the fact that the ODRPACK results agree with the fits by other authors there is, to date, no satisfactory explanation for the scatter and discrepancy in slopes. However, in Section 4.3 it will be shown that there is a dependency of some of these correlations on the properties of cooling flows.

### 3.2.2 *Interpolation of missing deprojection input data*

Returning to the issue of data required for the deprojection analysis, the ODRPACK correlations between $L_{\rm X,ref}$ - $T_{\rm X,ref}$ - $\sigma_{\rm opt,ref}$ can be used to predict unknown values according to the following procedure. Missing values of $\sigma_{\rm opt,ref}$ are determined from $T_{\rm X,ref}$ where available, or alternatively $L_{\rm X,ref}$. Conversely, missing values of $T_{\rm X,ref}$ are interpolated from $\sigma_{\rm opt,ref}$, or if necessary $L_{\rm X,ref}$. For the few cases where neither $T_{\rm X,ref}$ or $\sigma_{\rm opt,ref}$ are known, values are obtained by an iteration to a consistent deprojection solution. This is possible because of the interdependence of $T_{\rm X,ref}$, $L_{\rm X,ref}$ and $\sigma_{\rm opt,ref}$.

All input parameters obtained through this interpolation procedure are indicated as such in Table 1. The actual values of $\sigma_{\rm opt,ref}$ and $T_{\rm X}$ used in the deprojections are given in Table 2 and Table 3. The reference values shown in brackets have been interpolated if they are written in italic front, or iterated-to if the bracketed entry is blank. Certain velocity dispersion values also may be significantly different from the reference values, (*e.g.* A426) if $\sigma_{\rm opt,ref}$ gives a deprojected temperature which is far too high, and cannot be reconciled with $T_{\rm X,ref}$ by varying $P_{\rm out}$ and $R_{\rm core}$. In these cases $\sigma_{\rm opt,ref}$ is interpolated from $T_{\rm X,ref}$ (or $L_{\rm X,ref}$).

## 4 DEPROJECTION RESULTS

Having determined the required input data ($N_{\rm H}$, $T_{\rm X,ref}$ and $\sigma_{\rm opt,ref}$; see Table 2), each cluster surface-brightness profile is deprojected with the aim of obtaining a temperature profile that is isothermal (at least outside the size of a typical cooling flow) and consistent with the reference value, $T_{\rm X,ref}$. As noted previously this can be achieved through a suitable choice of $P_{\rm out}$ and $R_{\rm core}$; the resulting distributions of $R_{\rm core}$ are shown in Fig. 2, together with that for $\sigma_{\rm opt}$.

In Table 3 the average deprojected temperatures (*i.e.* reprojected spatial-median values with 10th and 90th percentile limits) are given for comparison with the reference values – they generally appear consistent. The quality of the sample's calibration is shown through a comparison of the deprojected and reference temperature distributions in Fig. 3(a). Both have a median value of 4.3 keV (from the IPC results), and the Kolmogrov-Smirnov (KS) test (Press et al. 1989) shows they are similarly distributed at the 96 percent confidence level (and this ignores uncertainties in the data). There are also no large systematic deviations in the temperature calibration as a function of cooling flow mass-deposition rate [Fig. 3(b); see below for details on cooling flow class definition]. These results indicate that the deprojection results are well calibrated with no systematic bias.

Table 3 also includes certain parameters, *e.g.* bolometric luminosity ($L_{\rm X}$), gas mass ($M_{\rm gas}$), gravitational mass ($M_{\rm grav}$), baryon fraction ($f_{\rm b} = M_{\rm gas}/M_{\rm grav}$; if $R_{\rm out} > R_{\rm core}$ as $f_{\rm b}$ varies rapidly within $R_{\rm core}$), integrated out to the maximum radius, $R_{\rm out}$, of each deprojection. The cooling flow properties, Thomson depth from the center of the cluster to $R_{\rm out}$, and the calculated Sunyaev-Zeldovich microwave decrement within 6 arcmin, are also given for each cluster. For ease in comparing $M_{\rm gas}$, $M_{\rm grav}$, $f_{\rm b}$, and $L_{\rm X}$ for each cluster, their integrated values within 0.5 and 1.0 Mpc are given in Table 4. The table also includes 'half-light radii', *i.e.* the radius which contains half the luminosity within 0.2, 0.5 and 1.0 Mpc.

The large number of clusters deprojected in this analysis provides an ideal opportunity for the study of cooling flow properties. However, it should be remembered that this sample is not homogeneously selected, *i.e.* flux-limited, especially when interpreting the estimated number of cooling flows in this sample. Nevertheless, given the large number of clusters it is possible to extract information on their properties as a function of cooling flow mass-deposition rate.

### 4.1 Fraction of clusters with cooling flows in the Einstein Observatory Sample

To detect a cooling flow in this deprojection analysis requires that the cooling time of the intracluster gas in the core of the cluster is less than a Hubble Time, *i.e.* $t_{\rm cool} < t_0$. The determination of the cooling time in the central bin of each deprojection is given in Table 3; and the distribution of their values are shown in Fig. 4. The corresponding integrated mass-deposition rates, $\dot{M}$, within the cooling radius, $r_{\rm cool}$, are given column (xiii) of the table, and their distribution shown in Fig. 5. For reference a catalogue of the cooling flows detected in the analysis are presented in Table 5, in decreasing order of $\dot{M}$, for both the IPC and HRI deprojections. These results are compared with those from previous analyses, in Fig. 6, using the compilation from various sources by Edge et al. (1992) (given in Table 6); there is a good agreement for both $\dot{M}$ and $t_{\rm cool}$. The slight systematic differences are partially due to: (i) changes in the code (an error in the density calculation in previous use of the code prior to ROSAT data analysis has been corrected), (ii) the use of a different critical timescale $t_0$, and (iii) a difference in methodology, *i.e.* the requirement for flat deprojected temperature profiles. Overall

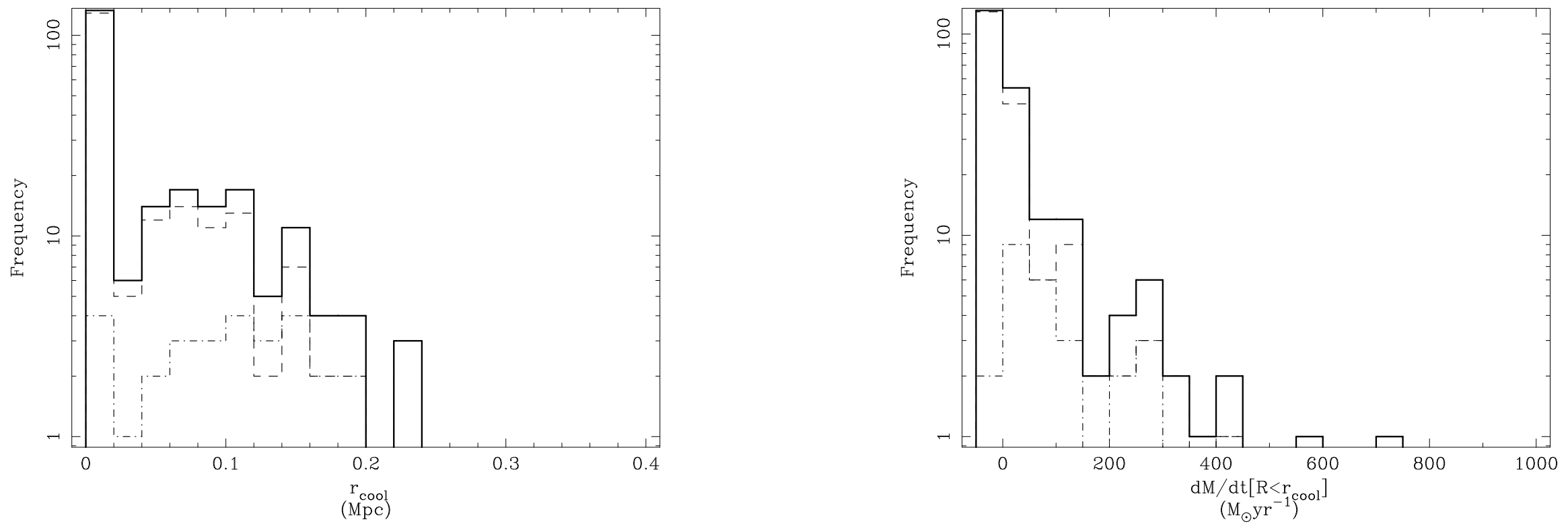

(a) (b)

**Figure 5.**

These figures show the distributions of (a) $r_{\rm cool}$ , and (b) $\dot{M}$ values. The IPC data are shown by the dashed lines, the HRI data by dot-dashed lines, and the combined sample by the solid lines. Note, in these plots the frequency is plotted in logarithmic units, and the bin with negative $\dot{M}$corresponds to non-cooling flow clusters where $\dot{M} = 0$.

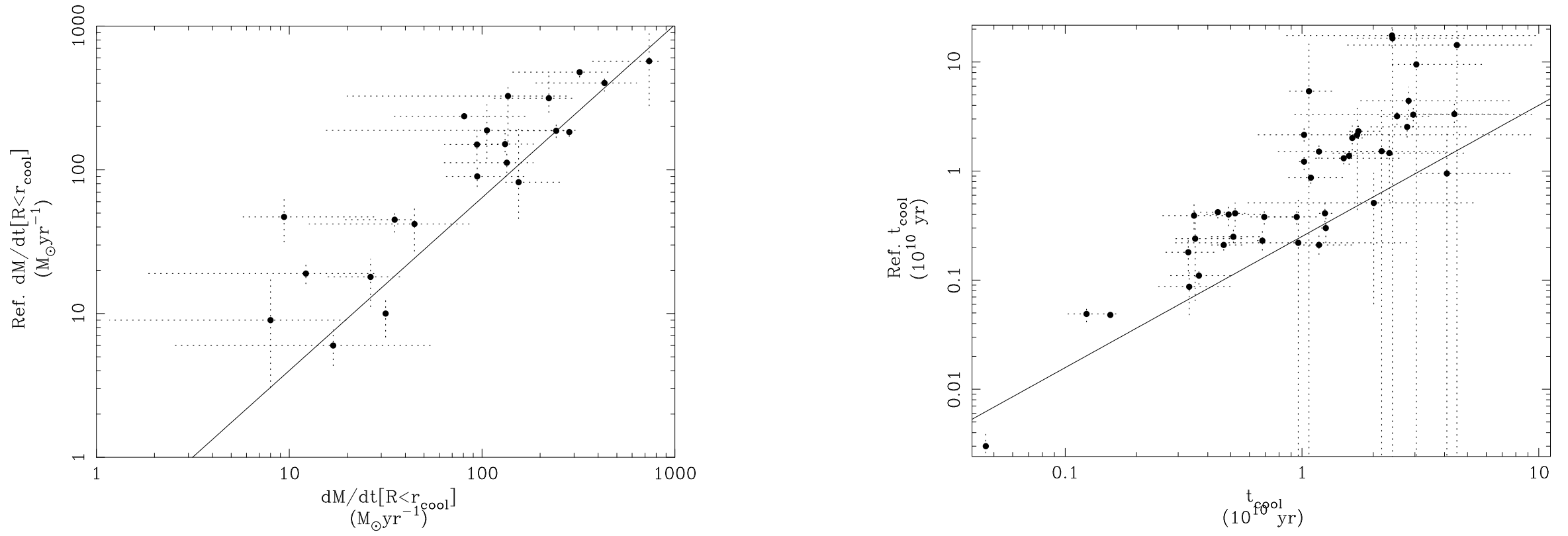

$\dot{M}_{\rm ref} = 0.25 \pm 0.14\ \dot{M}^{1.20\pm0.10}$ (22)
(a)

$t_{\rm cool,ref} = 0.69 \pm 0.06\ t_{\rm cool}{}^{1.49\pm0.05}$ (43)
(b)

**Figure 6.**

These plots compare the deprojection results against the values for objects in common with the compilation by Edge, Stewart, & Fabian (1992) in Tab. 6. Panel (a) compares $\dot{M}$ values, and (b) $t_{\rm cool}$ results. [Note, $\dot{M}$ values of zero do not show up in (a) as the data are plotted logarithmically.] The encapsulated tables present ODRPACK best fits results. There is generally reasonable agreement between these and the previous results (listed in Table 6). The errors plotted and quoted are $1\sigma$ standard deviations.

though, the results from the two samples appear consistent with each other.

According to the IPC results the number of cooling flows detected in the sample is 36 percent (71/200 IPC deprojections), however this is significantly different from the HRI detection rate of 93 percent (26/28). This discrepancy can also be seen, in Fig 4(a), as a deficit in the number of IPC deprojections with central cooling times shorter than the critical value of $t_0 = 1.3 \times 10^{10}$ yr. Although there are different selection effects in the IPC and HRI data (*i.e.* the HRI observations were generally on bright, nearby objects) there is real bias due to spatial resolution. Poor spatial resolution will affect the detection of cooling flows because the cooling time decreases with increasing density towards the centre of a cluster. Fig 4(b) shows that the IPC cooling flows results have better spatial resolution than the non-cooling flow results, *i.e.* $\overline{\Delta r}_{\rm CF} = 85{\pm}38$ kpc and $\overline{\Delta r}_{\rm NCF} = 158 \pm 69$ kpc respectively, and so it is clear that the central regions of some non-cooling flow clusters have not been resolved with the IPC data. However, assuming the HRI results are not affected by spatial-resolution, the bias in the IPC results can be estimated. If we look at the detected number of cooling flows in the HRI and IPC sample, Fig 4(d), we see that the HRI results produce detection cooling flows independent of redshift, while the IPC data show a clear decline with increasing redshift which we attribute to the decreasing angular resolution, and not evolutionary effects (note also that even the IPC detects all clusters as cooling flows in the nearest redshift bin). The HRI data indicate that the average size of a cooling flow is $\overline{r}_{\rm cool} = 105\,{\rm kpc} \pm 53\,{\rm kpc}$ [see Fig 5(a)]. There are 76 IPC deprojections which have sufficient spatial-resolution to resolve a region of this size (*i.e.* $\Delta r < \overline{r}_{\rm cool}$), and also have a bin-size greater than the IPC PSF of 60 arcsec. Thus, the proportion of detected cooling flows in this subsample of IPC results is $62^{+12}_{-15}$ percent (the errors come from the uncertainties in $\overline{r}_{\rm cool}$). This should not be interpreted as an accurate determination of the prevalence of cooling flows in clusters, but as a statement of the estimated number of cooling flows in this sample. In a flux-limited sample Edge, Stewart, & Fabian (1992) estimated that up to 90 percent of clusters could contain a cooling flow.

### 4.2 The effect of cooling flows on cluster properties

Given that many, if not most, clusters contain a cooling flow it is important to understand their effect on, and relation to, other clusters properties. Their relevance to the properties in the core region of clusters can be seen from the fact that 36 percent [see Fig. 7 (a)[‖]] of the X-ray luminosity within $r_{\rm cool}$ is due to complete cooling of the gas to zero-temperature rather than cooling due to gravitational work done [see eq. (6)]. Fig. 7 (b) shows a similar result for the fraction of the X-ray luminosity within $r_{\rm cool}$ compared to the luminosity within 0.5 Mpc and 1.0 Mpc. The significance of cooling flows on larger scales is apparent through 'half-light radii'. Fig. 8 shows that a larger fraction of the global luminosity arises from the central regions of cooling flows than non-cooling flows, *i.e.* the emission in cooling flows is more centrally concentrated. Thus, the effect of cooling flows on core properties appears to be strong enough to manifest itself in the properties on a global scale. In the following text the relationship between cooling flow properties and more general properties will be investigated.

In general, we find that larger cooling flows occupy larger volumes, and thus are naturally found in more massive clusters. Fig. 9 presents various correlations which support this. The correlation of $\dot{M}$ with $r_{\rm cool}$ shows the dependence is steeper than linear in volume, *i.e.* $\dot{M} \propto {r_{\rm cool}}^5$, while $L_{\rm X}$ also appears to be linearly related to $\dot{M}$. $T_{\rm X}$ and $f_{\rm b}$ are also well correlated with $\dot{M}$, with a shallower dependence than a power-law of index unity, that again confirms that larger cooling flows are generated in more massive clusters. Fig. 9(d) indicates that in the smallest cooling flows the baryon fraction within 0.5 Mpc may reach as low as the nucleosynthesis predictions of 5 percent, but not in larger cooling flows (see comments below). A trend between the optical $L_{\rm H\alpha}$ luminosity (Heckman et al. 1989) and $\dot{M}$ indicates there is some link, indirect or otherwise, between the line luminosity around central galaxies and the cooling flow (this was also shown in Edge & Stewart 1991a). A correlation between radio power (1.4 GHz; Owen et al. 1982) and $\dot{M}$ has also been investigated, although the scatter is too large to claim any significant trend.

To investigate the effect of cooling flows on more general cluster properties, the IPC deprojection results are divided into *cooling flow classes* according to the integrated mass-deposition rate ($\dot{M}$ – note only the median value is used and no account of the uncertainty is accounted for here), as follows:

- **All Cooling Flows (ACF):** 71 deprojections each value $\dot{M} > 0\,{\rm M_\odot\ yr^{-1}}$ subdivided as follows:
  – **Large Cooling Flows (LCF):** 26 deprojections with $\dot{M} > 50\,{\rm M_\odot\ yr^{-1}}$ ; star symbols; dashed lines.
  – **Moderate Cooling Flows (MCF):** 22 deprojections with $10 < \dot{M} \leq 50\,{\rm M_\odot}$; square symbols; dot-dash lines.
  – **Small Cooling Flows (SCF):** 23 deprojections with $0 < \dot{M} \leq 10\,{\rm M_\odot}$; triangle symbols; dot-dot-dot-dash lines.
- **Non-cooling flows (NCF):** 129 deprojections with $\dot{M} = 0\,{\rm M_\odot\ yr^{-1}}$ are divided according to whether the spatial-resolution, $\Delta r$, of the deprojection was sufficient to resolve a typical cooling flow of size $\overline{r}_{\rm cool}$ (see Section 4.1):
  – **Possible Cooling Flows (PCF):** 101 deprojections with $\Delta r > \overline{r}_{\rm cool}$.
  – **Excluded Cooling Flows (XCF):** 28 deprojections with $\Delta r < \overline{r}_{\rm cool}$; circle symbols; dotted lines.

As the actual number of detected cooling flows in the (IPC) sample is relatively low (36 percent), due to spatial-resolution bias, the most likely mass-deposition rate for the PCF class clusters is estimated using the correlation between $\dot{M}$ and $L_{\rm X}$ [a radius of 0.5 Mpc is chosen in order to obtain a large region of the cluster but not exclude most of the deprojection results; see Fig. 9(c)]. The division of the PCF results into the estimated CF classes (see Table 3) are: PLCF: 0, PMCF: 14, PSCF: 50, and PXCF: 26, with 11 left unassigned because they are not deprojected out to 0.5 Mpc (although there are no objects in the PXCF class, any cluster with an interpolated value of $\dot{M} < 1\,{\rm M_\odot\ yr^{-1}}$ would have been placed in this class). After this procedure there are sufficient clusters in each CF class (determined and probable) to enable cluster properties to be investigated as a function of the cooling flow properties. The bias introduced into the results by this procedure does not appear to be severe; most of the following correlations are similar to those obtained without the assignment of PCF data to possible cooling flow classes.

### 4.3 Cooling flows and the X-ray luminosity, temperature and optical velocity dispersion relations

In Section 3.2 correlations between reference data on $L_{\rm X,ref}$ - $T_{\rm X,ref}$ - $\sigma_{\rm opt,ref}$ were presented. These conformed to expectations from previous results (*e.g.* Edge & Stewart 1991a, Lubin & Bahcall 1993), but not the relations expected from scaling arguments. This analysis shows that when the correlations are performed on data from individual cooling flow classes, significant differences occur which depend on some parameter related to the cooling flow properties. Before this is discussed in detail it should be noted that the correlations obtained from the deprojection results as a whole are not independent of the reference data and correlations. This is because the deprojection $\sigma_{\rm opt}$ values *are* the reference values, while $T_{\rm X}$ is adjusted to be consistent with the reference value, and $L_{\rm X}$ is subject to the $T_{\rm X}$ determination. Thus, it should be expected that the general trends from the deprojection results are consistent with the reference correlations. However, their dependence on a third parameter, such as $\dot{M}$, should not be directly biased by these correlations.

The equivalent deprojection results for the $L_{\rm X}$ - $T_{\rm X}$ - $\sigma_{\rm opt}$ correlations are shown in Fig. 10. These appear reasonably consistent with the reference correlations which are overplotted as heavy solid-lines. (There is a slight difference in normalisation of the luminosity within 0.5 Mpc, but as this disappears in the comparison at 1 Mpc

[‖] Note all fits and plots exclude the HRI results to avoid duplication of cluster observations.

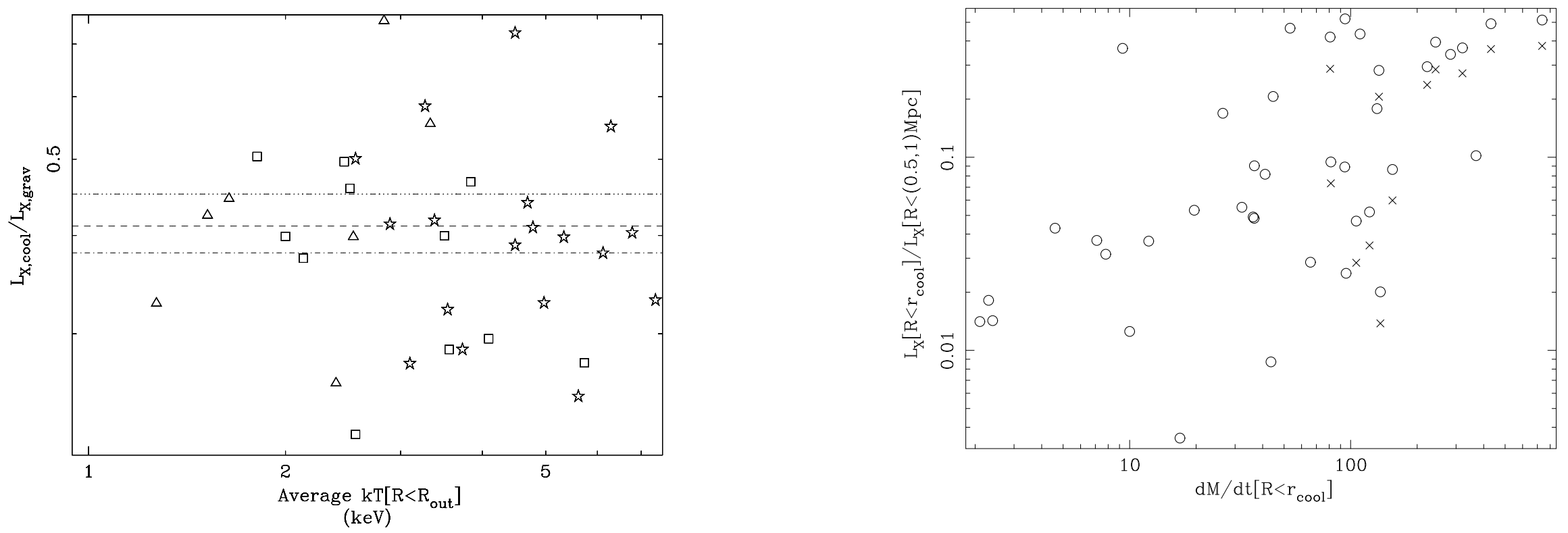

LCF: $L_{\rm CF} = 0.41\, T_{\rm X}^{\ [0.00]}$ (18)
MCF: $L_{\rm CF} = 0.38\, T_{\rm X}^{\ [0.00]}$ (11)
SCF: $L_{\rm CF} = 0.45\, T_{\rm X}^{\ [0.00]}$ (7)
XCF: Not applicable

(a)

(b)

**Figure 7.**

This shows (a) the fraction of the total luminosity within $r_{\rm cool}$ attributable to cooling, rather than gravitational work [see the $h$ terms compared with $\phi$ terms in eq. 6]. The fits show power-laws of fixed, flat slopes (*i.e.* essentially the results are weighted means of the ordinate data) fitted to the different cooling flow classes which are defined in Section 4. (The uncertainties in the temperature are not plotted.) In (b) we plot the X-ray luminosity within $r_{\rm cool}$ as as fraction of that within 0.5 Mpc (triangles) and 1.0 Mpc (crosses).

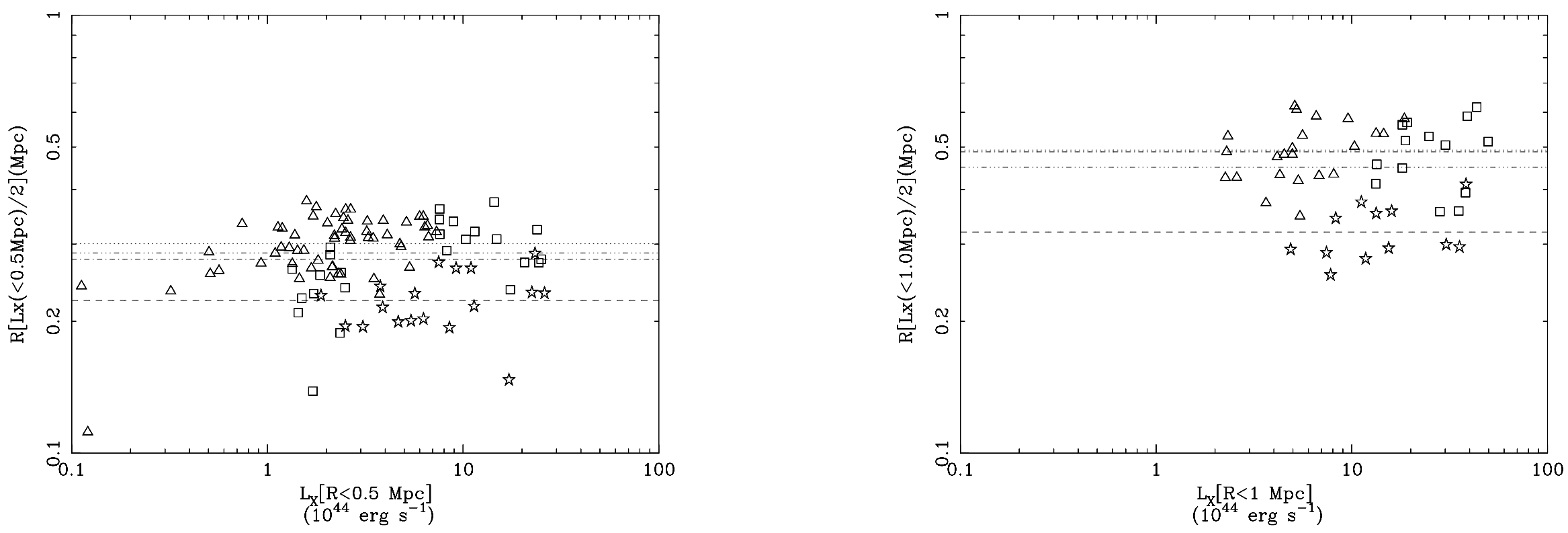

LCF: $R[L_{\rm X}\,(R < 0.5\,{\rm Mpc})] = 0.22\, L_{\rm X}\,[R < 0.5\,{\rm Mpc}]^{[0.00]}$ (18)
MCF: $R[L_{\rm X}\,(R < 0.5\,{\rm Mpc})] = 0.28\, L_{\rm X}\,[R < 0.5\,{\rm Mpc}]^{[0.00]}$ (25)
SCF: $R[L_{\rm X}\,(R < 0.5\,{\rm Mpc})] = 0.30\, L_{\rm X}\,[R < 0.5\,{\rm Mpc}]^{[0.00]}$ (58)
XCF: $R[L_{\rm X}\,(R < 0.5\,{\rm Mpc})] = 0.29\, L_{\rm X}\,[R < 0.5\,{\rm Mpc}]^{[0.00]}$ (34)

(a)

LCF: $R[L_{\rm X}\,(R < 1\,{\rm Mpc})] = 0.32\, L_{\rm X}\,[R < 1\,{\rm Mpc}]^{[0.00]}$ (12)
MCF: $R[L_{\rm X}\,(R < 1\,{\rm Mpc})] = 0.49\, L_{\rm X}\,[R < 1\,{\rm Mpc}]^{[0.00]}$ (14)
SCF: $R[L_{\rm X}\,(R < 1\,{\rm Mpc})] = 0.49\, L_{\rm X}\,[R < 1\,{\rm Mpc}]^{[0.00]}$ (23)
XCF: $R[L_{\rm X}\,(R < 1\,{\rm Mpc})] = 0.45\, L_{\rm X}\,[R < 1\,{\rm Mpc}]^{[0.00]}$ (2)

(b)

**Figure 8.**

These diagrams show 'half-light radii', *i.e.* radii which contain half the luminosity within (a) 0.5 Mpc and (b) 1 Mpc, for clusters in different ranges of mass-deposition rate. Note that luminosities are more centrally concentrated in larger cooling flow clusters, and that the fits are power-laws of flat slope (*i.e.* weighted means of the ordinate data).

this indicates that the discrepancy is because the broad-beam reference luminosities average over a larger radius than 0.5 Mpc. We also note that reference correlations are of observed, *i.e.* projected, temperatures and luminosities whereas those discussed below are correlations obtained from the deprojected values.) When the whole sample is divided by CF class there is significant segregation in the $L_{\rm X}$ - $\sigma_{\rm opt}$ correlations, and possibly a segregation of the larger CFs in the $T_{\rm X}$ - $L_{\rm X}$ correlation (as claimed by Fabian et al. 1994), but there are no real systematic differences in the $T_{\rm X}$ - $\sigma_{\rm opt}$ relation. Thus, it appears that segregation is mostly due to luminosity which primarily results from density differences ($L_{\rm X} \propto\!\!\!\sim n_{\rm e}^{\ 2} T_{\rm X}^{\ 1/2}$). Somewhat paradoxically the $L_{\rm X}$ versus $\sigma_{\rm opt}$ correlation is flatter than the standard correlation when the cooling flow $\dot{M}$ class is considered. This is due to the fact that a cluster with a smaller $\sigma_{\rm opt}$ requires a

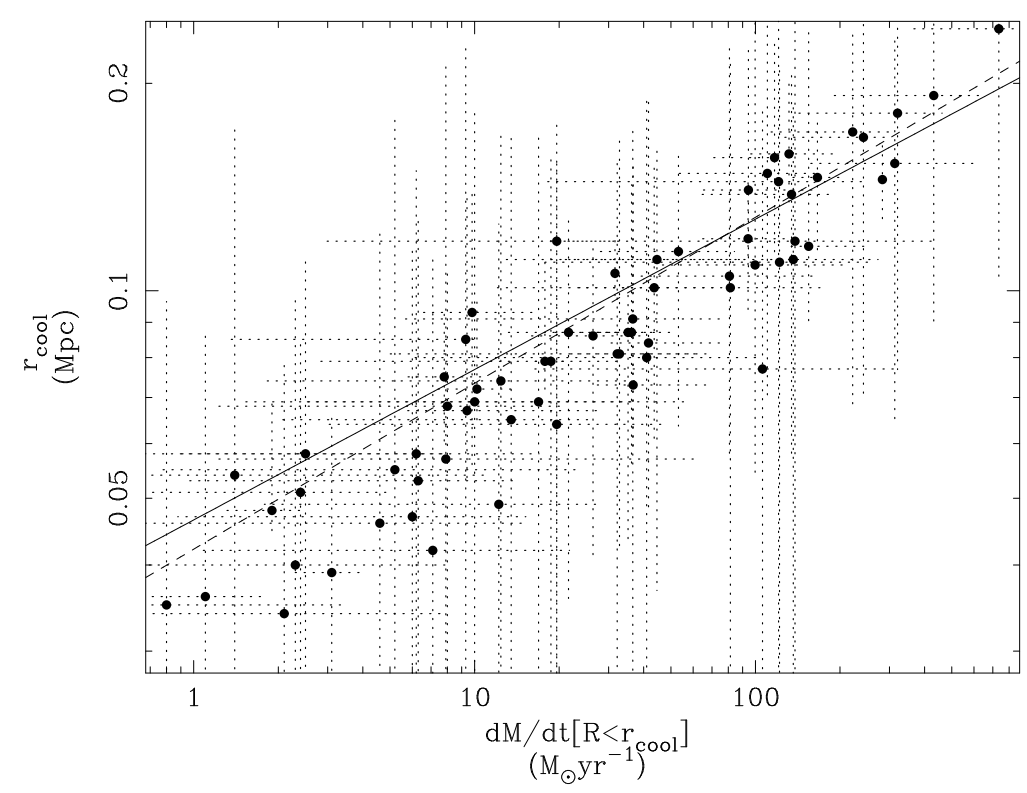

$r_{\rm cool} = (4.22 \pm 0.75) \times 10^{-2}\, \dot{M}^{0.24\pm0.05}$ (68)
(a)

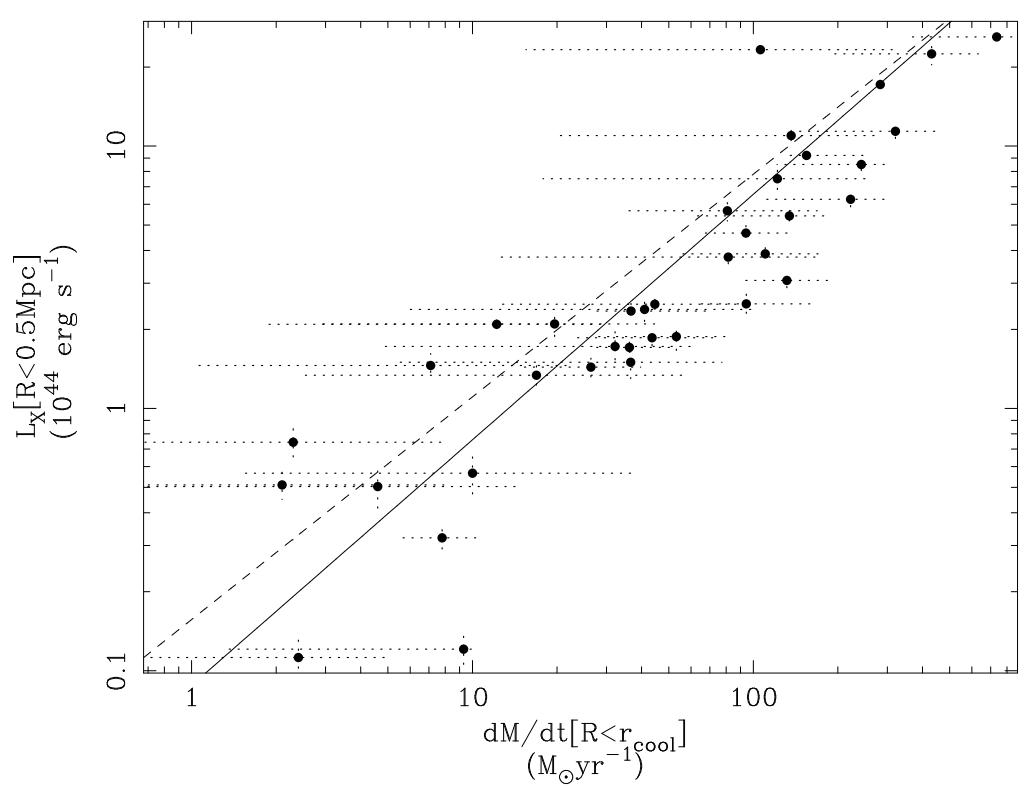

$L_{\rm X}\,[R < 0.5\,{\rm Mpc}] = 0.16 \pm 0.28\, \dot{M}^{0.85\pm1.18}$ (37)
(b)

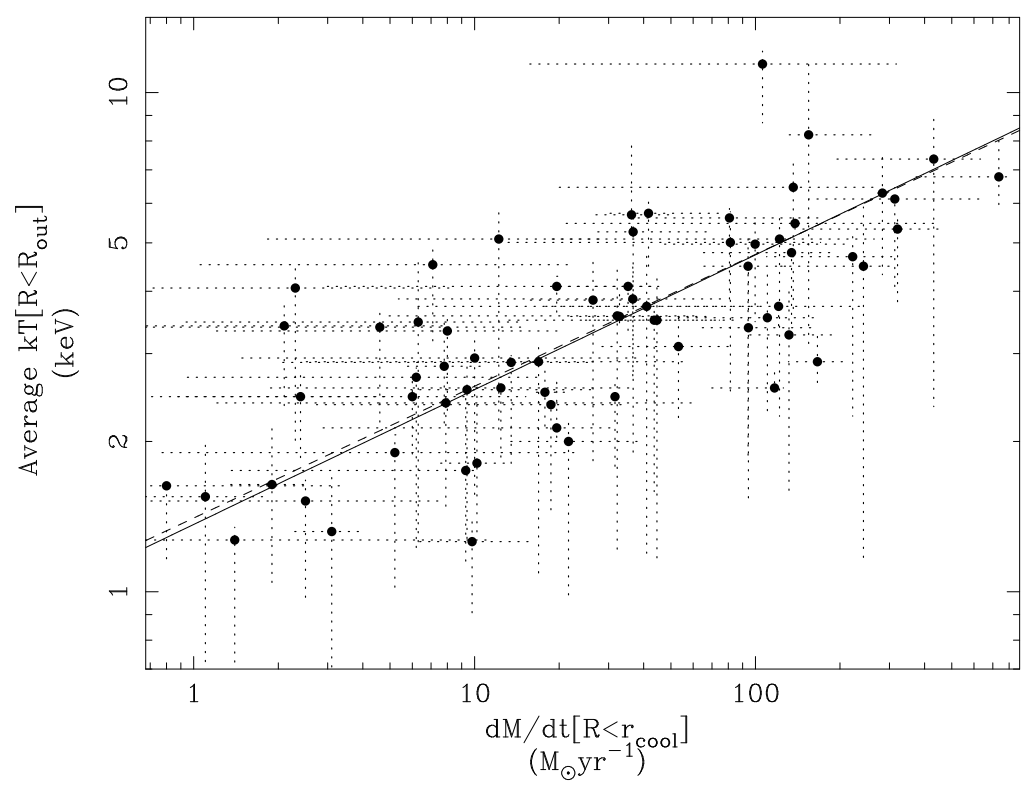

$kT = 1.41 \pm 0.29\, \dot{M}^{0.26\pm0.05}$ (66)
(c)

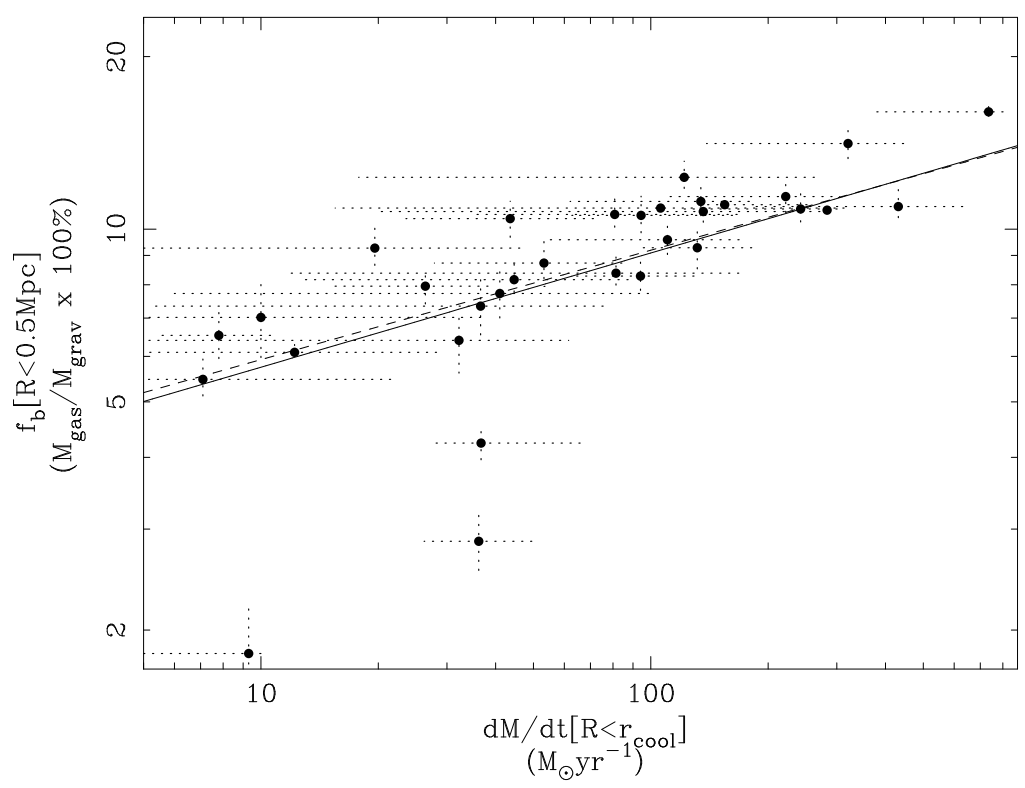

$f_{\rm b} = 3.61 \pm 4.00\, \dot{M}^{0.27\pm0.15}$ (52)
(d)

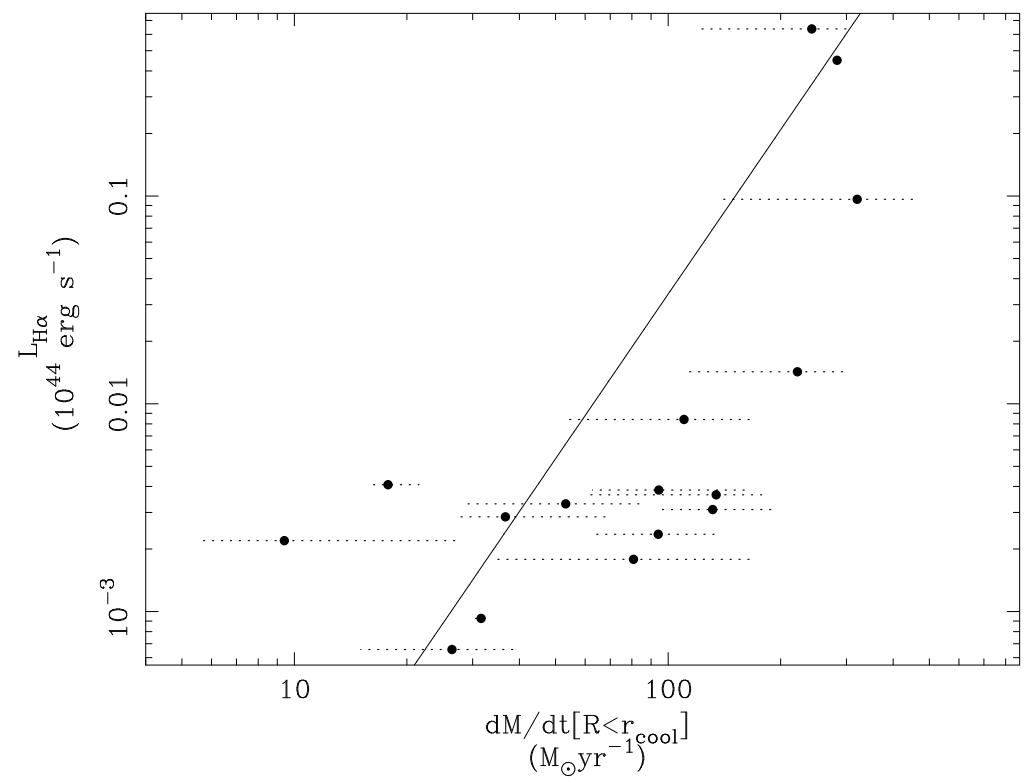

$\dot{M} = 363.21 \pm 22.01\, L_{\rm H\alpha}{}^{0.38\pm0.01}$ (16)
(e)

**Figure 9.**

These diagrams show the correlations of various parameters with cooling flow mass-deposition rate using data from individual deprojections. Power-law fits results are determined using the ODRPACK fitting package (see Section 3.2). [Note, the errors are $1\sigma$ standard deviations. The reciprocal relation is shown as a dashed line in all plots except (e), as the $L_{\rm H\alpha}$ data have no errors (and the fit is therefore conventional least-squares regression.]

steeper potential, *i.e.* a smaller $R_{\rm core}$ , in order to be in the same CF class as a cluster with a larger $\sigma_{\rm opt}$ . This will narrow the difference in luminosity between clusters of similar $\dot{M}$ class which have significantly different $\sigma_{\rm opt}$ and give flatter correlations in separate CF classes than from the sample as a whole.

### 4.4 Segregation of radial properties

To enhance differences in general radial properties as a function of mass-deposition rate the deprojected profiles can be *averaged* from results in each CF class. Individual profiles for each cluster in a particular CF class are grouped together into usually 20 data points (according to their order of the abscissae) and then averaged. The

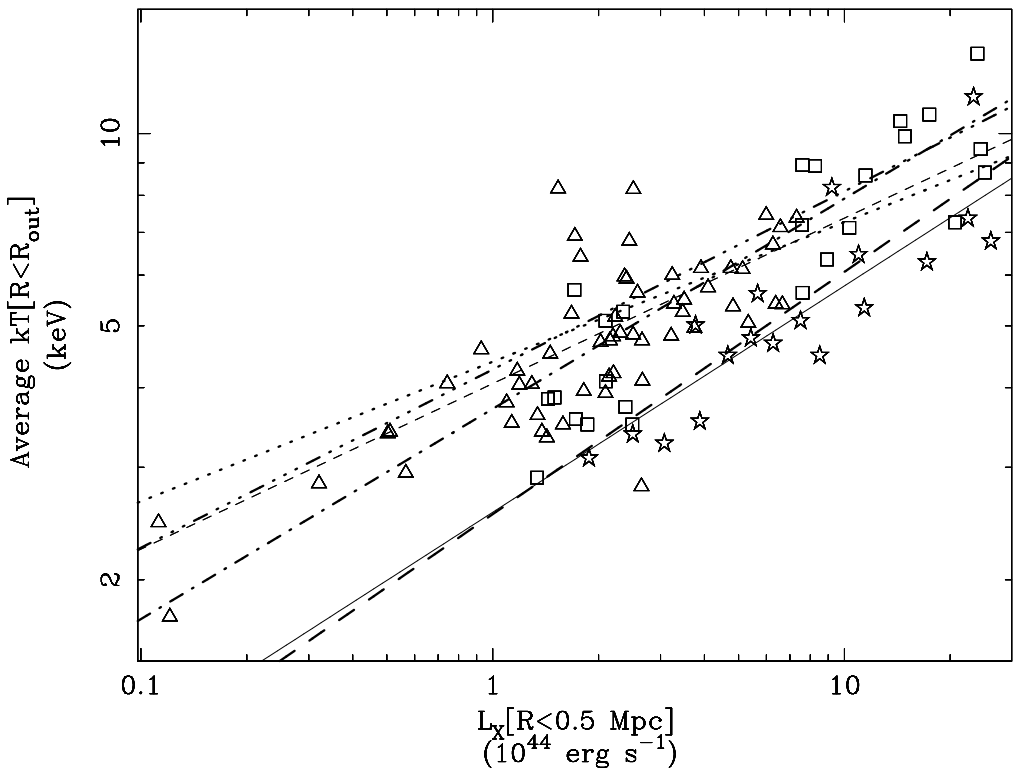

LCF: $kT = 2.53 \pm 0.57\, L_X\, [R < 0.5\,\mathrm{Mpc}]^{0.38\pm0.09}$ (18)
MCF: $kT = 3.70 \pm 0.59\, L_X\, [R < 0.5\,\mathrm{Mpc}]^{0.33\pm0.06}$ (25)
SCF: $kT = 4.39 \pm 0.77\, L_X\, [R < 0.5\,\mathrm{Mpc}]^{0.22\pm0.14}$ (58)
XCF: $kT = 4.27 \pm 0.63\, L_X\, [R < 0.5\,\mathrm{Mpc}]^{0.28\pm0.21}$ (40)
(ALL: $kT = 4.06 \pm 0.23\, L_X\, [R < 0.5\,\mathrm{Mpc}]^{0.26\pm0.03}$ (141) )
(a)

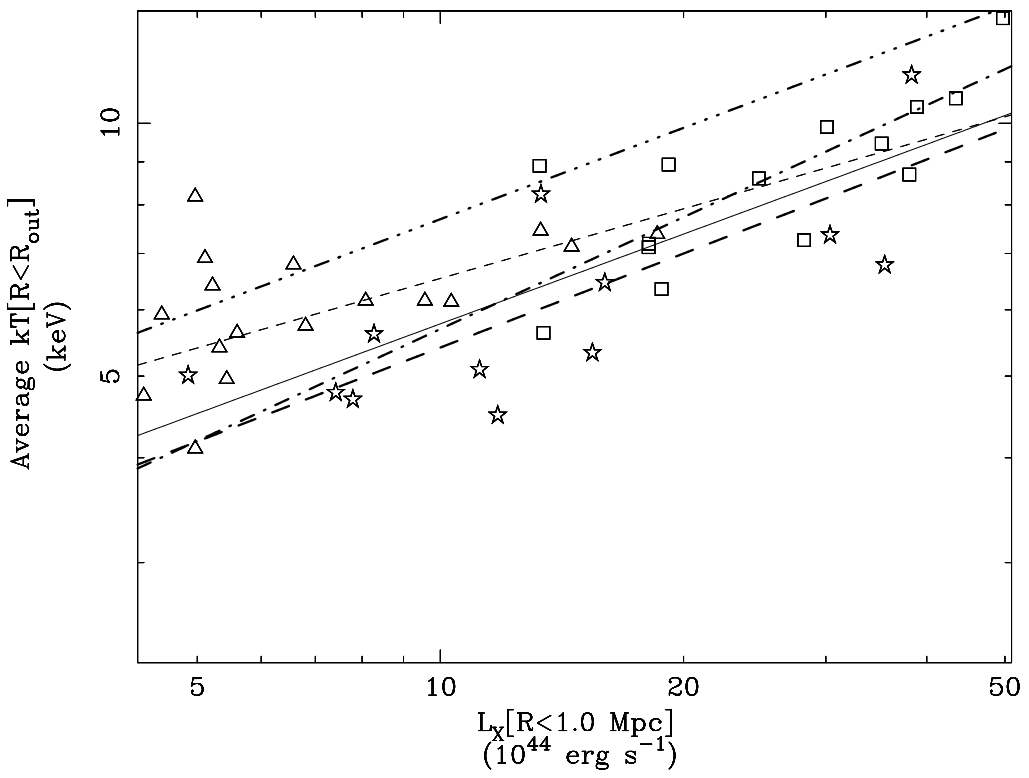

LCF: $kT = 2.30 \pm 0.66\, L_X\, [R < 1\,\mathrm{Mpc}]^{0.37\pm0.10}$ (12)
MCF: $kT = 2.05 \pm 0.62\, L_X\, [R < 1\,\mathrm{Mpc}]^{0.44\pm0.09}$ (14)
SCF:
XCF: $kT = 3.35 \pm 1.04\, L_X\, [R < 1\,\mathrm{Mpc}]^{0.36\pm0.21}$ (9)
(ALL: $kT = 3.47 \pm 0.32\, L_X\, [R < 1\,\mathrm{Mpc}]^{0.28\pm0.03}$ (58) )
(b)

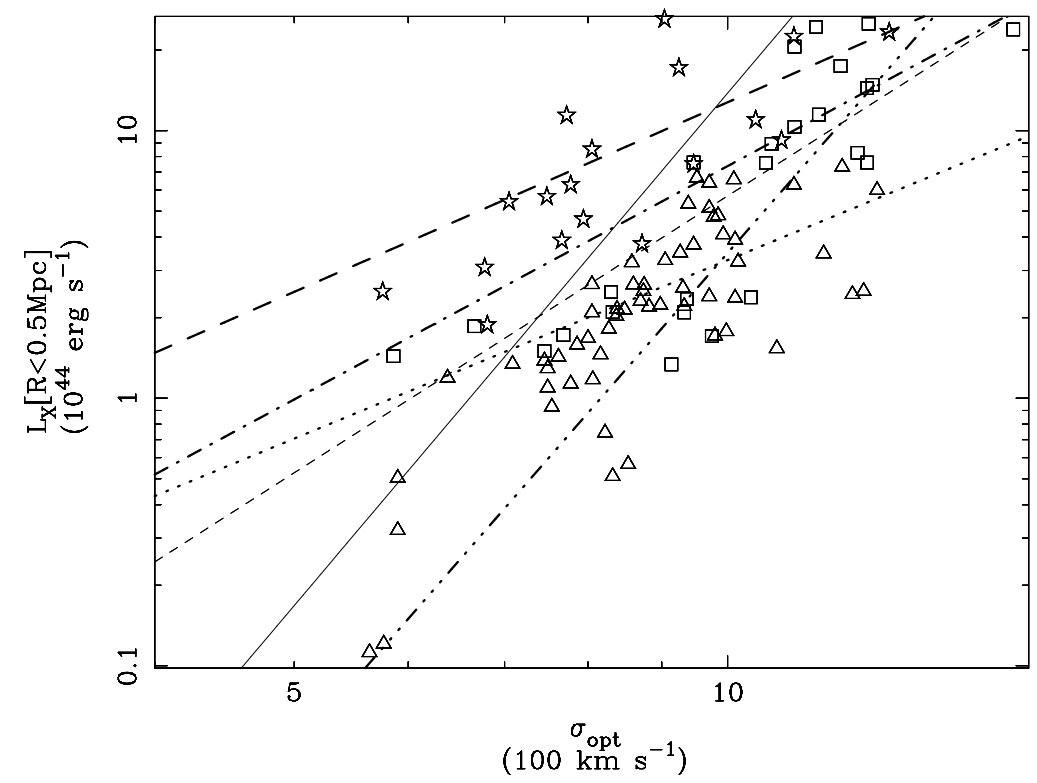

LCF: $L_X\, [R < 0.5\,\mathrm{Mpc}] = (5.66 \pm 1.79) \times 10^{-2}\, \sigma_{\rm opt}{}^{2.35\pm0.13}$ (18)
MCF: $L_X\, [R < 0.5\,\mathrm{Mpc}] = (9.34 \pm 2.60) \times 10^{-3}\, \sigma_{\rm opt}{}^{2.90\pm0.12}$ (25)
SCF: $L_X\, [R < 0.5\,\mathrm{Mpc}] = (1.99 \pm 1.47) \times 10^{-2}\, \sigma_{\rm opt}{}^{2.22\pm0.28}$ (58)
XCF: $L_X\, [R < 0.5\,\mathrm{Mpc}] = (2.40 \pm 50.26) \times 10^{-6}\, \sigma_{\rm opt}{}^{6.16\pm1.90}$ (40)
(ALL: $L_X\, [R < 0.5\,\mathrm{Mpc}] = (2.08 \pm 0.44) \times 10^{-3}\, \sigma_{\rm opt}{}^{3.44\pm0.08}$ (141) )
(c)

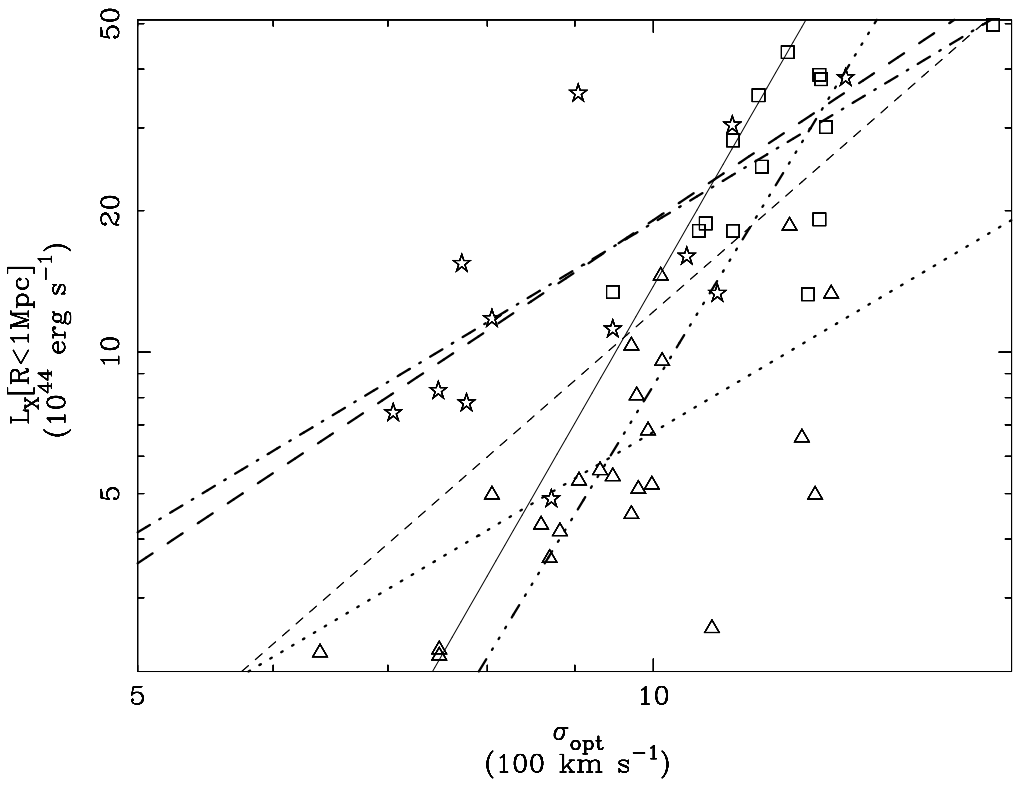

LCF: $L_X\, [R < 1\,\mathrm{Mpc}] = (7.16 \pm 2.13) \times 10^{-2}\, \sigma_{\rm opt}{}^{2.43\pm0.13}$ (12)
MCF: $L_X\, [R < 1\,\mathrm{Mpc}] = 0.12 \pm 0.03\, \sigma_{\rm opt}{}^{2.19\pm0.08}$ (14)
SCF: $L_X\, [R < 1\,\mathrm{Mpc}] = (4.71 \pm 2.72) \times 10^{-2}\, \sigma_{\rm opt}{}^{2.16\pm0.25}$ (23)
XCF: $L_X\, [R < 1\,\mathrm{Mpc}] = (9.09 \pm 113.05) \times 10^{-6}\, \sigma_{\rm opt}{}^{5.97\pm1.43}$ (9)
(ALL: $L_X\, [R < 1\,\mathrm{Mpc}] = (7.94 \pm 1.00) \times 10^{-3}\, \sigma_{\rm opt}{}^{3.19\pm0.05}$ (58) )
(d)

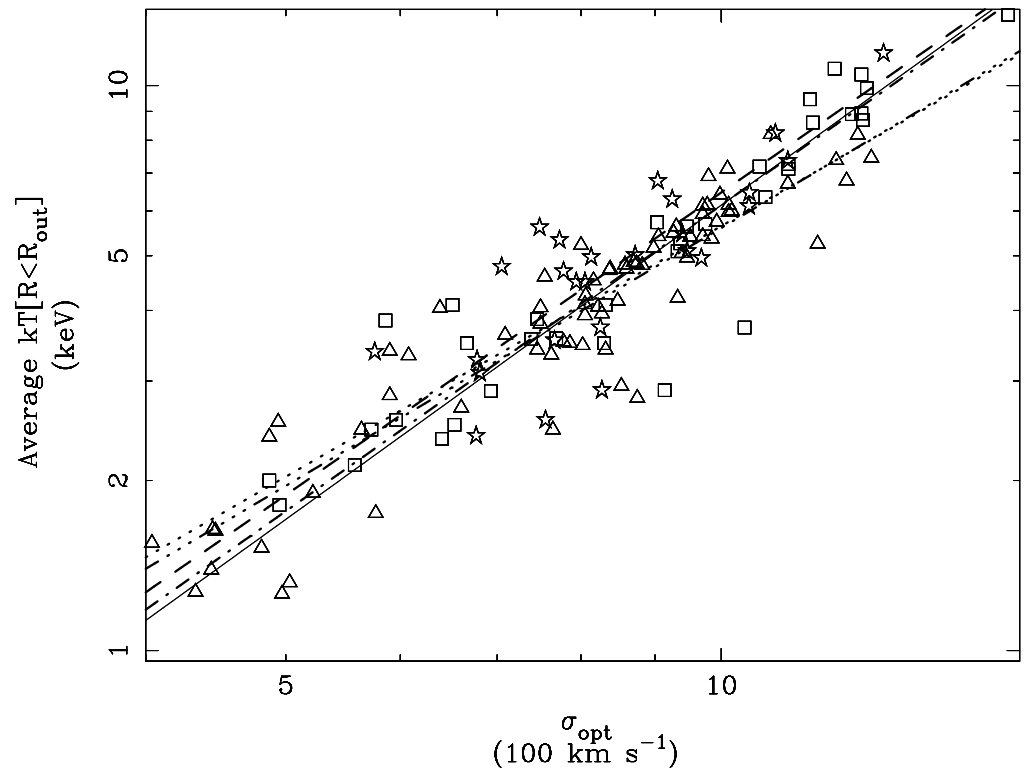

LCF: $kT = 0.11 \pm 0.07\, \sigma_{\rm opt}{}^{1.78\pm0.29}$ (26)
MCF: $kT = (9.76 \pm 4.21) \times 10^{-2}\, \sigma_{\rm opt}{}^{1.80\pm0.18}$ (35)
SCF: $kT = 0.19 \pm 0.06\, \sigma_{\rm opt}{}^{1.47\pm0.14}$ (72)
XCF: $kT = 0.17 \pm 0.10\, \sigma_{\rm opt}{}^{1.52\pm0.27}$ (53)
(ALL: $kT = 0.11 \pm 0.02\, \sigma_{\rm opt}{}^{1.72\pm0.09}$ (197) )
(e)

**Figure 10.**

These plots from the deprojection analysis are equivalent to the $L_X$ - $T_X$ - $\sigma_{\rm opt}$ reference correlations presented in Fig. 1 (a caveat here is that the reference correlations use observed, *i.e.* projected, rather than deprojected values which are plotted here). The results show there is clear segregation according to cooling flow class, *i.e.* $\dot{M}$, when luminosity is involved. The fits to the combined data are shown by lighter dotted lines, while the reference correlations are shown as the lighter solid lines. Star symbols with a dashed line refers to the LCF data and power-law fit; squares and dot-dashed line to MCF; triangles and dot-dot-dot-dashed line to SCF; circle and dotted line to XCF. A detailed discussion of these results is given in Section 4.3. Note the plots do not show the error bars for clarity, and a uniform weighting function has been used for all the data (otherwise the results are dominated by a couple of points with very small error bars).

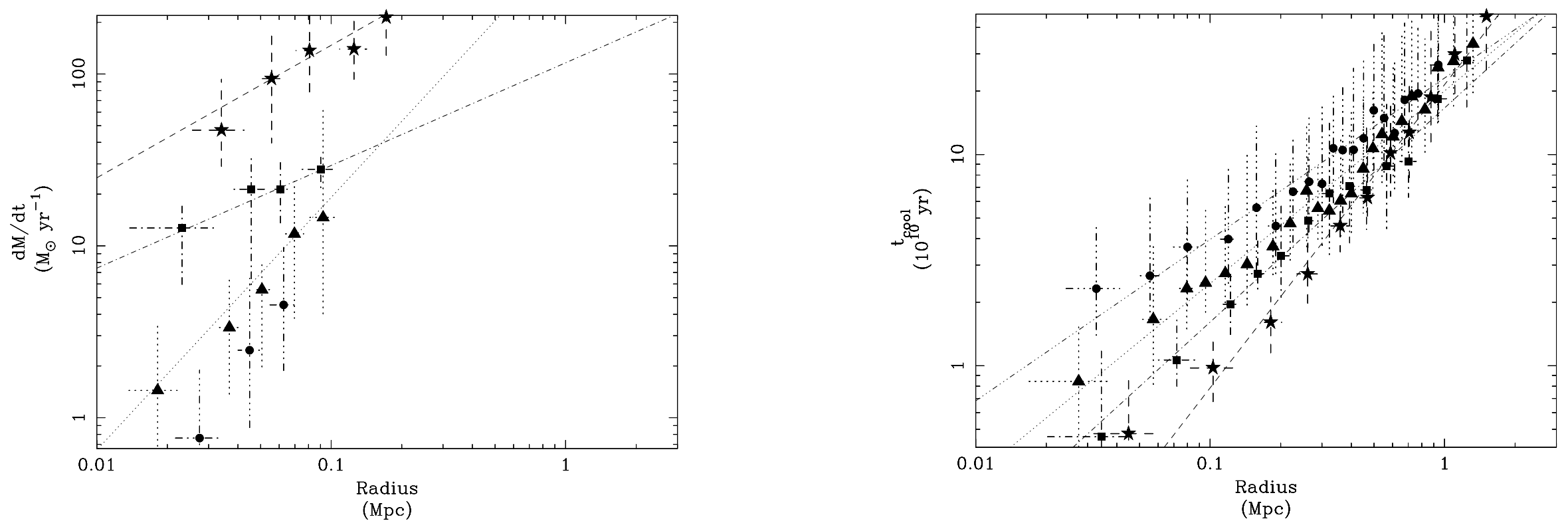

LCF: $\dot{M} = 868.17 \pm 1428.52\, R^{0.77 \pm 0.54}$ (5)
MCF: $\dot{M} = 116.22 \pm 422.83\, R^{0.60 \pm 0.65}$ (4)
SCF: $\dot{M} = 564.89 \pm 39377.18\, R^{1.47 \pm 1.46}$ (5)
XCF: Not applicable

(a)

LCF: $t_{\rm cool} = 21.44 \pm 7.95\, R^{1.44 \pm 0.32}$ (11)
MCF: $t_{\rm cool} = 16.63 \pm 5.77\, R^{1.01 \pm 0.22}$ (13)
SCF: $t_{\rm cool} = 20.02 \pm 7.39\, R^{0.91 \pm 0.32}$ (23)
XCF: $t_{\rm cool} = 23.04 \pm 12.58\, R^{0.76 \pm 0.35}$ (19)

(b)

**Figure 11.**

The diagrams show averaged radial profiles of (a) the integrated mass-deposition rate within the cooling radii, and (b) the cooling time at a radius, both as a function of cooling flow class (as defined in Section 4.2). Both these diagrams include data from the XCF class, *i.e.* non-cooling flow clusters, as $\dot{M}$ profiles are calculated for these clusters but the cooling time of the gas never falls below $t_0 = 1.3 \times 10^{10}$ yr, as seen in (b). Details of the ODRPACK fits results are given in the encapsulated tables.

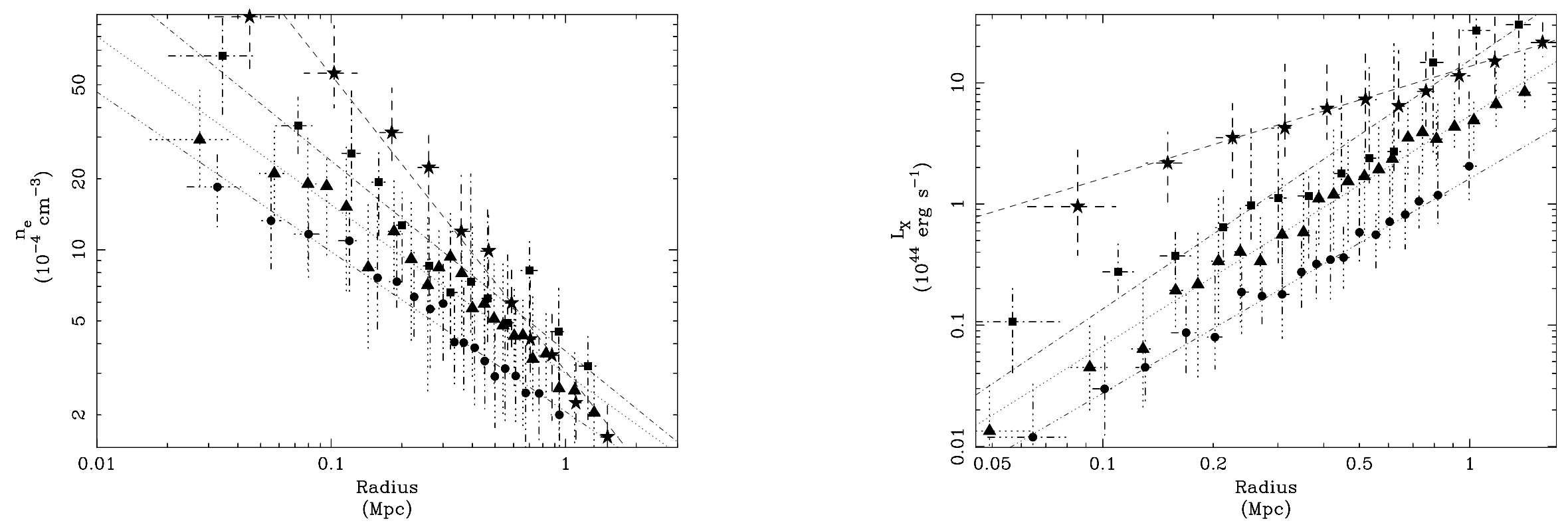

LCF: $n_{\rm e} = 3.06 \pm 0.58\, R^{-1.26 \pm 0.19}$ (11)
MCF: $n_{\rm e} = 3.72 \pm 1.16\, R^{-0.81 \pm 0.22}$ (13)
SCF: $n_{\rm e} = 3.00 \pm 0.57\, R^{-0.71 \pm 0.13}$ (23)
XCF: $n_{\rm e} = 2.06 \pm 0.53\, R^{-0.68 \pm 0.15}$ (19)

(a)

LCF: $L_{\rm X} = 13.67 \pm 4.55\, R^{0.92 \pm 0.40}$ (11)
MCF: $L_{\rm X} = 15.38 \pm 7.71\, R^{2.05 \pm 0.46}$ (13)
SCF: $L_{\rm X} = 5.35 \pm 1.38\, R^{1.90 \pm 0.26}$ (23)
XCF: $L_{\rm X} = 1.62 \pm 0.82\, R^{1.77 \pm 0.40}$ (19)

(b)

**Figure 12.**

The plots show averaged radial profiles of (a) electron density, and (b) bolometric X-ray luminosity. The segregation in density is responsible for that in many other parameters, including the luminosity profiles and the $L_{\rm X}$ - $T_{\rm X}$ - $\sigma_{\rm opt}$ relations (Fig.10).

data are presented as mean values with standard deviation errors if the original data have no statistical errors (*e.g.* $M_{\rm grav}$ ), or otherwise as median values (50th percentile) with pseudo-1$\sigma$ limits (16th and 84th percentiles). (Formally, these statistics should be identical if the data are distributed symmetrically.) This averaging procedure appears very good at segregating trends between different CF classes, even when the dispersion in the data within a class is larger than that between classes. (To avoid confusion between averaged profiles and individual deprojection results in figures presented hereafter, filled symbols and open symbols are used respectively.)

It is generally suggested that $\dot{M} \propto R$ (*e.g.* Arnaud 1988), and the ODRPACK fit results to the averaged radial profiles of $\sum \dot{M}(r)$, in Fig. 11(a), show that these deprojection results are consistent with the expected dependence. [Note, the $\sum \dot{M}(r)$ profiles are not the integrated $\dot{M}$ values within $r_{\rm cool}$ , but are the averages of all the radially integrated mass-deposition rate profiles out to approximately $r_{\rm cool}$ . Thus, although the mass-deposition rate profile can be defined for a non-cooling flow, each XCF clusters has zero $\dot{M}$ because the average $t_{\rm cool}$ profile never falls below $t_0 = 1.3 \times 10^{10}$ yr, as seen in Fig. 11(b).]

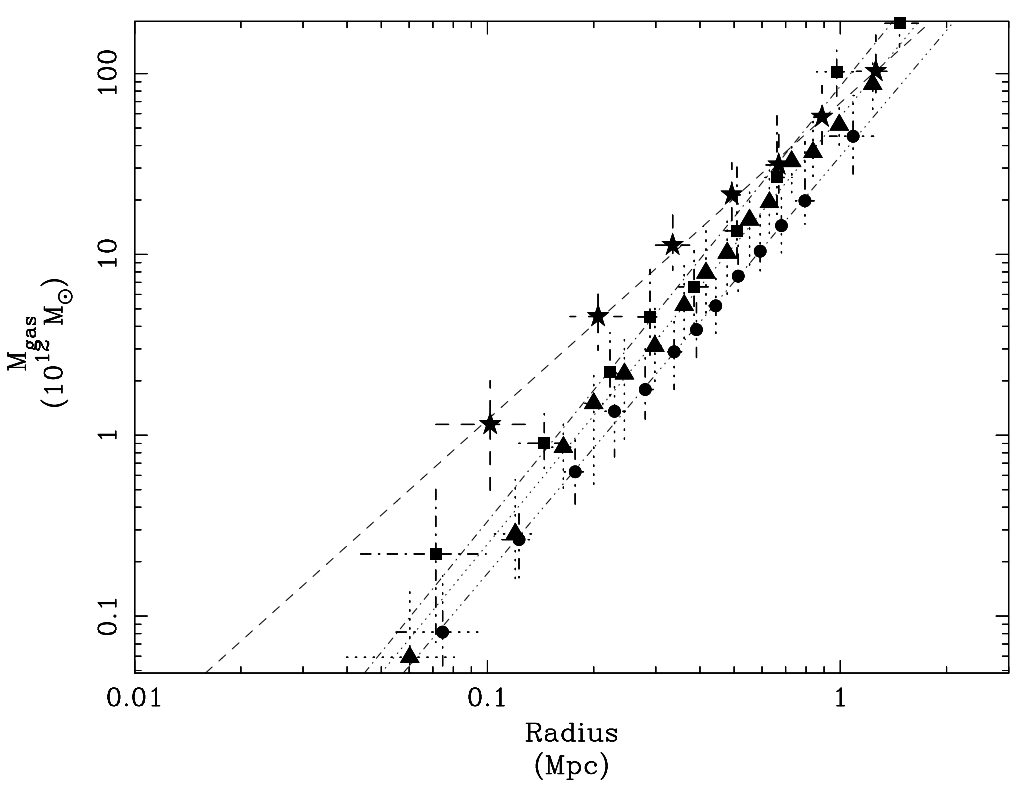

LCF: $M_{\rm gas} = 68.79 \pm 22.37\, R^{1.75\pm0.40}$ (7)
MCF: $M_{\rm gas} = 85.15 \pm 28.06\, R^{2.41\pm0.29}$ (9)
SCF: $M_{\rm gas} = 57.71 \pm 12.46\, R^{2.36\pm0.22}$ (15)
XCF: $M_{\rm gas} = 34.97 \pm 10.76\, R^{2.31\pm0.30}$ (13)
(a)

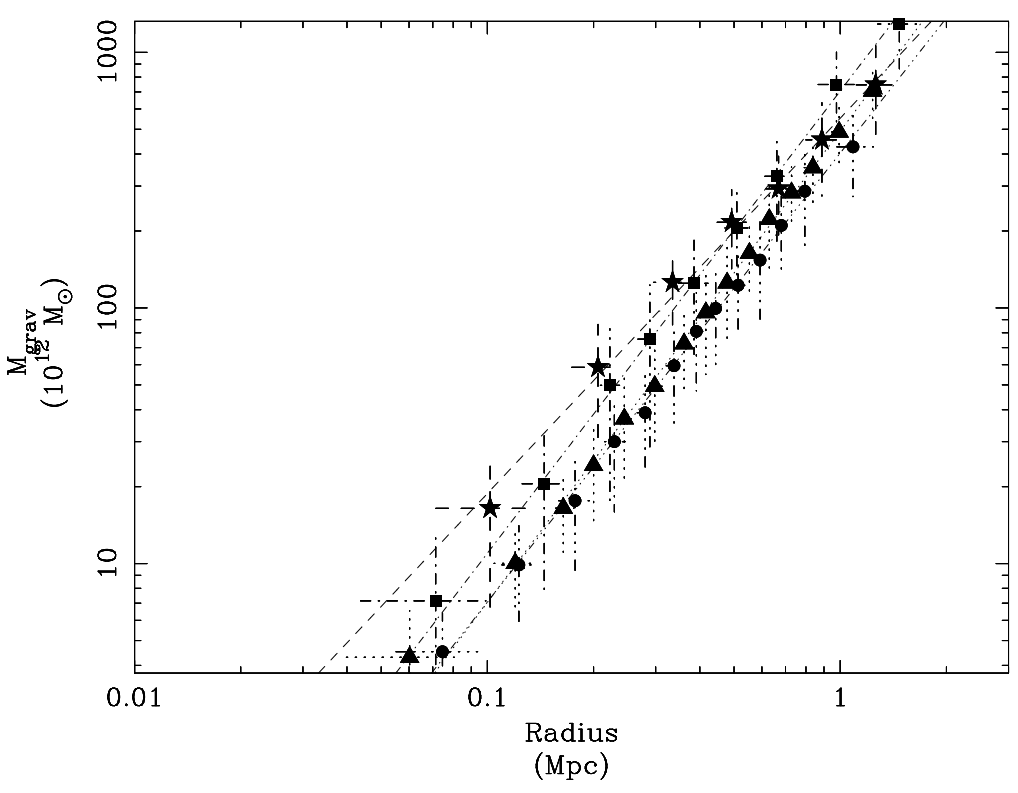

LCF: $M_{\rm grav} = 552.42 \pm 157.67\, R^{1.47\pm0.34}$ (7)
MCF: $M_{\rm grav} = 705.67 \pm 209.59\, R^{1.81\pm0.29}$ (9)
SCF: $M_{\rm grav} = 491.46 \pm 83.64\, R^{1.84\pm0.16}$ (15)
XCF: $M_{\rm grav} = 403.40 \pm 106.76\, R^{1.76\pm0.24}$ (13)
(b)

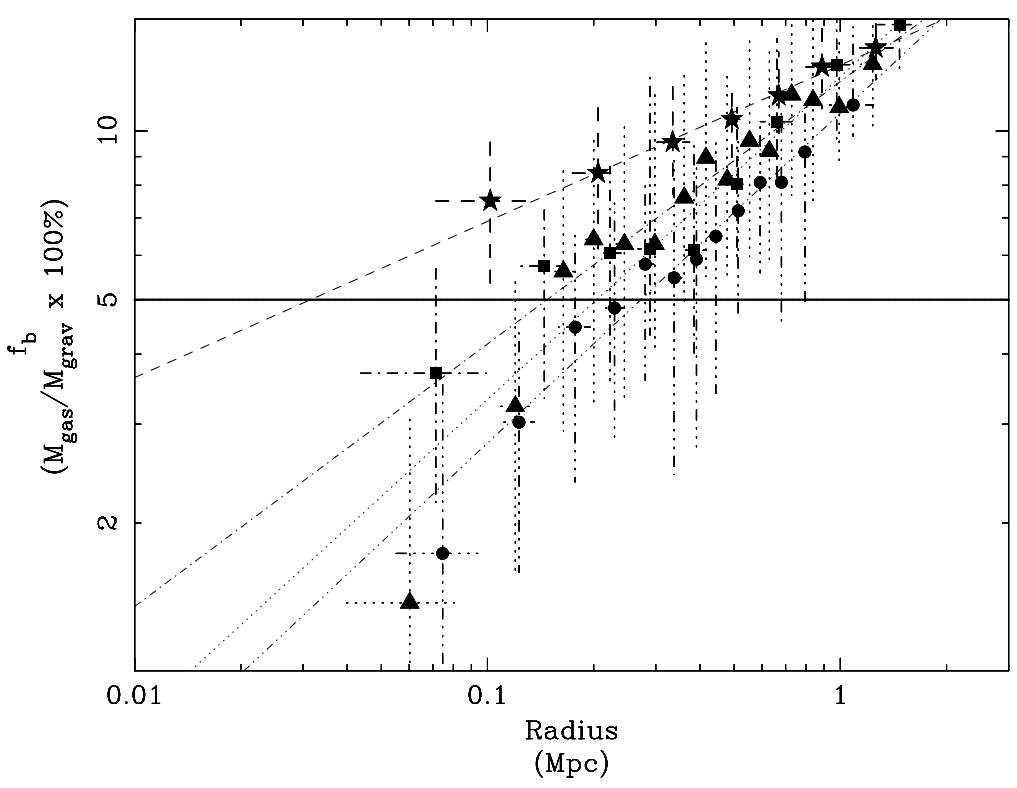

LCF: $f_{\rm b} = 13.10 \pm 2.59\, R^{0.28\pm0.20}$ (7)
MCF: $f_{\rm b} = 12.25 \pm 2.71\, R^{0.47\pm0.20}$ (9)
SCF: $f_{\rm b} = 12.49 \pm 1.85\, R^{0.58\pm0.16}$ (15)
XCF: $f_{\rm b} = 10.76 \pm 2.07\, R^{0.59\pm0.20}$ (13)
(c)

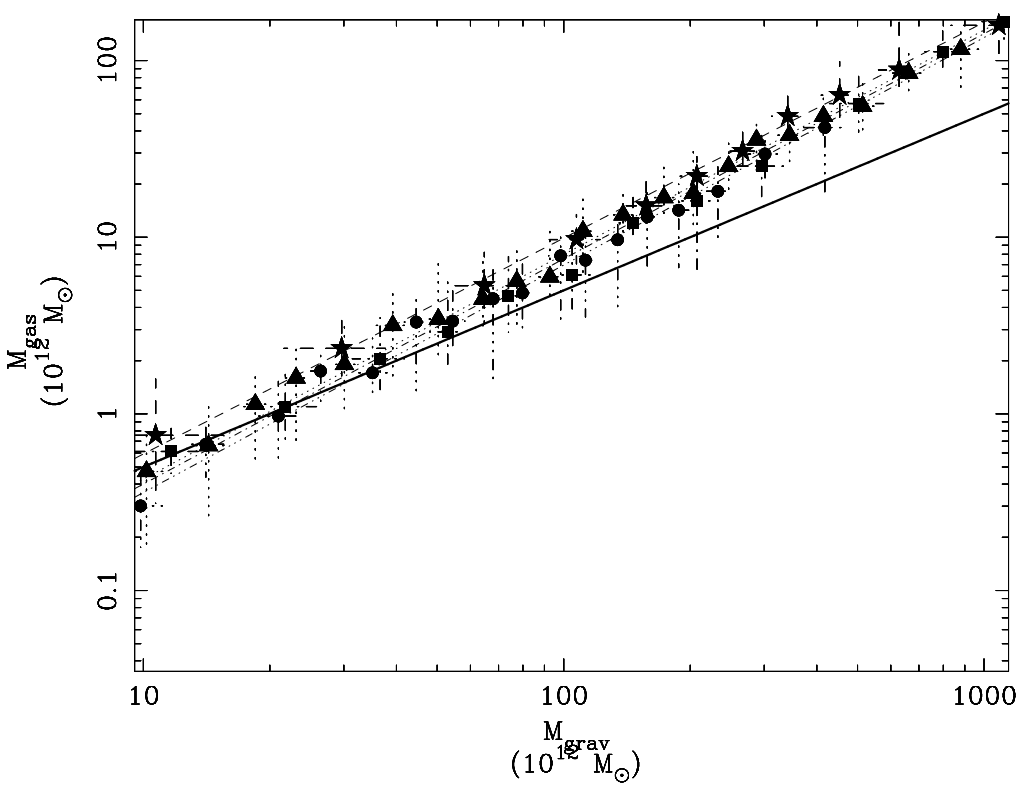

LCF: $M_{\rm gas} = (3.55 \pm 3.46) \times 10^{-2}\, M_{\rm grav}{}^{1.22\pm0.18}$ (11)
MCF: $M_{\rm gas} = (2.12 \pm 1.42) \times 10^{-2}\, M_{\rm grav}{}^{1.27\pm0.11}$ (13)
SCF: $M_{\rm gas} = (2.29 \pm 1.06) \times 10^{-2}\, M_{\rm grav}{}^{1.27\pm0.09}$ (23)
XCF: $M_{\rm gas} = (1.85 \pm 0.75) \times 10^{-2}\, M_{\rm grav}{}^{1.29\pm0.09}$ (19)
(d)

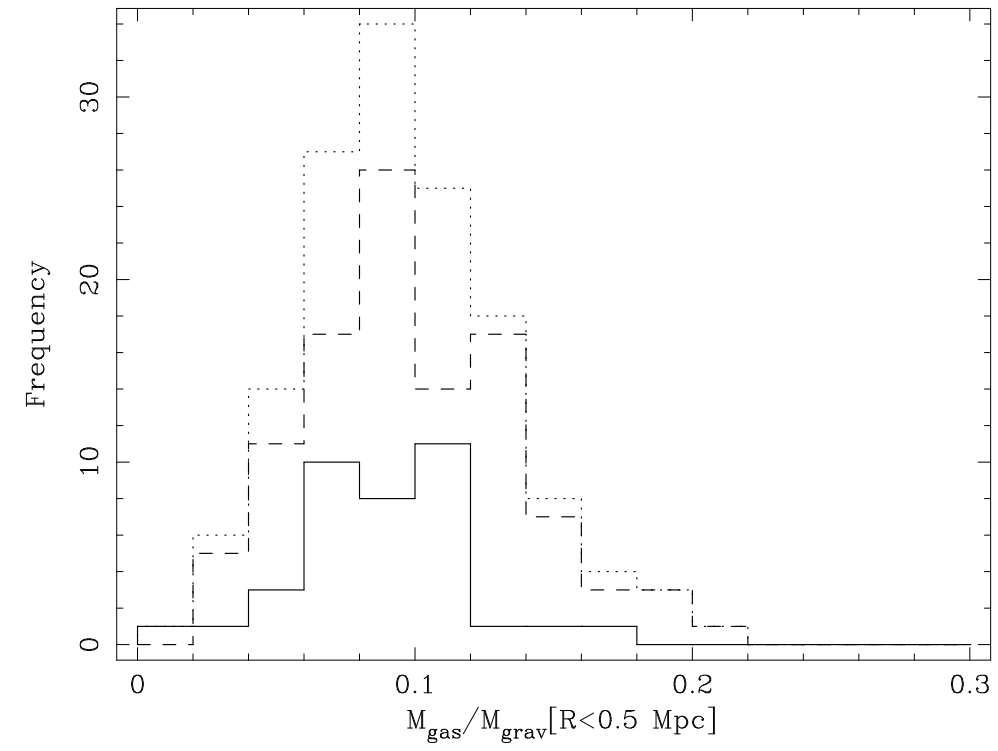

(e)

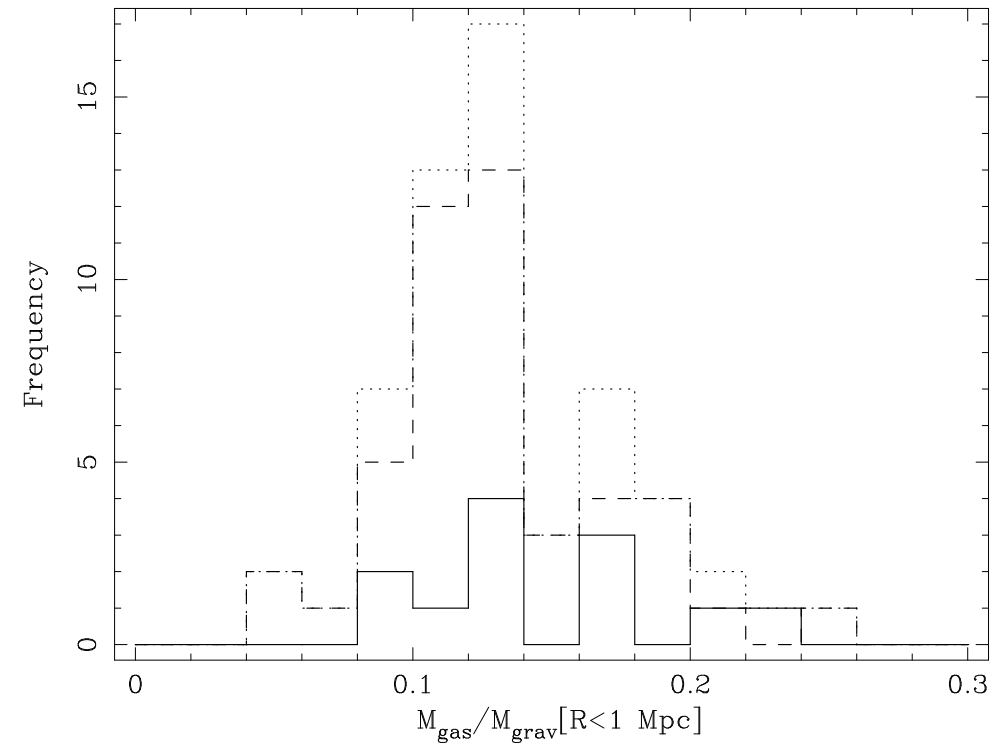

(f)

**Figure 13.**

These diagrams show averaged radial profiles of (a) gas mass, (b) total gravitational mass, (c) baryon fraction ($M_{\rm gas}/M_{\rm grav}$), (d) the gas versus gravitational mass, and histograms (from IPC data) of baryon fraction within (e) 0.5 Mpc and (f) 1 Mpc. The heavy solid lines in (c) and (d) represents the relation expected for a baryon fraction of 5 per cent which is given by primordial nucleosynthesis ($\Omega_0 = 1$) calculations.

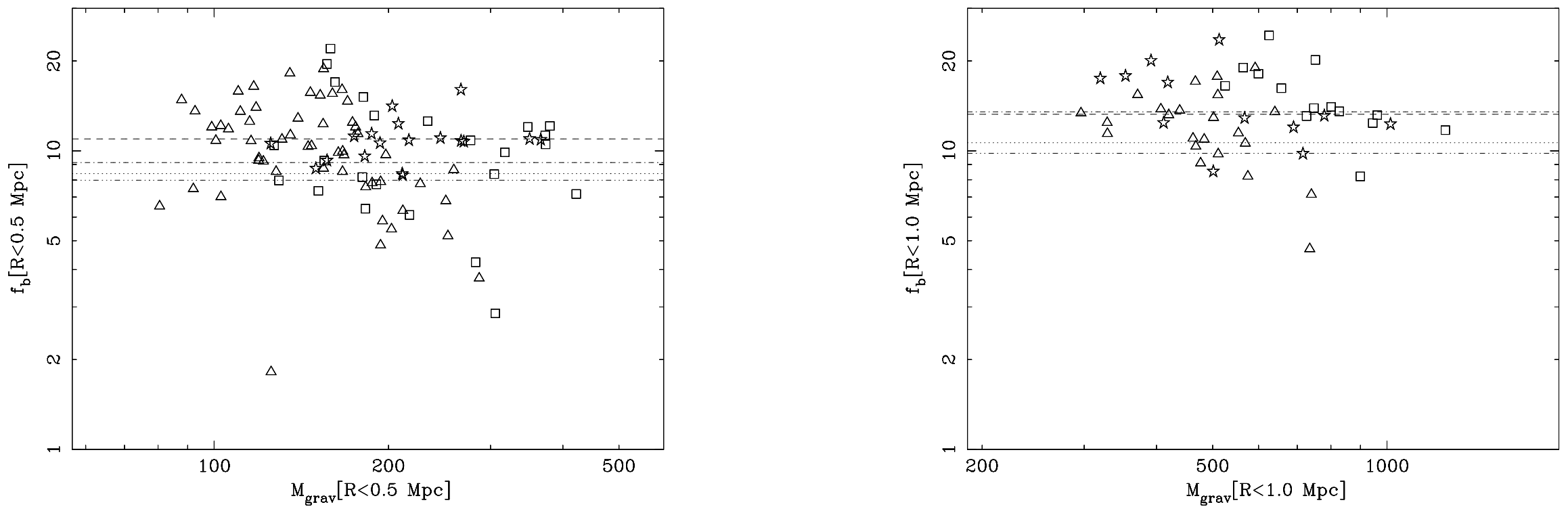

LCF: $f_{\rm b}[R<0.5\,{\rm Mpc}] = 10.96 \pm 0.08\, M_{\rm grav}[R<0.5\,{\rm Mpc}]^{[0.00\pm0.00]}$ (18)
MCF: $f_{\rm b}[R<0.5\,{\rm Mpc}] = 9.14 \pm 0.12\, M_{\rm grav}[R<0.5\,{\rm Mpc}]^{[0.00\pm0.00]}$ (24)
SCF: $f_{\rm b}[R<0.5\,{\rm Mpc}] = 8.39 \pm 0.15\, M_{\rm grav}[R<0.5\,{\rm Mpc}]^{[0.00\pm0.00]}$ (54)
XCF: $f_{\rm b}[R<0.5\,{\rm Mpc}] = 7.97 \pm 0.11\, M_{\rm grav}[R<0.5\,{\rm Mpc}]^{[0.00\pm0.00]}$ (35)
(a)

LCF: $f_{\rm b}[R<1.0\,{\rm Mpc}] = 13.26 \pm 0.22\, M_{\rm grav}[R<1.0\,{\rm Mpc}]^{[0.00\pm0.00]}$ (12)
MCF: $f_{\rm b}[R<1.0\,{\rm Mpc}] = 13.51 \pm 0.23\, M_{\rm grav}[R<1.0\,{\rm Mpc}]^{[0.00\pm0.00]}$ (14)
SCF: $f_{\rm b}[R<1.0\,{\rm Mpc}] = 10.65 \pm 0.23\, M_{\rm grav}[R<1.0\,{\rm Mpc}]^{[0.00\pm0.00]}$ (23)
XCF: $f_{\rm b}[R<1.0\,{\rm Mpc}] = 9.81 \pm 0.27\, M_{\rm grav}[R<1.0\,{\rm Mpc}]^{[0.00\pm0.00]}$ (9)
(b)

**Figure 14.**

These two plots show the dependence of cluster baryon fractions on the mass of the system [within 0.5 Mpc in (a) and 1.0 Mpc in (b)], as a function of cooling flow class (with line types and symbols as usual). Note the index of the power-law is fixed to zero, as indicated by the square-bracket notation (giving weighted mean fits to the ordinate data).

The hypothesis stated in Section 4.3, that the segregation in the $L_{\rm X}$ - $T_{\rm X}$ - $\sigma_{\rm opt}$ relations arises mainly from density segregation in different cooling flow classes is confirmed from the radial density profiles shown in Fig 12(a). It is seen that the central densities increase with mass-deposition rate, but also that the profiles approach convergence at around 1 Mpc. As density characteristics are well constrained ($L_{\rm X} \propto {n_{\rm e}}^2 {T_{\rm X}}^{1/2}$) this behaviour is translated to the many other related properties (including the radial mass-deposition rate profiles seen already). The averaged luminosity profiles have slopes which flatten with decreasing mass-deposition rate, *i.e.* a larger fraction of the luminosity in cooling flows arises from the central regions (which is consistent with the observation that cooling flow surface-brightness profiles are very sharply peaked). Gas mass profiles also show a significant segregation, see Fig 13(a), which is comparatively greater than that seen in the gravitational masses (b). This leads to differences in the baryon fraction ($f_{\rm b} = M_{\rm gas} / M_{\rm grav}$) profiles (c) with mass-deposition rate. In the core of smaller cooling flows, and non-cooling flows, the baryon fraction may be consistent with the expected mean baryon density of $\Omega_{\rm b} = 0.05 \pm 0.01\, h_{50}^{-2}$ in an $\Omega_0 = 1$ Universe (Walker et al. 1991). However, at a radius of $\sim$ 1 Mpc baryon contents exceed 10 percent for all clusters, irrespective of their cooling flow properties. This discrepancy between the primordial nucleosynthesis calculations and the X-ray determined baryon fractions is also apparent in a plot of $M_{\rm gas}$ against $M_{\rm grav}$ (d), and histograms of baryon fractions at (e) 0.5 Mpc and (f) 1 Mpc. These results are in good agreement with the determinations by White & Fabian (1995), and show that the 'baryon catastrophe' (White et al. 1993) is prevalent in clusters of galaxies.

David, Jones, & Forman (1995) have suggested that the baryon fraction may depend on the mass of the system in question, such that the baryon fraction is generally larger in more massive systems. Although this statement is supported by Fig 13(d), in Fig. 14 we show a plot of the baryon fraction (within 0.5 Mpc and 1.0 Mpc) against the mass of the system (within the same radii respectively), as a function of cooling flow class. There appears to be little variation in baryon fraction as a function of mass within any particular cooling flow class, or even for the sample as a whole. (We note our fits are fixed with a power-law index of zero, and if anything the data points follow an anticorrelation with mass rather than a correlation) . There is, however, a segregation between cooling flow class, such that larger cooling flows tend to have larger baryon fractions at any particular radius. There is clearly greater dispersion in the data (though not necessarily the average of each class) within 0.5 Mpc than 1.0 Mpc, as would be expected given that any effect resulting from cooling flow properties should be greater in the core.

Hypothetically, the differing baryon contents in the core of cooling flow and non-cooling flow clusters may be viewed as either a relative enhancement of baryons in cooling flow clusters, or a deficit in non-cooling flows. Although, the first option would be a natural consequence of mass deposition from the cooling flow, the discrepancy in baryon content appears over a larger volume than the typical cooling regions ($r_{\rm cool}$). Alternatively, the deficit of baryons in non-cooling flows could be due to heating and expansion of the intracluster gas, as might be expected during merger events which would disrupt any previously existing cooling flow. A plausible compromise would be that both these mechanisms are in operation. After intracluster gas is heated during a merger the cluster will relax and the cooling flow will be re-established. The central density will then increase as the gas cools, leading to a segregation in density related properties.

### 4.5 Reprojected results

Although this paper has concentrated on the cooling flow properties of clusters, other interesting properties may also be derived through the reprojection of the deprojection results. In particular, Sunyaev-Zeldovich (S-Z; d$T_{\rm mw}$) microwave decrements can be predicted from the deprojected results to produce a possible target list for future radio detections of the microwave decrement. The microwave decrements are calculated from the properties within the projected central 6 arcmin of each deprojection, see Table 3. Despite the noisiness of the averaged S-Z profiles, there are again indications of segregation with CF class, see Fig. 15(a). Larger decrements, of the order of just less than a mK, are seen in more massive (higher temperature) clusters.

The Thomson depth, also given in Table 3, indicates the probability that an optical photon, escaping from the centre of each cluster, will be scattered at least once by an electron in the hot ICM. Extrapolation of the fits shown in Fig. 15(b) indicates that for a photon originating from within 10 kpc of the centre of a LCF cluster the scattering probability may approach one percent.

Other reprojected parameters are projected X-ray luminosities [Fig. 15(c)], emission-weighted temperatures (which have been used to calibrate the deprojection results by comparison with $T_{\rm X,ref}$), and projected gravitational masses (*e.g.* for comparison with lensing studies) [Fig. 15(d)].

## 5 SUMMARY

This paper presents an X-ray deprojection analysis of EINSTEIN OBSERVATORY imaging data on 207 clusters of galaxies. The large number of clusters in the sample enables a detailed investigation into the properties of cooling flows, and their relation to general cluster properties. After correcting for a spatial-resolution bias, the detected proportion of cooling flows in this sample is estimated to be $62^{+12}_{-15}$ percent. This should not be taken as a quantitative statement of the prevalence of cooling flows as this sample is not homogeneously selected, *i.e.* flux-limited, but it does indicate that cooling flows are common in clusters. A catalogue of detected cooling flows has been compiled for reference, but again a literal interpretation of the $\dot{M}$ values given in this catalogue is cautioned against as these values are subject to many assumptions. It is suggested that any particular mass-deposition rate be considered accurate to within only a factor of two.

The deprojection results taken as a whole indicate that cooling flows are fundamentally related to the global properties of clusters, and play an important role in explaining the scatter and deviation in slopes of the scaling laws between $L_{\rm X}$ - $T_{\rm X}$ - $\sigma_{\rm opt}$. The resulting correlations between these parameters are consistent with previous determinations (*e.g.* Edge & Stewart 1991a; Lubin & Bahcall 1993), and indicate that on a global scale clusters are isothermal and consistent with $\beta$-values of unity. These results also confirm the scatter in the $L_{\rm X}$ - $T_{\rm X}$ relation, as a function of cooling flows mass-deposition rate, found by Fabian et al. (1994). This analysis additionally shows there is greater scatter in the $L_{\rm X}$ - $\sigma_{\rm opt}$ relation, and that the scatter in both relationships is due to differences in luminosity which results from segregation of the density profiles. This leads to segregation in many other related properties, notably baryon fraction profiles. In the core of non-cooling or smaller cooling flows the baryon fraction may be consistent with the primordial nucleosynthesis predictions of $\Omega_{\rm b} = 0.05 \pm 0.01 \Omega_0$ for the mean baryon content in a flat ($\Omega_0 = 1$) Universe, but this is *only* in the very core regions. At the centre of larger cooling flows, and *all* clusters at larger radii ($\sim 1$ Mpc), baryon fractions are greater than 10 to 15 percent. This is inconsistent with primordial nucleosynthesis calculations, and agrees with the results from a smaller analysis by White & Fabian (1995) and David, Jones, & Forman (1995).

Reprojected parameters such as Sunyaev-Zeldovich microwave decrements also are calculated for all the clusters in the sample, and may be useful in the search for new S-Z targets. Thompson depths indicate that the probability of a photon being scattered while escaping from the central 10 kpc of a large cooling flow cluster is around one per cent. Half-light radii (*i.e.* radii which contain the X-luminosity at particular reference radii of 0.2 Mpc, 0.5 Mpc and 1 Mpc) may be of use in numerical simulations of large scale structure formation in the Universe.

In summary, this deprojection analysis provides vast amount of reference information on the X-ray properties of clusters and reveals the significance of cooling flows in the determination of fundamental and global cluster properties.

## 6 ACKNOWLEDGEMENTS

D.A. White acknowledges financial support from the PPARC, and the Smithsonian Institute. C. Jones and W. Forman acknowledge support from the Smithsonian Institute.

Since the initial conception of the analysis procedure by A. Fabian, the deprojection code has been developed by many authors, including D. White, K. Arnaud, P. Thomas, G. Stewart, A. Edge. The version used in this analysis has been developed extensively by D.A. White with the help and advice from S. Allen and S. Daines. We also gratefully acknowledge K. Arnaud for supplying the EINSTEIN OBSERVATORY HRI data used in this analysis.

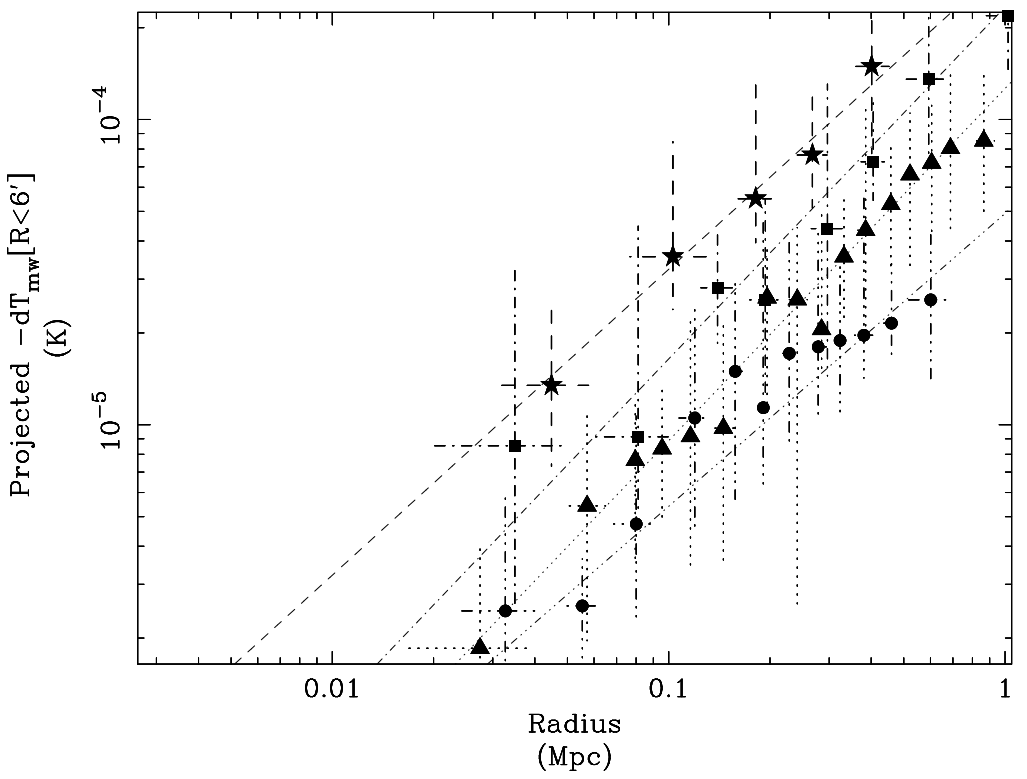

LCF: $\Delta T_{\rm mw} = (3.24 \pm 3.33) \times 10^{-4}\, R^{1.00\pm0.56}$ (5)
MCF: $\Delta T_{\rm mw} = (2.34 \pm 1.36) \times 10^{-4}\, R^{1.15\pm0.53}$ (8)
SCF: $\Delta T_{\rm mw} = (1.26 \pm 0.38) \times 10^{-4}\, R^{1.15\pm0.22}$ (16)
XCF: $\Delta T_{\rm mw} = (4.92 \pm 2.43) \times 10^{-5}\, R^{0.96\pm0.28}$ (12)
(a)

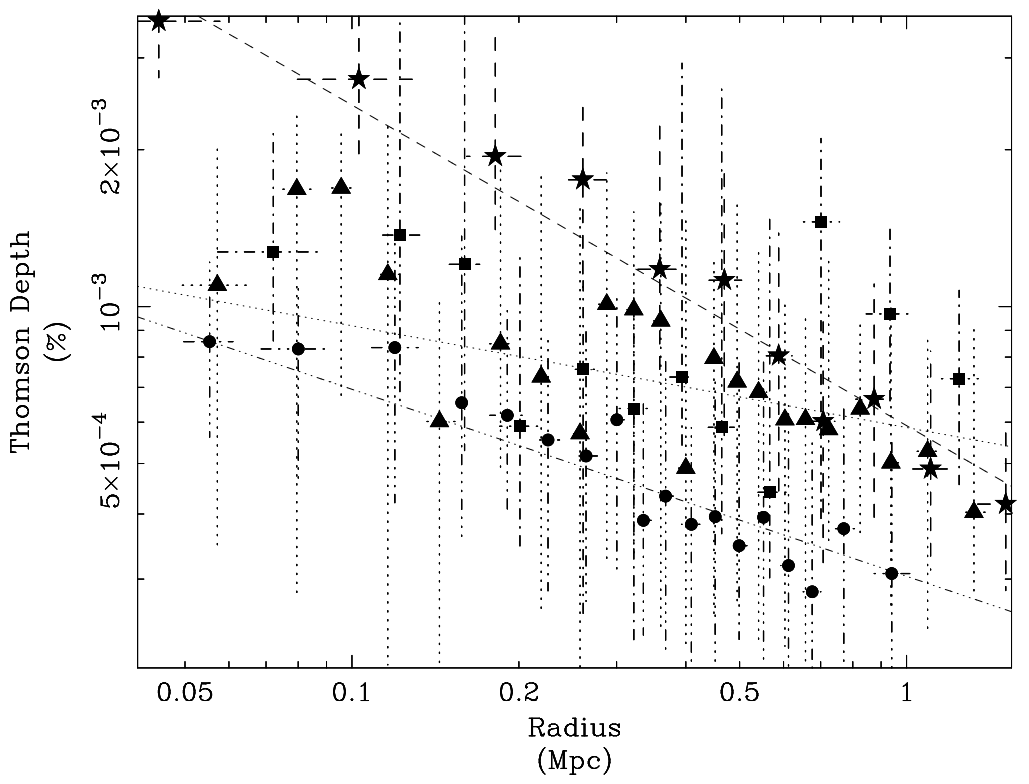

LCF: $d\tau = (5.91 \pm 1.18) \times 10^{-4}\, R^{-0.61\pm0.17}$ (11)
MCF:
SCF: $d\tau = (5.86 \pm 1.38) \times 10^{-4}\, R^{-0.20\pm0.16}$ (23)
XCF: $d\tau = (3.04 \pm 0.94) \times 10^{-4}\, R^{-0.36\pm0.17}$ (19)
(b)

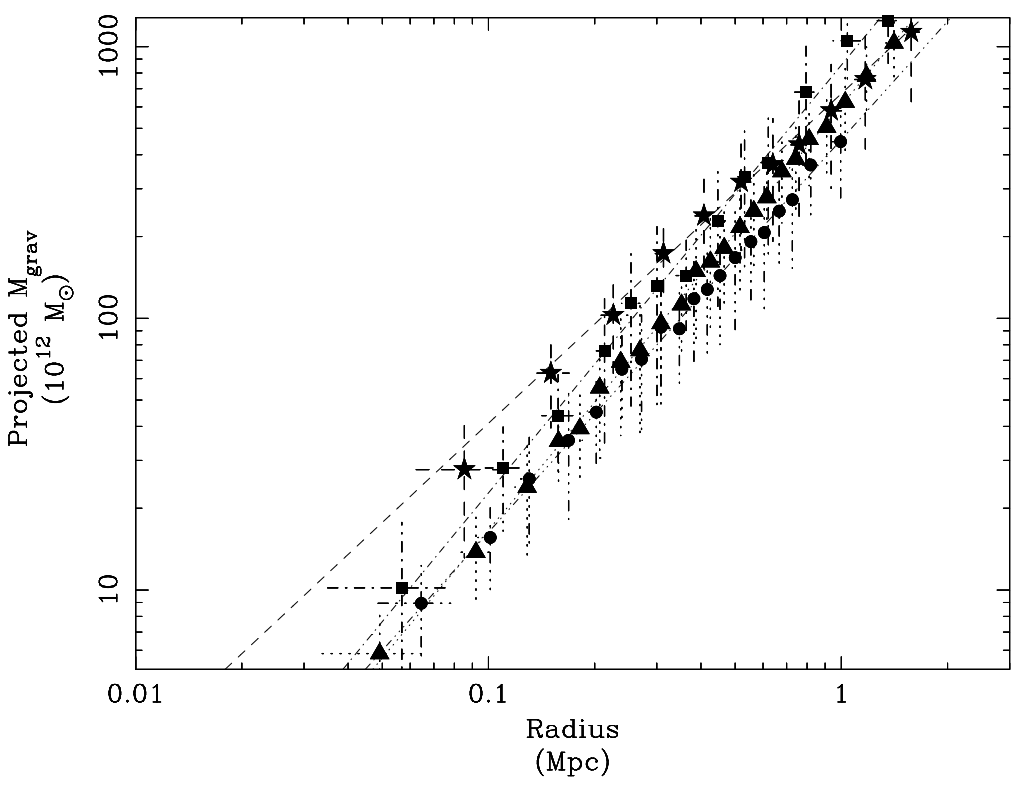

LCF: $M_{\rm grav,proj} = 671.15 \pm 137.62\, R^{1.21\pm0.20}$ (11)
MCF: $M_{\rm grav,proj} = 854.17 \pm 192.17\, R^{1.57\pm0.20}$ (13)
SCF: $M_{\rm grav,proj} = 607.42 \pm 78.35\, R^{1.56\pm0.12}$ (23)
XCF: $M_{\rm grav,proj} = 454.70 \pm 107.02\, R^{1.44\pm0.16}$ (19)
(c)

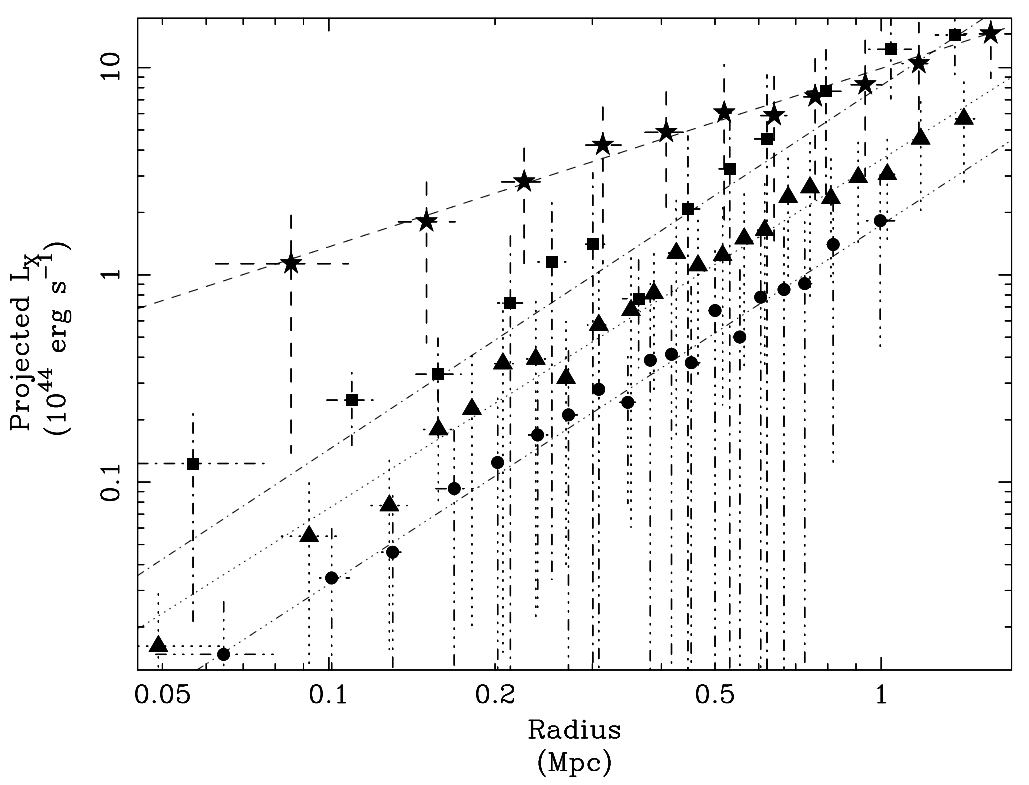

LCF: $L_{\rm X,proj} = 9.92 \pm 2.12\, R^{0.86\pm0.30}$ (11)
MCF: $L_{\rm X,proj} = 8.18 \pm 3.17\, R^{1.76\pm0.34}$ (13)
SCF: $L_{\rm X,proj} = 3.60 \pm 0.78\, R^{1.68\pm0.21}$ (23)
XCF: $L_{\rm X,proj} = 1.74 \pm 0.92\, R^{1.73\pm0.37}$ (19)
(d)

**Figure 15.**

These diagrams show various projected parameters from the deprojection results. (a) Shows the expected S-Z microwave decrement, (b) Thomson depth, (c) projected gravitational mass, and (d) projected bolometric X-ray luminosities.

**Table 1.** Reference Data

This table contains the reference data used to provide input data, both directly from reference sources and from correlations between $L_{\rm X,ref}$ -$T_{\rm X,ref}$ -$\sigma_{\rm opt,ref}$ determined from data in this table which have quoted uncertainties. Note, the X-ray luminosities have been converted from various band-passes to 'pseudo' bolometric luminosities [over $0.01 - 80$ keV – the range of the Raymond & Smith (1977) data], assuming an abundance of $Z = 0.4\,{\rm Z}_{\odot}$. The values actually used in the deprojections are given in Table 2. Literature references or the parameters from which the data has been interpolated (see Section 3.2), are indicated to the right of each data entry. The data in this table used to determine the $L_{\rm X}$ -$T_{\rm X}$ -$\sigma_{\rm opt}$ correlations (again see Section 3.2) are those $L_{\rm X}$ , $T_{\rm X}$ and $\sigma_{\rm opt}$ values which have uncertainties quoted. The reference abbreviations correspond as follows: ABL – Abell (1958) or Abell, Corwin, & Olowin (1989); ABL89 – Abell, Corwin, & Olowin (1989); APM – Dalton et al. (1994); BAH – Bahcall (1981); BAR – Bardelli et al. (1994); BEE – Beers et al. (1984); DAN – Danese, Zotti, & di Tullio (1980); LPD – David et al. (1993); DRE – Dressler (1978); EMSS – Stocke et al. (1991); ES – Edge & Stewart (1991b); ESP – Edge & Stewart private communication; FJM, FJC – Forman & Jones measured observationally or calculated value respectively (private communication); GMT – Garilli, Maccagni, & Tarenghi (1993); GGP – Green, Godwin, & Peach (1988); HIN – Hintzen & Scott (1979); ILT – Impey, Lawrence, & Tapia (1991); LEH – Lea & Henry (1988); LDC – Lucey, Currie, & Dickens (1986); M10 – redshift from 10th brightest galaxy distance estimator (*e.g.* see ABL); NIL – Nilsson et al. (1993); NOO – Noonan (1981); PHG – Postman, Huchra, & Geller (1992); QLA – Quintana & Lawrie (1982); QUD – Quintana & DeSouza (1993); QRM – Quintana et al. (1995); SPN – Spinrad et al. (1985); SR – Struble & Rood (1987), Struble & Rood (1991); TAD – Tadhunter et al. (1993); ZHG – Zabludoff, Huchra, & Geller (1990).

**Table 2.** Deprojection Input Data

This table contains the input data required in the deprojection analysis, the results of which are given in Table 3. The columns contain the data as follows: (i) Reference number cross-referencing against the results in Table 3. (ii) Cluster name (ordered with Abell clusters first, followed by clusters with alternative names following). (iii) EINSTEIN OBSERVATORY sequence number of the observation, prefixed 'I' to indicate IPC data and 'H' for HRI data. (iv)–(vi) IPC energy-range of the extracted data [which is empty for HRI data as this is essentially fixed $(0.4 - 4$ keV)], IPC gain, and the exposure time. (vii)–(viii) Optical redshift, and the luminosity distance ( $H_0 = 50$ km s$^{-1}$ Mpc$^{-1}$ and $q_0 = 0.5$) for the cluster. (ix) Galactic column density as determined by the 21 cm radio emission survey of Stark et al.. (x) The spatial-average X-ray temperature for calibration of the deprojection results. (xi)–(xii) Optical velocity dispersion, and core radius used to parameterise each cluster's gravitational potential. (xiii)–(xvi) The total number of bins, the bin-size (spatial and metric) and outer radius of each deprojection. (xvii) The gas pressure, at the outer radius which, in conjunction with the core radius used to give a deprojected temperature consistent with the reference value in column (x) (see Section 4).

**Table 3.** Deprojection Results

This table summarises the deprojection results. The columns contain the data as follows: (i)–(iii) Cluster reference number (for cross reference with data in Table 2). (iv)–(v) The core radius used and outer radius of the deprojection. (vi) The spatially-averaged, emission-weighted reprojected temperature (50th percentile with 10th and 90th percentile errors) from deprojected profile, for comparison with the reference temperature [given in the brackets and listed in Table 2]. (vii) The spatially-averaged bolometric X-ray luminosity from the deprojection compared to the reference value in the brackets. (Note interpolated reference values are given in italic font.) (viii) The integrated gas mass to the outer radius. (ix) Integrated gravitational mass at the outer radius (from the gravitational potential parameters $R_{\rm core}$ , $\sigma_{\rm opt}$ and $P_{\rm out}$ ). (Note, the Fornax-A deprojection uses only a single isothermal sphere for the central galaxy, unlike all others which model the cluster potential together with a standard central galaxy potential.) (x) The baryon fraction ($M_{\rm gas}/M_{\rm grav}$) at $R_{\rm out}$ is given only $R_{\rm core} > R_{\rm out}$ as it varies significantly within $R_{\rm core}$ . (xi)–(xiii) Cooling flow parameters $t_{\rm cool}$ , in the central bin (bin-size dependent), $r_{\rm cool}$ and $\dot{M}$. [These numbers are 50th percentiles with 10th and 90th percentile limits]. The occurrence of a '>' symbols indicates that $t_{\rm cool}$ is larger than this as $t_{\rm cool} < t_0$ even at $R_{\rm out}$ (this usually only happens for HRI deprojections); this leads to corresponding limits in the $\dot{M}$ values. (xiv) The cooling flow class as detailed in Section 4.3 which is used extensively in the analysis. (xv) The projected Sunyaev-Zeldovich microwave decrement (with respect to $T_{\rm cmb} = 2.735$ K) from the over 6 arcmin radius.

**Table 4.** Integrated Deprojection Results At Specific Radii

This table, where applicable, gives certain deprojection results at specific radii. Column (i) gives the deprojection reference number for comparison with data in Tables 2 and 3. (iv)–(vi) 'Half-light' radii, *i.e.* the radius which contains half the X-ray luminosity at the reference radius of 0.2, 0.5 or 1.0 Mpc. (vii)–(xiv) Integrated values of $L_{\rm X}$ , $M_{\rm gas}$ , $M_{\rm grav}$ and $M_{\rm gas}/M_{\rm grav}$ within 0.5 and 1.0 Mpc.

**Table 5.** A Catalogue Of Cooling Flows Detected In The Einstein Observatory Deprojections

This table summarises the cooling flow parameters for those clusters detected as a cooling flow in this deprojection analysis, *i.e.* $\dot{M} > 0\,{\rm M}_{\odot}\,{\rm yr}^{-1}$ ($t_{\rm cool} < t_0$). It does not include possible cooling flows missed due to the spatial-resolution bias discussed in Section 4.1. The entries are ordered by decreasing $\dot{M}$, and may contain multiple entries for a single cluster if both IPC and HRI data are available and indicate a cooling flow (the second entry is superscripted with a *b*). The occurrence of '>' symbols indicate (usually for HRI results) indicate that the outer radius of the deprojection is not large enough that $t_{\rm cool}$ increases above $t_0$, and therefore the numbers are quoted at the radius $R_{\rm out}$.

**Table 6.** Supplementary Reference Data From The Literature

This table gives values of $\dot{M}$ and $t_{\rm cool}$ values from Edge, Stewart, & Fabian (1992), which are compared with the deprojection results, as shown in Fig. 6. Values for $L_{\rm H\alpha}$ from Heckman et al. (1989) are also given.

**Table 1** Cluster Reference Data

| Cluster | $N_{\rm H}$ ($10^{20}$ cm$^{-2}$) | Redshift $z$ | $\sigma_{\rm opt,ref}$ (km s$^{-1}$) | $T_{\rm X,ref}$ (keV) | $L_{\rm X,ref}$ ($10^{44}$ erg s$^{-1}$) |
|---|---|---|---|---|---|
| (i) | (ii) | (iii) | (iv) | (v) | (vi) |
| A21 | 4.32 | 0.0948 [STR] | *882* [$T_{\rm X}$] | 4.8 [JFC] | $5.587^{+0.295}_{-0.295}$ [JFM] |
| A74 | 1.44 | 0.0672 [JOF] | *577* [$T_{\rm X}$] | 2.2 [JFC] | $0.479^{+0.092}_{-0.092}$ [JFM] |
| A76 | 3.85 | 0.0416[ABL89] | *462* [$T_{\rm X}$] | $1.5^{+1.1}_{-0.6}$[LPD] | $0.923^{+0.075}_{-0.075}$ [JFM] |
| A77 | 5.02 | 0.0719[ABL89] | *788* [$T_{\rm X}$] | 3.9 [JFC] | $2.138^{+0.068}_{-0.068}$ [JFM] |
| A84 | 3.42 | 0.1030[ABL89] | *769* [$T_{\rm X}$] | 3.8 [JFC] | $2.858^{+0.134}_{-0.134}$ [JFM] |
| A85 | 3.01 | 0.0521 [ZHG] | 749 [ABL89] | $6.2^{+0.2}_{-0.3}$[LPD] | $16.511^{+0.098}_{-0.098}$[JFM] |
| A98 | 3.40 | 0.1042 [ZHG] | $819^{+154}_{-49}$ [ZAB] | *4.2* [$\sigma_{\rm opt}$] | *4.059* [$\sigma_{\rm opt}$] |
| A98N | 3.41 | 0.1043[ABL89] | 829 [SRD] | 3.3 [JFC] | $1.389^{+0.116}_{-0.116}$ [JFM] |
| A98S | 3.39 | 0.1063 [FJE] | 919 [ABL89] | 3.2 [JFC] | $1.142^{+0.113}_{-0.113}$ [JFM] |
| A98SS | 3.39 | 0.1053 [FJE] | *602* [$T_{\rm X}$] | 2.4 [JFC] | $0.398^{+0.082}_{-0.082}$ [JFM] |
| A114 | 1.44 | 0.0566 [STR] | $517^{+146}_{-86}$ [ZHG] | *1.8* [$\sigma_{\rm opt}$] | *0.168* [$\sigma_{\rm opt}$] |
| A115 | ..... | 0.1971[ABL89] | 1167 [SRD] | *8.0* [$\sigma_{\rm opt}$] | *47.055* [$\sigma_{\rm opt}$] |
| A115N | 5.09 | 0.1971[ABL89] | *1045* [$T_{\rm X}$] | 6.6 [JFC] | 14.711 [FJE] |
| A115S | 5.09 | 0.1971[ABL89] | *969* [$T_{\rm X}$] | 5.7 [JFC] | 9.055 [FJE] |
| A117 | 3.67 | 0.0535[ABL89] | *601* [$T_{\rm X}$] | 2.4 [JFC] | $0.568^{+0.043}_{-0.043}$ [JFM] |
| A119 | 3.45 | 0.0443 [ZHG] | $863^{+178}_{-112}$ [ZAB] | $5.1^{+0.8}_{-0.6}$ [ES] | $5.781^{+0.147}_{-0.147}$ [ES] |
| A133 | 1.58 | 0.0604[ABL89] | *767* [$T_{\rm X}$] | $3.8^{+1.7}_{-0.8}$ [ES] | $5.763^{+0.310}_{-0.310}$ [ES] |
| A150N | 3.87 | 0.0596[ABL89] | *543* [$T_{\rm X}$] | 2.0 [JFC] | $0.213^{+0.038}_{-0.038}$ [JFM] |
| A150S | 3.87 | 0.0596[ABL89] | *632* [$T_{\rm X}$] | 2.6 [JFC] | $0.506^{+0.047}_{-0.047}$ [JFM] |
| A154 | 4.32 | 0.0652 [ZHG] | $833^{+274}_{-140}$ [ZAB] | 3.1 [JFC] | $1.278^{+0.085}_{-0.085}$ [JFM] |
| A160 | 4.17 | 0.0447[ABL89] | 572 [SRD] | *2.2* [$\sigma_{\rm opt}$] | *0.339* [$\sigma_{\rm opt}$] |
| A168 | 3.36 | 0.0438 [ZHG] | 571 [DAN] | $2.6^{+1.1}_{-0.6}$[LPD] | $1.125^{+0.027}_{-0.027}$ [JFM] |
| A189 | 3.17 | 0.0335[ABL89] | $259^{+94}_{-49}$ [ZHG] | *0.5* [$\sigma_{\rm opt}$] | *0.001* [$\sigma_{\rm opt}$] |
| A193 | 4.36 | 0.0482 [ESP] | *820* [$T_{\rm X}$] | $4.2^{+1.4}_{-0.7}$ [ES] | $3.239^{+0.186}_{-0.186}$ [ES] |
| A194 | 4.22 | 0.0180 [ZHG] | $480^{+48}_{-38}$ [ZAB] | 1.9 [JFC] | $0.157^{+0.004}_{-0.004}$ [JFM] |
| A195 | 40.74 | 0.0422[ABL89] | *678* [$T_{\rm X}$] | $3.0^{+1.0}_{-1.0}$ [ESP] | $0.286^{+0.029}_{-0.029}$ [JFM] |
| A240 | 4.41 | 0.0618[ABL89] | *591* [$T_{\rm X}$] | 2.3 [JFC] | $0.597^{+0.083}_{-0.083}$ [JFM] |
| A262 | 5.33 | 0.0163 [ZHG] | $494^{+63}_{-47}$ [ZAB] | $2.4^{+0.3}_{-0.2}$ [ES] | $0.915^{+0.030}_{-0.030}$ [ES] |
| A278 | 5.01 | 0.0896 [STR] | *739* [$T_{\rm X}$] | 3.5 [JFC] | $1.867^{+0.118}_{-0.118}$ [JFM] |
| A347 | 6.34 | 0.0190 [ZHG] | $590^{+134}_{-85}$ [ZAB] | *2.3* [$\sigma_{\rm opt}$] | *0.420* [$\sigma_{\rm opt}$] |
| A367 | 2.63 | 0.0891 [PHG] | *802* [$T_{\rm X}$] | 4.1 [JFC] | $2.294^{+0.220}_{-0.220}$ [JFM] |
| A376 | 6.30 | 0.0489 [STR] | *903* [$T_{\rm X}$] | $5.1^{+2.7}_{-1.6}$ [ES] | 2.366 [ES] |
| A389 | 1.84 | 0.1160[ABL89] | *848* [$T_{\rm X}$] | 4.5 [JFC] | $4.540^{+0.187}_{-0.187}$ [JFM] |
| A397 | 9.52 | 0.0325[ABL89] | 447 [SRD] | 1.6 [JFC] | $0.095^{+0.013}_{-0.013}$ [JFM] |
| A399 | 11.70 | 0.0715 [STR] | 1424 [ABL89] | $5.8^{+0.8}_{-0.7}$[LPD] | $8.774^{+0.147}_{-0.147}$ [JFM] |
| A400 | 8.82 | 0.0238 [ZHG] | $610^{+59}_{-46}$ [ZAB] | $2.1^{+1.1}_{-0.5}$ [ES] | $0.756^{+0.065}_{-0.065}$ [ES] |
| A401 | 11.10 | 0.0748 [STR] | $1290^{+351}_{-196}$[QLA] | $7.8^{+0.6}_{-0.6}$[LPD] | $19.317^{+0.214}_{-0.214}$[JFM] |
| A407 | 9.82 | 0.0472 [STR] | 590 [ABL89] | 2.8 [JFC] | $0.915^{+0.068}_{-0.068}$ [JFM] |
| A415 | 5.60 | 0.0788[ABL89] | *743* [$T_{\rm X}$] | 3.5 [JFC] | $2.159^{+0.119}_{-0.119}$ [JFM] |
| A419 | 1.74 | 0.0406 [STR] | *522* [$T_{\rm X}$] | 1.9 [JFC] | $0.549^{+0.067}_{-0.067}$ [JFM] |
| A426 | 14.50 | 0.0179 [ZHG] | $1277^{+95}_{-78}$ [ZAB] | $5.5^{+0.4}_{-0.4}$ [ES] | $21.434^{+1.031}_{-1.031}$ [ES] |
| A458 | 1.58 | 0.1054 [ZHG] | $709^{+116}_{-79}$ [ZAB] | 4.4 [JFC] | $4.350^{+0.245}_{-0.245}$ [JFM] |
| A478 | 13.60 | 0.0881 [ZHG] | $904^{+261}_{-140}$ [ZAB] | $6.8^{+0.9}_{-0.8}$ [ES] | $47.909^{+1.126}_{-1.126}$ [ES] |
| A496 | 4.41 | 0.0330 [ZHG] | $705^{+98}_{-69}$ [ZAB] | $4.7^{+0.8}_{-0.7}$ [ES] | $5.808^{+0.212}_{-0.212}$ [ES] |
| A500 | 2.92 | 0.0666 [STR] | *720* [$T_{\rm X}$] | 3.3 [JFC] | $1.688^{+0.064}_{-0.064}$ [JFM] |
| A520 | 7.50 | 0.2030 [NOO] | *1112* [$T_{\rm X}$] | 7.4 [JFC] | 21.927 [JFM] |
| A539 | 13.30 | 0.0288 [ZHG] | $832^{+76}_{-60}$ [ZAB] | $3.0^{+0.5}_{-0.4}$[LPD] | $1.299^{+0.028}_{-0.028}$ [JFM] |
| A545 | 11.40 | 0.1530 [STR] | *947* [$T_{\rm X}$] | $5.5^{+6.2}_{-2.1}$[LPD] | $17.347^{+0.567}_{-0.567}$[JFM] |
| A548S | 2.04 | 0.0415 [ZHG] | $853^{+62}_{-51}$ [ZAB] | $2.4^{+0.7}_{-0.5}$[LPD] | $1.174^{+0.026}_{-0.026}$ [JFM] |
| A566 | 5.09 | 0.0957 [STR] | *860* [$T_{\rm X}$] | 4.6 [JFC] | $4.456^{+0.297}_{-0.297}$ [JFM] |
| A569 | 8.18 | 0.0196[ABL89] | 444 [SRD] | *1.4* [$\sigma_{\rm opt}$] | *0.059* [$\sigma_{\rm opt}$] |
| A576 | 5.65 | 0.0381 [ZHG] | $914^{+113}_{-83}$ [ZAB] | $3.7^{+0.7}_{-0.5}$ [ES] | $2.761^{+0.096}_{-0.096}$ [ES] |
| A592 | 4.46 | 0.0624 [STR] | *711* [$T_{\rm X}$] | 3.3 [JFC] | $1.506^{+0.047}_{-0.047}$ [JFM] |
| A629 | 4.31 | 0.1380 [STR] | *713* [$T_{\rm X}$] | 3.3 [JFC] | $1.469^{+0.139}_{-0.139}$ [JFM] |
| A634 | ..... | 0.0267[ABL89] | $318^{+51}_{-46}$ [ZHG] | *0.8* [$\sigma_{\rm opt}$] | *0.006* [$\sigma_{\rm opt}$] |

**Table 1** Cluster Reference Data

| Cluster (i) | $N_{\rm H}$ ($10^{20}$ cm$^{-2}$) (ii) | Redshift $z$ (iii) | $\sigma_{\rm opt,ref}$ (km s$^{-1}$) (iv) | $T_{\rm X,ref}$ (keV) (v) | $L_{\rm X,ref}$ ($10^{44}$ erg s$^{-1}$) (vi) |
|---|---|---|---|---|---|
| A644 | 7.32 | 0.0704[ABL89] | *1046* [$T_{\rm X}$] | $6.6^{+0.1}_{-0.1}$ [LPD] | $14.652^{+0.148}_{-0.148}$ [JFM] |
| A646 | 4.50 | 0.1303 [STR] | *927* [$T_{\rm X}$] | 5.3 [JFC] | $7.355^{+0.584}_{-0.584}$ [JFM] |
| A665 | 4.21 | 0.1816[ABL89] | 1201 [SRD] | $8.3^{+0.6}_{-0.5}$ [LPD] | $25.395^{+0.719}_{-0.719}$ [JFM] |
| A671 | 3.94 | 0.0494 [STR] | *704* [$T_{\rm X}$] | 3.2 [JFC] | $1.435^{+0.082}_{-0.082}$ [JFM] |
| A690 | 4.15 | 0.0788[ABL89] | *546* [$T_{\rm X}$] | 2.0 [JFC] | $0.326^{+0.059}_{-0.059}$ [JFM] |
| A744 | 3.65 | 0.0732 [ZHG] | $814^{+173}_{-106}$ [ZAB] | 2.7 [JFC] | $0.553^{+0.026}_{-0.026}$ [JFM] |
| A754 | 4.74 | 0.0542 [ZHG] | $747^{+81}_{-61}$ [ZAB] | $8.7^{+1.5}_{-1.3}$ [ES] | $23.125^{+0.491}_{-0.491}$ [ES] |
| A779 | 1.74 | 0.0230 [ZHG] | $503^{+100}_{-63}$ [ZAB] | 1.5 [JFC] | $0.200^{+0.019}_{-0.019}$ [JFM] |
| A795 | 3.55 | 0.1357 [STR] | *978* [$T_{\rm X}$] | 5.8 [JFC] | $10.477^{+0.664}_{-0.664}$ [JFM] |
| A957 | 3.57 | 0.0440[ABL89] | *669* [$T_{\rm X}$] | 2.9 [JFC] | $1.062^{+0.070}_{-0.070}$ [JFM] |
| A963 | ..... | 0.2060 [STR] | *1099* [$T_{\rm X}$] | 7.2 [JFC] | $16.132^{+1.773}_{-1.773}$ [LEH] |
| A970 | 5.27 | 0.0595 | *832* [$T_{\rm X}$] | 4.3 [JFC] | $0.890^{+0.058}_{-0.058}$ [JFM] |
| A992 | 2.39 | 0.0880 [ESP] | *898* [$T_{\rm X}$] | $5.0^{+0.9}_{-0.9}$ [ESP] | 7.512 [ESP] |
| A999 | 3.91 | 0.0319 [ZHG] | $404^{+76}_{-49}$ [ZAB] | 1.2 [JFC] | $0.065^{+0.022}_{-0.022}$ [JFM] |
| A1016 | 3.43 | 0.0321 | $230^{+45}_{-31}$ [ZHG] | *0.4* [$\sigma_{\rm opt}$] | *0.001* [$\sigma_{\rm opt}$] |
| A1060 | 5.01 | 0.0124 [ZHG] | $608^{+47}_{-38}$ [ZAB] | $3.3^{+0.3}_{-0.3}$ [ES] | $0.781^{+0.018}_{-0.018}$ [ES] |
| A1126 | 2.40 | 0.0852[ABL89] | *713* [$T_{\rm X}$] | 3.3 [JFC] | $1.073^{+0.055}_{-0.055}$ [JFM] |
| A1142 | 2.21 | 0.0360 [ZHG] | $658^{+109}_{-73}$ [ZAB] | 3.7 [LPD] | $0.716^{+0.021}_{-0.021}$ [JFM] |
| A1146 | 5.26 | 0.1418 [ZHG] | $1166^{+186}_{-128}$[ZAB] | 5.0 [JFC] | $6.007^{+0.688}_{-0.688}$ [JFM] |
| A1185 | 1.77 | 0.0314 [ZHG] | $869^{+216}_{-125}$ [ZAB] | $3.9^{+2.0}_{-1.1}$ [LPD] | $0.563^{+0.017}_{-0.017}$ [JFM] |
| A1213 | 1.52 | 0.0468[ABL89] | 598 [SRD] | 2.0 [JFC] | $0.316^{+0.064}_{-0.064}$ [JFM] |
| A1254 | 1.70 | 0.0628 [STR] | *755* [$T_{\rm X}$] | 3.7 [JFC] | $0.527^{+0.080}_{-0.080}$ [JFM] |
| A1272 | 1.36 | 0.1505 [M10] | *800* [$T_{\rm X}$] | 4.1 [JFC] | $3.457^{+0.271}_{-0.271}$ [JFM] |
| A1285 | 3.92 | 0.1050 [EMSS] | *805* [$T_{\rm X}$] | $4.1^{+5.3}_{-1.7}$ [LPD] | $6.372^{+0.337}_{-0.337}$ [JFM] |
| A1291 | 9.66 | 0.0530[ABL89] | 975 [SRD] | 2.6 [JFC] | $0.676^{+0.046}_{-0.046}$ [JFM] |
| A1314 | 1.68 | 0.0338 [ZHG] | 664 [ZAB] | $5.0^{+4.5}_{-1.8}$ [LPD] | $0.922^{+0.030}_{-0.030}$ [JFM] |
| A1367 | 2.16 | 0.0214 [ZHG] | $822^{+69}_{-55}$ [ZAB] | $3.5^{+0.4}_{-0.4}$ [ES] | $1.760^{+0.052}_{-0.052}$ [ES] |
| A1377 | 1.02 | 0.0514[ABL89] | 488 [SRD] | 2.7 [JFC] | $0.918^{+0.046}_{-0.046}$ [JFM] |
| A1382 | 1.50 | 0.1053 [STR] | *689* [$T_{\rm X}$] | 3.1 [JFC] | $1.331^{+0.236}_{-0.236}$ [JFM] |
| A1413 | 1.98 | 0.1427 [STR] | *1231* [$T_{\rm X}$] | $8.9^{+0.3}_{-0.3}$ [LPD] | $23.623^{+1.060}_{-1.060}$ [JFM] |
| A1446 | 1.52 | 0.1028[ABL89] | *791* [$T_{\rm X}$] | 4.0 [JFC] | $3.058^{+0.106}_{-0.106}$ [JFM] |
| A1477 | 2.02 | 0.1104 [HHPG] | *633* [$T_{\rm X}$] | 2.7 [JFC] | $0.596^{+0.072}_{-0.072}$ [JFM] |
| A1569N | 2.11 | 0.0784[ABL89] | *599* [$T_{\rm X}$] | 2.4 [JFC] | 0.474 [FJE] |
| A1569S | 2.11 | 0.0784[ABL89] | *720* [$T_{\rm X}$] | 3.3 [JFC] | 1.454 [FJE] |
| A1589 | 1.90 | 0.0718[ABL89] | *786* [$T_{\rm X}$] | 3.9 [JFC] | $2.911^{+0.188}_{-0.188}$ [JFM] |
| A1631 | 3.57 | 0.0466 [ZHG] | $628^{+78}_{-56}$ [ZAB] | 2.8 [JFC] | $0.785^{+0.040}_{-0.040}$ [JFM] |
| A1631E | 3.57 | 0.0466 [ZHG] | *561* [$T_{\rm X}$] | 2.1 [JFC] | 0.273 [FJE] |
| A1644 | 4.68 | 0.0474 [ZHG] | $933^{+85}_{-67}$ [ZAB] | $4.7^{+0.5}_{-0.5}$ [LPD] | $5.156^{+0.062}_{-0.062}$ [JFM] |
| A1644NE | 4.68 | 0.0474 [ZHG] | *700* [$T_{\rm X}$] | 3.2 [JFC] | $2.630^{+0.039}_{-0.039}$ [JFM] |
| A1644SW | 4.68 | 0.0474 [ZHG] | *868* [$T_{\rm X}$] | 4.7 [JFM] | $1.124^{+0.029}_{-0.029}$ [JFM] |
| A1650 | 1.53 | 0.0840[ABL89] | *947* [$T_{\rm X}$] | $5.5^{+1.3}_{-1.0}$ [LPD] | $13.297^{+0.306}_{-0.306}$ [JFM] |
| A1651 | 1.69 | 0.0825 [ESP] | 965 [ZHG] | $7.0^{+1.7}_{-1.7}$ [ESP] | 23.434 [ESP] |
| A1656 | 0.90 | 0.0231 [ZHG] | $1010^{+51}_{-44}$ [ZAB] | $8.0^{+0.3}_{-0.3}$ [ES] | $15.563^{+0.830}_{-0.830}$ [ES] |
| A1689 | 1.87 | 0.1810[ABL89] | 1253 [SRD] | $10.1^{+5.4}_{-2.8}$[LPD] | $50.456^{+1.056}_{-1.056}$ [JFM] |
| A1709 | 8.03 | 0.0527 [PHG] | *609* [$T_{\rm X}$] | 2.5 [JFC] | $0.348^{+0.033}_{-0.033}$ [JFM] |
| A1736 | 4.99 | 0.0431 [STR] | *858* [$T_{\rm X}$] | $4.6^{+0.6}_{-0.5}$ [LPD] | 3.162 [ESP] |
| A1750N | 2.30 | 0.0855 [ZHG] | $778^{+97}_{-71}$ [ZAB] | 3.7 [JFC] | $1.973^{+0.196}_{-0.196}$ [JFM] |
| A1750S | 2.30 | 0.0855 [ZHG] | 778 [ZAB] | 4.4 [JFC] | $3.525^{+0.261}_{-0.261}$ [JFM] |
| A1750SS | 2.30 | 0.0855 [ZHG] | 778 [ZAB] | 2.5 [JFC] | $0.425^{+0.111}_{-0.111}$ [JFM] |
| A1763 | 0.93 | 0.1870 [NOO] | *1073* [$T_{\rm X}$] | 6.9 [LPD] | 19.973 [JFM] |
| A1767 | 1.72 | 0.0700 [ZHG] | $933^{+232}_{-134}$ [ZAB] | $4.1^{+1.8}_{-1.1}$ [LPD] | $4.805^{+0.145}_{-0.145}$ [JFM] |
| A1775 | 1.07 | 0.0696[ABL89] | $1571^{+666}_{-293}$[QLA] | $4.9^{+2.7}_{-1.4}$ [LPD] | $4.683^{+0.149}_{-0.149}$ [JFM] |
| A1795 | 1.16 | 0.0621 [ZHG] | $773^{+98}_{-71}$ [ZAB] | $5.1^{+0.3}_{-0.4}$ [ES] | $20.146^{+0.292}_{-0.292}$ [ES] |
| A1809 | 2.02 | 0.0789 [ZHG] | 249 [ZAB] | $3.7^{+1.0}_{-1.0}$ [ESP] | $3.230^{+0.169}_{-0.169}$ [JFM] |
| A1837 | 4.38 | 0.0376 [STR] | *596* [$T_{\rm X}$] | $2.4^{+0.7}_{-0.7}$ [ES] | $1.065^{+0.146}_{-0.146}$ [ES] |

**Table 1** Cluster Reference Data

| Cluster | $N_{\rm H}$ $(10^{20}\,{\rm cm}^{-2})$ | Redshift $z$ | $\sigma_{\rm opt,ref}$ $({\rm km\,s}^{-1})$ | $T_{\rm X,ref}$ $({\rm keV})$ | $L_{\rm X,ref}$ $(10^{44}\,{\rm erg\,s}^{-1})$ |
|---|---|---|---|---|---|
| (i) | (ii) | (iii) | (iv) | (v) | (vi) |
| A1839 | 2.69 | 0.1521 [M10] | *746* [$T_{\rm X}$] | 3.6 [JFC] | $2.445^{+0.300}_{-0.300}$ [JFM] |
| A1890 | 2.08 | 0.0579[ABL89] | *661* [$T_{\rm X}$] | 2.9 [JFC] | $1.070^{+0.077}_{-0.077}$ [JFM] |
| A1904 | 1.81 | 0.0714 [STR] | $730^{+151}_{-95}$ [QLA] | *3.4* [$\sigma_{\rm opt}$] | *1.831* [$\sigma_{\rm opt}$] |
| A1913 | 1.58 | 0.0527 [ZHG] | $454^{+128}_{-75}$ [ZAB] | 2.9 [JFC] | $1.079^{+0.033}_{-0.033}$ [JFM] |
| A1918 | 0.69 | 0.1415[ABL89] | *944* [$T_{\rm X}$] | 5.5 [JFC] | $8.731^{+0.319}_{-0.319}$ [JFM] |
| A1924 | 6.96 | 0.1353 [M10] | *718* [$T_{\rm X}$] | 3.3 [JFC] | $1.473^{+0.232}_{-0.232}$ [JFM] |
| A1940 | 1.33 | 0.1384 [ZHG] | $534^{+177}_{-92}$ [ZAB] | 4.3 [JFC] | $3.783^{+0.182}_{-0.182}$ [JFM] |
| A1983 | 2.10 | 0.0444 [ZHG] | $765^{+85}_{-64}$ [ZAB] | 2.5 [JFC] | $0.659^{+0.041}_{-0.041}$ [JFM] |
| A1991 | 2.54 | 0.0586[ABL89] | *937* [$T_{\rm X}$] | $5.4^{+5.9}_{-2.2}$ [LPD] | $3.291^{+0.107}_{-0.107}$ [JFM] |
| A2009 | 3.30 | 0.1530[ABL89] | 804 [SRD] | $7.8^{+4.4}_{-2.1}$ [LPD] | $21.872^{+0.560}_{-0.560}$[JFM] |
| A2022 | 2.24 | 0.0564 [STR] | *651* [$T_{\rm X}$] | 2.8 [JFC] | $0.939^{+0.121}_{-0.121}$ [JFM] |
| A2029 | 3.07 | 0.0765 [ZHG] | $786^{+205}_{-117}$ [ZAB] | $7.8^{+0.7}_{-0.6}$ [LPD] | 95.757 [ESP] |
| A2040 | 2.62 | 0.0456 [STR] | *609* [$T_{\rm X}$] | 2.5 [JFC] | $0.635^{+0.036}_{-0.036}$ [JFM] |
| A2050 | 4.39 | 0.1183 [STR] | *905* [$T_{\rm X}$] | 5.1 [JFC] | $6.455^{+0.358}_{-0.358}$ [JFM] |
| A2052 | 2.91 | 0.0348[ABL89] | 576 [QUI] | $3.4^{+0.5}_{-0.4}$ [ES] | $3.645^{+0.119}_{-0.119}$ [ES] |
| A2055 | 3.15 | 0.0530 [STR] | *975* [$T_{\rm X}$] | 5.8 [LPD] | $2.188^{+0.130}_{-0.130}$ [JFM] |
| A2063 | 2.91 | 0.0355 [ZHG] | $652^{+123}_{-80}$ [ZAB] | $4.1^{+0.6}_{-0.6}$ [LPD] | $3.642^{+0.066}_{-0.066}$ [JFM] |
| A2063SW | 2.91 | 0.0355 [ZHG] | *537* [$T_{\rm X}$] | 2.0 [JFC] | $0.113^{+0.011}_{-0.011}$ [JFM] |
| A2065 | 2.87 | 0.0722 [ZHG] | $1082^{+204}_{-132}$[ZAB] | $8.4^{+1.7}_{-1.2}$ [LPD] | $11.283^{+0.309}_{-0.309}$[JFM] |
| A2069 | 1.95 | 0.1160 [ABL] | 831 [STR] | *4.3* [$\sigma_{\rm opt}$] | *4.489* [$\sigma_{\rm opt}$] |
| A2079 | 2.42 | 0.0656 [ZHG] | $639^{+106}_{-71}$ [ZAB] | 3.2 [JFC] | $1.362^{+0.103}_{-0.103}$ [JFM] |
| A2089 | 2.24 | 0.0669 [STR] | $551^{+117}_{-72}$ [ZHG] | *2.1* [$\sigma_{\rm opt}$] | *0.261* [$\sigma_{\rm opt}$] |
| A2092 | 2.24 | 0.0670 [ZHG] | $504^{+115}_{-69}$ [ZAB] | 2.5 [JFC] | $0.706^{+0.088}_{-0.088}$ [JFM] |
| A2107 | 4.50 | 0.0421 [STR] | *816* [$T_{\rm X}$] | $4.2^{+1.9}_{-1.1}$ [LPD] | $2.714^{+0.096}_{-0.096}$ [JFM] |
| A2124 | 1.74 | 0.0654 [ZHG] | $848^{+300}_{-148}$ [ZAB] | 3.6 [JFC] | $2.237^{+0.097}_{-0.097}$ [JFM] |
| A2124a | 1.74 | 0.0654 [ZHG] | *617* [$T_{\rm X}$] | 2.5 [JFC] | 0.562 [FJE] |
| A2142 | 3.88 | 0.0899 [STR] | $1295^{+373}_{-207}$[QLA] | $11.0^{+1.7}_{-1.5}$ [ES] | $58.593^{+0.736}_{-0.736}$ [ES] |
| A2147 | 3.28 | 0.0356 [STR] | 1148 [TAR] | $4.4^{+1.9}_{-0.9}$ [ES] | $4.231^{+0.232}_{-0.232}$ [ES] |
| A2151 | 3.38 | 0.0370 [ZHG] | $827^{+69}_{-55}$ [ZAB] | $3.5^{+0.9}_{-0.9}$ [ES] | 1.460 [ES] |
| A2151E | 3.38 | 0.0370 [ZHG] | *596* [$T_{\rm X}$] | 2.4 [JFC] | $0.181^{+0.012}_{-0.012}$ [JFM] |
| A2151W | 3.38 | 0.0370 [ZHG] | *772* [$T_{\rm X}$] | 3.8 [JFM] | $0.759^{+0.022}_{-0.022}$ [JFM] |
| A2152 | 3.38 | 0.0374[ABL89] | *558* [$T_{\rm X}$] | 2.1 [JFC] | $0.381^{+0.039}_{-0.039}$ [JFM] |
| A2163 | 11.00 | 0.2030 [ARN] | *1579* [$T_{\rm X}$] | $13.9^{+0.7}_{-0.5}$[LPD] | 100.532 [JFM] |
| A2197 | 0.95 | 0.0305 [ZHG] | $564^{+84}_{-59}$ [ZAB] | 1.6 [JFC] | $0.155^{+0.018}_{-0.018}$ [JFM] |
| A2199 | 0.87 | 0.0299 [ZHG] | $794^{+77}_{-60}$ [ZAB] | $4.7^{+0.4}_{-0.3}$ [ES] | $6.295^{+0.075}_{-0.075}$ [ES] |
| A2204 | 5.61 | 0.1523 [ESP] | *1243* [$T_{\rm X}$] | $9.0^{+2.6}_{-2.6}$ [ESP] | 49.473 [ESP] |
| A2210 | 5.59 | 0.1465 [STR] | *778* [$T_{\rm X}$] | 3.8 [JFC] | $2.408^{+0.361}_{-0.361}$ [JFM] |
| A2218 | 3.34 | 0.1710 [STR] | *1056* [$T_{\rm X}$] | $6.7^{+2.8}_{-1.6}$ [LPD] | $18.093^{+0.750}_{-0.750}$[JFM] |
| A2220 | 2.52 | 0.1106 [STR] | *669* [$T_{\rm X}$] | 2.9 [JFC] | $1.174^{+0.096}_{-0.096}$ [JFM] |
| A2241 | 2.37 | 0.0635 [STR] | *685* [$T_{\rm X}$] | 3.1 [JFC] | $1.266^{+0.065}_{-0.065}$ [JFM] |
| A2244 | 1.96 | 0.0970 [STR] | *1090* [$T_{\rm X}$] | $7.1^{+2.4}_{-1.5}$ [LPD] | $7.011^{+0.160}_{-0.160}$ [JFM] |
| A2245 | 2.05 | 0.0843 [PHG] | *647* [$T_{\rm X}$] | 2.8 [JFC] | $0.527^{+0.040}_{-0.040}$ [JFM] |
| A2250 | 2.84 | 0.0654 [DAN] | 693 [ABL89] | 2.8 [JFC] | $0.926^{+0.107}_{-0.107}$ [JFM] |
| A2255 | 2.60 | 0.0808 [ZHG] | $1221^{+181}_{-126}$[ZAB] | $7.3^{+1.7}_{-1.1}$ [LPD] | $9.721^{+0.127}_{-0.127}$ [JFM] |
| A2255A | ..... | 0.0820 [DAN] | 385 [DAN] | *1.1* [$\sigma_{\rm opt}$] | *0.022* [$\sigma_{\rm opt}$] |
| A2256 | 4.30 | 0.0581 [ZHG] | $1270^{+107}_{-85}$[ZAB] | $7.5^{+0.1}_{-0.1}$ [LPD] | $15.777^{+0.115}_{-0.115}$[JFM] |
| A2271 | 4.01 | 0.0568[ABL89] | *666* [$T_{\rm X}$] | 2.9 [JFC] | $1.036^{+0.067}_{-0.067}$ [JFM] |
| A2312 | 5.65 | 0.0652 [M10] | *746* [$T_{\rm X}$] | 3.6 [JFC] | $1.967^{+0.096}_{-0.096}$ [JFM] |
| A2319 | 8.59 | 0.0559 [DAN] | 1580 [DAN] | $9.9^{+0.8}_{-0.7}$ [LPD] | $28.515^{+0.231}_{-0.231}$[JFM] |
| A2319A | ..... | 0.0528 [DAN] | 848 [DAN] | *4.5* [$\sigma_{\rm opt}$] | *5.164* [$\sigma_{\rm opt}$] |
| A2319B | ..... | 0.0508 [NOO] | 590 [NOO] | *2.3* [$\sigma_{\rm opt}$] | *0.420* [$\sigma_{\rm opt}$] |
| A2328 | 4.55 | 0.1470 [STR] | *931* [$T_{\rm X}$] | 5.3 [JFC] | $7.404^{+0.623}_{-0.623}$ [JFM] |
| A2356 | 4.17 | 0.1161 [STR] | *806* [$T_{\rm X}$] | 4.1 [JFC] | $3.174^{+0.262}_{-0.262}$ [JFM] |
| A2366 | 3.51 | 0.0542[ABL89] | *548* [$T_{\rm X}$] | 2.0 [JFC] | $0.340^{+0.037}_{-0.037}$ [JFM] |
| A2384N | 3.00 | 0.0943 [STR] | *875* [$T_{\rm X}$] | 4.8 [JFC] | $3.028^{+0.202}_{-0.202}$ [JFM] |

**Table 1** Cluster Reference Data

| Cluster | $N_{\rm H}$ | Redshift | | $\sigma_{\rm opt,ref}$ | | $T_{\rm X,ref}$ | | $L_{\rm X,ref}$ | |
|---|---|---|---|---|---|---|---|---|---|
| | ($10^{20}$ cm$^{-2}$) | $z$ | | (km s$^{-1}$) | | (keV) | | ($10^{44}$ erg s$^{-1}$) | |
| (i) | (ii) | (iii) | | (iv) | | (v) | | (vi) | |
| A2384S | 3.00 | 0.0943 | [STR] | *778* | [$T_{\rm X}$] | 3.8 | [JFC] | $1.470^{+0.149}_{-0.149}$ | [JFM] |
| A2384SS | 3.00 | 0.0943 | [STR] | *712* | [$T_{\rm X}$] | 3.3 | [JFC] | $0.858^{+0.122}_{-0.122}$ | [JFM] |
| A2415 | 4.84 | 0.0597 | [ABL89] | *769* | [$T_{\rm X}$] | 3.8 | [JFC] | $2.392^{+0.198}_{-0.198}$ | [JFM] |
| A2420 | 3.72 | 0.0838 | [ABL89] | *993* | [$T_{\rm X}$] | $6.0^{+2.3}_{-1.2}$ | [LPD] | $9.296^{+0.179}_{-0.179}$ | [JFM] |
| A2440 | 5.16 | 0.0904 | [ABL89] | *1243* | [$T_{\rm X}$] | 9.0 | [LPD] | $7.214^{+0.305}_{-0.305}$ | [JFM] |
| A2507 | 5.10 | 0.1960 | [LPD] | *1273* | [$T_{\rm X}$] | $9.4^{+1.6}_{-1.2}$ | [LPD] | *28.893* | [TX] |
| A2554 | 2.10 | 0.1108 | [ZHG] | $827^{+141}_{-94}$ | [ZAB] | 4.1 | [JFC] | $3.675^{+0.141}_{-0.141}$ | [JFM] |
| A2556 | 2.09 | 0.0865 | [STD] | *872* | [$T_{\rm X}$] | 4.7 | [JFC] | $5.524^{+0.128}_{-0.128}$ | [JFM] |
| A2580 | 1.96 | 0.1297 | [M10] | *907* | [$T_{\rm X}$] | 5.1 | [JFC] | $4.579^{+0.321}_{-0.321}$ | [JFM] |
| A2589 | 3.99 | 0.0416 | [ZHG] | $500^{+110}_{-67}$ | [ZAB] | $3.7^{+1.9}_{-1.0}$ | [ES] | $3.082^{+0.245}_{-0.245}$ | [ES] |
| A2593 | 4.07 | 0.0433 | [STR] | *690* | [$T_{\rm X}$] | $3.1^{+1.5}_{-0.9}$ | [LPD] | $1.846^{+0.057}_{-0.057}$ | [JFM] |
| A2597 | 2.48 | 0.0852 | [STR] | *1250* | [$T_{\rm X}$] | 9.1 | [LPD] | $15.311^{+0.522}_{-0.522}$ | [JFM] |
| A2625 | 4.16 | 0.0609 | [ABL89] | *641* | [$T_{\rm X}$] | 2.7 | [JFC] | $0.820^{+0.081}_{-0.081}$ | [JFM] |
| A2626 | 4.18 | 0.0573 | [ABL89] | 681 | [SRD] | $2.9^{+2.5}_{-1.0}$ | [LPD] | $4.782^{+0.120}_{-0.120}$ | [JFM] |
| A2634 | 4.88 | 0.0309 | [ZHG] | $744^{+127}_{-84}$ | [ZAB] | $3.4^{+0.2}_{-0.3}$ | [LPD] | $1.204^{+0.022}_{-0.022}$ | [JFM] |
| A2638 | 2.95 | 0.0825 | [STR] | *583* | [$T_{\rm X}$] | 2.3 | [JFC] | $0.483^{+0.080}_{-0.080}$ | [JFM] |
| A2657 | 6.00 | 0.0414 | [ABL89] | 667 | [SRD] | $3.4^{+0.5}_{-0.3}$ | [LPD] | $2.932^{+0.040}_{-0.040}$ | [JFM] |
| A2666 | 4.59 | 0.0270 | [ZHG] | $476^{+95}_{-60}$ | [ZAB] | *1.6* | [$\sigma_{\rm opt}$] | *0.095* | [$\sigma_{\rm opt}$] |
| A2670 | 2.69 | 0.0759 | [ZHG] | $1038^{+60}_{-52}$ | [ZAB] | $3.9^{+1.6}_{-0.9}$ | [LPD] | $3.885^{+0.178}_{-0.178}$ | [JFM] |
| A2877 | 1.80 | 0.0230 | [DRE] | *738* | [$T_{\rm X}$] | $3.5^{+1.1}_{-0.8}$ | [LPD] | $0.470^{+0.013}_{-0.013}$ | [JFM] |
| A2943 | 1.78 | 0.1710 | [M10] | *725* | [$T_{\rm X}$] | 3.4 | [JFC] | $4.826^{+0.780}_{-0.780}$ | [JFM] |
| A3021 | 30.56 | 0.1044 | [M10] | *699* | [$T_{\rm X}$] | 3.2 | [JFC] | $1.372^{+0.233}_{-0.233}$ | [JFM] |
| A3041 | 10.70 | 0.1089 | [M10] | *822* | [$T_{\rm X}$] | 4.3 | [JFC] | $3.832^{+0.562}_{-0.562}$ | [JFM] |
| A3112 | 4.00 | 0.0746 | [ESP] | 860 | [EDG] | $4.1^{+2.0}_{-1.0}$ | [ES] | $11.596^{+0.765}_{-0.765}$ | [ES] |
| A3128 | 1.30 | 0.0554 | [ABL] | *685* | [$T_{\rm X}$] | 3.1 | [JFC] | $2.756^{+0.117}_{-0.117}$ | [JFM] |
| A3128NE | 1.20 | 0.0554 | [ABL] | *685* | [$T_{\rm X}$] | 3.1 | [JFC] | $1.286^{+0.076}_{-0.076}$ | [JFM] |
| A3128SW | 1.20 | 0.0554 | [ABL] | *733* | [$T_{\rm X}$] | 3.5 | [JFC] | $0.902^{+0.066}_{-0.066}$ | [JFM] |
| A3158 | 1.20 | 0.0575 | [DAN] | $1058^{+241}_{-146}$ | [QLA] | $5.5^{+0.3}_{-0.4}$ | [LPD] | $6.970^{+0.137}_{-0.137}$ | [JFM] |
| A3186 | 6.01 | 0.1270 | [EMSS] | *985* | [$T_{\rm X}$] | 5.9 | [JFC] | $16.059^{+0.892}_{-0.892}$ | [JFM] |
| A3266 | 3.00 | 0.0594 | [ABL] | *1012* | [$T_{\rm X}$] | $6.2^{+0.5}_{-0.4}$ | [LPD] | $16.270^{+0.148}_{-0.148}$ | [JFM] |
| A3322 | 2.38 | 0.1044 | [M10] | *838* | [$T_{\rm X}$] | 4.4 | [JFC] | $2.841^{+0.221}_{-0.221}$ | [JFM] |
| A3376 | 4.41 | 0.0490 | [DRE] | *802* | [$T_{\rm X}$] | 4.1 | [JFC] | $3.125^{+0.058}_{-0.058}$ | [JFM] |
| A3389 | 4.80 | 0.0267 | [GMT] | $487^{+84}_{-57}$ | [GMT] | $2.1^{+0.9}_{-0.6}$ | [LPD] | $0.386^{+0.011}_{-0.011}$ | [JFM] |
| A3391 | 4.50 | 0.0531 | [ABL] | *918* | [$T_{\rm X}$] | $5.2^{+1.3}_{-0.9}$ | [LPD] | $3.372^{+0.071}_{-0.071}$ | [JFM] |
| A3395 | 4.50 | 0.0498 | [ABL] | 460 | [NOO] | $4.7^{+1.1}_{-0.7}$ | [LPD] | $3.658^{+0.066}_{-0.066}$ | [JFM] |
| A3395NE | 4.50 | ...... | ....... | *868* | [$T_{\rm X}$] | 4.7 | [JFM] | *7.614* | [TX] |
| A3395SW | 4.50 | ...... | ....... | *868* | [$T_{\rm X}$] | 4.7 | [JFM] | *7.614* | [TX] |
| A3528N | 6.17 | 0.0553 | [NOO] | *778* | [$T_{\rm X}$] | 3.8 | [JFC] | $0.960^{+0.015}_{-0.015}$ | [JFM] |
| A3528S | 6.17 | 0.0553 | [NOO] | *799* | [$T_{\rm X}$] | 4.0 | [JFC] | $0.833^{+0.014}_{-0.014}$ | [JFM] |
| A3530 | 6.20 | 0.0544 | [BRFJ] | *707* | [$T_{\rm X}$] | 3.2 | [JFC] | 0.959 | [FJE] |
| A3532 | 6.20 | 0.0537 | [QRM] | *837* | [$T_{\rm X}$] | $4.4^{+1.5}_{-1.5}$ | [ESP] | $2.882^{+0.103}_{-0.103}$ | [JFM] |
| A3556 | ..... | 0.4820 | [BAR] | $554^{+47}_{-47}$ | [BARD] | *2.1* | [$\sigma_{\rm opt}$] | *0.271* | [$\sigma_{\rm opt}$] |
| A3558 | 4.50 | 0.0476 | [BAR] | $986^{+60}_{-60}$ | [BARD] | $3.8^{+1.0}_{-0.5}$ | [LPD] | 10.415 | [ESP] |
| A3562 | 4.16 | 0.0499 | [LPD] | *767* | [$T_{\rm X}$] | $3.8^{+0.8}_{-0.7}$ | [ES] | $9.908^{+0.421}_{-0.421}$ | [ES] |
| SC1329-31 | 4.16 | 0.0499 | [ABL] | *684* | [$T_{\rm X}$] | 3.0 | [JFC] | $0.805^{+0.032}_{-0.032}$ | [JFM] |
| A3565 | 4.26 | 0.0109 | [ABL] | *334* | [$T_{\rm X}$] | 0.8 | [JFC] | $0.010^{+0.001}_{-0.001}$ | [JFM] |
| A3571 | 4.04 | 0.0390 | [ESP] | 1070 | [QUD] | $7.6^{+1.0}_{-0.8}$ | [ES] | $17.157^{+0.238}_{-0.238}$ | [ES] |
| A3581 | 4.78 | 0.0216 | [PHG] | *654* | [$T_{\rm X}$] | 2.8 | [JFC] | $0.845^{+0.025}_{-0.025}$ | [JFM] |
| A3602 | 11.40 | 0.1044 | [M10] | *805* | [$T_{\rm X}$] | 4.1 | [JFC] | $2.845^{+0.290}_{-0.290}$ | [JFM] |
| A3654 | 6.84 | 0.1044 | [M10] | *763* | [$T_{\rm X}$] | 3.7 | [JFC] | $1.461^{+0.092}_{-0.092}$ | [JFM] |
| A3656 | ..... | 0.0189 | [GMT] | $420^{+144}_{-76}$ | [GMT] | *1.3* | [$\sigma_{\rm opt}$] | *0.040* | [$\sigma_{\rm opt}$] |
| A3667 | 4.00 | 0.0542 | [QLA] | $1667^{+346}_{-221}$ | [QLA] | $6.5^{+1.0}_{-1.0}$ | [LPD] | $12.574^{+0.139}_{-0.139}$ | [JFM] |
| A3744 | 5.17 | 0.0387 | [GMT] | $925^{+169}_{-180}$ | [GMT] | *5.3* | [$\sigma_{\rm opt}$] | *9.423* | [$\sigma_{\rm opt}$] |
| A3880 | 1.08 | 0.0380 | [ODC] | *777* | [$T_{\rm X}$] | 3.8 | [JFC] | $1.428^{+0.060}_{-0.060}$ | [JFM] |
| A3888 | 1.22 | 0.1680 | [ABL] | *1157* | [$T_{\rm X}$] | 7.9 | [JFC] | $31.519^{+1.739}_{-1.739}$ | [JFM] |

**Table 1** Cluster Reference Data

| Cluster | $N_{\rm H}$ ($10^{20}$ cm$^{-2}$) | Redshift $z$ | | $\sigma_{\rm opt,ref}$ (km s$^{-1}$) | | $T_{\rm X,ref}$ (keV) | | $L_{\rm X,ref}$ ($10^{44}$ erg s$^{-1}$) | |
|---|---|---|---|---|---|---|---|---|---|
| (i) | (ii) | (iii) | | (iv) | | (v) | | (vi) | |
| A4039 | 1.53 | 0.0276 | [DAN] | *714* | [$T_{\rm X}$] | $3.3^{+1.6}_{-0.8}$ | [LPD] | *3.856* | [TX] |
| A4059 | ..... | 0.0478 | [BEE] | 845 | [GGP] | $3.5^{+0.5}_{-0.5}$ | [ES] | $4.936^{+0.179}_{-0.179}$ | [ES] |
| A4067 | 1.10 | 0.0959 | [APM] | *783* | [$T_{\rm X}$] | 3.9 | [JFC] | $3.147^{+0.165}_{-0.165}$ | [JFM] |
| A4067S | 1.10 | 0.0959 | [APM] | *639* | [$T_{\rm X}$] | 2.7 | [JFC] | $0.474^{+0.068}_{-0.068}$ | [JFM] |
| A4067SS | 1.10 | 0.0959 | [APM] | *626* | [$T_{\rm X}$] | 2.6 | [JFC] | $0.533^{+0.068}_{-0.068}$ | [JFM] |
| AWM4 | 4.93 | 0.0424 | [NOO] | *761* | [$T_{\rm X}$] | $3.7^{+2.0}_{-1.0}$ | [LPD] | $1.673^{+0.022}_{-0.022}$ | [JFM] |
| AWM5 | 4.87 | 0.0345 | [DFG] | *622* | [$T_{\rm X}$] | 2.6 | [JFC] | $0.583^{+0.041}_{-0.041}$ | [JFM] |
| AWM7 | 9.19 | 0.0176 | [NOO] | $830^{+161}_{-104}$ | [QLA] | $3.6^{+0.2}_{-0.2}$ | [ES] | $3.119^{+0.041}_{-0.041}$ | [ES] |
| CENTAURUS | 7.96 | 0.0107 | [NOO] | 586 | [LDC] | $3.6^{+0.1}_{-0.3}$ | [ES] | $1.406^{+0.032}_{-0.032}$ | [ES] |
| CYGNUS-A | 36.10 | 0.0570 | [ESP] | *805* | [$T_{\rm X}$] | $4.1^{+2.6}_{-0.8}$ | [LPD] | $10.658^{+0.153}_{-0.153}$ | [JFM] |
| FORNAX-A | 1.35 | 0.0048 | [DAN] | 240 | [DAN] | *0.5* | [$\sigma_{\rm opt}$] | *0.001* | [$\sigma_{\rm opt}$] |
| HERCULES-A | 6.26 | 0.1540 | [TAD] | *939* | [$T_{\rm X}$] | 5.4 | [JFC] | $7.547^{+0.134}_{-0.134}$ | [JFM] |
| HYDRA-A | 4.80 | 0.0522 | [ABL89] | *778* | [$T_{\rm X}$] | $3.8^{+0.8}_{-0.8}$ | [ES] | $7.234^{+0.607}_{-0.607}$ | [ES] |
| MKW3S | 2.89 | 0.0434 | [ESP] | *678* | [$T_{\rm X}$] | $3.0^{+0.3}_{-0.3}$ | [LPD] | $4.259^{+0.110}_{-0.110}$ | [JFM] |
| MKW4 | 1.85 | 0.0196 | [NOO] | *495* | [$T_{\rm X}$] | $1.7^{+1.7}_{-0.7}$ | [LPD] | $0.445^{+0.007}_{-0.007}$ | [JFM] |
| MKW4S | 1.66 | 0.0288 | [DHG] | *533* | [$T_{\rm X}$] | 1.9 | [JFC] | $0.323^{+0.032}_{-0.032}$ | [JFM] |
| OPHIUCHUS | 19.70 | 0.0280 | [ESP] | *1242* | [$T_{\rm X}$] | $9.0^{+0.7}_{-0.6}$ | [ES] | $31.410^{+2.761}_{-2.761}$ | [ES] |
| PKS0745-19 | 46.60 | 0.1028 | [ESP] | *1207* | [$T_{\rm X}$] | $8.5^{+1.6}_{-1.2}$ | [ES] | $57.175^{+4.360}_{-4.360}$ | [ES] |
| SERSIC159-03 | 2.09 | 0.0580 | [EMSS] | *677* | [$T_{\rm X}$] | $3.0^{+1.0}_{-0.6}$ | [ES] | $4.060^{+0.328}_{-0.328}$ | [ES] |
| SC0247-31 | 2.17 | 0.0210 | [DRE] | *528* | [$T_{\rm X}$] | 1.9 | [JFC] | $0.264^{+0.014}_{-0.014}$ | [JFM] |
| SC1842-63 | 6.60 | 0.0141 | [NOO] | *445* | [$T_{\rm X}$] | $1.4^{+0.3}_{-0.3}$ | [LPD] | $0.126^{+0.003}_{-0.003}$ | [JFM] |
| SC2316-3632 | 1.65 | 0.0121 | [Z120] | *874* | [$T_{\rm X}$] | 4.8 | [JFC] | $0.038^{+0.003}_{-0.003}$ | [JFM] |
| TRIANGULUM-A | 19.80 | 0.0510 | [ESP] | *1160* | [$T_{\rm X}$] | $7.9^{+1.2}_{-1.1}$ | [ES] | 26.421 | [ES] |
| M87 | 2.50 | 0.0037 | [DAN] | 573 | [BTS] | $2.4^{+0.3}_{-0.3}$ | [ES] | 0.665 | [ES] |
| WP23 | 5.02 | 0.0087 | [NOO] | *370* | [$T_{\rm X}$] | $1.0^{+0.6}_{-0.4}$ | [LPD] | $0.205^{+0.006}_{-0.006}$ | [JFM] |
| SC2059-25 | ..... | ...... | ....... | *1082* | [$T_{\rm X}$] | $7.0^{+4.2}_{-1.3}$ | [LPD] | *16.386* | [TX] |
| S617 | ..... | 0.0344 | [GMT] | $862^{+293}_{-146}$ | [GMT] | *4.6* | [$\sigma_{\rm opt}$] | *5.783* | [$\sigma_{\rm opt}$] |
| S639 | ..... | 0.0211 | [GMT] | $523^{+162}_{-87}$ | [GMT] | *1.9* | [$\sigma_{\rm opt}$] | *0.182* | [$\sigma_{\rm opt}$] |
| S963 | ..... | 0.0328 | [GMT] | $531^{+141}_{-181}$ | [GMT] | *1.9* | [$\sigma_{\rm opt}$] | *0.202* | [$\sigma_{\rm opt}$] |
| K38 | ..... | 0.0310 | [GMT] | $422^{+183}_{-85}$ | [GMT] | *1.3* | [$\sigma_{\rm opt}$] | *0.041* | [$\sigma_{\rm opt}$] |
| CL0422-09 | 6.02 | 0.0390 | [ESP] | *669* | [$T_{\rm X}$] | $2.9^{+0.7}_{-0.5}$ | [ES] | $3.103^{+0.199}_{-0.199}$ | [ES] |
| 2A0336+09 | 17.20 | 0.0350 | [ESP] | *684* | [$T_{\rm X}$] | $3.0^{+0.2}_{-0.2}$ | [ES] | $7.067^{+0.156}_{-0.156}$ | [ES] |
| 3C129 | 57.10 | 0.0218 | [NOO] | *959* | [$T_{\rm X}$] | $5.6^{+0.6}_{-0.5}$ | [ES] | $4.010^{+0.107}_{-0.107}$ | [ES] |
| 3C130 | 51.70 | 0.1090 | [SPN] | *898* | [$T_{\rm X}$] | 5.0 | [JFC] | $3.806^{+0.326}_{-0.326}$ | [JFM] |
| 3C370 | 3.45 | 0.0540 | [BAH] | *618* | [$T_{\rm X}$] | 2.5 | [JFC] | $0.675^{+0.034}_{-0.034}$ | [JFM] |
| 3C449 | 10.80 | 0.0171 | [ILT] | *481* | [$T_{\rm X}$] | 1.6 | [JFC] | $0.119^{+0.010}_{-0.010}$ | [JFM] |
| 3C66B | 7.43 | 0.0214 | [7SAM] | *467* | [$T_{\rm X}$] | 1.5 | [JFC] | $0.089^{+0.015}_{-0.015}$ | [JFM] |
| ZW0628+25 | 31.20 | 0.0810 | [LPD] | *1012* | [$T_{\rm X}$] | $6.2^{+3.6}_{-1.7}$ | [LPD] | $5.156^{+0.209}_{-0.209}$ | [JFM] |
| ZW0712+534 | 6.60 | 0.0644 | [NIL] | *656* | [$T_{\rm X}$] | 2.8 | [JFC] | $0.911^{+0.104}_{-0.104}$ | [JFM] |
| ZW1615+35 | 1.45 | 0.0321 | [NOO] | *665* | [$T_{\rm X}$] | $2.9^{+2.6}_{-1.1}$ | [LPD] | $0.589^{+0.017}_{-0.017}$ | [JFM] |

Table 2: Deprojection Input Data

| # | Name | Alternative Name | Sequence | IPC Energy ( keV) | Gain (ALP) | Exposure ( s) | Redshift $z$ | $D_{\rm L}$ ( Mpc) | $N_{\rm H}$ ($10^{21}$ cm$^{-2}$ ) | k$T$ ( keV) | $\sigma_{\rm clust}$ ( km s$^{-1}$ ) | $R_{\rm core}$ ( Mpc) | $N$ Bins | d$R$ ( arcsec) | d$R$ ( Mpc) | $R_{\rm out}$ ( Mpc) | $P_{\rm out}/10^4$ ( K cm$^{-3}$ ) |
|---|---|---|---|---|---|---|---|---|---|---|---|---|---|---|---|---|---|
| (i) | (ii) | (iii) | (iv) | (v) | (vi) | (vii) | (viii) | (ix) | (x) | (xi) | (xii) | (xiii) | (xiv) | (xv) | (xvi) | (xvii) | (xviii) |
| 1 | A21 | | I6012 | 0.5 − 3.0 | 14.3 | 1959 | 0.095 | 581 | 0.43 | 4.8 | 882 (*882*) | 0.60 | 5 | 96 | 0.226 | 1.130 | 1.00 |
| 2 | A74 | | I8989 | 0.6 − 4.5 | 11.5 | 2113 | 0.067 | 409 | 0.14 | 2.2 | 577 (*577*) | 0.20 | 3 | 96 | 0.167 | 0.502 | 0.20 |
| 3 | A76 | | I1817 | 0.6 − 4.5 | 15.1 | 1511 | 0.042 | 252 | 0.38 | 1.5 | 462 (*462*) | 0.80 | 5 | 96 | 0.108 | 0.540 | 0.50 |
| 4 | A84 | | I7640 | 0.6 − 4.5 | 14.4 | 6512 | 0.103 | 633 | 0.34 | 3.8 | 769 (*769*) | 0.80 | 5 | 72 | 0.182 | 0.908 | 0.50 |
| 5 | A85 | | I292 | 0.6 − 4.5 | 14.8 | 14274 | 0.052 | 316 | 0.30 | 6.2 | 749 (749) | 0.15 | 18 | 72 | 0.100 | 1.800 | 0.90 |
| | | | H6013 | ............ | ..... | 7156 | 0.052 | 316 | 0.30 | 6.2 | 749 (749) | 0.20 | 11 | 20 | 0.028 | 0.305 | 21.00 |
| 6 | A98N | | I208 | 0.6 − 4.5 | 22.2 | 4308 | 0.104 | 641 | 0.34 | 3.3 | 829 (829) | 0.50 | 3 | 72 | 0.183 | 0.550 | 1.00 |
| 7 | A98S | | I208 | 0.6 − 4.5 | 22.2 | 4308 | 0.106 | 653 | 0.34 | 3.2 | 694 (919) | 1.00 | 4 | 48 | 0.124 | 0.497 | 3.00 |
| 8 | A115N | | I209 | 0.5 − 3.0 | 14.3 | 2564 | 0.197 | 1235 | 0.51 | 6.6 | 1045 (*1045*) | 0.20 | 3 | 24 | 0.100 | 0.301 | 25.00 |
| 9 | A115S | | I209 | 0.5 − 3.0 | 14.3 | 2564 | 0.197 | 1235 | 0.51 | 5.7 | 969 (*969*) | 0.25 | 3 | 16 | 0.067 | 0.201 | 35.00 |
| 10 | A117 | | I8992 | 0.6 − 4.5 | 12.1 | 5019 | 0.054 | 325 | 0.37 | 2.4 | 601 (*601*) | 0.50 | 5 | 96 | 0.136 | 0.681 | 0.30 |
| 11 | A119 | | I1770 | 0.6 − 4.5 | 14.5 | 3706 | 0.044 | 268 | 0.34 | 5.1 | 863 (863) | 0.80 | 18 | 72 | 0.086 | 1.550 | 0.80 |
| 12 | A133 | | I2333 | 0.6 − 4.5 | 18.2 | 4312 | 0.060 | 367 | 0.16 | 3.8 | 767 (*767*) | 0.30 | 7 | 72 | 0.114 | 0.798 | 1.50 |
| 13 | A150N | | I10766 | 0.6 − 4.5 | 17.1 | 2587 | 0.060 | 363 | 0.39 | 2.0 | 543 (*543*) | 0.40 | 7 | 72 | 0.113 | 0.789 | 0.30 |
| 14 | A154 | | I6135 | 0.6 − 4.5 | 15.6 | 4399 | 0.065 | 397 | 0.43 | 3.1 | 833 (833) | 0.70 | 6 | 72 | 0.122 | 0.733 | 1.00 |
| 15 | A160 | | I154 | 0.6 − 4.5 | 24.1 | 2610 | 0.045 | 271 | 0.42 | 2.2 | 572 (572) | 0.40 | 5 | 96 | 0.116 | 0.578 | 0.50 |
| 16 | A168 | | I6083 | 0.6 − 4.5 | 14.7 | 12081 | 0.045 | 274 | 0.34 | 2.6 | 571 (571) | 1.00 | 8 | 96 | 0.117 | 0.934 | 0.80 |
| 17 | A194 | | I6084 | 0.6 − 4.5 | 14.8 | 9444 | 0.018 | 108 | 0.42 | 1.9 | 480 (480) | 0.60 | 12 | 72 | 0.037 | 0.438 | 0.50 |
| 18 | A240 | | I189 | 0.6 − 4.5 | 17.1 | 2279 | 0.062 | 376 | 0.44 | 2.3 | 591 (*591*) | 0.50 | 4 | 72 | 0.116 | 0.466 | 0.50 |
| 19 | A262 | | I295 | 0.6 − 4.5 | 17.8 | 3838 | 0.016 | 98 | 0.53 | 2.4 | 494 (494) | 0.20 | 15 | 72 | 0.033 | 0.497 | 0.70 |
| 20 | A278 | | I7698 | 0.5 − 3.5 | 15.0 | 4658 | 0.090 | 549 | 0.50 | 3.5 | 739 (*739*) | 0.15 | 3 | 32 | 0.072 | 0.215 | 9.00 |
| 21 | A347 | | I302 | 0.5 − 3.0 | 15.0 | 4737 | 0.019 | 114 | 0.63 | 2.3 | 590 (590) | 0.60 | 17 | 72 | 0.038 | 0.654 | 0.60 |
| 22 | A367 | | I3445 | 0.6 − 4.5 | 17.5 | 1637 | 0.089 | 546 | 0.26 | 4.1 | 802 (*802*) | 0.50 | 3 | 96 | 0.214 | 0.643 | 0.70 |
| 23 | A397 | | I7699 | 0.6 − 4.5 | 14.4 | 5047 | 0.032 | 196 | 0.95 | 1.6 | 447 (447) | 0.20 | 4 | 48 | 0.043 | 0.172 | 0.80 |
| 24 | A399 | | I185 | 0.6 − 4.5 | 17.1 | 6942 | 0.072 | 436 | 1.17 | 5.8 | 971 (1424) | 0.70 | 11 | 72 | 0.133 | 1.460 | 1.00 |
| 25 | A400 | | I6085 | 0.6 − 4.5 | 14.5 | 10474 | 0.024 | 144 | 0.88 | 2.1 | 610 (610) | 0.70 | 14 | 72 | 0.048 | 0.669 | 0.50 |
| 26 | A401 | | H1777 | ............ | ..... | 3964 | 0.075 | 457 | 1.11 | 7.8 | 1113 (1290) | 0.55 | 9 | 50 | 0.096 | 0.862 | 5.00 |
| | | | I1776 | 0.5 − 4.0 | 17.5 | 8621 | 0.075 | 457 | 1.11 | 7.8 | 1113 (1290) | 0.55 | 10 | 72 | 0.138 | 1.380 | 2.00 |
| 27 | A407 | | I1825 | 0.6 − 4.5 | 11.9 | 2040 | 0.047 | 285 | 0.98 | 2.8 | 590 (590) | 0.50 | 8 | 72 | 0.091 | 0.726 | 1.00 |
| 28 | A419 | | I8993 | 0.6 − 4.5 | 11.6 | 1143 | 0.041 | 246 | 0.17 | 1.9 | 522 (*522*) | 0.30 | 5 | 72 | 0.079 | 0.396 | 0.50 |
| 29 | A426 | PERSEUS | H285 | ............ | ..... | 15040 | 0.018 | 108 | 1.45 | 5.5 | 925 (1277) | 0.15 | 13 | 20 | 0.010 | 0.131 | 150.00 |
| | | | I283 | 0.5 − 4.0 | 15.0 | 14231 | 0.018 | 108 | 1.45 | 5.5 | 925 (1277) | 0.10 | 16 | 96 | 0.048 | 0.775 | 4.00 |
| 30 | A458 | | I6018 | 0.5 − 3.0 | 14.3 | 2962 | 0.105 | 648 | 0.16 | 4.4 | 709 (709) | 0.50 | 4 | 72 | 0.185 | 0.740 | 2.00 |
| 31 | A478 | | H4198 | ............ | ..... | 11543 | 0.088 | 539 | 1.36 | 6.8 | 904 (904) | 0.20 | 11 | 20 | 0.044 | 0.486 | 15.00 |

Table 2: Deprojection Input Data

| # | Name | Alternative Name | Sequence | IPC Energy (keV) | Gain (ALP) | Exposure (s) | Redshift $z$ | $D_{\rm L}$ (Mpc) | $N_{\rm H}$ ($10^{21}$ cm$^{-2}$) | k$T$ (keV) | $\sigma_{\rm clust}$ (km s$^{-1}$) | $R_{\rm core}$ (Mpc) | $N$ Bins | d$R$ (arcsec) | d$R$ (Mpc) | $R_{\rm out}$ (Mpc) | $P_{\rm out}/10^4$ (K cm$^{-3}$) |
|---|---|---|---|---|---|---|---|---|---|---|---|---|---|---|---|---|---|
| (i) | (ii) | (iii) | (iv) | (v) | (vi) | (vii) | (viii) | (ix) | (x) | (xi) | (xii) | (xiii) | (xiv) | (xv) | (xvi) | (xvii) | (xviii) |
| | | | I303 | 0.5 − 3.0 | 14.3 | 3509 | 0.088 | 539 | 1.36 | 6.8 | 904 (904) | 0.20 | 12 | 72 | 0.159 | 1.910 | 1.50 |
| 32 | A496 | | I2348 | 0.6 − 4.5 | 11.8 | 3678 | 0.033 | 199 | 0.44 | 4.7 | 705 (705) | 0.15 | 14 | 96 | 0.087 | 1.220 | 1.20 |
| | | | H10401 | ………… | ….. | 17990 | 0.033 | 199 | 0.44 | 4.7 | 705 (705) | 0.15 | 8 | 25 | 0.023 | 0.181 | 30.00 |
| 33 | A500 | | I6232 | 0.6 − 4.5 | 16.5 | 9065 | 0.067 | 406 | 0.29 | 3.3 | 720 (*720*) | 0.60 | 6 | 72 | 0.125 | 0.747 | 1.20 |
| 34 | A520 | | I6841 | 0.5 − 3.0 | 14.3 | 4781 | 0.203 | 1273 | 0.75 | 7.4 | 1112 (*1112*) | 0.80 | 4 | 48 | 0.205 | 0.819 | 7.00 |
| 35 | A539 | | I2353 | 0.6 − 4.5 | 15.0 | 4336 | 0.029 | 174 | 1.33 | 3.0 | 832 (832) | 0.70 | 12 | 72 | 0.057 | 0.688 | 1.00 |
| 36 | A545 | | I310 | 0.5 − 4.0 | 15.0 | 4139 | 0.153 | 950 | 1.14 | 5.5 | 947 (*947*) | 0.60 | 8 | 72 | 0.249 | 2.000 | 0.50 |
| 37 | A548S | | I7860 | 0.5 − 4.0 | 15.0 | 10431 | 0.041 | 251 | 0.20 | 2.4 | 853 (853) | 0.80 | 10 | 72 | 0.081 | 0.809 | 0.40 |
| 38 | A566 | | I3553 | 0.6 − 4.5 | 13.6 | 1747 | 0.098 | 604 | 0.51 | 4.6 | 860 (*860*) | 0.50 | 7 | 72 | 0.175 | 1.220 | 0.70 |
| 39 | A569 | | I1836 | 0.6 − 4.5 | 16.5 | 735 | 0.019 | 116 | 0.82 | 1.4 | 444 (444) | 0.01 | 3 | 48 | 0.026 | 0.078 | 1.50 |
| 40 | A576 | | I3455 | 0.6 − 4.5 | 16.4 | 10378 | 0.038 | 231 | 0.56 | 3.7 | 914 (914) | 0.60 | 8 | 72 | 0.075 | 0.597 | 0.60 |
| 41 | A586 | | I211 | 0.5 − 3.0 | 14.3 | 2080 | 0.172 | 1072 | 0.57 | 5.8 | 940 (…..) | 0.45 | 4 | 32 | 0.121 | 0.484 | 12.00 |
| 42 | A592 | | I5170 | 0.6 − 4.5 | 17.2 | 12979 | 0.062 | 380 | 0.45 | 3.3 | 711 (*711*) | 0.40 | 4 | 72 | 0.117 | 0.470 | 2.00 |
| 43 | A629 | | I317 | 0.6 − 4.5 | 16.7 | 7721 | 0.138 | 854 | 0.43 | 3.3 | 713 (*713*) | 0.30 | 3 | 72 | 0.230 | 0.691 | 0.70 |
| 44 | A644 | | I5728 | 0.6 − 4.5 | 15.3 | 11272 | 0.070 | 429 | 0.73 | 6.6 | 1046 (*1046*) | 0.40 | 9 | 72 | 0.131 | 1.180 | 2.00 |
| 45 | A646 | | I1839 | 0.6 − 4.5 | 17.0 | 1425 | 0.130 | 805 | 0.45 | 5.3 | 927 (*927*) | 0.30 | 3 | 72 | 0.220 | 0.660 | 3.50 |
| 46 | A665 | | I305 | 0.5 − 4.0 | 15.0 | 6431 | 0.182 | 1134 | 0.42 | 8.3 | 1201 (1201) | 1.00 | 7 | 96 | 0.378 | 2.650 | 0.50 |
| 47 | A671 | | I7337 | 0.6 − 4.5 | 12.6 | 21895 | 0.049 | 300 | 0.39 | 3.2 | 704 (*704*) | 0.60 | 8 | 72 | 0.095 | 0.760 | 1.00 |
| 48 | A690 | | I6020 | 0.6 − 4.5 | 12.6 | 6061 | 0.079 | 481 | 0.41 | 2.0 | 546 (*546*) | 0.40 | 5 | 72 | 0.144 | 0.722 | 0.40 |
| 49 | A732 | | I6118 | 0.6 − 4.5 | 15.7 | 4925 | 0.203 | 1273 | 0.39 | 7.7 | 1063 (…..) | 0.70 | 4 | 72 | 0.307 | 1.230 | 4.00 |
| 50 | A744 | | I481 | 0.6 − 4.5 | 17.6 | 13459 | 0.073 | 447 | 0.37 | 2.7 | 814 (814) | 0.60 | 5 | 72 | 0.135 | 0.677 | 0.50 |
| 51 | A754 | | I1784 | 0.6 − 4.5 | 16.3 | 3114 | 0.054 | 329 | 0.47 | 8.7 | 1250 (747) | 0.80 | 7 | 120 | 0.172 | 1.210 | 4.50 |
| | | | H1786 | ………… | ….. | 5513 | 0.054 | 329 | 0.47 | 8.7 | 1250 (747) | 0.80 | 4 | 68 | 0.097 | 0.388 | 18.00 |
| 52 | A779 | | I1841 | 0.6 − 4.5 | 17.3 | 1551 | 0.023 | 139 | 0.17 | 1.5 | 503 (503) | 0.20 | 4 | 72 | 0.046 | 0.185 | 0.50 |
| 53 | A795 | | I212 | 0.6 − 4.5 | 12.3 | 1555 | 0.136 | 840 | 0.35 | 5.8 | 978 (*978*) | 0.60 | 6 | 72 | 0.227 | 1.360 | 1.50 |
| 54 | A838 | | I6097 | 0.6 − 4.5 | 15.5 | 11039 | 0.051 | 308 | 0.35 | 1.3 | 497 (…..) | 0.40 | 3 | 72 | 0.097 | 0.292 | 0.50 |
| 55 | A910 | | I1788 | 0.5 − 4.0 | 15.0 | 4443 | 0.206 | 1290 | 0.39 | 11.4 | 1250 (…..) | 1.00 | 7 | 72 | 0.310 | 2.170 | 3.00 |
| 56 | A957 | | I6023 | 0.6 − 4.5 | 17.6 | 2125 | 0.044 | 267 | 0.36 | 2.9 | 669 (*669*) | 0.60 | 9 | 72 | 0.085 | 0.769 | 0.50 |
| 57 | A970 | | I7791 | 0.6 − 4.5 | 15.2 | 2092 | 0.060 | 362 | 0.53 | 4.3 | 832 (*832*) | 0.50 | 6 | 72 | 0.113 | 0.675 | 2.50 |
| 58 | A979 | | I6098 | 0.6 − 4.5 | 12.7 | 12087 | 0.055 | 334 | 0.46 | 0.9 | 433 (…..) | 0.60 | 8 | 24 | 0.035 | 0.279 | 0.30 |
| 59 | A999 | | I7700 | 0.6 − 4.5 | 15.2 | 4921 | 0.032 | 193 | 0.39 | 1.2 | 404 (404) | 0.10 | 4 | 72 | 0.063 | 0.253 | 0.20 |
| 60 | A1060 | HYDRA | I6114 | 0.6 − 4.5 | 15.1 | 10441 | 0.012 | 75 | 0.50 | 3.3 | 608 (608) | 0.20 | 15 | 72 | 0.025 | 0.381 | 2.00 |
| 61 | A1142 | | I6079 | 0.6 − 4.5 | 11.9 | 9433 | 0.036 | 218 | 0.22 | 3.7 | 658 (658) | 0.50 | 6 | 72 | 0.071 | 0.425 | 1.50 |
| 62 | A1146 | | I217 | 0.6 − 4.5 | 16.4 | 949 | 0.142 | 878 | 0.53 | 5.0 | 1166 (1166) | 0.70 | 3 | 72 | 0.235 | 0.706 | 3.00 |
| 63 | A1185 | | I6100 | 0.6 − 4.5 | 15.5 | 11459 | 0.031 | 190 | 0.18 | 3.9 | 869 (869) | 0.30 | 7 | 72 | 0.062 | 0.436 | 1.00 |

Table 2: Deprojection Input Data

| # | Name | Alternative Name | Sequence | IPC Energy ( keV) | Gain (ALP) | Exposure ( s) | Redshift $z$ | $D_{\rm L}$ ( Mpc) | $N_{\rm H}$ ($10^{21}$ cm$^{-2}$ ) | k$T$ ( keV) | $\sigma_{\rm clust}$ ( km s$^{-1}$ ) | $R_{\rm core}$ ( Mpc) | $N$ Bins | d$R$ ( arcsec) | d$R$ ( Mpc) | $R_{\rm out}$ ( Mpc) | $P_{\rm out}/10^4$ ( K cm$^{-3}$ ) |
|---|---|---|---|---|---|---|---|---|---|---|---|---|---|---|---|---|---|
| (i) | (ii) | (iii) | (iv) | (v) | (vi) | (vii) | (viii) | (ix) | (x) | (xi) | (xii) | (xiii) | (xiv) | (xv) | (xvi) | (xvii) | (xviii) |
| 64 | A1213 | | I1844 | 0.6 − 4.5 | 18.0 | 1533 | 0.047 | 284 | 0.15 | 2.0 | 598 (598) | 0.80 | 3 | 72 | 0.090 | 0.271 | 1.50 |
| 65 | A1246 | | I233 | 0.5 − 3.0 | 14.3 | 1795 | 0.216 | 1358 | 0.14 | 6.2 | 971 (.....) | 0.70 | 4 | 64 | 0.285 | 1.140 | 3.00 |
| 66 | A1254 | | I172 | 0.6 − 4.5 | 16.2 | 2834 | 0.153 | 947 | 0.17 | 3.7 | 755 (*755*) | 0.20 | 3 | 72 | 0.249 | 0.746 | 1.00 |
| 67 | A1272 | | I331 | 0.6 − 4.5 | 11.9 | 3492 | 0.152 | 944 | 0.14 | 4.1 | 800 (*800*) | 0.80 | 5 | 96 | 0.331 | 1.650 | 0.40 |
| 68 | A1285 | | I331 | 0.6 − 4.5 | 16.4 | 3492 | 0.100 | 614 | 0.39 | 4.1 | 805 (*805*) | 0.70 | 8 | 72 | 0.177 | 1.420 | 0.50 |
| 69 | A1291 | | I6293 | 0.6 − 4.5 | 15.2 | 6588 | 0.053 | 322 | 0.97 | 2.6 | 614 (975) | 0.40 | 4 | 72 | 0.101 | 0.405 | 1.50 |
| 70 | A1314 | | I6120 | 0.6 − 4.5 | 14.0 | 6407 | 0.034 | 204 | 0.17 | 5.0 | 664 (664) | 0.50 | 6 | 72 | 0.067 | 0.400 | 3.50 |
| 71 | A1367 | | I296 | 0.6 − 4.5 | 16.2 | 23777 | 0.021 | 129 | 0.22 | 3.5 | 822 (822) | 0.80 | 9 | 120 | 0.072 | 0.647 | 2.00 |
| 72 | A1377 | | I6101 | 0.6 − 4.5 | 17.2 | 5997 | 0.051 | 312 | 0.10 | 2.7 | 488 (488) | 0.50 | 5 | 96 | 0.131 | 0.657 | 0.70 |
| 73 | A1413 | | I308 | 0.6 − 4.5 | 16.9 | 1738 | 0.143 | 884 | 0.20 | 8.9 | 1231 (*1231*) | 0.50 | 7 | 72 | 0.236 | 1.650 | 1.50 |
| 74 | A1546 | | I6868 | 0.5 − 3.0 | 14.3 | 3552 | 0.146 | 905 | 0.20 | 4.2 | 825 (.....) | 0.40 | 3 | 32 | 0.107 | 0.321 | 10.00 |
| 75 | A1569S | | I1849 | 0.6 − 4.5 | 17.7 | 2929 | 0.078 | 479 | 0.21 | 3.3 | 720 (*720*) | 0.40 | 4 | 72 | 0.144 | 0.575 | 1.00 |
| 76 | A1576 | | I6871 | 0.5 − 3.0 | 14.3 | 3559 | 0.302 | 1930 | 0.17 | 6.2 | 971 (.....) | 0.60 | 4 | 32 | 0.177 | 0.707 | 5.00 |
| 77 | A1589 | | I145 | 0.6 − 4.5 | 17.7 | 1555 | 0.072 | 438 | 0.19 | 3.9 | 786 (*786*) | 0.80 | 5 | 96 | 0.177 | 0.887 | 2.00 |
| 78 | A1617 | | I6875 | 0.5 − 3.0 | 14.3 | 4234 | 0.152 | 944 | 0.12 | 2.2 | 629 (.....) | 0.60 | 3 | 32 | 0.110 | 0.331 | 1.00 |
| 79 | A1631 | | I1900 | 0.6 − 4.5 | 16.7 | 5511 | 0.047 | 283 | 0.36 | 2.8 | 628 (628) | 0.20 | 3 | 120 | 0.150 | 0.450 | 0.50 |
| 80 | A1644 | | I7654 | 0.5 − 3.5 | 15.0 | 10689 | 0.047 | 287 | 0.47 | 4.7 | 933 (933) | 0.35 | 6 | 72 | 0.091 | 0.549 | 3.00 |
| 81 | A1650 | | I6034 | 0.6 − 4.5 | 14.6 | 3327 | 0.084 | 514 | 0.15 | 5.5 | 947 (*947*) | 0.45 | 8 | 72 | 0.153 | 1.220 | 1.00 |
| 82 | A1656 | COMA BERENICES | I1793 | 0.5 − 4.5 | 14.3 | 8292 | 0.023 | 139 | 0.09 | 8.0 | 1010 (1010) | 0.50 | 24 | 72 | 0.046 | 1.110 | 2.50 |
| 83 | A1689 | | I6123 | 0.5 − 3.0 | 14.3 | 6335 | 0.181 | 1130 | 0.19 | 10.1 | 1253 (1253) | 0.40 | 6 | 72 | 0.283 | 1.700 | 1.50 |
| 84 | A1704 | | I6877 | 0.5 − 3.0 | 14.3 | 5851 | 0.220 | 1385 | 0.18 | 5.9 | 952 (.....) | 0.60 | 3 | 48 | 0.216 | 0.649 | 7.00 |
| 85 | A1709 | | I8996 | 0.6 − 4.5 | 12.2 | 3938 | 0.052 | 315 | 0.80 | 2.5 | 609 (*609*) | 0.40 | 5 | 72 | 0.099 | 0.497 | 0.50 |
| 86 | A1736 | | I7653 | 0.6 − 4.5 | 14.4 | 10272 | 0.043 | 261 | 0.50 | 4.6 | 858 (*858*) | 0.80 | 12 | 72 | 0.084 | 1.010 | 1.50 |
| 87 | A1750N | | I144 | 0.6 − 4.5 | 17.1 | 1281 | 0.086 | 523 | 0.23 | 3.7 | 778 (778) | 0.60 | 5 | 72 | 0.155 | 0.775 | 1.00 |
| 88 | A1763 | | I3930 | 0.6 − 4.5 | 11.7 | 4292 | 0.187 | 1169 | 0.09 | 6.9 | 1073 (*1073*) | 0.80 | 7 | 72 | 0.290 | 2.030 | 1.00 |
| 89 | A1767 | | I5731 | 0.6 − 4.5 | 18.4 | 4244 | 0.070 | 427 | 0.17 | 4.1 | 933 (933) | 0.70 | 6 | 72 | 0.130 | 0.781 | 2.00 |
| 90 | A1775 | | I320 | 0.6 − 4.5 | 20.0 | 3507 | 0.070 | 424 | 0.11 | 4.9 | 870 (1571) | 0.50 | 8 | 72 | 0.129 | 1.040 | 1.20 |
| 91 | A1795 | | I293 | 0.6 − 4.5 | 16.6 | 6498 | 0.062 | 378 | 0.12 | 5.1 | 773 (773) | 0.20 | 16 | 72 | 0.117 | 1.870 | 0.80 |
| | | | H7881 | ............ | ..... | 5041 | 0.062 | 378 | 0.12 | 5.1 | 773 (773) | 0.25 | 10 | 20 | 0.032 | 0.325 | 22.00 |
| 92 | A1809 | | I142 | 0.6 − 4.5 | 15.1 | 2444 | 0.079 | 482 | 0.20 | 3.7 | 750 (249) | 0.70 | 8 | 72 | 0.145 | 1.160 | 0.80 |
| 93 | A1837 | | I141 | 0.6 − 4.5 | 15.7 | 2349 | 0.038 | 228 | 0.44 | 2.4 | 596 (*596*) | 0.30 | 6 | 72 | 0.074 | 0.443 | 1.00 |
| 94 | A1839 | | I6037 | 0.5 − 4.0 | 15.0 | 2243 | 0.131 | 810 | 0.27 | 3.6 | 746 (*746*) | 0.80 | 4 | 72 | 0.221 | 0.884 | 1.50 |
| 95 | A1877 | | I6883 | 0.5 − 3.0 | 14.3 | 3147 | 0.124 | 766 | 0.16 | 3.1 | 756 (.....) | 0.40 | 3 | 24 | 0.071 | 0.212 | 10.00 |
| 96 | A1890 | | I165 | 0.5 − 3.0 | 15.5 | 3085 | 0.058 | 352 | 0.21 | 2.9 | 661 (*661*) | 0.40 | 4 | 72 | 0.110 | 0.439 | 1.00 |
| 97 | A1913 | | I6077 | 0.6 − 4.5 | 15.6 | 14136 | 0.053 | 320 | 0.16 | 2.9 | 454 (454) | 0.80 | 6 | 72 | 0.101 | 0.605 | 1.00 |

Table 2: Deprojection Input Data

| # | Name | Alternative Name | Sequence | IPC Energy ( keV) | Gain (ALP) | Exposure ( s) | Redshift $z$ | $D_{\rm L}$ ( Mpc) | $N_{\rm H}$ ($10^{21}$ cm$^{-2}$ ) | k$T$ ( keV) | $\sigma_{\rm clust}$ ( km s$^{-1}$ ) | $R_{\rm core}$ ( Mpc) | $N$ Bins | d$R$ ( arcsec) | d$R$ ( Mpc) | $R_{\rm out}$ ( Mpc) | $P_{\rm out}/10^4$ ( K cm$^{-3}$ ) |
|---|---|---|---|---|---|---|---|---|---|---|---|---|---|---|---|---|---|
| (i) | (ii) | (iii) | (iv) | (v) | (vi) | (vii) | (viii) | (ix) | (x) | (xi) | (xii) | (xiii) | (xiv) | (xv) | (xvi) | (xvii) | (xviii) |
| 98 | A1940 | | I6124 | 0.5 − 3.5 | 15.0 | 8819 | 0.138 | 857 | 0.13 | 4.3 | 534 (534) | 0.40 | 5 | 72 | 0.231 | 1.150 | 1.00 |
| 99 | A1942 | | I6847 | 0.5 − 3.0 | 14.3 | 7158 | 0.224 | 1411 | 0.26 | 3.4 | 756 (.....) | 0.60 | 3 | 32 | 0.146 | 0.438 | 2.00 |
| 100 | A1983 | | I4190 | 0.6 − 4.5 | 14.4 | 4079 | 0.044 | 269 | 0.21 | 2.5 | 765 (765) | 0.40 | 4 | 72 | 0.086 | 0.344 | 1.50 |
| 101 | A1991 | | I6039 | 0.6 − 4.5 | 14.1 | 4158 | 0.059 | 356 | 0.25 | 5.4 | 937 (*937*) | 0.20 | 6 | 72 | 0.111 | 0.666 | 1.00 |
| 102 | A2009 | | I3582 | 0.6 − 4.5 | 10.8 | 6092 | 0.153 | 950 | 0.33 | 7.8 | 1113 (804) | 0.10 | 6 | 72 | 0.249 | 1.500 | 2.00 |
| 103 | A2022 | | I4191 | 0.6 − 4.5 | 13.8 | 1034 | 0.056 | 343 | 0.22 | 2.8 | 651 (*651*) | 0.60 | 4 | 72 | 0.107 | 0.429 | 1.00 |
| 104 | A2029 | | I138 | 0.6 − 4.5 | 16.9 | 474 | 0.076 | 467 | 0.31 | 7.8 | 1112 (786) | 0.30 | 11 | 72 | 0.141 | 1.550 | 1.00 |
| | | | H7882 | ............ | ..... | 8881 | 0.076 | 467 | 0.31 | 7.8 | 1112 (786) | 0.30 | 6 | 25 | 0.049 | 0.293 | 40.00 |
| 105 | A2040 | | I6104 | 0.6 − 4.5 | 16.3 | 9226 | 0.046 | 276 | 0.26 | 2.5 | 609 (*609*) | 0.70 | 7 | 72 | 0.088 | 0.618 | 1.00 |
| 106 | A2050 | | I3444 | 0.6 − 4.5 | 11.8 | 2583 | 0.118 | 729 | 0.44 | 5.1 | 905 (*905*) | 0.60 | 7 | 72 | 0.204 | 1.420 | 1.00 |
| 107 | A2052 | | I1853 | 0.6 − 4.5 | 16.9 | 1425 | 0.035 | 210 | 0.29 | 3.4 | 576 (576) | 0.20 | 13 | 72 | 0.069 | 0.892 | 1.00 |
| | | | H5728 | ............ | ..... | 11272 | 0.035 | 210 | 0.29 | 3.4 | 576 (576) | 0.15 | 8 | 20 | 0.019 | 0.152 | 17.00 |
| 108 | A2055 | | I137 | 0.6 − 4.5 | 11.9 | 1854 | 0.053 | 322 | 0.31 | 5.8 | 975 (*975*) | 0.15 | 5 | 72 | 0.101 | 0.507 | 3.00 |
| 109 | A2063 | | H4595 | ............ | ..... | 9791 | 0.036 | 215 | 0.29 | 4.1 | 652 (652) | 0.25 | 4 | 20 | 0.019 | 0.078 | 32.00 |
| | | | I162 | 0.5 − 4.0 | 12.5 | 3781 | 0.036 | 215 | 0.29 | 4.1 | 652 (652) | 0.25 | 4 | 72 | 0.070 | 0.280 | 10.00 |
| 110 | A2065 | CORONA BOREALIS | I1795 | 0.6 − 4.5 | 18.0 | 2445 | 0.072 | 440 | 0.29 | 8.4 | 1082 (1082) | 0.40 | 9 | 72 | 0.134 | 1.200 | 1.00 |
| 111 | A2069E | | I10404 | 0.6 − 4.5 | 12.0 | 5566 | 0.116 | 715 | 0.19 | 6.1 | 998 (.....) | 1.00 | 13 | 72 | 0.200 | 2.600 | 0.50 |
| 112 | A2069W | | I10404 | 0.6 − 4.5 | 12.0 | 5566 | 0.116 | 715 | 0.19 | 5.9 | 980 (.....) | 1.00 | 12 | 72 | 0.200 | 2.400 | 0.60 |
| 113 | A2079 | | I1854 | 0.6 − 4.5 | 17.5 | 2536 | 0.066 | 400 | 0.24 | 3.2 | 639 (639) | 0.60 | 6 | 72 | 0.123 | 0.737 | 1.00 |
| 114 | A2092 | | I135 | 0.6 − 4.5 | 18.2 | 2529 | 0.067 | 408 | 0.22 | 2.5 | 504 (504) | 0.25 | 5 | 72 | 0.125 | 0.626 | 0.50 |
| 115 | A2107 | | I134 | 0.6 − 4.5 | 14.8 | 1836 | 0.042 | 255 | 0.45 | 4.2 | 816 (*816*) | 0.30 | 8 | 72 | 0.082 | 0.656 | 1.50 |
| 116 | A2111 | | I239 | 0.5 − 3.0 | 14.3 | 2191 | 0.229 | 1444 | 0.19 | 6.3 | 1017 (.....) | 0.80 | 5 | 32 | 0.148 | 0.741 | 5.00 |
| 117 | A2124 | | I4192 | 0.6 − 4.5 | 15.9 | 4084 | 0.065 | 398 | 0.17 | 3.6 | 848 (848) | 0.50 | 4 | 72 | 0.122 | 0.490 | 1.50 |
| 118 | A2142 | | H1800 | ............ | ..... | 8182 | 0.090 | 551 | 0.39 | 11.0 | 1295 (1295) | 0.50 | 8 | 40 | 0.090 | 0.719 | 18.00 |
| | | | I1798 | 0.5 − 4.0 | 15.0 | 2951 | 0.090 | 551 | 0.39 | 11.0 | 1295 (1295) | 0.45 | 16 | 60 | 0.135 | 2.160 | 1.50 |
| 119 | A2147 | | I297 | 0.6 − 4.5 | 12.1 | 10076 | 0.036 | 215 | 0.33 | 4.4 | 837 (1148) | 0.50 | 10 | 72 | 0.070 | 0.701 | 2.00 |
| 120 | A2151 | | H9264 | ............ | ..... | 17760 | 0.037 | 224 | 0.34 | 3.5 | 827 (827) | 0.10 | 5 | 45 | 0.045 | 0.227 | 1.00 |
| | | | I1801 | 0.5 − 4.0 | 15.0 | 5763 | 0.037 | 224 | 0.34 | 3.5 | 827 (827) | 0.25 | 5 | 72 | 0.073 | 0.363 | 4.00 |
| 121 | A2152 | | I1855 | 0.6 − 4.5 | 16.4 | 3304 | 0.037 | 226 | 0.34 | 2.1 | 558 (*558*) | 0.30 | 5 | 72 | 0.073 | 0.367 | 1.00 |
| 122 | A2163 | | I4526 | 0.6 − 4.5 | 11.9 | 1517 | 0.170 | 1059 | 1.10 | 13.9 | 1579 (*1579*) | 0.70 | 12 | 72 | 0.270 | 3.240 | 1.00 |
| 123 | A2197 | | I1857 | 0.6 − 4.5 | 12.2 | 5503 | 0.031 | 184 | 0.09 | 1.6 | 564 (564) | 0.70 | 9 | 72 | 0.061 | 0.545 | 0.60 |
| 124 | A2199 | | I4193 | 0.6 − 4.5 | 18.1 | 4708 | 0.030 | 181 | 0.09 | 4.7 | 794 (794) | 0.20 | 12 | 72 | 0.059 | 0.713 | 1.50 |
| | | | H4597 | ............ | ..... | 1823 | 0.030 | 181 | 0.09 | 4.7 | 794 (794) | 0.20 | 7 | 30 | 0.025 | 0.173 | 35.00 |
| 125 | A2218 | | H3160 | ............ | ..... | 36000 | 0.171 | 1066 | 0.33 | 6.7 | 1056 (*1056*) | 0.40 | 6 | 30 | 0.113 | 0.678 | 5.00 |
| 126 | A2220 | | I8351 | 0.6 − 4.5 | 15.3 | 8386 | 0.111 | 681 | 0.25 | 2.9 | 669 (*669*) | 0.60 | 3 | 48 | 0.128 | 0.385 | 1.00 |

Table 2: Deprojection Input Data

| # | Name | Alternative Name | Sequence | IPC Energy (keV) | Gain (ALP) | Exposure (s) | Redshift $z$ | $D_{\rm L}$ (Mpc) | $N_{\rm H}$ ($10^{21}$ cm$^{-2}$) | k$T$ (keV) | $\sigma_{\rm clust}$ (km s$^{-1}$) | $R_{\rm core}$ (Mpc) | $N$ Bins | d$R$ (arcsec) | d$R$ (Mpc) | $R_{\rm out}$ (Mpc) | $P_{\rm out}/10^4$ (K cm$^{-3}$) |
|---|---|---|---|---|---|---|---|---|---|---|---|---|---|---|---|---|---|
| (i) | (ii) | (iii) | (iv) | (v) | (vi) | (vii) | (viii) | (ix) | (x) | (xi) | (xii) | (xiii) | (xiv) | (xv) | (xvi) | (xvii) | (xviii) |
| 127 | A2244 | | H10190 | ........... | ..... | 25458 | 0.097 | 595 | 0.20 | 7.1 | 1090 (*1090*) | 0.50 | 12 | 15 | 0.036 | 0.432 | 20.00 |
| | | | | 0.5 − 3.0 | 14.3 | 7154 | 0.097 | 595 | 0.20 | 7.1 | 1090 (*1090*) | 0.50 | 12 | 72 | 0.173 | 2.070 | 1.00 |
| 128 | A2250 | | I3090 | 0.6 − 4.5 | 17.2 | 2005 | 0.065 | 398 | 0.28 | 2.8 | 693 (693) | 0.30 | 4 | 72 | 0.122 | 0.490 | 1.00 |
| 129 | A2255 | | I160 | 0.6 − 4.5 | 16.5 | 15287 | 0.081 | 494 | 0.26 | 7.3 | 1221 (1221) | 1.20 | 11 | 72 | 0.148 | 1.620 | 1.50 |
| 130 | A2256 | | I300 | 0.6 − 4.5 | 11.4 | 13702 | 0.058 | 353 | 0.43 | 7.5 | 1270 (1270) | 1.00 | 15 | 72 | 0.110 | 1.650 | 1.20 |
| | | | H10189 | ........... | ..... | 48523 | 0.058 | 353 | 0.43 | 7.5 | 1270 (1270) | 1.00 | 11 | 20 | 0.031 | 0.337 | 24.00 |
| 131 | A2271 | | I6042 | 0.6 − 4.5 | 15.1 | 3374 | 0.057 | 345 | 0.40 | 2.9 | 666 (*666*) | 0.55 | 8 | 72 | 0.108 | 0.863 | 0.50 |
| 132 | A2306 | | I7696 | 0.6 − 4.5 | 14.6 | 19604 | 0.116 | 715 | 0.62 | 3.8 | 767 (.....) | 0.80 | 9 | 72 | 0.200 | 1.800 | 0.40 |
| 133 | A2312 | | I6269 | 0.6 − 4.5 | 14.8 | 3159 | 0.065 | 397 | 0.56 | 3.6 | 746 (*746*) | 0.40 | 7 | 72 | 0.122 | 0.855 | 1.00 |
| 134 | A2319 | | I3456 | 0.6 − 4.5 | 14.0 | 6360 | 0.056 | 340 | 0.86 | 9.9 | 1261 (1580) | 0.50 | 15 | 72 | 0.106 | 1.600 | 2.00 |
| 135 | A2328 | | I6271 | 0.6 − 4.5 | 14.6 | 1557 | 0.147 | 912 | 0.46 | 5.3 | 931 (*931*) | 0.80 | 5 | 72 | 0.242 | 1.210 | 2.00 |
| 136 | A2356 | | I7802 | 0.6 − 4.5 | 16.7 | 2507 | 0.116 | 715 | 0.42 | 4.1 | 806 (*806*) | 0.60 | 4 | 72 | 0.200 | 0.802 | 1.50 |
| 137 | A2366 | | I133 | 0.6 − 4.5 | 16.2 | 7307 | 0.054 | 329 | 0.35 | 2.0 | 548 (*548*) | 0.80 | 4 | 72 | 0.103 | 0.414 | 0.50 |
| 138 | A2384N | | I7805 | 0.6 − 4.5 | 15.0 | 1566 | 0.094 | 578 | 0.30 | 4.8 | 875 (*875*) | 0.50 | 3 | 72 | 0.169 | 0.506 | 2.50 |
| 139 | A2390 | | I9125 | 0.6 − 4.5 | 12.9 | 1154 | 0.124 | 763 | 0.70 | 9.3 | 1152 (.....) | 0.30 | 7 | 72 | 0.211 | 1.480 | 3.50 |
| 140 | A2397 | | I242 | 0.5 − 3.0 | 14.3 | 6968 | 0.224 | 1411 | 0.54 | 4.6 | 858 (.....) | 0.60 | 5 | 32 | 0.146 | 0.731 | 3.00 |
| 141 | A2410 | | I6071 | 0.6 − 4.5 | 15.8 | 12852 | 0.081 | 493 | 0.39 | 2.6 | 670 (.....) | 0.80 | 8 | 72 | 0.147 | 1.180 | 0.20 |
| 142 | A2415 | | I130 | 0.6 − 4.5 | 11.8 | 866 | 0.060 | 363 | 0.48 | 3.8 | 769 (*769*) | 0.30 | 7 | 72 | 0.113 | 0.790 | 0.50 |
| 143 | A2420 | | I6045 | 0.5 − 4.0 | 15.0 | 7135 | 0.084 | 513 | 0.37 | 6.0 | 993 (*993*) | 0.60 | 7 | 72 | 0.152 | 1.070 | 2.00 |
| 144 | A2440 | | I129 | 0.6 − 4.5 | 12.1 | 2822 | 0.090 | 554 | 0.52 | 9.0 | 1243 (*1243*) | 0.80 | 6 | 120 | 0.271 | 1.630 | 1.00 |
| 145 | A2554 | | I336 | 0.6 − 4.5 | 11.8 | 8015 | 0.111 | 682 | 0.21 | 4.1 | 827 (827) | 0.50 | 5 | 72 | 0.193 | 0.964 | 1.00 |
| 146 | A2556 | | I336 | 0.6 − 4.5 | 11.8 | 8015 | 0.086 | 529 | 0.21 | 4.7 | 872 (*872*) | 0.35 | 7 | 72 | 0.157 | 1.100 | 1.00 |
| 147 | A2577 | | I1875 | 0.6 − 4.5 | 11.3 | 1771 | 0.130 | 801 | 0.20 | 5.9 | 947 (.....) | 0.40 | 7 | 72 | 0.219 | 1.530 | 0.40 |
| 148 | A2580 | | H5751 | ........... | ..... | 12744 | 0.130 | 801 | 0.20 | 5.1 | 907 (*907*) | 0.30 | 7 | 36 | 0.110 | 0.767 | 1.00 |
| 149 | A2593 | | I6134 | 0.6 − 4.5 | 17.5 | 5689 | 0.043 | 262 | 0.41 | 3.1 | 690 (*690*) | 0.70 | 13 | 72 | 0.084 | 1.090 | 0.50 |
| 150 | A2625 | | I156 | 0.6 − 4.5 | 18.6 | 2656 | 0.061 | 371 | 0.42 | 2.7 | 641 (*641*) | 0.20 | 3 | 72 | 0.115 | 0.345 | 1.00 |
| 151 | A2626 | | I201 | 0.6 − 4.5 | 18.6 | 2050 | 0.057 | 348 | 0.42 | 2.9 | 681 (681) | 0.30 | 7 | 72 | 0.109 | 0.761 | 0.80 |
| 152 | A2634 | | I199 | 0.5 − 4.0 | 15.0 | 10056 | 0.031 | 187 | 0.49 | 3.4 | 744 (744) | 0.70 | 12 | 72 | 0.061 | 0.736 | 1.00 |
| 153 | A2657 | | I290 | 0.6 − 4.5 | 16.7 | 10499 | 0.041 | 251 | 0.60 | 3.4 | 667 (667) | 0.40 | 8 | 96 | 0.108 | 0.861 | 1.00 |
| 154 | A2666 | | I294 | 0.6 − 4.5 | 24.3 | 11517 | 0.027 | 163 | 0.46 | 1.6 | 476 (476) | 0.15 | 2 | 72 | 0.054 | 0.108 | 1.40 |
| 155 | A2670 | | I314 | 0.6 − 4.5 | 16.5 | 2558 | 0.076 | 463 | 0.27 | 3.9 | 1038 (1038) | 0.60 | 7 | 72 | 0.140 | 0.978 | 0.30 |
| 156 | A2703 | | I5360 | 0.6 − 4.5 | 13.4 | 5166 | 0.114 | 704 | 0.39 | 3.6 | 750 (.....) | 0.70 | 6 | 72 | 0.198 | 1.190 | 1.00 |
| 157 | A2715 | | I4517 | 0.5 − 3.5 | 15.0 | 4188 | 0.114 | 702 | 0.11 | 4.1 | 813 (.....) | 0.10 | 4 | 32 | 0.088 | 0.351 | 5.00 |
| 158 | A2877 | | I6088 | 0.6 − 4.5 | 14.7 | 5461 | 0.023 | 139 | 0.18 | 3.5 | 738 (*738*) | 0.20 | 10 | 72 | 0.046 | 0.463 | 0.30 |
| 159 | A3158 | SC0340-538, CA0340-53 | H5753 | ........... | ..... | 11622 | 0.058 | 350 | 0.12 | 5.5 | 1058 (1058) | 0.65 | 11 | 20 | 0.030 | 0.333 | 16.00 |

Table 2: Deprojection Input Data

| # | Name | Alternative Name | Sequence | IPC Energy (keV) | Gain (ALP) | Exposure (s) | Redshift $z$ | $D_L$ (Mpc) | $N_H$ ($10^{21}$ cm$^{-2}$) | k$T$ (keV) | $\sigma_{clust}$ (km s$^{-1}$) | $R_{core}$ (Mpc) | $N$ Bins | d$R$ (arcsec) | d$R$ (Mpc) | $R_{out}$ (Mpc) | $P_{out}/10^4$ (K cm$^{-3}$) |
|---|---|---|---|---|---|---|---|---|---|---|---|---|---|---|---|---|---|
| (i) | (ii) | (iii) | (iv) | (v) | (vi) | (vii) | (viii) | (ix) | (x) | (xi) | (xii) | (xiii) | (xiv) | (xv) | (xvi) | (xvii) | (xviii) |
| | | | I1829 | 0.5 – 4.0 | 17.0 | 4360 | 0.058 | 350 | 0.12 | 5.5 | 1058 (1058) | 0.65 | 8 | 64 | 0.097 | 0.776 | 3.00 |
| 160 | A3186 | | I8385 | 0.6 – 4.5 | 12.9 | 1136 | 0.104 | 642 | 0.60 | 5.9 | 985 (*985*) | 0.60 | 5 | 72 | 0.184 | 0.918 | 3.00 |
| 161 | A3266 | SC0430-61 | I1831 | 0.5 – 3.5 | 15.0 | 8244 | 0.059 | 361 | 0.30 | 6.2 | 1012 (*1012*) | 0.90 | 13 | 72 | 0.112 | 1.460 | 2.00 |
| 162 | A3322 | | I7705 | 0.6 – 4.5 | 15.6 | 1819 | 0.104 | 642 | 0.24 | 4.4 | 838 (*838*) | 0.30 | 5 | 48 | 0.122 | 0.612 | 2.00 |
| 163 | A3376 | SC0559-40 | I5167 | 0.6 – 4.5 | 16.6 | 7253 | 0.049 | 297 | 0.44 | 4.1 | 802 (*802*) | 0.50 | 5 | 72 | 0.094 | 0.472 | 1.50 |
| 164 | A3389E | SC0622-64 | I5169 | 0.6 – 4.5 | 17.8 | 8652 | 0.026 | 160 | 0.48 | 2.1 | 487 (.....) | 0.10 | 4 | 72 | 0.053 | 0.212 | 2.00 |
| 165 | A3389W | SC0622-64 | I5169 | 0.6 – 4.5 | 17.8 | 8652 | 0.026 | 160 | 0.48 | 2.1 | 487 (.....) | 0.60 | 9 | 72 | 0.053 | 0.477 | 2.00 |
| 166 | A3391 | SC0624-53 | I8309 | 0.5 – 4.0 | 15.0 | 5877 | 0.053 | 322 | 0.45 | 5.2 | 918 (*918*) | 0.50 | 8 | 72 | 0.102 | 0.812 | 1.00 |
| 167 | A3532 | | I6173 | 0.8 – 3.5 | 14.4 | 3322 | 0.069 | 421 | 0.62 | 4.4 | 837 (*837*) | 0.70 | 7 | 72 | 0.128 | 0.899 | 2.50 |
| 168 | A3562W | | I5730 | 0.6 – 4.5 | 14.4 | 6223 | 0.050 | 303 | 0.42 | 3.0 | 677 (.....) | 0.70 | 5 | 72 | 0.096 | 0.479 | 1.50 |
| 169 | A3581 | PKS1404-267 | I9980 | 0.6 – 4.5 | 15.5 | 2542 | 0.021 | 126 | 0.48 | 2.8 | 654 (*654*) | 0.10 | 5 | 80 | 0.047 | 0.234 | 2.00 |
| 170 | A3602 | | I5252 | 0.6 – 4.5 | 15.7 | 1541 | 0.104 | 642 | 0.50 | 4.1 | 805 (*805*) | 0.30 | 3 | 72 | 0.184 | 0.551 | 2.50 |
| 171 | A3654 | | I3289 | 0.6 – 4.5 | 16.1 | 5153 | 0.104 | 642 | 0.68 | 3.7 | 763 (*763*) | 0.60 | 4 | 96 | 0.245 | 0.979 | 1.00 |
| 172 | A3667 | SC2009-56, 2A2009 | I5735 | 0.6 – 4.5 | 13.3 | 6249 | 0.058 | 356 | 0.40 | 6.5 | 1010 (1667) | 0.80 | 12 | 96 | 0.148 | 1.770 | 2.00 |
| 173 | A370 | | I245 | 0.6 – 4.5 | 16.7 | 4078 | 0.373 | 2413 | 0.30 | 10.3 | 1198 (.....) | 0.10 | 5 | 48 | 0.298 | 1.490 | 7.00 |
| 174 | A3744 | SC2009-56, 2A2009 | I3044 | 0.6 – 4.5 | 17.3 | 1499 | 0.026 | 157 | 0.52 | 5.4 | 925 (925) | 0.50 | 9 | 72 | 0.052 | 0.468 | 2.00 |
| 175 | A376 | | I1773 | 0.5 – 3.5 | 15.0 | 1565 | 0.049 | 297 | 0.63 | 5.1 | 903 (*903*) | 0.15 | 4 | 64 | 0.084 | 0.335 | 5.00 |
| 176 | A3888 | | I1872 | 0.6 – 4.5 | 11.4 | 1122 | 0.168 | 1046 | 0.12 | 7.9 | 1157 (*1157*) | 0.50 | 6 | 72 | 0.268 | 1.610 | 2.00 |
| 177 | A389 | | I6128 | 0.5 – 4.0 | 15.0 | 6039 | 0.116 | 715 | 0.18 | 4.5 | 848 (*848*) | 0.40 | 5 | 72 | 0.200 | 1.000 | 1.00 |
| 178 | A3998 | | I6385 | 0.6 – 4.5 | 14.8 | 9768 | 0.104 | 642 | 0.44 | 4.1 | 800 (.....) | 0.30 | 5 | 72 | 0.184 | 0.918 | 1.50 |
| 179 | A4067S | | I5745 | 0.6 – 4.5 | 12.2 | 4036 | 0.100 | 615 | 0.11 | 3.9 | 639 (*639*) | 0.40 | 6 | 72 | 0.177 | 1.060 | 1.00 |
| 180 | AWM4 | | I10543 | 0.6 – 4.5 | 16.7 | 22762 | 0.042 | 257 | 0.49 | 3.7 | 761 (*761*) | 0.40 | 10 | 96 | 0.110 | 1.100 | 0.20 |
| 181 | AWM5 | | I3302 | 0.6 – 4.5 | 11.8 | 1534 | 0.034 | 209 | 0.49 | 2.6 | 622 (*622*) | 0.40 | 7 | 72 | 0.068 | 0.476 | 0.50 |
| 182 | AWM7 | | I6698 | 0.6 – 4.5 | 17.5 | 748 | 0.018 | 107 | 0.92 | 3.6 | 830 (830) | 0.40 | 13 | 96 | 0.048 | 0.623 | 1.00 |
| | | | H6638 | ........... | ..... | 11623 | 0.018 | 106 | 0.92 | 3.6 | 830 (830) | 0.40 | 11 | 20 | 0.010 | 0.109 | 20.00 |
| 183 | CENTAURUS | | I298 | 0.6 – 4.5 | 11.6 | 7843 | 0.011 | 64 | 0.80 | 3.6 | 586 (586) | 0.15 | 29 | 72 | 0.022 | 0.637 | 1.00 |
| | | | H4341 | ........... | ..... | 9216 | 0.011 | 64 | 0.80 | 3.6 | 586 (586) | 0.15 | 16 | 20 | 0.006 | 0.098 | 24.00 |
| 184 | CL0016+16 | | H7755 | ........... | ..... | 60068 | 0.550 | 3658 | 0.32 | 7.3 | 1039 (.....) | 0.50 | 6 | 10 | 0.074 | 0.443 | 24.00 |
| 185 | CYGNUS-A | | I1807 | 0.6 – 4.5 | 12.2 | 4283 | 0.057 | 346 | 3.61 | 4.1 | 805 (*805*) | 0.20 | 10 | 72 | 0.108 | 1.080 | 0.70 |
| | | | H10760 | ........... | ..... | 21868 | 0.057 | 346 | 3.61 | 4.1 | 805 (*805*) | 0.20 | 8 | 20 | 0.030 | 0.241 | 25.00 |
| 186 | FORNAX-A | | H1885 | ........... | ..... | 23274 | 0.005 | 29 | 0.44 | 0.4 | 259 (240) | 0.01 | 3 | 40 | 0.006 | 0.017 | 1.00 |
| 187 | HERCULES-A | | I10533 | 0.6 – 4.5 | 16.9 | 41238 | 0.154 | 956 | 0.63 | 5.4 | 939 (*939*) | 0.30 | 3 | 72 | 0.251 | 0.752 | 2.50 |
| 188 | HYDRA-A | | I1894 | 0.6 – 4.5 | 15.3 | 10845 | 0.052 | 317 | 0.48 | 3.8 | 778 (*778*) | 0.30 | 15 | 72 | 0.100 | 1.500 | 0.40 |
| 189 | M87 | VIRGO | H282 | ........... | ..... | 69000 | 0.004 | 22 | 0.25 | 2.4 | 573 (573) | 0.30 | 6 | 20 | 0.002 | 0.013 | 62.00 |
| | | | I10362 | 0.5 – 4.0 | 13.0 | 9253 | 0.004 | 22 | 0.25 | 2.4 | 573 (573) | 0.30 | 23 | 120 | 0.013 | 0.295 | 3.00 |

Table 2: Deprojection Input Data

| # | Name | Alternative Name | Sequence | IPC Energy ( keV) | Gain (ALP) | Exposure ( s) | Redshift $z$ | $D_{\rm L}$ ( Mpc) | $N_{\rm H}$ ($10^{21}$ cm$^{-2}$ ) | k$T$ ( keV) | $\sigma_{\rm clust}$ ( km s$^{-1}$ ) | $R_{\rm core}$ ( Mpc) | $N$ Bins | d$R$ ( arcsec) | d$R$ ( Mpc) | $R_{\rm out}$ ( Mpc) | $P_{\rm out}/10^4$ ( K cm$^{-3}$ ) |
|---|---|---|---|---|---|---|---|---|---|---|---|---|---|---|---|---|---|
| (i) | (ii) | (iii) | (iv) | (v) | (vi) | (vii) | (viii) | (ix) | (x) | (xi) | (xii) | (xiii) | (xiv) | (xv) | (xvi) | (xvii) | (xviii) |
| 190 | MKW3S | | I2604 | 0.6 − 4.5 | 12.3 | 2087 | 0.043 | 263 | 0.29 | 3.0 | 678 (*678*) | 0.25 | 9 | 72 | 0.084 | 0.759 | 1.00 |
| | | | H4359 | ........... | ..... | 10306 | 0.043 | 263 | 0.29 | 3.0 | 678 (*678*) | 0.25 | 9 | 20 | 0.023 | 0.211 | 14.00 |
| 191 | MKW4 | | I2601 | 0.6 − 4.5 | 13.2 | 10359 | 0.020 | 123 | 0.19 | 1.7 | 495 (*495*) | 0.20 | 11 | 72 | 0.041 | 0.453 | 0.50 |
| 192 | OPHIUCHUS | | H6553 | ........... | ..... | 8495 | 0.028 | 169 | 1.97 | 9.0 | 1242 (*1242*) | 0.25 | 9 | 60 | 0.047 | 0.419 | 10.00 |
| 193 | PKS0745-19 | | H6541 | ........... | ..... | 5133 | 0.103 | 631 | 4.66 | 8.5 | 1207 (*1207*) | 0.20 | 8 | 15 | 0.038 | 0.302 | 25.00 |
| 194 | SC1842-63 | S0805 | I6105 | 0.6 − 4.5 | 15.7 | 16954 | 0.014 | 85 | 0.66 | 1.4 | 445 (*445*) | 0.30 | 9 | 72 | 0.029 | 0.259 | 0.80 |
| 195 | SC2316-3632 | S1109 | I7569 | 0.6 − 4.5 | 14.8 | 1947 | 0.140 | 867 | 0.17 | 4.8 | 874 (*874*) | 0.50 | 4 | 72 | 0.233 | 0.931 | 1.50 |
| 196 | SERSIC159-03 | SC2311-43, S1101 | I1874 | 0.5 − 3.5 | 15.0 | 1899 | 0.056 | 338 | 0.21 | 3.0 | 677 (*677*) | 0.05 | 6 | 24 | 0.035 | 0.212 | 1.20 |
| 197 | ZW0258+43 | | I4611 | 0.6 − 4.5 | 15.6 | 4298 | 0.065 | 396 | 1.38 | 2.1 | 620 (.....) | 0.60 | 6 | 72 | 0.122 | 0.731 | 0.50 |
| 198 | ZW0628+25 | | I4613 | 0.6 − 4.5 | 15.5 | 3733 | 0.081 | 495 | 3.12 | 6.2 | 1012 (*1012*) | 0.40 | 6 | 72 | 0.148 | 0.887 | 1.50 |
| 199 | ZW0712+53 | | I4620 | 0.6 − 4.5 | 16.8 | 1816 | 0.064 | 390 | 0.66 | 2.8 | 656 (*656*) | 0.40 | 4 | 72 | 0.120 | 0.481 | 1.00 |
| 200 | ZW1615+35 | | I322 | 0.6 − 4.5 | 15.8 | 9432 | 0.032 | 194 | 0.14 | 2.9 | 665 (*665*) | 0.20 | 6 | 72 | 0.064 | 0.381 | 0.50 |
| 201 | 3C129 | | I1832 | 0.6 − 4.5 | 12.8 | 7121 | 0.022 | 131 | 5.71 | 5.6 | 959 (*959*) | 0.45 | 19 | 72 | 0.044 | 0.835 | 0.98 |
| 202 | 3C130 | | I3924 | 0.6 − 4.5 | 15.7 | 1818 | 0.110 | 677 | 5.17 | 5.0 | 898 (*898*) | 0.70 | 5 | 72 | 0.192 | 0.959 | 2.00 |
| 203 | 3C295 | | H4290 | ........... | ..... | 94365 | 0.461 | 3028 | 0.14 | 7.5 | 1049 (.....) | 0.10 | 5 | 10 | 0.069 | 0.344 | 16.00 |
| 204 | 3C330 | | I272 | 0.5 − 3.0 | 14.3 | 5948 | 0.550 | 3658 | 0.29 | 3.6 | 750 (.....) | 0.50 | 4 | 24 | 0.177 | 0.709 | 1.00 |
| 205 | 3C370 | ZW1501+26 | I1907 | 0.6 − 4.5 | 14.4 | 9624 | 0.054 | 328 | 0.34 | 2.5 | 618 (*618*) | 0.50 | 4 | 96 | 0.137 | 0.550 | 0.40 |
| 206 | 3C411 | | I6833 | 0.5 − 3.0 | 14.3 | 11991 | 0.467 | 3068 | 1.16 | 1.7 | 563 (.....) | 0.80 | 3 | 24 | 0.166 | 0.498 | 0.40 |
| 207 | 3C449 | | I3916 | 0.6 − 4.5 | 15.0 | 2135 | 0.017 | 103 | 1.08 | 1.6 | 481 (*481*) | 0.25 | 5 | 72 | 0.035 | 0.174 | 1.00 |

Table 3: Deprojection Results

| # | Name | Sequence Number | $R_{\rm core}$ | $R_{\rm out}$ | k$T$ | $L_{\rm X}$ | $M_{\rm gas}$ | $M_{\rm grav}$ | $\frac{M_{\rm gas}}{M_{\rm grav}}$ | $t_{\rm cool}$ | $R_{\rm cool}$ | $\dot{M}$ | CF | S-Z -d$T$ | T-Depth |
|---|---|---|---|---|---|---|---|---|---|---|---|---|---|---|---|
| | | | ( Mpc) | | ( keV) | ($10^{44}$ erg s$^{-1}$ ) | ($10^{12}$ M$_\odot$) | | (%) | ( $H_0^{-1}$) | ( kpc) | ( M$_\odot$ yr$^{-1}$ ) | (Class) | (mK) | (%) |
| (i) | (ii) | (iii) | (iv) | (v) | (vi) | (vii) | (viii) | (ix) | (x) | (xi) | (xii) | (xiii) | (xiv) | (xv) | (xvi) |
| 1 | A21 | I6012 | 0.60 | 1.130 | $4.8^{+0.2}_{-3.0}$ (4.8) | 4.49 (5.59) | $61.7 \pm 7.6$ | 545.0 | $11.3 \pm 1.4$ | $2.55^{+1.01}_{-0.58}$ | $0^{+113}_{-0}$ | $0.0^{+43.8}_{-0.0}$ | PSCF | 0.09 | 0.19 |
| 2 | A74 | I8989 | 0.20 | 0.502 | $1.8^{+1.0}_{-0.7}$ (2.2) | 0.12 (0.48) | $2.3 \pm 0.5$ | 126.0 | $1.8 \pm 0.4$ | $0.87^{+0.29}_{-0.28}$ | $85^{+166}_{-85}$ | $9.3^{+1.0}_{-9.3}$ | SCF | 0.00 | 0.05 |
| 3 | A76 | I1817 | 0.80 | 0.540 | $1.6^{+0.4}_{-1.0}$ (1.5) | 0.32 (0.92) | $6.5 \pm 1.0$ | 57.5 | .......... | $1.26^{+4.66}_{-0.66}$ | $0^{+162}_{-0}$ | $0.0^{+18.6}_{-0.0}$ | PXCF | 0.01 | 0.07 |
| 4 | A84 | I7640 | 0.80 | 0.908 | $3.7^{+0.2}_{-2.9}$ (3.8) | 1.16 (2.86) | $25.4 \pm 3.9$ | 271.0 | $9.4 \pm 1.5$ | $4.08^{+7.42}_{-2.06}$ | $0^{+91}_{-0}$ | $0.0^{+13.7}_{-0.0}$ | PXCF | 0.03 | 0.10 |
| 5 | A85 | I292 | 0.15 | 1.800 | $5.6^{+0.3}_{-2.2}$ (6.2) | 9.92 (16.51) | $142.0 \pm 9.4$ | 574.0 | $24.7 \pm 1.6$ | $0.73^{+0.22}_{-0.17}$ | $105^{+145}_{-55}$ | $81^{+103}_{-54}$ | LCF | 0.22 | 0.37 |
| | | H6013 | 0.20 | 0.305 | $6.5^{+1.9}_{-3.2}$ (6.2) | 4.05 (16.51) | $7.6 \pm 0.8$ | 102.0 | $7.5 \pm 0.8$ | $0.10^{+0.09}_{-0.03}$ | $131^{+21}_{-62}$ | $108^{+58}_{-61}$ | LCF | 0.14 | 0.41 |
| 6 | A98N | I208 | 0.50 | 0.550 | $3.0^{+0.1}_{-1.3}$ (3.3) | 0.68 (1.39) | $9.1 \pm 1.6$ | 181.0 | $5.0 \pm 0.9$ | $1.54^{+0.72}_{-0.45}$ | $0^{+92}_{-0}$ | $0.0^{+24.3}_{-0.0}$ | PXCF | 0.02 | 0.10 |
| 7 | A98S | I208 | 1.00 | 0.497 | $3.6^{+0.5}_{-0.2}$ (3.2) | 0.64 (1.14) | $8.5 \pm 1.4$ | 62.6 | .......... | $2.06^{+12.27}_{-1.31}$ | $0^{+186}_{-0}$ | $0.0^{+21.4}_{-0.0}$ | PCF | 0.03 | 0.08 |
| 8 | A115N | I209 | 0.20 | 0.301 | $6.1^{+0.4}_{-2.5}$ (6.6) | 14.50 (14.71) | $15.4 \pm 1.4$ | 179.0 | $8.6 \pm 0.8$ | $0.35^{+0.06}_{-0.05}$ | $153^{+97}_{-103}$ | $313^{+369}_{-186}$ | LCF | 0.28 | 0.55 |
| 9 | A115S | I209 | 0.25 | 0.201 | $4.9^{+0.0}_{-0.1}$ (5.7) | 5.33 (9.05) | $5.8 \pm 0.9$ | 63.9 | .......... | $0.35^{+0.26}_{-0.12}$ | $> 234^{+0}_{->134}$ | $> 294^{+0}_{->182}$ | LCF | 0.18 | 0.38 |
| 10 | A117 | I8992 | 0.50 | 0.681 | $2.2^{+0.4}_{-1.5}$ (2.4) | 0.35 (0.57) | $9.1 \pm 1.3$ | 154.0 | $5.9 \pm 0.9$ | $4.89^{+10.28}_{-3.65}$ | $0^{+68}_{-0}$ | $0.0^{+6.9}_{-0.0}$ | PXCF | 0.02 | 0.07 |
| 11 | A119 | I1770 | 0.80 | 1.550 | $5.3^{+0.4}_{-2.3}$ (5.1) | 6.06 (5.78) | $113.0 \pm 8.6$ | 723.0 | $15.7 \pm 1.2$ | $1.79^{+2.42}_{-0.93}$ | $0^{+129}_{-0}$ | $0.0^{+23.7}_{-0.0}$ | XCF | 0.17 | 0.20 |
| 12 | A133 | I2333 | 0.30 | 0.798 | $3.5^{+0.9}_{-1.5}$ (3.8) | 5.19 (5.76) | $38.7 \pm 2.5$ | 320.0 | $12.1 \pm 0.8$ | $0.50^{+0.18}_{-0.11}$ | $148^{+137}_{-91}$ | $110^{+71}_{-67}$ | LCF | 0.11 | 0.30 |
| 13 | A150N | I10766 | 0.40 | 0.789 | $2.6^{+0.4}_{-1.9}$ (2.0) | 0.43 (0.21) | $11.4 \pm 1.6$ | 175.0 | $6.5 \pm 0.9$ | $1.33^{+2.07}_{-0.55}$ | $0^{+169}_{-0}$ | $0.0^{+16.6}_{-0.0}$ | PXCF | 0.02 | 0.08 |
| 14 | A154 | I6135 | 0.70 | 0.733 | $3.6^{+0.1}_{-1.8}$ (3.1) | 1.07 (1.28) | $18.0 \pm 2.1$ | 234.0 | $7.7 \pm 0.9$ | $2.50^{+3.67}_{-1.35}$ | $0^{+61}_{-0}$ | $0.0^{+10.8}_{-0.0}$ | PXCF | 0.04 | 0.11 |
| 15 | A160 | I154 | 0.40 | 0.578 | $2.7^{+0.4}_{-1.3}$ (*2.2*) | 0.27 (*1.78*) | $4.8 \pm 0.6$ | 125.0 | $3.8 \pm 0.5$ | $3.68^{+11.58}_{-2.95}$ | $0^{+173}_{-0}$ | $0.0^{+25.5}_{-0.0}$ | PXCF | 0.01 | 0.06 |
| 16 | A168 | I6083 | 1.00 | 0.934 | $2.8^{+0.7}_{-1.3}$ (2.6) | 1.10 (1.12) | $28.8 \pm 2.6$ | 157.0 | .......... | $3.90^{+7.34}_{-2.12}$ | $0^{+58}_{-0}$ | $0.0^{+2.7}_{-0.0}$ | PXCF | 0.04 | 0.09 |
| 17 | A194 | I6084 | 0.60 | 0.438 | $2.0^{+0.3}_{-1.2}$ (1.9) | 0.13 (0.16) | $2.9 \pm 0.3$ | 50.6 | .......... | $1.10^{+2.33}_{-0.75}$ | $0^{+55}_{-0}$ | $0.0^{+1.4}_{-0.0}$ | XCF | 0.01 | 0.05 |
| 18 | A240 | I189 | 0.50 | 0.466 | $2.1^{+0.1}_{-1.0}$ (2.3) | 0.19 (0.60) | $3.7 \pm 0.6$ | 81.5 | .......... | $1.29^{+3.01}_{-0.56}$ | $0^{+175}_{-0}$ | $0.0^{+14.5}_{-0.0}$ | PCF | 0.01 | 0.06 |
| 19 | A262 | I295 | 0.20 | 0.497 | $2.5^{+0.4}_{-1.5}$ (2.4) | 0.52 (0.91) | $6.5 \pm 0.5$ | 99.2 | $6.6 \pm 0.5$ | $0.26^{+0.16}_{-0.08}$ | $67^{+82}_{-17}$ | $9.4^{+21.2}_{-4.4}$ | SCF | 0.04 | 0.14 |
| 20 | A278 | I7698 | 0.15 | 0.215 | $3.6^{+0.5}_{-0.1}$ (3.5) | 0.99 (1.87) | $2.7 \pm 0.4$ | 68.3 | $3.9 \pm 0.5$ | $0.36^{+0.08}_{-0.08}$ | $81^{+99}_{-45}$ | $33^{+44}_{-14}$ | MCF | 0.06 | 0.19 |
| 21 | A347 | I302 | 0.60 | 0.654 | $2.8^{+0.8}_{-0.8}$ (*2.3*) | 0.35 (*1.98*) | $7.2 \pm 0.8$ | 127.0 | $5.6 \pm 0.6$ | $0.57^{+0.23}_{-0.15}$ | $75^{+21}_{-18}$ | $7.8^{+3.5}_{-2.7}$ | SCF | 0.03 | 0.10 |
| 22 | A367 | I3445 | 0.50 | 0.643 | $3.8^{+1.0}_{-0.9}$ (4.1) | 0.20 (2.29) | $5.6 \pm 1.4$ | 223.0 | $2.5 \pm 0.6$ | $4.90^{+6.06}_{-2.63}$ | $0^{+107}_{-0}$ | $0.0^{+8.3}_{-0.0}$ | PXCF | 0.01 | 0.05 |
| 23 | A397 | I7699 | 0.20 | 0.172 | $1.6^{+0.0}_{-0.6}$ (1.6) | 0.03 (0.09) | $0.3 \pm 0.1$ | 19.8 | .......... | $0.56^{+2.06}_{-0.13}$ | $35^{+72}_{-35}$ | $0.8^{+3.1}_{-0.8}$ | SCF | 0.00 | 0.03 |
| 24 | A399 | I185 | 0.70 | 1.460 | $5.9^{+1.3}_{-3.6}$ (5.8) | 6.12 (8.77) | $102.0 \pm 7.2$ | 856.0 | $12.0 \pm 0.8$ | $2.13^{+2.16}_{-0.77}$ | $0^{+66}_{-0}$ | $0.0^{+17.1}_{-0.0}$ | PSCF | 0.17 | 0.21 |
| 25 | A400 | I6085 | 0.70 | 0.669 | $2.2^{+0.2}_{-0.6}$ (2.1) | 0.57 (0.76) | $11.1 \pm 0.8$ | 125.0 | .......... | $1.79^{+5.13}_{-1.55}$ | $0^{+167}_{-0}$ | $0.0^{+28.3}_{-0.0}$ | XCF | 0.03 | 0.10 |
| 26 | A401 | H1777 | 0.55 | 0.862 | $7.6^{+0.5}_{-3.8}$ (7.8) | 16.30 (19.32) | $71.0 \pm 11.5$ | 590.0 | $12.0 \pm 2.0$ | $1.12^{+1.56}_{-0.78}$ | $0^{+240}_{-0}$ | $0.0^{+262.1}_{-0.0}$ | XCF | 0.26 | 0.38 |
| | | I1776 | 0.55 | 1.380 | $7.1^{+0.7}_{-3.5}$ (7.8) | 20.50 (19.32) | $146.0 \pm 5.5$ | 1070.0 | $13.6 \pm 0.5$ | $1.25^{+0.16}_{-0.10}$ | $0^{+69}_{-0}$ | $0.0^{+50.1}_{-0.0}$ | PMCF | 0.35 | 0.39 |
| 27 | A407 | I1825 | 0.50 | 0.726 | $3.4^{+0.4}_{-1.5}$ (2.8) | 0.75 (0.91) | $14.5 \pm 2.0$ | 165.0 | $8.8 \pm 1.2$ | $0.99^{+0.94}_{-0.41}$ | $46^{+90}_{-46}$ | $4.6^{+11.8}_{-4.6}$ | SCF | 0.04 | 0.10 |
| 28 | A419 | I8993 | 0.30 | 0.396 | $1.9^{+0.2}_{-1.0}$ (1.9) | 0.19 (0.55) | $2.8 \pm 0.5$ | 72.1 | $4.0 \pm 0.6$ | $0.78^{+1.09}_{-0.30}$ | $55^{+143}_{-55}$ | $5.2^{+21.2}_{-5.2}$ | SCF | 0.01 | 0.07 |
| 29 | A426 | H285 | 0.15 | 0.131 | $6.7^{+3.0}_{-4.3}$ (5.5) | 10.20 (21.43) | $3.5 \pm 0.1$ | 42.0 | .......... | $0.00^{+0.00}_{-0.00}$ | $123^{+>14}_{-7}$ | $291^{+>-7}_{-58}$ | SCF | 0.37 | 1.13 |
| | | I283 | 0.10 | 0.775 | $6.3^{+1.5}_{-2.3}$ (5.5) | 19.70 (21.43) | $55.0 \pm 1.0$ | 383.0 | $14.4 \pm 0.3$ | $0.12^{+0.00}_{-0.00}$ | $145^{+25}_{-24}$ | $283^{+14}_{-12}$ | LCF | 0.63 | 0.78 |
| 30 | A458 | I6018 | 0.50 | 0.740 | $3.6^{+1.0}_{-0.9}$ (4.4) | 2.50 (4.35) | $28.8 \pm 2.3$ | 226.0 | $12.7 \pm 1.0$ | $1.44^{+0.51}_{-0.36}$ | $0^{+93}_{-0}$ | $0.0^{+39.8}_{-0.0}$ | PSCF | 0.05 | 0.16 |
| 31 | A478 | H4198 | 0.20 | 0.486 | $7.1^{+0.8}_{-4.9}$ (6.8) | 24.80 (47.91) | $38.3 \pm 3.4$ | 257.0 | $14.9 \pm 1.3$ | $0.10^{+0.05}_{-0.02}$ | $144^{+143}_{-33}$ | $299^{+498}_{-117}$ | PSCF | 0.27 | 0.77 |

Table 3: Deprojection Results

| # | Name | Sequence Number | $R_{\rm core}$ | $R_{\rm out}$ | k$T$ | $L_{\rm X}$ | $M_{\rm gas}$ | $M_{\rm grav}$ | $\frac{M_{\rm gas}}{M_{\rm grav}}$ | $t_{\rm cool}$ | $R_{\rm cool}$ | $\dot{M}$ | CF | S-Z -d$T$ | T-Depth |
|---|---|---|---|---|---|---|---|---|---|---|---|---|---|---|---|
| | | | (Mpc) | | (keV) | ($10^{44}$ erg s$^{-1}$) | ($10^{12}$ M$_\odot$) | | (%) | ($H_0^{-1}$) | (kpc) | (M$_\odot$ yr$^{-1}$) | (Class) | (mK) | (%) |
| (i) | (ii) | (iii) | (iv) | (v) | (vi) | (vii) | (viii) | (ix) | (x) | (xi) | (xii) | (xiii) | (xiv) | (xv) | (xvi) |
| | | I303 | 0.20 | 1.910 | $6.8^{+1.5}_{-1.0}$ (6.8) | 38.90 (47.91) | $232.0 \pm 20.0$ | 870.0 | $26.7 \pm 2.3$ | $0.52^{+0.03}_{-0.03}$ | $240^{+158}_{-161}$ | $736^{+114}_{-434}$ | LCF | 0.32 | 0.67 |
| 32 | A496 | I2348 | 0.15 | 1.220 | $4.8^{+0.2}_{-1.7}$ (4.7) | 7.91 (5.81) | $74.1 \pm 4.5$ | 376.0 | $19.7 \pm 1.2$ | $0.36^{+0.10}_{-0.06}$ | $138^{+79}_{-95}$ | $134^{+58}_{-85}$ | LCF | 0.22 | 0.39 |
| | | H10401 | 0.15 | 0.181 | $5.6^{+0.7}_{-2.7}$ (4.7) | 1.92 (5.81) | $2.7 \pm 0.2$ | 48.4 | $5.5 \pm 0.5$ | $0.07^{+0.02}_{-0.02}$ | $101^{+69}_{-22}$ | $56^{+76}_{-13}$ | LCF | 0.12 | 0.36 |
| 33 | A500 | I6232 | 0.60 | 0.747 | $3.5^{+0.2}_{-1.4}$ (3.3) | 1.25 (1.69) | $19.2 \pm 1.9$ | 213.0 | $9.0 \pm 0.9$ | $2.01^{+1.56}_{-0.83}$ | $0^{+62}_{-0}$ | $0.0^{+9.3}_{-0.0}$ | PXCF | 0.04 | 0.12 |
| 34 | A520 | I6841 | 0.80 | 0.819 | $6.7^{+0.2}_{-0.5}$ (7.4) | 11.90 (21.93) | $65.1 \pm 10.2$ | 421.0 | $15.5 \pm 2.4$ | $3.04^{+2.69}_{-1.05}$ | $0^{+102}_{-0}$ | $0.0^{+50.4}_{-0.0}$ | PSCF | 0.15 | 0.26 |
| 35 | A539 | I2353 | 0.70 | 0.688 | $3.4^{+0.5}_{-1.3}$ (3.0) | 0.67 (1.30) | $12.2 \pm 1.0$ | 209.0 | ......... | $0.82^{+1.42}_{-0.30}$ | $34^{+52}_{-34}$ | $2.1^{+6.8}_{-2.1}$ | SCF | 0.05 | 0.11 |
| 36 | A545 | I310 | 0.60 | 2.000 | $5.6^{+1.1}_{-3.8}$ (5.5) | 17.90 (17.35) | $224.0 \pm 38.2$ | 1160.0 | $19.3 \pm 3.3$ | $1.46^{+0.26}_{-0.17}$ | $0^{+125}_{-0}$ | $0.0^{+158.4}_{-0.0}$ | PMCF | 0.15 | 0.35 |
| 37 | A548S | I7860 | 0.80 | 0.809 | $2.9^{+0.4}_{-1.7}$ (2.4) | 0.80 (1.17) | $15.2 \pm 2.6$ | 262.0 | $5.8 \pm 1.0$ | $0.71^{+0.74}_{-0.25}$ | $69^{+134}_{-69}$ | $10^{+32}_{-10}$ | SCF | 0.04 | 0.12 |
| 38 | A566 | I3553 | 0.50 | 1.220 | $4.7^{+0.3}_{-2.8}$ (4.6) | 4.59 (4.46) | $61.9 \pm 8.4$ | 594.0 | $10.4 \pm 1.4$ | $1.10^{+0.75}_{-0.31}$ | $0^{+262}_{-0}$ | $0.0^{+117.7}_{-0.0}$ | PSCF | 0.08 | 0.22 |
| 39 | A569 | I1836 | 0.01 | 0.078 | $1.4^{+0.1}_{-0.4}$ (*1.4*) | 0.05 (*0.73*) | $0.1 \pm 0.0$ | 11.9 | $0.9 \pm 0.2$ | $0.19^{+0.21}_{-0.08}$ | $> 91^{+0}_{->78}$ | $> 5.2^{+0.0}_{->4.2}$ | SCF | 0.01 | 0.07 |
| 40 | A576 | I3455 | 0.60 | 0.597 | $2.9^{+0.3}_{-2.1}$ (3.7) | 1.58 (2.76) | $15.9 \pm 1.3$ | 214.0 | ......... | $0.91^{+0.97}_{-0.35}$ | $69^{+117}_{-69}$ | $17^{+47}_{-17}$ | MCF | 0.06 | 0.16 |
| 41 | A586 | I211 | 0.45 | 0.484 | $5.5^{+0.1}_{-0.3}$ (.....) | 9.81 (.....) | $28.4 \pm 3.5$ | 194.0 | $14.7 \pm 1.8$ | $0.93^{+0.94}_{-0.32}$ | $118^{+185}_{-118}$ | $138^{+347}_{-138}$ | LCF | 0.15 | 0.34 |
| 42 | A592 | I5170 | 0.40 | 0.470 | $3.4^{+0.1}_{-0.8}$ (3.3) | 0.57 (1.51) | $6.9 \pm 0.5$ | 126.0 | $5.5 \pm 0.4$ | $1.76^{+0.85}_{-0.59}$ | $0^{+59}_{-0}$ | $0.0^{+8.8}_{-0.0}$ | PCF | 0.03 | 0.10 |
| 43 | A629 | I317 | 0.30 | 0.691 | $3.3^{+0.3}_{-1.1}$ (3.3) | 0.95 (1.47) | $13.6 \pm 2.5$ | 242.0 | $5.7 \pm 1.0$ | $2.56^{+1.19}_{-0.80}$ | $0^{+115}_{-0}$ | $0.0^{+33.0}_{-0.0}$ | PXCF | 0.02 | 0.10 |
| 44 | A644 | I5728 | 0.40 | 1.180 | $6.5^{+0.9}_{-3.3}$ (6.6) | 16.60 (14.65) | $98.5 \pm 5.9$ | 821.0 | $12.0 \pm 0.7$ | $0.84^{+0.41}_{-0.20}$ | $111^{+86}_{-111}$ | $136^{+161}_{-136}$ | LCF | 0.28 | 0.41 |
| 45 | A646 | I1839 | 0.30 | 0.660 | $5.5^{+0.6}_{-1.2}$ (5.3) | 3.93 (7.36) | $23.3 \pm 3.3$ | 361.0 | $6.5 \pm 0.9$ | $1.55^{+0.61}_{-0.37}$ | $0^{+110}_{-0}$ | $0.0^{+94.5}_{-0.0}$ | PSCF | 0.08 | 0.21 |
| 46 | A665 | I305 | 1.00 | 2.650 | $7.4^{+0.5}_{-4.3}$ (8.3) | 31.30 (25.40) | $407.0 \pm 56.6$ | 2370.0 | $17.2 \pm 2.4$ | $2.64^{+0.55}_{-0.32}$ | $0^{+189}_{-0}$ | $0.0^{+205.6}_{-0.0}$ | PSCF | 0.23 | 0.35 |
| 47 | A671 | I7337 | 0.60 | 0.760 | $3.1^{+0.4}_{-1.3}$ (3.2) | 1.03 (1.44) | $18.0 \pm 1.5$ | 211.0 | $8.5 \pm 0.7$ | $1.17^{+0.69}_{-0.38}$ | $0^{+143}_{-0}$ | $0.0^{+28.9}_{-0.0}$ | XCF | 0.05 | 0.13 |
| 48 | A690 | I6020 | 0.40 | 0.722 | $2.8^{+0.8}_{-1.5}$ (2.0) | 0.20 (0.33) | $6.6 \pm 1.4$ | 158.0 | $4.2 \pm 0.9$ | $1.50^{+1.74}_{-0.53}$ | $0^{+217}_{-0}$ | $0.0^{+15.3}_{-0.0}$ | PXCF | 0.01 | 0.06 |
| 49 | A732 | I6118 | 0.70 | 1.230 | $7.2^{+0.2}_{-1.1}$ (.....) | 20.50 (.....) | $144.0 \pm 15.6$ | 809.0 | $17.9 \pm 1.9$ | $2.54^{+0.92}_{-0.57}$ | $0^{+154}_{-0}$ | $0.0^{+156.8}_{-0.0}$ | PMCF | 0.16 | 0.32 |
| 50 | A744 | I481 | 0.60 | 0.677 | $2.7^{+0.7}_{-1.2}$ (2.7) | 0.63 (0.55) | $11.4 \pm 1.4$ | 220.0 | $5.2 \pm 0.7$ | $2.75^{+3.53}_{-1.02}$ | $0^{+68}_{-0}$ | $0.0^{+9.7}_{-0.0}$ | PXCF | 0.02 | 0.09 |
| 51 | A754 | I1784 | 0.80 | 1.210 | $8.9^{+0.6}_{-5.2}$ (8.7) | 21.60 (23.13) | $137.0 \pm 7.7$ | 1000.0 | $13.7 \pm 0.8$ | $3.14^{+3.37}_{-1.00}$ | $0^{+86}_{-0}$ | $0.0^{+30.7}_{-0.0}$ | PMCF | 0.50 | 0.32 |
| | | H1786 | 0.80 | 0.388 | $7.0^{+0.8}_{-0.7}$ (8.7) | 2.86 (23.13) | $10.0 \pm 1.2$ | 104.0 | ......... | $1.60^{+2.06}_{-0.69}$ | $0^{+145}_{-0}$ | $0.0^{+79.9}_{-0.0}$ | PMCF | 0.14 | 0.20 |
| 52 | A779 | I1841 | 0.20 | 0.185 | $1.3^{+0.5}_{-0.8}$ (1.5) | 0.04 (0.20) | $0.3 \pm 0.1$ | 25.7 | ......... | $0.16^{+0.07}_{-0.05}$ | $39^{+31}_{-16}$ | $3.1^{+1.1}_{-1.1}$ | SCF | 0.00 | 0.05 |
| 53 | A795 | I212 | 0.60 | 1.360 | $6.2^{+1.1}_{-3.1}$ (5.8) | 10.10 (10.48) | $113.0 \pm 13.9$ | 824.0 | $13.7 \pm 1.7$ | $1.56^{+0.70}_{-0.46}$ | $0^{+114}_{-0}$ | $0.0^{+116.9}_{-0.0}$ | PSCF | 0.14 | 0.27 |
| 54 | A838 | I6097 | 0.40 | 0.292 | $1.3^{+0.2}_{-0.4}$ (.....) | 0.10 (.....) | $1.2 \pm 0.2$ | 34.6 | ......... | $0.57^{+0.18}_{-0.12}$ | $93^{+53}_{-45}$ | $9.8^{+7.3}_{-4.2}$ | SCF | 0.00 | 0.05 |
| 55 | A910 | I1788 | 1.00 | 2.170 | $10.5^{+1.4}_{-4.0}$ (.....) | 69.20 (.....) | $521.0 \pm 84.1$ | 2040.0 | $25.5 \pm 4.1$ | $3.48^{+3.90}_{-1.22}$ | $0^{+155}_{-0}$ | $0.0^{+177.6}_{-0.0}$ | PMCF | 0.45 | 0.47 |
| 56 | A957 | I6023 | 0.60 | 0.769 | $2.7^{+0.5}_{-1.7}$ (2.9) | 0.80 (1.06) | $16.4 \pm 1.7$ | 199.0 | $8.3 \pm 0.9$ | $1.35^{+1.98}_{-0.62}$ | $0^{+128}_{-0}$ | $0.0^{+19.4}_{-0.0}$ | XCF | 0.04 | 0.11 |
| 57 | A970 | I7791 | 0.50 | 0.675 | $4.1^{+0.2}_{-0.6}$ (4.3) | 2.67 (0.89) | $23.7 \pm 2.1$ | 256.0 | $9.3 \pm 0.8$ | $0.81^{+0.36}_{-0.28}$ | $64^{+105}_{-64}$ | $20^{+32}_{-20}$ | MCF | 0.08 | 0.19 |
| 58 | A979 | I6098 | 0.60 | 0.279 | $1.3^{+0.1}_{-0.7}$ (.....) | 0.04 (.....) | $0.7 \pm 0.1$ | 22.5 | ......... | $0.51^{+1.33}_{-0.25}$ | $54^{+138}_{-54}$ | $1.4^{+11.3}_{-1.4}$ | SCF | 0.00 | 0.04 |
| 59 | A999 | I7700 | 0.10 | 0.253 | $1.5^{+0.5}_{-1.0}$ (1.2) | 0.02 (0.07) | $0.4 \pm 0.1$ | 38.2 | $1.1 \pm 0.2$ | $0.65^{+0.49}_{-0.26}$ | $36^{+59}_{-36}$ | $1.1^{+0.8}_{-1.1}$ | SCF | 0.00 | 0.03 |
| 60 | A1060 | I6114 | 0.20 | 0.381 | $3.3^{+0.2}_{-1.2}$ (3.3) | 0.62 (0.78) | $4.8 \pm 0.3$ | 99.0 | $4.9 \pm 0.3$ | $0.74^{+1.73}_{-0.62}$ | $68^{+46}_{-68}$ | $8.0^{+14.9}_{-8.0}$ | SCF | 0.07 | 0.14 |
| 61 | A1142 | I6079 | 0.50 | 0.425 | $3.5^{+0.2}_{-1.2}$ (3.7) | 0.15 (0.72) | $3.1 \pm 0.4$ | 80.1 | ......... | $3.14^{+13.96}_{-2.39}$ | $0^{+106}_{-0}$ | $0.0^{+3.8}_{-0.0}$ | XCF | 0.02 | 0.05 |
| 62 | A1146 | I217 | 0.70 | 0.706 | $5.3^{+0.1}_{-1.6}$ (5.0) | 4.51 (6.01) | $31.8 \pm 4.8$ | 389.0 | $8.2 \pm 1.2$ | $2.12^{+1.66}_{-0.73}$ | $0^{+118}_{-0}$ | $0.0^{+90.4}_{-0.0}$ | PSCF | 0.07 | 0.20 |
| 63 | A1185 | I6100 | 0.30 | 0.436 | $4.4^{+1.3}_{-3.1}$ (3.9) | 0.34 (0.56) | $4.6 \pm 0.4$ | 185.0 | $2.5 \pm 0.2$ | $1.73^{+1.35}_{-0.56}$ | $0^{+31}_{-0}$ | $0.0^{+1.5}_{-0.0}$ | XCF | 0.04 | 0.08 |

Table 3: Deprojection Results

| # | Name | Sequence Number | $R_{\rm core}$ ( Mpc) | $R_{\rm out}$ | k$T$ ( keV) | $L_{\rm X}$ ($10^{44}$ erg s$^{-1}$ ) | $M_{\rm gas}$ ($10^{12}$ M$_\odot$) | $M_{\rm grav}$ | $\frac{M_{\rm gas}}{M_{\rm grav}}$ (%) | $t_{\rm cool}$ ( $H_0^{-1}$) | $R_{\rm cool}$ ( kpc) | $\dot{M}$ ( M$_\odot$ yr$^{-1}$ ) | CF (Class) | S-Z -d$T$ (mK) | T-Depth (%) |
|---|---|---|---|---|---|---|---|---|---|---|---|---|---|---|---|
| (i) | (ii) | (iii) | (iv) | (v) | (vi) | (vii) | (viii) | (ix) | (x) | (xi) | (xii) | (xiii) | (xiv) | (xv) | (xvi) |
| 64 | A1213 | I1844 | 0.80 | 0.271 | $1.8^{+0.4}_{-0.4}$ (2.0) | 0.10 (0.32) | $1.3 \pm 0.3$ | 22.4 | .......... | $1.66^{+7.46}_{-1.04}$ | $0^{+136}_{-0}$ | $0.0^{+11.5}_{-0.0}$ | XCF | 0.01 | 0.05 |
| 65 | A1246 | I233 | 0.70 | 1.140 | $6.1^{+0.6}_{-1.4}$ (.....) | 11.10 (.....) | $92.5 \pm 18.6$ | 620.0 | $14.9 \pm 3.0$ | $2.32^{+1.72}_{-0.80}$ | $0^{+143}_{-0}$ | $0.0^{+138.8}_{-0.0}$ | PSCF | 0.10 | 0.26 |
| 66 | A1254 | I172 | 0.20 | 0.746 | $4.6^{+0.6}_{-1.9}$ (3.7) | 1.08 (0.53) | $14.9 \pm 3.6$ | 289.0 | $5.1 \pm 1.3$ | $3.68^{+2.58}_{-1.92}$ | $0^{+124}_{-0}$ | $0.0^{+30.0}_{-0.0}$ | PSCF | 0.03 | 0.10 |
| 67 | A1272 | I331 | 0.80 | 1.650 | $3.2^{+2.3}_{-2.1}$ (4.1) | 4.45 (3.46) | $120.0 \pm 15.8$ | 689.0 | $17.5 \pm 2.3$ | $5.84^{+6.24}_{-2.11}$ | $0^{+165}_{-0}$ | $0.0^{+32.8}_{-0.0}$ | PXCF | 0.07 | 0.14 |
| 68 | A1285 | I331 | 0.70 | 1.420 | $4.1^{+0.3}_{-2.9}$ (4.1) | 5.81 (6.37) | $92.6 \pm 10.7$ | 594.0 | $15.6 \pm 1.8$ | $3.10^{+9.20}_{-1.86}$ | $0^{+89}_{-0}$ | $0.0^{+31.5}_{-0.0}$ | PSCF | 0.09 | 0.20 |
| 69 | A1291 | I6293 | 0.40 | 0.405 | $2.6^{+0.1}_{-1.2}$ (2.6) | 0.57 (0.68) | $5.1 \pm 0.4$ | 79.2 | $6.5 \pm 0.5$ | $1.20^{+1.07}_{-0.50}$ | $0^{+253}_{-0}$ | $0.0^{+62.1}_{-0.0}$ | XCF | 0.02 | 0.11 |
| 70 | A1314 | I6120 | 0.50 | 0.400 | $4.9^{+0.3}_{-1.6}$ (5.0) | 0.28 (0.92) | $3.8 \pm 0.4$ | 72.7 | .......... | $3.64^{+13.19}_{-2.88}$ | $0^{+100}_{-0}$ | $0.0^{+3.0}_{-0.0}$ | XCF | 0.04 | 0.06 |
| 71 | A1367 | I296 | 0.80 | 0.647 | $4.1^{+0.6}_{-2.5}$ (3.5) | 1.17 (1.76) | $16.4 \pm 1.2$ | 162.0 | .......... | $0.78^{+7.61}_{-0.35}$ | $40^{+68}_{-40}$ | $2.3^{+6.8}_{-2.3}$ | SCF | 0.08 | 0.11 |
| 72 | A1377 | I6101 | 0.50 | 0.657 | $2.6^{+0.1}_{-1.2}$ (2.7) | 0.42 (0.92) | $9.5 \pm 1.1$ | 110.0 | $8.7 \pm 1.0$ | $4.56^{+8.67}_{-3.56}$ | $0^{+66}_{-0}$ | $0.0^{+7.2}_{-0.0}$ | PXCF | 0.02 | 0.07 |
| 73 | A1413 | I308 | 0.50 | 1.650 | $8.9^{+2.0}_{-4.7}$ (8.9) | 16.60 (23.62) | $156.0 \pm 16.6$ | 1560.0 | $10.1 \pm 1.1$ | $1.63^{+0.50}_{-0.31}$ | $0^{+118}_{-0}$ | $0.0^{+133.0}_{-0.0}$ | PMCF | 0.24 | 0.33 |
| 74 | A1546 | I6868 | 0.40 | 0.321 | $3.7^{+0.1}_{-0.4}$ (.....) | 2.96 (.....) | $9.0 \pm 0.7$ | 79.7 | .......... | $0.71^{+0.38}_{-0.21}$ | $144^{+123}_{-144}$ | $121^{+141}_{-121}$ | LCF | 0.09 | 0.25 |
| 75 | A1569S | I1849 | 0.40 | 0.575 | $2.9^{+0.1}_{-0.8}$ (3.3) | 0.99 (1.45) | $12.3 \pm 1.9$ | 177.0 | $6.9 \pm 1.1$ | $1.15^{+9.51}_{-0.16}$ | $0^{+216}_{-0}$ | $0.0^{+57.8}_{-0.0}$ | PXCF | 0.02 | 0.11 |
| 76 | A1576 | I6871 | 0.60 | 0.707 | $5.4^{+0.3}_{-2.2}$ (.....) | 9.32 (.....) | $47.2 \pm 8.8$ | 320.0 | $14.8 \pm 2.8$ | $1.64^{+2.29}_{-0.66}$ | $0^{+265}_{-0}$ | $0.0^{+278.4}_{-0.0}$ | PSCF | 0.12 | 0.28 |
| 77 | A1589 | I145 | 0.80 | 0.887 | $3.5^{+1.1}_{-1.0}$ (3.9) | 4.17 (2.91) | $50.1 \pm 5.5$ | 269.0 | $18.6 \pm 2.0$ | $2.25^{+2.82}_{-0.92}$ | $0^{+89}_{-0}$ | $0.0^{+23.4}_{-0.0}$ | PSCF | 0.09 | 0.17 |
| 78 | A1617 | I6875 | 0.60 | 0.331 | $1.4^{+0.5}_{-0.6}$ (.....) | 0.33 (.....) | $3.2 \pm 0.7$ | 40.7 | .......... | $1.09^{+1.86}_{-0.65}$ | $0^{+165}_{-0}$ | $0.0^{+41.9}_{-0.0}$ | PCF | 0.01 | 0.08 |
| 79 | A1631 | I1900 | 0.20 | 0.450 | $2.8^{+0.4}_{-1.8}$ (2.8) | 0.13 (0.78) | $2.4 \pm 0.5$ | 127.0 | $1.9 \pm 0.4$ | $1.30^{+0.42}_{-0.31}$ | $0^{+225}_{-0}$ | $0.0^{+7.0}_{-0.0}$ | PCF | 0.02 | 0.06 |
| 80 | A1644 | I7654 | 0.35 | 0.549 | $5.1^{+0.8}_{-3.5}$ (4.7) | 2.29 (5.16) | $15.7 \pm 0.5$ | 251.0 | $6.3 \pm 0.2$ | $0.96^{+0.09}_{-0.10}$ | $49^{+89}_{-49}$ | $12^{+19}_{-12}$ | MCF | 0.11 | 0.19 |
| 81 | A1650 | I6034 | 0.45 | 1.220 | $5.1^{+0.6}_{-3.3}$ (5.5) | 11.80 (13.30) | $93.3 \pm 7.2$ | 709.0 | $13.2 \pm 1.0$ | $0.82^{+0.27}_{-0.20}$ | $110^{+119}_{-110}$ | $122^{+168}_{-122}$ | LCF | 0.17 | 0.36 |
| 82 | A1656 | I1793 | 0.50 | 1.110 | $7.3^{+0.5}_{-0.7}$ (8.0) | 9.14 (15.56) | $75.6 \pm 2.1$ | 712.0 | $10.6 \pm 0.3$ | $2.33^{+2.63}_{-0.61}$ | $0^{+23}_{-0}$ | $0.0^{+1.0}_{-0.0}$ | XCF | 0.37 | 0.26 |
| 83 | A1689 | I6123 | 0.40 | 1.700 | $8.7^{+0.0}_{-3.3}$ (10.1) | 42.30 (50.46) | $219.0 \pm 20.2$ | 1610.0 | $13.6 \pm 1.3$ | $1.21^{+0.07}_{-0.05}$ | $0^{+141}_{-0}$ | $0.0^{+398.4}_{-0.0}$ | PMCF | 0.23 | 0.52 |
| 84 | A1704 | I6877 | 0.60 | 0.649 | $5.4^{+0.0}_{-0.2}$ (.....) | 8.91 (.....) | $42.4 \pm 5.1$ | 267.0 | $15.9 \pm 1.9$ | $1.58^{+0.39}_{-0.33}$ | $0^{+108}_{-0}$ | $0.0^{+113.7}_{-0.0}$ | PSCF | 0.12 | 0.28 |
| 85 | A1709 | I8996 | 0.40 | 0.497 | $2.6^{+0.3}_{-1.6}$ (2.5) | 0.15 (0.35) | $3.7 \pm 0.6$ | 109.0 | $3.3 \pm 0.6$ | $1.78^{+3.01}_{-0.81}$ | $0^{+149}_{-0}$ | $0.0^{+7.0}_{-0.0}$ | XCF | 0.01 | 0.05 |
| 86 | A1736 | I7653 | 0.80 | 1.010 | $5.1^{+0.6}_{-3.4}$ (4.6) | 2.70 (3.16) | $42.8 \pm 3.4$ | 384.0 | $11.1 \pm 0.9$ | $2.26^{+5.68}_{-1.83}$ | $0^{+126}_{-0}$ | $0.0^{+15.7}_{-0.0}$ | XCF | 0.11 | 0.15 |
| 87 | A1750N | I144 | 0.60 | 0.775 | $3.5^{+0.5}_{-2.2}$ (3.7) | 1.45 (1.97) | $20.3 \pm 2.6$ | 256.0 | $7.9 \pm 1.0$ | $1.55^{+1.22}_{-0.59}$ | $0^{+232}_{-0}$ | $0.0^{+63.5}_{-0.0}$ | PSCF | 0.04 | 0.14 |
| 88 | A1763 | I3930 | 0.80 | 2.030 | $6.3^{+1.1}_{-3.3}$ (6.9) | 27.80 (19.97) | $316.0 \pm 32.3$ | 1470.0 | $21.5 \pm 2.2$ | $1.91^{+0.39}_{-0.26}$ | $0^{+145}_{-0}$ | $0.0^{+181.0}_{-0.0}$ | PMCF | 0.22 | 0.38 |
| 89 | A1767 | I5731 | 0.70 | 0.781 | $4.2^{+0.3}_{-1.2}$ (4.1) | 3.07 (4.81) | $30.2 \pm 2.4$ | 317.0 | $9.5 \pm 0.8$ | $1.63^{+1.25}_{-0.64}$ | $0^{+195}_{-0}$ | $0.0^{+82.6}_{-0.0}$ | PSCF | 0.08 | 0.18 |
| 90 | A1775 | I320 | 0.50 | 1.040 | $4.9^{+0.4}_{-1.5}$ (4.9) | 3.70 (4.68) | $45.5 \pm 4.9$ | 500.0 | $9.1 \pm 1.0$ | $1.70^{+1.30}_{-0.58}$ | $0^{+65}_{-0}$ | $0.0^{+25.4}_{-0.0}$ | PSCF | 0.10 | 0.20 |
| 91 | A1795 | I293 | 0.20 | 1.870 | $5.3^{+0.4}_{-1.9}$ (5.1) | 17.60 (20.15) | $176.0 \pm 10.9$ | 655.0 | $26.8 \pm 1.7$ | $0.39^{+0.15}_{-0.10}$ | $181^{+112}_{-122}$ | $321^{+166}_{-213}$ | LCF | 0.24 | 0.52 |
| | | H7881 | 0.25 | 0.325 | $5.3^{+0.6}_{-2.1}$ (5.1) | 9.01 (20.15) | $13.8 \pm 1.2$ | 106.0 | $13.1 \pm 1.1$ | $0.19^{+0.19}_{-0.09}$ | $155^{+121}_{-41}$ | $238^{+344}_{-84}$ | LCF | 0.17 | 0.54 |
| 92 | A1809 | I142 | 0.70 | 1.160 | $4.1^{+0.3}_{-1.6}$ (3.7) | 2.42 (3.23) | $45.9 \pm 5.8$ | 404.0 | $11.3 \pm 1.4$ | $2.06^{+2.93}_{-0.94}$ | $0^{+72}_{-0}$ | $0.0^{+19.7}_{-0.0}$ | PSCF | 0.06 | 0.15 |
| 93 | A1837 | I141 | 0.30 | 0.443 | $2.6^{+0.1}_{-0.8}$ (2.4) | 0.46 (1.07) | $5.1 \pm 0.6$ | 103.0 | $5.0 \pm 0.6$ | $0.77^{+0.48}_{-0.30}$ | $74^{+110}_{-74}$ | $12^{+29}_{-12}$ | MCF | 0.02 | 0.11 |
| 94 | A1839 | I6037 | 0.80 | 0.884 | $3.4^{+0.6}_{-0.7}$ (3.6) | 2.30 (2.45) | $33.0 \pm 5.6$ | 246.0 | $13.4 \pm 2.3$ | $2.33^{+4.18}_{-1.14}$ | $0^{+110}_{-0}$ | $0.0^{+52.5}_{-0.0}$ | PSCF | 0.04 | 0.14 |
| 95 | A1877 | I6883 | 0.40 | 0.212 | $2.6^{+0.3}_{-0.1}$ (.....) | 1.23 (.....) | $3.3 \pm 0.3$ | 30.5 | .......... | $0.61^{+0.64}_{-0.26}$ | $156^{+>91}_{-50}$ | $117^{+>11}_{-58}$ | LCF | 0.05 | 0.19 |
| 96 | A1890 | I165 | 0.40 | 0.439 | $2.7^{+0.1}_{-1.8}$ (2.9) | 0.43 (1.07) | $5.5 \pm 0.9$ | 101.0 | $5.4 \pm 0.9$ | $0.85^{+5.16}_{-0.12}$ | $58^{+107}_{-58}$ | $6.2^{+12.6}_{-6.2}$ | SCF | 0.02 | 0.09 |
| 97 | A1913 | I6077 | 0.80 | 0.605 | $2.6^{+0.7}_{-1.5}$ (2.9) | 0.29 (1.08) | $7.6 \pm 1.2$ | 68.0 | .......... | $3.16^{+8.44}_{-2.39}$ | $0^{+151}_{-0}$ | $0.0^{+5.6}_{-0.0}$ | XCF | 0.02 | 0.06 |

Table 3: Deprojection Results

| # | Name | Sequence Number | $R_{\rm core}$ | $R_{\rm out}$ | k$T$ | $L_{\rm X}$ | $M_{\rm gas}$ | $M_{\rm grav}$ | $\frac{M_{\rm gas}}{M_{\rm grav}}$ | $t_{\rm cool}$ | $R_{\rm cool}$ | $\dot{M}$ | CF | S-Z -d$T$ | T-Depth |
|---|---|---|---|---|---|---|---|---|---|---|---|---|---|---|---|
| | | | (Mpc) | | (keV) | ($10^{44}$ erg s$^{-1}$) | ($10^{12}$ M$_\odot$) | | (%) | ($H_0^{-1}$) | (kpc) | (M$_\odot$ yr$^{-1}$) | (Class) | (mK) | (%) |
| (i) | (ii) | (iii) | (iv) | (v) | (vi) | (vii) | (viii) | (ix) | (x) | (xi) | (xii) | (xiii) | (xiv) | (xv) | (xvi) |
| 98 | A1940 | I6124 | 0.40 | 1.150 | $3.8^{+1.1}_{-0.3}$ (4.3) | 2.19 (3.78) | $47.2 \pm 8.5$ | 260.0 | $18.2 \pm 3.3$ | $2.42^{+0.91}_{-0.71}$ | $0^{+115}_{-0}$ | $0.0^{+23.5}_{-0.0}$ | PXCF | 0.04 | 0.12 |
| 99 | A1942 | I6847 | 0.60 | 0.438 | $2.2^{+0.9}_{-1.0}$ (.....) | 1.15 (.....) | $9.2 \pm 2.1$ | 87.9 | ......... | $1.40^{+2.28}_{-0.69}$ | $0^{+219}_{-0}$ | $0.0^{+82.8}_{-0.0}$ | PCF | 0.03 | 0.13 |
| 100 | A1983 | I4190 | 0.40 | 0.344 | $2.5^{+0.3}_{-0.7}$ (2.5) | 0.28 (0.66) | $3.1 \pm 0.2$ | 81.5 | ......... | $0.96^{+0.46}_{-0.30}$ | $47^{+82}_{-47}$ | $6.0^{+10.8}_{-6.0}$ | SCF | 0.02 | 0.09 |
| 101 | A1991 | I6039 | 0.20 | 0.666 | $5.3^{+0.4}_{-4.0}$ (5.4) | 2.55 (3.29) | $16.7 \pm 1.4$ | 380.0 | $4.4 \pm 0.4$ | $0.79^{+0.18}_{-0.18}$ | $73^{+94}_{-17}$ | $37^{+36}_{-11}$ | MCF | 0.08 | 0.22 |
| 102 | A2009 | I3582 | 0.10 | 1.500 | $7.3^{+1.6}_{-0.9}$ (7.8) | 30.30 (21.87) | $163.0 \pm 14.4$ | 922.0 | $17.7 \pm 1.6$ | $1.76^{+0.12}_{-0.13}$ | $0^{+125}_{-0}$ | $0.0^{+203.5}_{-0.0}$ | PMCF | 0.20 | 0.46 |
| 103 | A2022 | I4191 | 0.60 | 0.429 | $2.2^{+0.2}_{-0.6}$ (2.8) | 0.29 (0.94) | $4.3 \pm 0.9$ | 68.9 | ......... | $1.23^{+3.88}_{-0.56}$ | $0^{+161}_{-0}$ | $0.0^{+27.0}_{-0.0}$ | PCF | 0.01 | 0.07 |
| 104 | A2029 | I138 | 0.30 | 1.550 | $7.4^{+1.9}_{-5.9}$ (7.8) | 33.30 (95.76) | $176.0 \pm 12.7$ | 1150.0 | $15.4 \pm 1.1$ | $0.53^{+0.22}_{-0.10}$ | $192^{+160}_{-121}$ | $431^{+236}_{-288}$ | LCF | 0.41 | 0.61 |
| | | H7882 | 0.30 | 0.293 | $7.2^{+0.9}_{-3.0}$ (7.8) | 12.60 (95.76) | $13.1 \pm 1.2$ | 144.0 | ......... | $0.13^{+0.04}_{-0.03}$ | $144^{+27}_{-22}$ | $298^{+85}_{-98}$ | LCF | 0.28 | 0.63 |
| 105 | A2040 | I6104 | 0.70 | 0.618 | $2.8^{+0.5}_{-1.3}$ (2.5) | 0.35 (0.64) | $8.3 \pm 1.0$ | 108.0 | ......... | $2.07^{+3.54}_{-1.07}$ | $0^{+44}_{-0}$ | $0.0^{+3.3}_{-0.0}$ | XCF | 0.03 | 0.08 |
| 106 | A2050 | I3444 | 0.60 | 1.420 | $5.4^{+0.8}_{-2.7}$ (5.1) | 6.44 (6.46) | $91.2 \pm 10.1$ | 754.0 | $12.1 \pm 1.3$ | $2.09^{+1.65}_{-0.63}$ | $0^{+102}_{-0}$ | $0.0^{+57.8}_{-0.0}$ | PSCF | 0.11 | 0.22 |
| 107 | A2052 | I1853 | 0.20 | 0.892 | $3.4^{+1.2}_{-2.2}$ (3.4) | 3.13 (3.64) | $30.8 \pm 2.7$ | 220.0 | $14.0 \pm 1.2$ | $0.28^{+0.14}_{-0.09}$ | $140^{+100}_{-38}$ | $94^{+84}_{-37}$ | LCF | 0.09 | 0.29 |
| | | H5728 | 0.15 | 0.152 | $3.3^{+1.2}_{-0.9}$ (3.4) | 0.93 (3.64) | $1.5 \pm 0.1$ | 26.8 | $5.4 \pm 0.3$ | $0.08^{+0.03}_{-0.02}$ | $94^{+11}_{-8}$ | $54^{+17}_{-16}$ | LCF | 0.05 | 0.28 |
| 108 | A2055 | I137 | 0.15 | 0.507 | $5.7^{+2.6}_{-0.5}$ (5.8) | 1.72 (2.19) | $9.0 \pm 1.1$ | 310.0 | $2.9 \pm 0.4$ | $0.67^{+0.19}_{-0.12}$ | $87^{+65}_{-37}$ | $36^{+17}_{-12}$ | MCF | 0.07 | 0.20 |
| 109 | A2063 | H4595 | 0.25 | 0.078 | $3.4^{+0.4}_{-1.0}$ (4.1) | 0.20 (3.64) | $0.3 \pm 0.0$ | 6.1 | ......... | $0.23^{+0.87}_{-0.16}$ | $65^{+>22}_{-17}$ | $16^{+>3}_{-11}$ | MCF | 0.05 | 0.14 |
| | | I162 | 0.25 | 0.280 | $4.1^{+0.2}_{-0.9}$ (4.1) | 1.31 (3.64) | $4.7 \pm 0.2$ | 63.4 | $7.4 \pm 0.3$ | $0.40^{+0.05}_{-0.04}$ | $87^{+18}_{-52}$ | $35^{+10}_{-19}$ | MCF | 0.07 | 0.20 |
| 110 | A2065 | I1795 | 0.40 | 1.200 | $8.2^{+0.9}_{-5.4}$ (8.4) | 2.76 (11.28) | $43.8 \pm 3.6$ | 893.0 | $4.9 \pm 0.4$ | $3.37^{+2.62}_{-1.04}$ | $0^{+67}_{-0}$ | $0.0^{+9.0}_{-0.0}$ | PSCF | 0.14 | 0.15 |
| 111 | A2069E | I10404 | 1.00 | 2.600 | $6.4^{+2.0}_{-4.8}$ (.....) | 12.20 (.....) | $318.0 \pm 27.3$ | 1650.0 | $19.2 \pm 1.6$ | $5.41^{+7.12}_{-4.02}$ | $0^{+100}_{-0}$ | $0.0^{+19.6}_{-0.0}$ | PSCF | 0.22 | 0.21 |
| 112 | A2069W | I10404 | 1.00 | 2.400 | $6.9^{+0.7}_{-4.8}$ (.....) | 10.80 (.....) | $275.0 \pm 25.3$ | 1470.0 | $18.8 \pm 1.7$ | $4.33^{+7.98}_{-3.04}$ | $0^{+100}_{-0}$ | $0.0^{+29.0}_{-0.0}$ | PSCF | 0.20 | 0.21 |
| 113 | A2079 | I1854 | 0.60 | 0.737 | $3.2^{+0.7}_{-1.6}$ (3.2) | 0.82 (1.36) | $16.4 \pm 2.3$ | 173.0 | $9.4 \pm 1.3$ | $3.36^{+6.60}_{-2.42}$ | $0^{+184}_{-0}$ | $0.0^{+15.6}_{-0.0}$ | PXCF | 0.04 | 0.09 |
| 114 | A2092 | I135 | 0.25 | 0.626 | $2.5^{+0.5}_{-1.1}$ (2.5) | 0.43 (0.71) | $7.4 \pm 1.1$ | 129.0 | $5.8 \pm 0.8$ | $2.28^{+3.81}_{-1.02}$ | $0^{+63}_{-0}$ | $0.0^{+9.9}_{-0.0}$ | PXCF | 0.01 | 0.08 |
| 115 | A2107 | I134 | 0.30 | 0.656 | $4.5^{+0.5}_{-1.7}$ (4.2) | 1.62 (2.71) | $15.6 \pm 1.5$ | 286.0 | $5.5 \pm 0.5$ | $0.97^{+0.43}_{-0.31}$ | $42^{+81}_{-42}$ | $7.1^{+17.2}_{-7.1}$ | SCF | 0.08 | 0.16 |
| 116 | A2111 | I239 | 0.80 | 0.741 | $6.0^{+0.3}_{-2.4}$ (.....) | 4.98 (.....) | $35.5 \pm 6.6$ | 298.0 | ......... | $1.58^{+4.43}_{-0.92}$ | $0^{+222}_{-0}$ | $0.0^{+99.2}_{-0.0}$ | PSCF | 0.10 | 0.21 |
| 117 | A2124 | I4192 | 0.50 | 0.490 | $3.4^{+0.0}_{-2.1}$ (3.6) | 0.98 (2.24) | $9.6 \pm 1.0$ | 153.0 | ......... | $2.22^{+2.12}_{-1.04}$ | $0^{+61}_{-0}$ | $0.0^{+13.2}_{-0.0}$ | PCF | 0.03 | 0.12 |
| 118 | A2142 | H1800 | 0.50 | 0.719 | $10.1^{+1.7}_{-3.9}$ (11.0) | 33.70 (58.59) | $73.8 \pm 4.8$ | 622.0 | $11.9 \pm 0.8$ | $0.33^{+0.05}_{-0.03}$ | $172^{+52}_{-38}$ | $369^{+130}_{-87}$ | PCF | 0.36 | 0.64 |
| | | I1798 | 0.45 | 2.160 | $11.4^{+0.8}_{-3.2}$ (11.0) | 47.00 (58.59) | $321.0 \pm 25.1$ | 2140.0 | $15.0 \pm 1.2$ | $0.97^{+0.09}_{-0.09}$ | $77^{+125}_{-77}$ | $106^{+248}_{-106}$ | LCF | 0.64 | 0.58 |
| 119 | A2147 | I297 | 0.50 | 0.701 | $5.4^{+0.8}_{-4.2}$ (4.4) | 2.06 (4.23) | $22.3 \pm 2.1$ | 274.0 | $8.2 \pm 0.8$ | $1.66^{+1.75}_{-0.73}$ | $0^{+105}_{-0}$ | $0.0^{+14.5}_{-0.0}$ | XCF | 0.12 | 0.15 |
| 120 | A2151 | H9264 | 0.10 | 0.227 | $3.7^{+0.2}_{-2.7}$ (3.5) | 0.29 (1.46) | $1.3 \pm 0.3$ | 101.0 | $1.3 \pm 0.3$ | $0.52^{+0.13}_{-0.15}$ | $50^{+154}_{-27}$ | $6.3^{+26.3}_{-3.2}$ | XCF | 0.02 | 0.11 |
| | | I1801 | 0.25 | 0.363 | $2.9^{+0.5}_{-0.4}$ (3.5) | 4.06 (1.46) | $11.4 \pm 1.1$ | 141.0 | $8.1 \pm 0.8$ | $0.25^{+0.03}_{-0.03}$ | $146^{+36}_{-37}$ | $166^{+51}_{-41}$ | LCF | 0.08 | 0.34 |
| 121 | A2152 | I1855 | 0.30 | 0.367 | $2.1^{+0.6}_{-0.2}$ (2.1) | 0.26 (0.38) | $2.8 \pm 0.5$ | 70.4 | $4.0 \pm 0.7$ | $0.81^{+0.87}_{-0.29}$ | $118^{+66}_{-118}$ | $20^{+13}_{-20}$ | MCF | 0.01 | 0.08 |
| 122 | A2163 | I4526 | 0.70 | 3.240 | $13.3^{+6.0}_{-9.6}$ (13.9) | 85.00 (100.53) | $843.0 \pm 89.3$ | 4720.0 | $17.8 \pm 1.9$ | $1.79^{+1.03}_{-0.33}$ | $0^{+135}_{-0}$ | $0.0^{+256.2}_{-0.0}$ | PMCF | 0.90 | 0.56 |
| 123 | A2197 | I1857 | 0.70 | 0.545 | $2.5^{+0.4}_{-0.7}$ (1.6) | 0.13 (0.15) | $3.9 \pm 0.5$ | 78.8 | ......... | $0.72^{+0.71}_{-0.19}$ | $51^{+40}_{-51}$ | $2.4^{+3.0}_{-2.4}$ | SCF | 0.01 | 0.06 |
| 124 | A2199 | I4193 | 0.20 | 0.713 | $4.5^{+0.8}_{-3.1}$ (4.7) | 5.39 (6.30) | $29.4 \pm 1.6$ | 303.0 | $9.7 \pm 0.5$ | $0.27^{+0.19}_{-0.06}$ | $119^{+30}_{-30}$ | $94^{+45}_{-35}$ | LCF | 0.19 | 0.36 |
| | | H4597 | 0.20 | 0.173 | $5.4^{+2.3}_{-2.2}$ (4.7) | 2.08 (6.30) | $2.7 \pm 0.2$ | 43.2 | ......... | $0.12^{+0.01}_{-0.02}$ | $124^{+37}_{-37}$ | $97^{+9}_{-31}$ | LCF | 0.12 | 0.33 |
| 125 | A2218 | H3160 | 0.40 | 0.678 | $6.2^{+1.7}_{-3.8}$ (6.7) | 10.90 (18.09) | $43.8 \pm 4.4$ | 436.0 | $10.1 \pm 1.0$ | $0.74^{+0.24}_{-0.13}$ | $85^{+84}_{-29}$ | $66^{+76}_{-30}$ | LCF | 0.15 | 0.35 |
| 126 | A2220 | I8351 | 0.60 | 0.385 | $2.2^{+0.3}_{-1.2}$ (2.9) | 0.19 (1.17) | $3.0 \pm 0.5$ | 58.2 | ......... | $2.35^{+6.30}_{-1.46}$ | $0^{+193}_{-0}$ | $0.0^{+24.6}_{-0.0}$ | PCF | 0.01 | 0.06 |

Table 3: Deprojection Results

| # | Name | Sequence | $R_{\rm core}$ | $R_{\rm out}$ | k$T$ | $L_{\rm X}$ | $M_{\rm gas}$ | $M_{\rm grav}$ | $\frac{M_{\rm gas}}{M_{\rm grav}}$ | $t_{\rm cool}$ | $R_{\rm cool}$ | $\dot{M}$ | CF | S-Z -d$T$ | T-Depth |
|---|---|---|---|---|---|---|---|---|---|---|---|---|---|---|---|
| | | Number | ( Mpc) | | ( keV) | ($10^{44}$ erg s$^{-1}$ ) | ($10^{12}$ M$_\odot$) | | (%) | ( $H_0^{-1}$) | ( kpc) | ( M$_\odot$ yr$^{-1}$ ) | (Class) | (mK) | (%) |
| (i) | (ii) | (iii) | (iv) | (v) | (vi) | (vii) | (viii) | (ix) | (x) | (xi) | (xii) | (xiii) | (xiv) | (xv) | (xvi) |
| 127 | A2244 | H10190 | 0.50 | 0.432 | $7.3^{+3.6}_{-2.8}$ (7.1) | 8.26 (7.01) | $18.6 \pm 1.8$ | 182.0 | .......... | $0.27^{+0.39}_{-0.10}$ | $182^{+51}_{-164}$ | $205^{+160}_{-202}$ | PCF | 0.18 | 0.40 |
| | | | 0.50 | 2.070 | $8.2^{+3.8}_{-6.1}$ (7.1) | 19.50 (7.01) | $232.0 \pm 17.7$ | 1520.0 | $15.2 \pm 1.2$ | $0.78^{+0.06}_{-0.04}$ | $116^{+143}_{-30}$ | $155^{+121}_{-31}$ | LCF | 0.33 | 0.41 |
| 128 | A2250 | I3090 | 0.30 | 0.490 | $2.9^{+0.1}_{-1.3}$ (2.8) | 0.66 (0.93) | $7.6 \pm 1.1$ | 150.0 | $5.1 \pm 0.7$ | $0.88^{+0.47}_{-0.33}$ | $65^{+119}_{-65}$ | $14^{+15}_{-14}$ | MCF | 0.02 | 0.11 |
| 129 | A2255 | I160 | 1.20 | 1.620 | $6.8^{+0.6}_{-1.1}$ (7.3) | 8.91 (9.72) | $137.0 \pm 8.6$ | 1210.0 | $11.3 \pm 0.7$ | $3.46^{+4.46}_{-2.68}$ | $0^{+221}_{-0}$ | $0.0^{+69.2}_{-0.0}$ | PSCF | 0.21 | 0.22 |
| 130 | A2256 | I300 | 1.00 | 1.650 | $7.4^{+0.8}_{-3.8}$ (7.5) | 17.60 (15.78) | $186.0 \pm 10.3$ | 1470.0 | $12.6 \pm 0.7$ | $1.84^{+6.92}_{-1.42}$ | $0^{+165}_{-0}$ | $0.0^{+76.9}_{-0.0}$ | PSCF | 0.37 | 0.32 |
| | | H10189 | 1.00 | 0.337 | $6.9^{+1.6}_{-3.2}$ (7.5) | 2.68 (15.78) | $8.4 \pm 0.9$ | 58.1 | .......... | $0.45^{+0.73}_{-0.20}$ | $36^{+40}_{-36}$ | $3.3^{+7.8}_{-3.3}$ | PSCF | 0.14 | 0.20 |
| 131 | A2271 | I6042 | 0.55 | 0.863 | $3.0^{+0.4}_{-1.9}$ (2.9) | 0.86 (1.04) | $18.8 \pm 2.5$ | 244.0 | $7.7 \pm 1.0$ | $1.25^{+1.11}_{-0.51}$ | $0^{+162}_{-0}$ | $0.0^{+34.5}_{-0.0}$ | PXCF | 0.04 | 0.11 |
| 132 | A2306 | I7696 | 0.80 | 1.800 | $4.5^{+1.7}_{-2.7}$ (.....) | 2.95 (.....) | $108.0 \pm 11.7$ | 709.0 | $15.3 \pm 1.6$ | $4.02^{+4.33}_{-1.43}$ | $0^{+100}_{-0}$ | $0.0^{+14.4}_{-0.0}$ | PXCF | 0.08 | 0.13 |
| 133 | A2312 | I6269 | 0.40 | 0.855 | $3.9^{+0.8}_{-1.4}$ (3.6) | 2.15 (1.97) | $27.0 \pm 3.2$ | 319.0 | $8.5 \pm 1.0$ | $0.81^{+0.43}_{-0.22}$ | $91^{+93}_{-91}$ | $37^{+48}_{-37}$ | MCF | 0.06 | 0.18 |
| 134 | A2319 | I3456 | 0.50 | 1.600 | $9.9^{+2.7}_{-4.6}$ (9.9) | 38.50 (28.52) | $242.0 \pm 14.5$ | 1570.0 | $15.4 \pm 0.9$ | $1.15^{+1.01}_{-0.36}$ | $0^{+160}_{-0}$ | $0.0^{+149.9}_{-0.0}$ | PMCF | 0.79 | 0.48 |
| 135 | A2328 | I6271 | 0.80 | 1.210 | $5.6^{+0.7}_{-2.2}$ (5.3) | 6.56 (7.40) | $83.3 \pm 12.7$ | 589.0 | $14.1 \pm 2.2$ | $4.16^{+5.35}_{-2.49}$ | $0^{+121}_{-0}$ | $0.0^{+53.7}_{-0.0}$ | PSCF | 0.10 | 0.19 |
| 136 | A2356 | I7802 | 0.60 | 0.802 | $4.3^{+0.3}_{-1.5}$ (4.1) | 2.31 (3.17) | $31.0 \pm 4.0$ | 286.0 | $10.8 \pm 1.4$ | $2.51^{+2.46}_{-1.09}$ | $0^{+100}_{-0}$ | $0.0^{+32.9}_{-0.0}$ | PSCF | 0.04 | 0.14 |
| 137 | A2366 | I133 | 0.80 | 0.414 | $1.6^{+0.3}_{-0.5}$ (2.0) | 0.11 (0.34) | $2.6 \pm 0.5$ | 42.6 | .......... | $1.60^{+6.90}_{-0.80}$ | $0^{+155}_{-0}$ | $0.0^{+9.4}_{-0.0}$ | XCF | 0.01 | 0.04 |
| 138 | A2384N | I7805 | 0.50 | 0.506 | $2.8^{+1.0}_{-1.2}$ (4.8) | 2.69 (3.03) | $17.1 \pm 1.8$ | 170.0 | $10.0 \pm 1.1$ | $1.65^{+1.03}_{-0.54}$ | $0^{+84}_{-0}$ | $0.0^{+43.2}_{-0.0}$ | PSCF | 0.05 | 0.18 |
| 139 | A2390 | I9125 | 0.30 | 1.480 | $9.5^{+1.3}_{-3.4}$ (.....) | 40.50 (.....) | $194.0 \pm 22.0$ | 1180.0 | $16.5 \pm 1.9$ | $1.03^{+0.19}_{-0.16}$ | $0^{+317}_{-0}$ | $0.0^{+653.5}_{-0.0}$ | PMCF | 0.38 | 0.55 |
| 140 | A2397 | I242 | 0.60 | 0.731 | $4.8^{+0.1}_{-1.6}$ (.....) | 4.59 (.....) | $33.5 \pm 7.6$ | 274.0 | $12.2 \pm 2.8$ | $1.03^{+0.83}_{-0.34}$ | $0^{+219}_{-0}$ | $0.0^{+125.4}_{-0.0}$ | PSCF | 0.07 | 0.22 |
| 141 | A2410 | I6071 | 0.80 | 1.180 | $2.5^{+1.3}_{-2.1}$ (.....) | 1.10 (.....) | $36.6 \pm 4.5$ | 327.0 | $11.2 \pm 1.4$ | $1.36^{+0.76}_{-0.44}$ | $0^{+221}_{-0}$ | $0.0^{+28.2}_{-0.0}$ | PXCF | 0.04 | 0.11 |
| 142 | A2415 | I130 | 0.30 | 0.790 | $3.6^{+1.1}_{-2.8}$ (3.8) | 2.38 (2.39) | $25.9 \pm 2.6$ | 318.0 | $8.2 \pm 0.8$ | $0.70^{+0.34}_{-0.18}$ | $81^{+88}_{-81}$ | $32^{+34}_{-32}$ | MCF | 0.07 | 0.20 |
| 143 | A2420 | I6045 | 0.60 | 1.070 | $5.7^{+0.7}_{-0.3}$ (6.0) | 6.91 (9.30) | $63.7 \pm 5.4$ | 624.0 | $10.2 \pm 0.9$ | $1.70^{+0.30}_{-0.25}$ | $0^{+76}_{-0}$ | $0.0^{+32.0}_{-0.0}$ | PSCF | 0.13 | 0.24 |
| 144 | A2440 | I129 | 0.80 | 1.630 | $8.2^{+1.4}_{-3.9}$ (9.0) | 6.55 (7.21) | $110.0 \pm 12.3$ | 1490.0 | $7.3 \pm 0.8$ | $7.52^{+6.03}_{-2.72}$ | $0^{+135}_{-0}$ | $0.0^{+28.4}_{-0.0}$ | PSCF | 0.20 | 0.17 |
| 145 | A2554 | I336 | 0.50 | 0.964 | $4.0^{+0.7}_{-2.2}$ (4.1) | 2.97 (3.67) | $39.7 \pm 4.2$ | 419.0 | $9.5 \pm 1.0$ | $1.57^{+0.65}_{-0.36}$ | $0^{+96}_{-0}$ | $0.0^{+48.9}_{-0.0}$ | PSCF | 0.06 | 0.18 |
| 146 | A2556 | I336 | 0.35 | 1.100 | $5.0^{+0.9}_{-2.9}$ (4.7) | 4.97 (5.52) | $47.3 \pm 3.7$ | 551.0 | $8.6 \pm 0.7$ | $0.86^{+0.31}_{-0.17}$ | $101^{+134}_{-101}$ | $81^{+105}_{-81}$ | LCF | 0.09 | 0.26 |
| 147 | A2577 | I1875 | 0.40 | 1.530 | $5.0^{+2.0}_{-3.9}$ (.....) | 6.23 (.....) | $86.0 \pm 11.6$ | 882.0 | $9.8 \pm 1.3$ | $1.05^{+0.36}_{-0.23}$ | $0^{+329}_{-0}$ | $0.0^{+177.2}_{-0.0}$ | PSCF | 0.09 | 0.25 |
| 148 | A2580 | H5751 | 0.30 | 0.767 | $4.3^{+0.7}_{-3.5}$ (5.1) | 11.90 (4.58) | $50.7 \pm 5.0$ | 411.0 | $12.3 \pm 1.2$ | $0.43^{+0.08}_{-0.05}$ | $91^{+73}_{-36}$ | $95^{+71}_{-34}$ | PSCF | 0.08 | 0.41 |
| 149 | A2593 | I6134 | 0.70 | 1.090 | $3.3^{+0.8}_{-2.1}$ (3.1) | 1.43 (1.85) | $33.9 \pm 3.1$ | 327.0 | $10.4 \pm 0.9$ | $2.30^{+4.03}_{-1.89}$ | $0^{+126}_{-0}$ | $0.0^{+21.3}_{-0.0}$ | XCF | 0.06 | 0.13 |
| 150 | A2625 | I156 | 0.20 | 0.345 | $2.4^{+0.0}_{-1.1}$ (2.7) | 0.42 (0.82) | $3.4 \pm 0.5$ | 94.6 | $3.6 \pm 0.5$ | $0.74^{+0.33}_{-0.21}$ | $79^{+93}_{-79}$ | $19^{+18}_{-19}$ | MCF | 0.02 | 0.11 |
| 151 | A2626 | I201 | 0.30 | 0.761 | $3.1^{+0.5}_{-1.0}$ (2.9) | 2.24 (4.78) | $22.4 \pm 2.4$ | 249.0 | $9.0 \pm 1.0$ | $0.52^{+0.21}_{-0.13}$ | $114^{+50}_{-59}$ | $53^{+36}_{-30}$ | LCF | 0.05 | 0.21 |
| 152 | A2634 | I199 | 0.70 | 0.736 | $3.5^{+0.4}_{-1.3}$ (3.4) | 0.65 (1.20) | $13.9 \pm 1.2$ | 197.0 | $7.1 \pm 0.6$ | $1.55^{+0.83}_{-0.47}$ | $0^{+31}_{-0}$ | $0.0^{+1.5}_{-0.0}$ | XCF | 0.05 | 0.09 |
| 153 | A2657 | I290 | 0.40 | 0.861 | $3.5^{+0.4}_{-2.2}$ (3.4) | 2.70 (2.93) | $32.6 \pm 2.4$ | 267.0 | $12.2 \pm 0.9$ | $0.65^{+0.22}_{-0.15}$ | $101^{+61}_{-47}$ | $44^{+36}_{-24}$ | MCF | 0.09 | 0.21 |
| 154 | A2666 | I294 | 0.15 | 0.108 | $1.7^{+0.0}_{-0.1}$ (*1.6*) | 0.01 (*0.94*) | $0.1 \pm 0.0$ | 11.7 | .......... | $1.07^{+0.78}_{-0.40}$ | $0^{+81}_{-0}$ | $0.0^{+2.6}_{-0.0}$ | XCF | 0.00 | 0.03 |
| 155 | A2670 | I314 | 0.60 | 0.978 | $3.7^{+1.1}_{-3.0}$ (3.9) | 3.51 (3.88) | $40.7 \pm 4.3$ | 599.0 | $6.8 \pm 0.7$ | $0.91^{+0.42}_{-0.24}$ | $80^{+129}_{-80}$ | $41^{+71}_{-41}$ | MCF | 0.08 | 0.21 |
| 156 | A2703 | I5360 | 0.70 | 1.190 | $3.8^{+1.4}_{-1.0}$ (.....) | 2.65 (.....) | $54.3 \pm 7.5$ | 420.0 | $12.9 \pm 1.8$ | $1.73^{+0.86}_{-0.53}$ | $0^{+99}_{-0}$ | $0.0^{+38.5}_{-0.0}$ | PSCF | 0.06 | 0.15 |
| 157 | A2715 | I4517 | 0.10 | 0.351 | $5.0^{+0.8}_{-1.8}$ (.....) | 3.91 (.....) | $9.4 \pm 1.7$ | 155.0 | $6.0 \pm 1.1$ | $0.56^{+0.09}_{-0.07}$ | $109^{+198}_{-65}$ | $100^{+126}_{-64}$ | LCF | 0.07 | 0.30 |
| 158 | A2877 | I6088 | 0.20 | 0.463 | $4.4^{+1.3}_{-4.1}$ (3.5) | 0.30 (0.47) | $4.6 \pm 0.4$ | 171.0 | $2.7 \pm 0.2$ | $1.43^{+1.18}_{-0.52}$ | $0^{+69}_{-0}$ | $0.0^{+2.9}_{-0.0}$ | XCF | 0.05 | 0.08 |
| 159 | A3158 | H5753 | 0.65 | 0.333 | $5.3^{+0.6}_{-2.8}$ (5.5) | 2.87 (6.97) | $8.9 \pm 0.9$ | 74.2 | .......... | $0.23^{+1.11}_{-0.13}$ | $53^{+113}_{-53}$ | $9.6^{+67.8}_{-9.6}$ | XCF | 0.11 | 0.25 |

Table 3: Deprojection Results

| # | Name | Sequence Number | $R_{\rm core}$ | $R_{\rm out}$ | k$T$ | $L_{\rm X}$ | $M_{\rm gas}$ | $M_{\rm grav}$ | $\frac{M_{\rm gas}}{M_{\rm grav}}$ | $t_{\rm cool}$ | $R_{\rm cool}$ | $\dot{M}$ | CF | S-Z -d$T$ | T-Depth |
|---|---|---|---|---|---|---|---|---|---|---|---|---|---|---|---|
| | | | ( Mpc) | | ( keV) | ($10^{44}$ erg s$^{-1}$ ) | ($10^{12}$ M$_{\odot}$) | | (%) | ( $H_0^{-1}$) | ( kpc) | ( M$_{\odot}$ yr$^{-1}$ ) | (Class) | (mK) | (%) |
| (i) | (ii) | (iii) | (iv) | (v) | (vi) | (vii) | (viii) | (ix) | (x) | (xi) | (xii) | (xiii) | (xiv) | (xv) | (xvi) |
| | | I1829 | 0.65 | 0.776 | $5.3^{+0.2}_{-2.8}$ (5.5) | 6.53 (6.97) | $43.2 \pm 1.6$ | 412.0 | $10.5 \pm 0.4$ | $1.33^{+0.44}_{-0.18}$ | $0^{+48}_{-0}$ | $0.0^{+14.5}_{-0.0}$ | XCF | 0.15 | 0.26 |
| 160 | A3186 | I8385 | 0.60 | 0.918 | $5.4^{+1.3}_{-3.1}$ (5.9) | 8.40 (16.06) | $59.8 \pm 8.0$ | 498.0 | $12.0 \pm 1.6$ | $1.30^{+0.84}_{-0.35}$ | $0^{+275}_{-0}$ | $0.0^{+207.3}_{-0.0}$ | PSCF | 0.14 | 0.28 |
| 161 | A3266 | I1831 | 0.90 | 1.460 | $6.2^{+0.5}_{-1.1}$ (6.2) | 12.10 (16.27) | $141.0 \pm 6.8$ | 852.0 | $16.6 \pm 0.8$ | $1.31^{+0.22}_{-0.15}$ | $0^{+56}_{-0}$ | $0.0^{+16.9}_{-0.0}$ | PSCF | 0.26 | 0.28 |
| 162 | A3322 | I7705 | 0.30 | 0.612 | $4.7^{+2.2}_{-1.4}$ (4.4) | 2.21 (2.84) | $15.1 \pm 2.5$ | 276.0 | $5.5 \pm 0.9$ | $1.33^{+1.34}_{-0.51}$ | $0^{+184}_{-0}$ | $0.0^{+90.2}_{-0.0}$ | PSCF | 0.04 | 0.17 |
| 163 | A3376 | I5167 | 0.50 | 0.472 | $3.5^{+1.3}_{-2.6}$ (4.1) | 0.76 (3.13) | $8.0 \pm 1.2$ | 130.0 | .......... | $0.64^{+8.36}_{-0.09}$ | $53^{+88}_{-53}$ | $6.3^{+9.3}_{-6.3}$ | SCF | 0.04 | 0.11 |
| 164 | A3389E | I5169 | 0.10 | 0.212 | $2.0^{+0.2}_{-1.2}$ (.....) | 0.47 (.....) | $1.8 \pm 0.2$ | 40.1 | $4.6 \pm 0.6$ | $0.18^{+0.08}_{-0.06}$ | $87^{+46}_{-60}$ | $22^{+21}_{-10}$ | MCF | 0.02 | 0.16 |
| 165 | A3389W | I5169 | 0.60 | 0.477 | $2.4^{+0.2}_{-1.1}$ (.....) | 0.84 (.....) | $8.4 \pm 1.0$ | 59.2 | .......... | $0.48^{+2.04}_{-0.31}$ | $57^{+182}_{-57}$ | $7.9^{+63.0}_{-7.9}$ | SCF | 0.04 | 0.13 |
| 166 | A3391 | I8309 | 0.50 | 0.812 | $5.0^{+0.2}_{-3.5}$ (5.2) | 2.29 (3.37) | $27.3 \pm 2.1$ | 399.0 | $6.8 \pm 0.5$ | $1.93^{+0.95}_{-0.40}$ | $0^{+51}_{-0}$ | $0.0^{+7.7}_{-0.0}$ | XCF | 0.09 | 0.16 |
| 167 | A3532 | I6173 | 0.70 | 0.899 | $4.7^{+0.5}_{-0.6}$ (4.4) | 4.01 (2.88) | $46.5 \pm 3.7$ | 334.0 | $13.9 \pm 1.1$ | $2.16^{+4.30}_{-0.97}$ | $0^{+64}_{-0}$ | $0.0^{+12.6}_{-0.0}$ | PSCF | 0.11 | 0.17 |
| 168 | A3562W | I5730 | 0.70 | 0.479 | $2.7^{+0.6}_{-1.3}$ (.....) | 0.37 (.....) | $6.1 \pm 0.9$ | 78.1 | .......... | $2.46^{+4.89}_{-1.20}$ | $0^{+48}_{-0}$ | $0.0^{+2.9}_{-0.0}$ | XCF | 0.02 | 0.07 |
| 169 | A3581 | I9980 | 0.10 | 0.234 | $2.5^{+0.4}_{-0.5}$ (2.8) | 0.36 (0.85) | $1.6 \pm 0.1$ | 70.7 | $2.3 \pm 0.1$ | $0.27^{+0.03}_{-0.03}$ | $79^{+38}_{-9}$ | $18^{+4}_{-3}$ | MCF | 0.03 | 0.14 |
| 170 | A3602 | I5252 | 0.30 | 0.551 | $3.9^{+0.2}_{-0.1}$ (4.1) | 2.22 (2.84) | $15.7 \pm 2.7$ | 209.0 | $7.5 \pm 1.3$ | $1.40^{+0.66}_{-0.38}$ | $0^{+92}_{-0}$ | $0.0^{+62.6}_{-0.0}$ | PSCF | 0.04 | 0.17 |
| 171 | A3654 | I3289 | 0.60 | 0.979 | $3.3^{+0.6}_{-0.6}$ (3.7) | 2.51 (1.46) | $39.8 \pm 5.9$ | 351.0 | $11.3 \pm 1.7$ | $1.83^{+0.68}_{-0.53}$ | $0^{+122}_{-0}$ | $0.0^{+60.3}_{-0.0}$ | PSCF | 0.05 | 0.15 |
| 172 | A3667 | I5735 | 0.80 | 1.770 | $7.1^{+1.8}_{-4.0}$ (6.5) | 24.70 (12.57) | $272.0 \pm 15.1$ | 1130.0 | $24.1 \pm 1.3$ | $1.85^{+2.18}_{-0.63}$ | $0^{+74}_{-0}$ | $0.0^{+41.2}_{-0.0}$ | PSCF | 0.41 | 0.35 |
| 173 | A370 | I245 | 0.10 | 1.490 | $10.7^{+5.2}_{-1.4}$ (.....) | 55.80 (.....) | $270.0 \pm 40.4$ | 1050.0 | $25.7 \pm 3.8$ | $2.26^{+0.72}_{-0.56}$ | $0^{+149}_{-0}$ | $0.0^{+199.9}_{-0.0}$ | PMCF | 0.32 | 0.48 |
| 174 | A3744 | I3044 | 0.50 | 0.468 | $5.8^{+1.5}_{-2.5}$ (*5.3*) | 0.17 (*9.51*) | $3.2 \pm 0.5$ | 162.0 | .......... | $1.66^{+5.45}_{-1.10}$ | $0^{+78}_{-0}$ | $0.0^{+4.3}_{-0.0}$ | XCF | 0.04 | 0.06 |
| 175 | A376 | I1773 | 0.15 | 0.335 | $5.7^{+0.4}_{-1.5}$ (5.1) | 1.65 (2.37) | $5.1 \pm 1.0$ | 175.0 | $2.9 \pm 0.6$ | $0.40^{+0.13}_{-0.13}$ | $84^{+125}_{-42}$ | $42^{+42}_{-14}$ | MCF | 0.06 | 0.22 |
| 176 | A3888 | I1872 | 0.50 | 1.610 | $8.6^{+1.5}_{-5.5}$ (7.9) | 30.20 (31.52) | $214.0 \pm 23.7$ | 1350.0 | $15.9 \pm 1.8$ | $2.25^{+0.91}_{-0.67}$ | $0^{+134}_{-0}$ | $0.0^{+168.7}_{-0.0}$ | PMCF | 0.27 | 0.39 |
| 177 | A389 | I6128 | 0.40 | 1.000 | $4.2^{+1.2}_{-1.1}$ (4.5) | 2.97 (4.54) | $37.0 \pm 4.8$ | 475.0 | $7.8 \pm 1.0$ | $1.56^{+0.24}_{-0.17}$ | $0^{+100}_{-0}$ | $0.0^{+54.0}_{-0.0}$ | PSCF | 0.05 | 0.18 |
| 178 | A3998 | I6385 | 0.30 | 0.918 | $5.2^{+1.2}_{-3.4}$ (.....) | 3.11 (.....) | $37.9 \pm 4.3$ | 397.0 | $9.5 \pm 1.1$ | $1.61^{+0.48}_{-0.30}$ | $0^{+92}_{-0}$ | $0.0^{+34.5}_{-0.0}$ | PSCF | 0.08 | 0.18 |
| 179 | A4067S | I5745 | 0.40 | 1.060 | $4.0^{+0.6}_{-0.8}$ (2.7) | 2.38 (0.47) | $43.2 \pm 6.4$ | 316.0 | $13.7 \pm 2.0$ | $1.24^{+9.74}_{-0.11}$ | $0^{+89}_{-0}$ | $0.0^{+30.9}_{-0.0}$ | PSCF | 0.05 | 0.15 |
| 180 | AWM4 | I10543 | 0.40 | 1.100 | $4.0^{+1.2}_{-3.0}$ (3.7) | 0.75 (1.67) | $23.3 \pm 2.5$ | 436.0 | $5.3 \pm 0.6$ | $1.51^{+0.51}_{-0.43}$ | $0^{+55}_{-0}$ | $0.0^{+8.8}_{-0.0}$ | PXCF | 0.05 | 0.10 |
| 181 | AWM5 | I3302 | 0.40 | 0.476 | $2.5^{+0.5}_{-1.9}$ (2.6) | 0.29 (0.58) | $5.0 \pm 0.7$ | 105.0 | $4.7 \pm 0.7$ | $1.10^{+2.90}_{-0.59}$ | $0^{+102}_{-0}$ | $0.0^{+8.2}_{-0.0}$ | XCF | 0.02 | 0.08 |
| 182 | AWM7 | I6698 | 0.40 | 0.623 | $3.5^{+0.9}_{-2.7}$ (3.6) | 2.87 (3.12) | $21.0 \pm 1.6$ | 252.0 | $8.3 \pm 0.6$ | $0.27^{+0.16}_{-0.08}$ | $111^{+57}_{-87}$ | $45^{+50}_{-37}$ | MCF | 0.14 | 0.26 |
| | | H6638 | 0.40 | 0.109 | $3.0^{+0.4}_{-1.6}$ (3.6) | 0.36 (3.12) | $0.6 \pm 0.1$ | 9.4 | .......... | $0.15^{+0.44}_{-0.12}$ | $> 114^{+0}_{->89}$ | $> 48^{+0}_{->47}$ | MCF | 0.04 | 0.17 |
| 183 | CENTAURUS | I298 | 0.15 | 0.637 | $3.8^{+1.0}_{-2.4}$ (3.6) | 1.90 (1.41) | $17.7 \pm 1.1$ | 162.0 | $10.9 \pm 0.7$ | $0.09^{+0.04}_{-0.02}$ | $86^{+35}_{-53}$ | $26^{+16}_{-13}$ | MCF | 0.13 | 0.28 |
| | | H4341 | 0.15 | 0.098 | $3.7^{+0.5}_{-1.5}$ (3.6) | 0.37 (1.41) | $0.5 \pm 0.0$ | 12.3 | .......... | $0.02^{+0.01}_{-0.00}$ | $72^{+>29}_{-32}$ | $15^{+>14}_{-10}$ | MCF | 0.04 | 0.27 |
| 184 | CL0016+16 | H7755 | 0.50 | 0.443 | $7.4^{+0.9}_{-1.8}$ (.....) | 12.30 (.....) | $24.5 \pm 4.8$ | 177.0 | .......... | $0.37^{+1.43}_{-0.20}$ | $106^{+226}_{-106}$ | $88^{+497}_{-88}$ | MCF | 0.24 | 0.42 |
| 185 | CYGNUS-A | I1807 | 0.20 | 1.080 | $4.5^{+1.6}_{-3.9}$ (4.1) | 12.00 (10.66) | $79.1 \pm 6.9$ | 447.0 | $17.7 \pm 1.5$ | $0.34^{+0.08}_{-0.08}$ | $167^{+104}_{-113}$ | $242^{+69}_{-142}$ | LCF | 0.20 | 0.47 |
| | | H10760 | 0.20 | 0.241 | $6.4^{+0.8}_{-3.8}$ (4.1) | 4.62 (10.66) | $5.5 \pm 0.7$ | 79.8 | $6.9 \pm 0.9$ | $0.05^{+0.01}_{-0.01}$ | $113^{+23}_{-37}$ | $150^{+42}_{-22}$ | LCF | 0.13 | 0.54 |
| 186 | FORNAX-A | H1885 | 0.01 | 0.017 | $0.3^{+0.1}_{-0.1}$ (*0.5*) | 0.00 (*0.09*) | $0.0 \pm 0.0$ | 0.6 | $0.4 \pm 0.1$ | $0.00^{+0.00}_{-0.00}$ | $> 19^{+0}_{->6}$ | $> 1.9^{+0.0}_{->2.0}$ | LCF | 0.00 | 0.03 |
| 187 | HERCULES-A | I10533 | 0.30 | 0.752 | $5.1^{+0.3}_{-0.5}$ (5.4) | 6.39 (7.55) | $39.2 \pm 2.1$ | 428.0 | $9.2 \pm 0.5$ | $1.56^{+0.36}_{-0.26}$ | $0^{+125}_{-0}$ | $0.0^{+135.8}_{-0.0}$ | PSCF | 0.07 | 0.24 |
| 188 | HYDRA-A | I1894 | 0.30 | 1.500 | $4.7^{+0.7}_{-2.9}$ (3.8) | 8.56 (7.23) | $91.0 \pm 6.7$ | 590.0 | $15.4 \pm 1.1$ | $0.25^{+0.09}_{-0.06}$ | $170^{+80}_{-120}$ | $222^{+98}_{-132}$ | LCF | 0.16 | 0.42 |
| 189 | M87 | H282 | 0.30 | 0.013 | $1.5^{+0.1}_{-0.2}$ (2.4) | 0.03 (0.67) | $0.0 \pm 0.0$ | 0.8 | .......... | $0.01^{+0.00}_{-0.00}$ | $> 14^{+0}_{->2}$ | $> 6.2^{+0.0}_{->0.3}$ | LCF | 0.02 | 0.18 |
| | | I10362 | 0.30 | 0.295 | $2.5^{+0.2}_{-0.8}$ (2.4) | 0.56 (0.67) | $3.1 \pm 0.0$ | 51.0 | .......... | $0.04^{+0.00}_{-0.00}$ | $106^{+16}_{-10}$ | $32^{+1}_{-2}$ | MCF | 0.07 | 0.23 |

Table 3: Deprojection Results

| # | Name | Sequence Number | $R_{\rm core}$ (Mpc) | $R_{\rm out}$ | k$T$ (keV) | $L_{\rm X}$ ($10^{44}$ erg s$^{-1}$) | $M_{\rm gas}$ ($10^{12}$ M$_\odot$) | $M_{\rm grav}$ | $\frac{M_{\rm gas}}{M_{\rm grav}}$ (%) | $t_{\rm cool}$ ($H_0^{-1}$) | $R_{\rm cool}$ (kpc) | $\dot{M}$ (M$_\odot$ yr$^{-1}$) | CF (Class) | S-Z -d$T$ (mK) | T-Depth (%) |
|---|---|---|---|---|---|---|---|---|---|---|---|---|---|---|---|
| (i) | (ii) | (iii) | (iv) | (v) | (vi) | (vii) | (viii) | (ix) | (x) | (xi) | (xii) | (xiii) | (xiv) | (xv) | (xvi) |
| 190 | MKW3S | I2604 | 0.25 | 0.759 | $3.3^{+0.9}_{-2.0}$ (3.0) | 3.69 (4.26) | $27.9 \pm 2.5$ | 248.0 | $11.2 \pm 1.0$ | $0.38^{+0.19}_{-0.10}$ | $158^{+53}_{-31}$ | $132^{+68}_{-45}$ | LCF | 0.09 | 0.29 |
| | | H4359 | 0.25 | 0.211 | $3.4^{+0.1}_{-1.3}$ (3.0) | 2.01 (4.26) | $3.8 \pm 0.4$ | 40.6 | .......... | $0.12^{+0.14}_{-0.05}$ | $148^{+>75}_{-89}$ | $121^{+>54}_{-79}$ | LCF | 0.07 | 0.32 |
| 191 | MKW4 | I2601 | 0.20 | 0.453 | $1.8^{+0.9}_{-0.6}$ (1.7) | 0.29 (0.44) | $4.1 \pm 0.3$ | 89.9 | $4.6 \pm 0.3$ | $0.26^{+0.15}_{-0.07}$ | $72^{+31}_{-10}$ | $10^{+4}_{-3}$ | MCF | 0.03 | 0.11 |
| 192 | OPHIUCHUS | H6553 | 0.25 | 0.419 | $8.6^{+2.6}_{-6.9}$ (9.0) | 16.00 (31.41) | $26.5 \pm 1.5$ | 360.0 | $7.4 \pm 0.4$ | $0.39^{+0.07}_{-0.06}$ | $68^{+95}_{-45}$ | $41^{+110}_{-28}$ | MCF | 0.56 | 0.48 |
| 193 | PKS0745-19 | H6541 | 0.20 | 0.302 | $8.3^{+0.5}_{-5.8}$ (8.5) | 30.50 (57.18) | $21.1 \pm 3.1$ | 235.0 | $9.0 \pm 1.3$ | $0.18^{+0.40}_{-0.09}$ | $177^{+31}_{-44}$ | $579^{+399}_{-215}$ | MCF | 0.37 | 0.82 |
| 194 | SC1842-63 | I6105 | 0.30 | 0.259 | $1.6^{+0.6}_{-0.7}$ (1.4) | 0.06 (0.13) | $0.9 \pm 0.1$ | 30.4 | .......... | $0.20^{+0.07}_{-0.04}$ | $48^{+24}_{-4}$ | $1.9^{+0.7}_{-0.7}$ | SCF | 0.01 | 0.06 |
| 195 | SC2316-3632 | I7569 | 0.50 | 0.931 | $4.8^{+0.4}_{-0.3}$ (4.8) | 4.12 (0.04) | $42.9 \pm 6.6$ | 441.0 | $9.7 \pm 1.5$ | $2.75^{+2.33}_{-1.14}$ | $0^{+116}_{-0}$ | $0.0^{+58.0}_{-0.0}$ | PSCF | 0.05 | 0.18 |
| 196 | SERSIC159-03 | I1874 | 0.05 | 0.212 | $2.4^{+0.2}_{-1.9}$ (3.0) | 3.95 (4.06) | $3.8 \pm 0.3$ | 66.9 | $5.7 \pm 0.4$ | $0.07^{+0.01}_{-0.01}$ | $> 229^{+0}_{->141}$ | $> 288^{+0}_{->130}$ | LCF | 0.04 | 0.55 |
| 197 | ZW0258+43 | I4611 | 0.60 | 0.731 | $2.3^{+1.1}_{-1.2}$ (.....) | 0.53 (.....) | $12.6 \pm 2.0$ | 145.0 | $8.7 \pm 1.4$ | $1.73^{+9.59}_{-1.00}$ | $0^{+183}_{-0}$ | $0.0^{+10.2}_{-0.0}$ | PXCF | 0.02 | 0.07 |
| 198 | ZW0628+25 | I4613 | 0.40 | 0.887 | $6.0^{+1.4}_{-4.1}$ (6.2) | 3.75 (5.16) | $35.7 \pm 3.6$ | 568.0 | $6.3 \pm 0.6$ | $1.41^{+0.49}_{-0.41}$ | $0^{+74}_{-0}$ | $0.0^{+37.0}_{-0.0}$ | PSCF | 0.12 | 0.20 |
| 199 | ZW0712+53 | I4620 | 0.40 | 0.481 | $2.7^{+0.1}_{-0.9}$ (2.8) | 0.43 (0.91) | $5.8 \pm 1.1$ | 116.0 | $5.0 \pm 0.9$ | $1.41^{+2.25}_{-0.70}$ | $0^{+180}_{-0}$ | $0.0^{+34.0}_{-0.0}$ | PCF | 0.02 | 0.09 |
| 200 | ZW1615+35 | I322 | 0.20 | 0.381 | $3.3^{+1.3}_{-2.6}$ (2.9) | 0.10 (0.59) | $2.0 \pm 0.3$ | 114.0 | $1.7 \pm 0.2$ | $1.40^{+0.85}_{-0.51}$ | $0^{+95}_{-0}$ | $0.0^{+2.4}_{-0.0}$ | XCF | 0.02 | 0.05 |
| 201 | 3C129 | I1832 | 0.45 | 0.835 | $6.1^{+1.0}_{-2.2}$ (5.6) | 1.17 (4.01) | $20.2 \pm 1.7$ | 445.0 | $4.5 \pm 0.4$ | $1.54^{+3.17}_{-1.29}$ | $0^{+66}_{-0}$ | $0.0^{+4.2}_{-0.0}$ | XCF | 0.11 | 0.11 |
| 202 | 3C130 | I3924 | 0.70 | 0.959 | $5.2^{+0.6}_{-3.0}$ (5.0) | 6.16 (3.81) | $64.4 \pm 8.4$ | 417.0 | $15.4 \pm 2.0$ | $2.19^{+2.96}_{-0.97}$ | $0^{+96}_{-0}$ | $0.0^{+37.2}_{-0.0}$ | PSCF | 0.10 | 0.20 |
| 203 | 3C295 | H4290 | 0.10 | 0.344 | $6.4^{+1.5}_{-1.6}$ (.....) | 17.60 (.....) | $16.4 \pm 2.4$ | 240.0 | $6.9 \pm 1.0$ | $0.16^{+0.02}_{-0.02}$ | $181^{+60}_{-78}$ | $410^{+125}_{-130}$ | PSCF | 0.22 | 0.72 |
| 204 | 3C330 | I272 | 0.50 | 0.709 | $3.5^{+0.1}_{-1.7}$ (.....) | 1.06 (.....) | $17.3 \pm 1.9$ | 232.0 | $7.5 \pm 0.8$ | $7.46^{+19.79}_{-5.91}$ | $0^{+89}_{-0}$ | $0.0^{+15.4}_{-0.0}$ | PXCF | 0.03 | 0.10 |
| 205 | 3C370 | I1907 | 0.50 | 0.550 | $2.0^{+0.1}_{-1.2}$ (2.5) | 0.32 (0.67) | $6.1 \pm 0.8$ | 115.0 | $5.3 \pm 0.7$ | $1.21^{+0.51}_{-0.38}$ | $0^{+206}_{-0}$ | $0.0^{+31.4}_{-0.0}$ | PXCF | 0.01 | 0.08 |
| 206 | 3C411 | I6833 | 0.80 | 0.498 | $1.4^{+0.5}_{-0.4}$ (.....) | 0.17 (.....) | $4.4 \pm 0.6$ | 60.4 | .......... | $3.01^{+3.29}_{-1.20}$ | $0^{+83}_{-0}$ | $0.0^{+5.5}_{-0.0}$ | PCF | 0.01 | 0.05 |
| 207 | 3C449 | I3916 | 0.25 | 0.174 | $1.5^{+0.2}_{-0.7}$ (1.6) | 0.04 (0.12) | $0.4 \pm 0.1$ | 18.9 | .......... | $0.56^{+1.06}_{-0.24}$ | $58^{+63}_{-58}$ | $2.5^{+6.0}_{-2.5}$ | SCF | 0.01 | 0.05 |

Table 4: Deprojection Results Summed To Different Radii

| # | Name | Sequence Number | $R[L_{\rm X}/2]$ ($10^{44}$ erg s$^{-1}$) | | | $L_{\rm X}$ ($10^{44}$ erg s$^{-1}$) | | $M_{\rm gas}$ ($10^{12}$ M$_{\odot}$) | | $M_{\rm grav}$ ($10^{12}$ M$_{\odot}$) | | $M_{\rm gas}/M_{\rm grav}$ (%) | |
|---|---|---|---|---|---|---|---|---|---|---|---|---|---|
| | | | ($R < 0.2$ Mpc) | ($R < 0.5$ Mpc) | ($R < 1.0$ Mpc) | ($R < 0.5$ Mpc) | ($R < 1.0$ Mpc) | ($R < 0.5$ Mpc) | ($R < 1.0$ Mpc) | ($R < 0.5$ Mpc) | ($R < 1.0$ Mpc) | ($R < 0.5$ Mpc) | ($R < 1.0$ Mpc) |
| (i) | (ii) | (iii) | (iv) | (v) | (vi) | (vii) | (viii) | (ix) | (x) | (xi) | (xii) | | |
| 1 | A21 | I6012 | ..... | 0.314 | 0.475 | 2.20 | 4.15 | $15.2 \pm 1.3$ | $51.1 \pm 6.2$ | 147.3 | 462.5 | $10.4 \pm 0.9$ | $11.0 \pm 1.3$ |
| 2 | A74 | I8989 | 0.087 | 0.112 | ..... | 0.12 | ..... | $2.3 \pm 0.5$ | .......... | 125.4 | ..... | $1.8 \pm 0.4$ | .......... |
| 3 | A76 | I1817 | 0.117 | 0.305 | ..... | 0.30 | ..... | $5.6 \pm 0.9$ | .......... | 51.0 | ..... | $10.9 \pm 1.7$ | .......... |
| 4 | A84 | I7640 | 0.123 | 0.320 | ..... | 0.62 | ..... | $8.5 \pm 1.3$ | .......... | 90.7 | ..... | $9.2 \pm 1.4$ | .......... |
| 5 | A85 | I292 | 0.127 | 0.231 | 0.344 | 5.67 | 8.27 | $20.5 \pm 1.4$ | $62.8 \pm 3.7$ | 193.4 | 353.6 | $10.6 \pm 0.7$ | $17.8 \pm 1.0$ |
| | | H6013 | 0.108 | ..... | ..... | ..... | ..... | .......... | .......... | ..... | ..... | .......... | .......... |
| 6 | A98N | I208 | 0.101 | 0.252 | ..... | 0.62 | ..... | $7.7 \pm 1.3$ | .......... | 154.6 | ..... | $5.0 \pm 0.8$ | .......... |
| 7 | A98S | I208 | 0.127 | ..... | ..... | ..... | ..... | .......... | .......... | ..... | ..... | .......... | .......... |
| 8 | A115N | I209 | 0.116 | ..... | ..... | ..... | ..... | .......... | .......... | ..... | ..... | .......... | .......... |
| 9 | A115S | I209 | 0.130 | ..... | ..... | ..... | ..... | .......... | .......... | ..... | ..... | .......... | .......... |
| 10 | A117 | I8992 | 0.153 | 0.261 | ..... | 0.26 | ..... | $5.2 \pm 0.8$ | .......... | 94.8 | ..... | $5.5 \pm 0.8$ | .......... |
| 11 | A119 | I1770 | 0.149 | 0.343 | 0.530 | 1.99 | 4.35 | $13.9 \pm 1.6$ | $52.5 \pm 4.1$ | 104.5 | 385.9 | $13.3 \pm 1.5$ | $13.6 \pm 1.1$ |
| 12 | A133 | I2333 | 0.124 | 0.215 | ..... | 3.88 | ..... | $17.4 \pm 1.4$ | .......... | 182.0 | ..... | $9.6 \pm 0.7$ | .......... |
| 13 | A150N | I10766 | 0.117 | 0.273 | ..... | 0.32 | ..... | $5.5 \pm 0.8$ | .......... | 93.9 | ..... | $5.8 \pm 0.9$ | .......... |
| 14 | A154 | I6135 | 0.146 | 0.268 | ..... | 0.68 | ..... | $8.1 \pm 1.1$ | .......... | 113.4 | ..... | $7.2 \pm 1.0$ | .......... |
| 15 | A160 | I154 | 0.137 | 0.251 | ..... | 0.26 | ..... | $4.1 \pm 0.6$ | .......... | 100.9 | ..... | $4.1 \pm 0.6$ | .......... |
| 16 | A168 | I6083 | 0.137 | 0.339 | ..... | 0.39 | ..... | $6.8 \pm 1.0$ | .......... | 52.9 | ..... | $12.8 \pm 1.8$ | .......... |
| 17 | A194 | I6084 | 0.142 | ..... | ..... | ..... | ..... | .......... | .......... | ..... | ..... | .......... | .......... |
| 18 | A240 | I189 | 0.107 | ..... | ..... | ..... | ..... | .......... | .......... | ..... | ..... | .......... | .......... |
| 19 | A262 | I295 | 0.115 | ..... | ..... | ..... | ..... | .......... | .......... | ..... | ..... | .......... | .......... |
| 20 | A278 | I7698 | 0.102 | ..... | ..... | ..... | ..... | .......... | .......... | ..... | ..... | .......... | .......... |
| 21 | A347 | I302 | 0.122 | 0.234 | ..... | 0.32 | ..... | $5.3 \pm 0.6$ | .......... | 80.5 | ..... | $6.5 \pm 0.7$ | .......... |
| 22 | A367 | I3445 | ..... | 0.269 | ..... | 0.17 | ..... | $3.9 \pm 1.0$ | .......... | 146.8 | ..... | $2.8 \pm 0.7$ | .......... |
| 23 | A397 | I7699 | ..... | ..... | ..... | ..... | ..... | .......... | .......... | ..... | ..... | .......... | .......... |
| 24 | A399 | I185 | 0.149 | 0.325 | 0.481 | 2.40 | 4.52 | $15.2 \pm 1.4$ | $50.0 \pm 3.9$ | 145.3 | 511.2 | $10.3 \pm 1.0$ | $9.8 \pm 0.8$ |
| 25 | A400 | I6085 | 0.135 | 0.280 | ..... | 0.43 | ..... | $6.5 \pm 0.5$ | .......... | 74.4 | ..... | $8.8 \pm 0.7$ | .......... |
| 26 | A401 | H1777 | 0.137 | 0.311 | ..... | 10.54 | ..... | $29.6 \pm 4.0$ | .......... | 231.5 | ..... | $12.8 \pm 1.7$ | .......... |
| | | I1776 | 0.149 | 0.308 | 0.448 | 10.33 | 18.10 | $29.4 \pm 0.5$ | $94.9 \pm 2.1$ | 233.7 | 726.2 | $12.6 \pm 0.2$ | $13.1 \pm 0.3$ |
| 27 | A407 | I1825 | 0.122 | 0.288 | ..... | 0.50 | ..... | $6.9 \pm 1.0$ | .......... | 92.1 | ..... | $7.5 \pm 1.1$ | .......... |
| 28 | A419 | I8993 | 0.119 | ..... | ..... | ..... | ..... | .......... | .......... | ..... | ..... | .......... | .......... |
| 29 | A426 | H285 | ..... | ..... | ..... | ..... | ..... | .......... | .......... | ..... | ..... | .......... | .......... |
| | | I283 | 0.090 | 0.147 | ..... | 17.16 | ..... | $28.7 \pm 0.2$ | .......... | 266.9 | ..... | $10.8 \pm 0.1$ | .......... |
| 30 | A458 | I6018 | 0.104 | 0.271 | ..... | 1.34 | ..... | $11.2 \pm 1.3$ | .......... | 121.6 | ..... | $9.2 \pm 1.1$ | .......... |
| 31 | A478 | H4198 | 0.113 | ..... | ..... | ..... | ..... | .......... | .......... | ..... | ..... | .......... | .......... |

Table 4: Deprojection Results Summed To Different Radii

| # | Name | Sequence Number | $R[L_X/2]$ ($10^{44}$ erg s$^{-1}$) ($R<0.2$ Mpc) | ($R<0.5$ Mpc) | ($R<1.0$ Mpc) | $L_X$ ($10^{44}$ erg s$^{-1}$) ($R<0.5$ Mpc) | ($R<1.0$ Mpc) | $M_{gas}$ ($10^{12}$ M$_\odot$) ($R<0.5$ Mpc) | ($R<1.0$ Mpc) | $M_{grav}$ ($10^{12}$ M$_\odot$) ($R<0.5$ Mpc) | ($R<1.0$ Mpc) | $M_{gas}/M_{grav}$ (%) ($R<0.5$ Mpc) | ($R<1.0$ Mpc) |
|---|---|---|---|---|---|---|---|---|---|---|---|---|---|
| (i) | (ii) | (iii) | (iv) | (v) | (vi) | (vii) | (viii) | (ix) | (x) | (xi) | (xii) | | |
| | | I303 | 0.111 | 0.232 | 0.296 | 26.08 | 35.49 | $43.0 \pm 0.9$ | $120.9 \pm 3.9$ | 266.7 | 513.1 | $16.0 \pm 0.3$ | $23.5 \pm 0.7$ |
| 32 | A496 | I2348 | 0.112 | 0.200 | 0.287 | 5.42 | 7.42 | $19.5 \pm 1.3$ | $56.0 \pm 3.2$ | 174.4 | 319.9 | $11.2 \pm 0.7$ | $17.5 \pm 1.0$ |
| | | H10401 | ..... | ..... | ..... | ..... | ..... | .......... | .......... | ..... | ..... | .......... | .......... |
| 33 | A500 | I6232 | 0.135 | 0.310 | ..... | 0.84 | ..... | $9.4 \pm 1.0$ | .......... | 105.8 | ..... | $8.9 \pm 1.0$ | .......... |
| 34 | A520 | I6841 | 0.100 | 0.347 | ..... | 6.26 | ..... | $25.5 \pm 2.6$ | .......... | 160.0 | ..... | $15.6 \pm 1.6$ | .......... |
| 35 | A539 | I2353 | 0.141 | 0.257 | ..... | 0.51 | ..... | $6.9 \pm 0.6$ | .......... | 113.0 | ..... | $6.1 \pm 0.5$ | .......... |
| 36 | A545 | I310 | ..... | 0.316 | 0.456 | 7.61 | 13.42 | $27.5 \pm 1.0$ | $86.9 \pm 7.5$ | 161.7 | 525.4 | $17.0 \pm 0.6$ | $16.5 \pm 1.4$ |
| 37 | A548S | I7860 | 0.125 | 0.261 | ..... | 0.57 | ..... | $7.2 \pm 1.2$ | .......... | 102.7 | ..... | $7.0 \pm 1.1$ | .......... |
| 38 | A566 | I3553 | 0.104 | 0.311 | 0.432 | 2.66 | 4.29 | $16.4 \pm 2.0$ | $48.5 \pm 5.9$ | 163.8 | 467.7 | $9.9 \pm 1.2$ | $10.4 \pm 1.3$ |
| 39 | A569 | I1836 | ..... | ..... | ..... | ..... | ..... | .......... | .......... | ..... | ..... | .......... | .......... |
| 40 | A576 | I3455 | 0.128 | 0.263 | ..... | 1.34 | ..... | $11.4 \pm 1.1$ | .......... | 152.5 | ..... | $7.5 \pm 0.7$ | .......... |
| 41 | A586 | I211 | 0.145 | ..... | ..... | ..... | ..... | .......... | .......... | ..... | ..... | .......... | .......... |
| 42 | A592 | I5170 | 0.130 | ..... | ..... | ..... | ..... | .......... | .......... | ..... | ..... | .......... | .......... |
| 43 | A629 | I317 | ..... | 0.270 | ..... | 0.83 | ..... | $9.3 \pm 1.3$ | .......... | 161.0 | ..... | $5.8 \pm 0.8$ | .......... |
| 44 | A644 | I5728 | 0.142 | 0.264 | 0.357 | 10.97 | 15.98 | $29.1 \pm 1.9$ | $82.5 \pm 4.4$ | 269.8 | 689.7 | $10.7 \pm 0.7$ | $12.0 \pm 0.6$ |
| 45 | A646 | I1839 | ..... | 0.250 | ..... | 3.50 | ..... | $16.9 \pm 2.0$ | .......... | 251.2 | ..... | $6.8 \pm 0.8$ | .......... |
| 46 | A665 | I305 | ..... | 0.320 | 0.580 | 7.32 | 18.56 | $29.3 \pm 1.5$ | $112.0 \pm 5.4$ | 154.5 | 592.0 | $18.8 \pm 1.1$ | $19.0 \pm 0.9$ |
| 47 | A671 | I7337 | 0.130 | 0.275 | ..... | 0.71 | ..... | $8.5 \pm 0.8$ | .......... | 102.6 | ..... | $8.3 \pm 0.7$ | .......... |
| 48 | A690 | I6020 | 0.107 | 0.222 | ..... | 0.15 | ..... | $3.6 \pm 0.7$ | .......... | 94.7 | ..... | $3.8 \pm 0.8$ | .......... |
| 49 | A732 | I6118 | ..... | 0.342 | 0.562 | 7.57 | 18.12 | $28.4 \pm 2.2$ | $108.3 \pm 9.0$ | 181.0 | 599.8 | $15.1 \pm 1.4$ | $18.1 \pm 1.4$ |
| 50 | A744 | I481 | 0.155 | 0.229 | ..... | 0.50 | ..... | $6.7 \pm 0.8$ | .......... | 128.6 | ..... | $5.2 \pm 0.7$ | .......... |
| 51 | A754 | I1784 | 0.145 | 0.361 | 0.569 | 7.60 | 19.16 | $24.9 \pm 2.3$ | $103.7 \pm 5.4$ | 189.0 | 747.8 | $13.1 \pm 1.2$ | $13.9 \pm 0.7$ |
| | | H1786 | 0.139 | ..... | ..... | ..... | ..... | .......... | .......... | ..... | ..... | .......... | .......... |
| 52 | A779 | I1841 | ..... | ..... | ..... | ..... | ..... | .......... | .......... | ..... | ..... | .......... | .......... |
| 53 | A795 | I212 | 0.100 | 0.300 | 0.433 | 4.76 | 8.08 | $21.6 \pm 2.2$ | $63.9 \pm 7.8$ | 173.3 | 553.9 | $12.5 \pm 1.3$ | $11.5 \pm 1.4$ |
| 54 | A838 | I6097 | 0.103 | ..... | ..... | ..... | ..... | .......... | .......... | ..... | ..... | .......... | .......... |
| 55 | A910 | I1788 | ..... | 0.374 | 0.588 | 14.41 | 38.91 | $39.0 \pm 4.0$ | $154.3 \pm 15.5$ | 158.8 | 625.8 | $22.0 \pm 3.3$ | $24.4 \pm 2.5$ |
| 56 | A957 | I6023 | 0.127 | 0.284 | ..... | 0.50 | ..... | $7.1 \pm 0.8$ | .......... | 95.5 | ..... | $7.4 \pm 0.9$ | .......... |
| 57 | A970 | I7791 | 0.110 | 0.295 | ..... | 2.10 | ..... | $14.4 \pm 1.3$ | .......... | 154.9 | ..... | $9.3 \pm 0.8$ | .......... |
| 58 | A979 | I6098 | 0.122 | ..... | ..... | ..... | ..... | .......... | .......... | ..... | ..... | .......... | .......... |
| 59 | A999 | I7700 | 0.061 | ..... | ..... | ..... | ..... | .......... | .......... | ..... | ..... | .......... | .......... |
| 60 | A1060 | I6114 | 0.121 | ..... | ..... | ..... | ..... | .......... | .......... | ..... | ..... | .......... | .......... |
| 61 | A1142 | I6079 | 0.150 | ..... | ..... | ..... | ..... | .......... | .......... | ..... | ..... | .......... | .......... |
| 62 | A1146 | I217 | ..... | 0.310 | ..... | 3.47 | ..... | $18.7 \pm 2.6$ | .......... | 197.8 | ..... | $9.7 \pm 1.3$ | .......... |
| 63 | A1185 | I6100 | 0.130 | ..... | ..... | ..... | ..... | .......... | .......... | ..... | ..... | .......... | .......... |

Table 4: Deprojection Results Summed To Different Radii

| # | Name | Sequence Number | $R[L_X/2]$ ($10^{44}$ erg s$^{-1}$) ($R < 0.2$ Mpc) | ($R < 0.5$ Mpc) | ($R < 1.0$ Mpc) | $L_X$ ($10^{44}$ erg s$^{-1}$) ($R < 0.5$ Mpc) | ($R < 1.0$ Mpc) | $M_{gas}$ ($10^{12}$ M$_\odot$) ($R < 0.5$ Mpc) | ($R < 1.0$ Mpc) | $M_{grav}$ ($10^{12}$ M$_\odot$) ($R < 0.5$ Mpc) | ($R < 1.0$ Mpc) | $M_{gas}/M_{grav}$ (%) ($R < 0.5$ Mpc) | ($R < 1.0$ Mpc) |
|---|---|---|---|---|---|---|---|---|---|---|---|---|---|
| (i) | (ii) | (iii) | (iv) | (v) | (vi) | (vii) | (viii) | (ix) | (x) | (xi) | (xii) | | |
| 64 | A1213 | I1844 | 0.129 | ..... | ..... | ..... | ..... | .......... | .......... | ..... | ..... | .......... | .......... |
| 65 | A1246 | I233 | ..... | 0.337 | 0.501 | 5.14 | 10.31 | $24.2 \pm 3.5$ | $78.2 \pm 15.3$ | 152.4 | 510.5 | $15.4 \pm 2.4$ | $15.4 \pm 3.0$ |
| 66 | A1254 | I172 | 0.100 | 0.272 | ..... | 0.93 | ..... | $9.4 \pm 2.2$ | .......... | 193.8 | ..... | $4.8 \pm 1.1$ | .......... |
| 67 | A1272 | I331 | ..... | 0.349 | 0.575 | 0.86 | 2.14 | $10.3 \pm 1.7$ | $37.4 \pm 6.8$ | 103.3 | 339.9 | $9.3 \pm 1.8$ | $10.9 \pm 2.0$ |
| 68 | A1285 | I331 | 0.134 | 0.360 | 0.481 | 2.66 | 4.97 | $17.8 \pm 2.1$ | $57.1 \pm 6.1$ | 110.1 | 371.1 | $15.8 \pm 2.0$ | $15.4 \pm 1.7$ |
| 69 | A1291 | I6293 | 0.135 | ..... | ..... | ..... | ..... | .......... | .......... | ..... | ..... | .......... | .......... |
| 70 | A1314 | I6120 | 0.142 | ..... | ..... | ..... | ..... | .......... | .......... | ..... | ..... | .......... | .......... |
| 71 | A1367 | I296 | 0.129 | 0.334 | ..... | 0.74 | ..... | $8.7 \pm 0.8$ | .......... | 96.8 | ..... | $9.0 \pm 0.8$ | .......... |
| 72 | A1377 | I6101 | 0.148 | 0.335 | ..... | 0.34 | ..... | $6.4 \pm 0.7$ | .......... | 72.3 | ..... | $8.8 \pm 1.0$ | .......... |
| 73 | A1413 | I308 | 0.100 | 0.290 | 0.412 | 8.26 | 13.27 | $25.4 \pm 2.2$ | $74.3 \pm 7.1$ | 304.5 | 899.3 | $8.4 \pm 0.7$ | $8.2 \pm 0.8$ |
| 74 | A1546 | I6868 | 0.134 | ..... | ..... | ..... | ..... | .......... | .......... | ..... | ..... | .......... | .......... |
| 75 | A1569S | I1849 | 0.128 | 0.288 | ..... | 0.85 | ..... | $9.5 \pm 1.5$ | .......... | 143.1 | ..... | $6.6 \pm 1.0$ | .......... |
| 76 | A1576 | I6871 | 0.125 | 0.328 | ..... | 6.38 | ..... | $25.2 \pm 3.7$ | .......... | 169.9 | ..... | $14.7 \pm 2.2$ | .......... |
| 77 | A1589 | I145 | 0.110 | 0.376 | ..... | 1.58 | ..... | $13.0 \pm 2.1$ | .......... | 92.7 | ..... | $13.6 \pm 2.3$ | .......... |
| 78 | A1617 | I6875 | 0.124 | ..... | ..... | ..... | ..... | .......... | .......... | ..... | ..... | .......... | .......... |
| 79 | A1631 | I1900 | 0.083 | ..... | ..... | ..... | ..... | .......... | .......... | ..... | ..... | .......... | .......... |
| 80 | A1644 | I7654 | 0.128 | 0.284 | ..... | 2.09 | ..... | $13.3 \pm 0.4$ | .......... | 217.4 | ..... | $6.1 \pm 0.2$ | .......... |
| 81 | A1650 | I6034 | 0.127 | 0.273 | 0.374 | 7.50 | 11.18 | $25.7 \pm 1.7$ | $73.4 \pm 5.3$ | 208.0 | 567.6 | $12.3 \pm 0.8$ | $12.9 \pm 0.9$ |
| 82 | A1656 | I1793 | 0.151 | 0.319 | 0.446 | 5.08 | 8.85 | $20.5 \pm 0.3$ | $67.1 \pm 1.6$ | 213.2 | 625.6 | $9.6 \pm 0.1$ | $10.8 \pm 0.3$ |
| 83 | A1689 | I6123 | 0.100 | 0.277 | 0.393 | 25.07 | 38.08 | $46.4 \pm 0.8$ | $127.2 \pm 4.7$ | 379.9 | 962.1 | $12.1 \pm 0.2$ | $13.2 \pm 0.5$ |
| 84 | A1704 | I6877 | ..... | 0.312 | ..... | 6.66 | ..... | $26.6 \pm 2.5$ | .......... | 166.3 | ..... | $16.0 \pm 1.4$ | .......... |
| 85 | A1709 | I8996 | 0.112 | ..... | ..... | ..... | ..... | .......... | .......... | ..... | ..... | .......... | .......... |
| 86 | A1736 | I7653 | 0.146 | 0.342 | 0.525 | 1.28 | 2.68 | $11.4 \pm 1.0$ | $42.1 \pm 3.3$ | 102.8 | 378.1 | $11.1 \pm 1.0$ | $11.1 \pm 0.9$ |
| 87 | A1750N | I144 | 0.117 | 0.328 | ..... | 1.13 | ..... | $11.1 \pm 1.2$ | .......... | 119.5 | ..... | $9.5 \pm 1.0$ | .......... |
| 88 | A1763 | I3930 | ..... | 0.338 | 0.517 | 8.92 | 18.79 | $31.5 \pm 1.4$ | $106.5 \pm 6.2$ | 156.6 | 564.4 | $19.5 \pm 1.0$ | $19.0 \pm 1.1$ |
| 89 | A1767 | I5731 | 0.144 | 0.309 | ..... | 2.21 | ..... | $15.2 \pm 1.2$ | .......... | 135.4 | ..... | $11.3 \pm 0.9$ | .......... |
| 90 | A1775 | I320 | 0.149 | 0.257 | 0.373 | 2.31 | 3.64 | $14.1 \pm 1.4$ | $43.4 \pm 4.5$ | 166.6 | 476.7 | $8.5 \pm 0.8$ | $9.1 \pm 0.9$ |
| 91 | A1795 | I293 | 0.120 | 0.216 | 0.294 | 11.38 | 15.42 | $28.8 \pm 1.9$ | $78.5 \pm 4.5$ | 202.9 | 391.3 | $14.1 \pm 0.9$ | $20.0 \pm 1.1$ |
| | | H7881 | 0.108 | ..... | ..... | ..... | ..... | .......... | .......... | ..... | ..... | .......... | .......... |
| 92 | A1809 | I142 | 0.154 | 0.295 | 0.426 | 1.29 | 2.26 | $11.7 \pm 1.4$ | $37.7 \pm 4.5$ | 99.0 | 329.0 | $12.0 \pm 1.4$ | $11.4 \pm 1.4$ |
| 93 | A1837 | I141 | 0.122 | ..... | ..... | ..... | ..... | .......... | .......... | ..... | ..... | .......... | .......... |
| 94 | A1839 | I6037 | ..... | 0.315 | ..... | 1.38 | ..... | $12.9 \pm 2.0$ | .......... | 87.8 | ..... | $14.8 \pm 2.3$ | .......... |
| 95 | A1877 | I6883 | 0.121 | ..... | ..... | ..... | ..... | .......... | .......... | ..... | ..... | .......... | .......... |
| 96 | A1890 | I165 | 0.099 | ..... | ..... | ..... | ..... | .......... | .......... | ..... | ..... | .......... | .......... |
| 97 | A1913 | I6077 | 0.120 | 0.351 | ..... | 0.21 | ..... | $4.8 \pm 0.8$ | .......... | 49.9 | ..... | $9.6 \pm 1.6$ | .......... |

Table 4: Deprojection Results Summed To Different Radii

| # | Name | Sequence Number | $R[L_X/2]$ ($10^{44}$ erg s$^{-1}$) ($R < 0.2$ Mpc) | $R[L_X/2]$ ($R < 0.5$ Mpc) | $R[L_X/2]$ ($R < 1.0$ Mpc) | $L_X$ ($10^{44}$ erg s$^{-1}$) ($R < 0.5$ Mpc) | $L_X$ ($R < 1.0$ Mpc) | $M_{gas}$ ($10^{12}$ M$_\odot$) ($R < 0.5$ Mpc) | $M_{gas}$ ($R < 1.0$ Mpc) | $M_{grav}$ ($10^{12}$ M$_\odot$) ($R < 0.5$ Mpc) | $M_{grav}$ ($R < 1.0$ Mpc) | $M_{gas}/M_{grav}$ (%) ($R < 0.5$ Mpc) | $M_{gas}/M_{grav}$ (%) ($R < 1.0$ Mpc) |
|---|---|---|---|---|---|---|---|---|---|---|---|---|---|
| (i) | (ii) | (iii) | (iv) | (v) | (vi) | (vii) | (viii) | (ix) | (x) | (xi) | (xii) | | |
| 98 | A1940 | I6124 | ..... | 0.288 | 0.624 | 0.66 | 2.06 | $8.2 \pm 1.3$ | $40.3 \pm 7.1$ | 91.8 | 223.7 | $8.5 \pm 1.4$ | $18.0 \pm 3.2$ |
| 99 | A1942 | I6847 | 0.129 | ..... | ..... | ..... | ..... | .......... | .......... | ..... | ..... | .......... | .......... |
| 100 | A1983 | I4190 | 0.116 | ..... | ..... | ..... | ..... | .......... | .......... | ..... | ..... | .......... | .......... |
| 101 | A1991 | I6039 | 0.113 | 0.188 | ..... | 2.35 | ..... | $12.0 \pm 0.9$ | .......... | 282.8 | ..... | $4.2 \pm 0.3$ | .......... |
| 102 | A2009 | I3582 | ..... | 0.273 | 0.356 | 20.62 | 28.21 | $39.4 \pm 0.9$ | $106.3 \pm 4.3$ | 373.6 | 657.1 | $10.5 \pm 0.2$ | $16.2 \pm 0.7$ |
| 103 | A2022 | I4191 | 0.128 | ..... | ..... | ..... | ..... | .......... | .......... | ..... | ..... | .......... | .......... |
| 104 | A2029 | I138 | 0.127 | 0.233 | 0.299 | 22.46 | 30.37 | $38.8 \pm 2.7$ | $102.4 \pm 6.6$ | 350.4 | 779.4 | $11.0 \pm 0.8$ | $13.1 \pm 0.8$ |
| | | H7882 | 0.105 | ..... | ..... | ..... | ..... | .......... | .......... | ..... | ..... | .......... | .......... |
| 105 | A2040 | I6104 | 0.128 | 0.294 | ..... | 0.27 | ..... | $5.3 \pm 0.6$ | .......... | 74.3 | ..... | $7.0 \pm 0.8$ | .......... |
| 106 | A2050 | I3444 | ..... | 0.310 | 0.419 | 3.28 | 5.32 | $18.9 \pm 1.8$ | $52.9 \pm 5.4$ | 154.2 | 483.5 | $12.3 \pm 1.2$ | $10.9 \pm 1.1$ |
| 107 | A2052 | I1853 | 0.112 | 0.195 | ..... | 2.50 | ..... | $13.2 \pm 1.1$ | .......... | 125.2 | ..... | $10.6 \pm 0.8$ | .......... |
| | | H5728 | ..... | ..... | ..... | ..... | ..... | .......... | .......... | ..... | ..... | .......... | .......... |
| 108 | A2055 | I137 | 0.103 | 0.139 | ..... | 1.71 | ..... | $8.8 \pm 1.1$ | .......... | 305.7 | ..... | $2.9 \pm 0.3$ | .......... |
| 109 | A2063 | H4595 | ..... | ..... | ..... | ..... | ..... | .......... | .......... | ..... | ..... | .......... | .......... |
| | | I162 | 0.122 | ..... | ..... | ..... | ..... | .......... | .......... | ..... | ..... | .......... | .......... |
| 110 | A2065 | I1795 | 0.147 | 0.290 | 0.426 | 1.54 | 2.58 | $10.8 \pm 0.9$ | $34.6 \pm 2.6$ | 286.8 | 735.6 | $3.7 \pm 0.3$ | $4.7 \pm 0.4$ |
| 111 | A2069E | I10404 | ..... | 0.365 | 0.609 | 1.78 | 5.22 | $13.1 \pm 2.2$ | $55.4 \pm 6.4$ | 105.9 | 420.0 | $11.8 \pm 2.2$ | $13.2 \pm 1.5$ |
| 112 | A2069W | I10404 | ..... | 0.348 | 0.620 | 1.71 | 5.11 | $12.8 \pm 2.1$ | $56.0 \pm 5.9$ | 102.7 | 407.0 | $12.1 \pm 2.1$ | $13.8 \pm 1.4$ |
| 113 | A2079 | I1854 | 0.134 | 0.343 | ..... | 0.50 | ..... | $7.5 \pm 1.2$ | .......... | 89.6 | ..... | $8.3 \pm 1.3$ | .......... |
| 114 | A2092 | I135 | 0.144 | 0.233 | ..... | 0.40 | ..... | $5.8 \pm 0.9$ | .......... | 99.7 | ..... | $5.8 \pm 0.9$ | .......... |
| 115 | A2107 | I134 | 0.141 | 0.250 | ..... | 1.46 | ..... | $11.1 \pm 0.9$ | .......... | 202.4 | ..... | $5.5 \pm 0.5$ | .......... |
| 116 | A2111 | I239 | 0.128 | 0.338 | ..... | 3.24 | ..... | $17.7 \pm 2.9$ | .......... | 135.7 | ..... | $13.2 \pm 2.1$ | .......... |
| 117 | A2124 | I4192 | 0.142 | ..... | ..... | ..... | ..... | .......... | .......... | ..... | ..... | .......... | .......... |
| 118 | A2142 | H1800 | 0.128 | 0.258 | ..... | 25.50 | ..... | $41.3 \pm 1.7$ | .......... | 332.6 | ..... | $12.3 \pm 0.5$ | .......... |
| | | I1798 | 0.147 | 0.286 | 0.411 | 23.30 | 38.32 | $39.8 \pm 1.0$ | $124.1 \pm 3.8$ | 366.4 | 1014.1 | $10.9 \pm 0.3$ | $12.3 \pm 0.4$ |
| 119 | A2147 | I297 | 0.148 | 0.296 | ..... | 1.28 | ..... | $10.4 \pm 1.1$ | .......... | 155.1 | ..... | $6.7 \pm 0.7$ | .......... |
| 120 | A2151 | H9264 | 0.091 | ..... | ..... | ..... | ..... | .......... | .......... | ..... | ..... | .......... | .......... |
| | | I1801 | 0.107 | ..... | ..... | ..... | ..... | .......... | .......... | ..... | ..... | .......... | .......... |
| 121 | A2152 | I1855 | 0.117 | ..... | ..... | ..... | ..... | .......... | .......... | ..... | ..... | .......... | .......... |
| 122 | A2163 | I4526 | ..... | 0.324 | 0.514 | 23.89 | 49.72 | $41.6 \pm 4.0$ | $148.7 \pm 12.3$ | 347.9 | 1261.2 | $12.0 \pm 1.2$ | $11.7 \pm 1.0$ |
| 123 | A2197 | I1857 | 0.103 | 0.241 | ..... | 0.11 | ..... | $3.2 \pm 0.4$ | .......... | 67.7 | ..... | $4.7 \pm 0.6$ | .......... |
| 124 | A2199 | I4193 | 0.112 | 0.199 | ..... | 4.66 | ..... | $17.6 \pm 1.2$ | .......... | 211.7 | ..... | $8.3 \pm 0.5$ | .......... |
| | | H4597 | ..... | ..... | ..... | ..... | ..... | .......... | .......... | ..... | ..... | .......... | .......... |
| 125 | A2218 | H3160 | 0.130 | 0.281 | ..... | 7.97 | ..... | $24.8 \pm 2.5$ | .......... | 273.8 | ..... | $8.9 \pm 0.9$ | .......... |
| 126 | A2220 | I8351 | 0.135 | ..... | ..... | ..... | ..... | .......... | .......... | ..... | ..... | .......... | .......... |

Table 4: Deprojection Results Summed To Different Radii

| # | Name | Sequence Number | $R[L_X/2]$ ($10^{44}$ erg s$^{-1}$) ($R<0.2$ Mpc) | ($R<0.5$ Mpc) | ($R<1.0$ Mpc) | $L_X$ ($10^{44}$ erg s$^{-1}$) ($R<0.5$ Mpc) | ($R<1.0$ Mpc) | $M_{gas}$ ($10^{12}$ M$_\odot$) ($R<0.5$ Mpc) | ($R<1.0$ Mpc) | $M_{grav}$ ($10^{12}$ M$_\odot$) ($R<0.5$ Mpc) | ($R<1.0$ Mpc) | $M_{gas}/M_{grav}$ (%) ($R<0.5$ Mpc) | ($R<1.0$ Mpc) |
|---|---|---|---|---|---|---|---|---|---|---|---|---|---|
| (i) | (ii) | (iii) | (iv) | (v) | (vi) | (vii) | (viii) | (ix) | (x) | (xi) | (xii) | | |
| 127 | A2244 | H10190 | 0.139 | ..... | ..... | ..... | ..... | .......... | .......... | ..... | ..... | .......... | .......... |
| | | | 0.110 | 0.264 | 0.352 | 9.21 | 13.32 | $26.8 \pm 0.6$ | $69.9 \pm 4.2$ | 245.8 | 716.1 | $11.0 \pm 0.3$ | $9.8 \pm 0.6$ |
| 128 | A2250 | I3090 | 0.095 | ..... | ..... | ..... | ..... | .......... | .......... | ..... | ..... | .......... | .......... |
| 129 | A2255 | I160 | 0.153 | 0.344 | 0.588 | 2.45 | 6.58 | $15.2 \pm 2.0$ | $64.8 \pm 4.2$ | 111.0 | 501.8 | $13.5 \pm 1.9$ | $12.9 \pm 0.9$ |
| 130 | A2256 | I300 | 0.129 | 0.347 | 0.537 | 5.99 | 13.31 | $22.8 \pm 2.3$ | $86.4 \pm 5.9$ | 146.6 | 640.3 | $15.7 \pm 1.6$ | $13.5 \pm 0.9$ |
| | | H10189 | 0.150 | ..... | ..... | ..... | ..... | .......... | .......... | ..... | ..... | .......... | .......... |
| 131 | A2271 | I6042 | 0.133 | 0.260 | ..... | 0.56 | ..... | $7.4 \pm 1.0$ | .......... | 101.8 | ..... | $7.3 \pm 1.0$ | .......... |
| 132 | A2306 | I7696 | ..... | 0.298 | 0.565 | 0.66 | 1.46 | $8.0 \pm 1.2$ | $29.9 \pm 4.1$ | 92.2 | 318.0 | $9.1 \pm 1.2$ | $9.4 \pm 1.3$ |
| 133 | A2312 | I6269 | 0.130 | 0.226 | ..... | 1.50 | ..... | $11.1 \pm 1.3$ | .......... | 151.3 | ..... | $7.3 \pm 0.9$ | .......... |
| 134 | A2319 | I3456 | 0.139 | 0.308 | 0.505 | 14.84 | 30.10 | $31.5 \pm 2.8$ | $117.2 \pm 7.8$ | 317.6 | 944.9 | $9.9 \pm 0.9$ | $12.4 \pm 0.8$ |
| 135 | A2328 | I6271 | ..... | 0.340 | 0.532 | 2.58 | 5.60 | $16.4 \pm 2.4$ | $59.9 \pm 9.2$ | 118.1 | 438.6 | $14.0 \pm 2.0$ | $13.7 \pm 2.1$ |
| 136 | A2356 | I7802 | ..... | 0.295 | ..... | 1.18 | ..... | $10.8 \pm 1.9$ | .......... | 127.9 | ..... | $8.5 \pm 1.5$ | .......... |
| 137 | A2366 | I133 | 0.116 | ..... | ..... | ..... | ..... | .......... | .......... | ..... | ..... | .......... | .......... |
| 138 | A2384N | I7805 | 0.124 | 0.305 | ..... | 2.65 | ..... | $16.7 \pm 1.8$ | .......... | 166.7 | ..... | $10.0 \pm 1.1$ | .......... |
| 139 | A2390 | I9125 | ..... | 0.272 | 0.357 | 24.40 | 35.19 | $42.4 \pm 2.6$ | $112.4 \pm 9.1$ | 373.0 | 827.5 | $11.3 \pm 0.7$ | $13.6 \pm 1.1$ |
| 140 | A2397 | I242 | 0.128 | 0.321 | ..... | 3.22 | ..... | $18.1 \pm 3.8$ | .......... | 139.5 | ..... | $12.9 \pm 2.7$ | .......... |
| 141 | A2410 | I6071 | 0.112 | 0.284 | 0.576 | 0.42 | 0.95 | $6.5 \pm 0.9$ | $27.3 \pm 3.3$ | 75.5 | 255.9 | $8.7 \pm 1.2$ | $10.7 \pm 1.3$ |
| 142 | A2415 | I130 | 0.118 | 0.231 | ..... | 1.72 | ..... | $11.7 \pm 1.4$ | .......... | 182.4 | ..... | $6.4 \pm 0.8$ | .......... |
| 143 | A2420 | I6045 | 0.152 | 0.315 | 0.430 | 4.10 | 6.81 | $20.1 \pm 0.6$ | $60.1 \pm 4.6$ | 177.1 | 569.3 | $11.4 \pm 0.3$ | $10.6 \pm 0.8$ |
| 144 | A2440 | I129 | ..... | 0.361 | 0.497 | 2.51 | 4.97 | $16.0 \pm 1.2$ | $53.0 \pm 4.0$ | 193.8 | 740.5 | $7.9 \pm 0.7$ | $7.1 \pm 0.5$ |
| 145 | A2554 | I336 | 0.103 | 0.276 | ..... | 1.82 | ..... | $13.4 \pm 1.1$ | .......... | 154.5 | ..... | $8.7 \pm 0.7$ | .......... |
| 146 | A2556 | I336 | 0.114 | 0.240 | 0.292 | 3.77 | 4.87 | $17.7 \pm 1.2$ | $42.8 \pm 3.1$ | 211.2 | 501.3 | $8.4 \pm 0.6$ | $8.5 \pm 0.6$ |
| 147 | A2577 | I1875 | ..... | 0.230 | 0.348 | 3.74 | 5.44 | $17.9 \pm 2.0$ | $47.4 \pm 6.2$ | 226.9 | 575.2 | $7.8 \pm 0.9$ | $8.2 \pm 1.1$ |
| 148 | A2580 | H5751 | 0.104 | 0.272 | ..... | 9.99 | ..... | $28.8 \pm 2.4$ | .......... | 242.7 | ..... | $11.8 \pm 1.0$ | .......... |
| 149 | A2593 | I6134 | 0.137 | 0.283 | 0.432 | 0.74 | 1.35 | $8.3 \pm 0.9$ | $29.1 \pm 2.7$ | 87.1 | 286.7 | $9.6 \pm 1.0$ | $10.2 \pm 0.9$ |
| 150 | A2625 | I156 | 0.102 | ..... | ..... | ..... | ..... | .......... | .......... | ..... | ..... | .......... | .......... |
| 151 | A2626 | I201 | 0.118 | 0.229 | ..... | 1.88 | ..... | $13.2 \pm 1.2$ | .......... | 149.8 | ..... | $8.7 \pm 0.8$ | .......... |
| 152 | A2634 | I199 | 0.132 | 0.320 | ..... | 0.40 | ..... | $6.4 \pm 0.5$ | .......... | 96.1 | ..... | $6.7 \pm 0.5$ | .......... |
| 153 | A2657 | I290 | 0.127 | 0.255 | ..... | 1.86 | ..... | $13.3 \pm 1.0$ | .......... | 126.9 | ..... | $10.4 \pm 0.8$ | .......... |
| 154 | A2666 | I294 | ..... | ..... | ..... | ..... | ..... | .......... | .......... | ..... | ..... | .......... | .......... |
| 155 | A2670 | I314 | 0.127 | 0.258 | ..... | 2.38 | ..... | $14.4 \pm 1.5$ | .......... | 190.6 | ..... | $7.7 \pm 0.8$ | .......... |
| 156 | A2703 | I5360 | 0.101 | 0.286 | 0.530 | 1.10 | 2.33 | $10.8 \pm 1.4$ | $40.7 \pm 4.8$ | 100.7 | 328.8 | $10.8 \pm 1.4$ | $12.4 \pm 1.4$ |
| 157 | A2715 | I4517 | 0.113 | ..... | ..... | ..... | ..... | .......... | .......... | ..... | ..... | .......... | .......... |
| 158 | A2877 | I6088 | 0.118 | ..... | ..... | ..... | ..... | .......... | .......... | ..... | ..... | .......... | .......... |
| 159 | A3158 | H5753 | 0.137 | ..... | ..... | ..... | ..... | .......... | .......... | ..... | ..... | .......... | .......... |

Table 4: Deprojection Results Summed To Different Radii

| # | Name | Sequence Number | $R[L_X/2]$ ($10^{44}$ erg s$^{-1}$) ($R<0.2$ Mpc) | ($R<0.5$ Mpc) | ($R<1.0$ Mpc) | $L_X$ ($10^{44}$ erg s$^{-1}$) ($R<0.5$ Mpc) | ($R<1.0$ Mpc) | $M_{gas}$ ($10^{12}$ M$_\odot$) ($R<0.5$ Mpc) | ($R<1.0$ Mpc) | $M_{grav}$ ($10^{12}$ M$_\odot$) ($R<0.5$ Mpc) | ($R<1.0$ Mpc) | $M_{gas}/M_{grav}$ (%) ($R<0.5$ Mpc) | ($R<1.0$ Mpc) |
|---|---|---|---|---|---|---|---|---|---|---|---|---|---|
| (i) | (ii) | (iii) | (iv) | (v) | (vi) | (vii) | (viii) | (ix) | (x) | (xi) | (xii) | | |
| | | I1829 | 0.138 | 0.304 | ..... | 4.61 | ..... | $20.8 \pm 0.6$ | .......... | 178.8 | ..... | $11.6 \pm 0.3$ | .......... |
| 160 | A3186 | I8385 | 0.110 | 0.296 | ..... | 4.82 | ..... | $21.0 \pm 2.4$ | .......... | 175.5 | ..... | $12.0 \pm 1.4$ | .......... |
| 161 | A3266 | I1831 | 0.137 | 0.340 | 0.580 | 3.91 | 9.57 | $19.2 \pm 0.5$ | $79.6 \pm 1.9$ | 117.1 | 466.8 | $16.4 \pm 0.4$ | $17.1 \pm 0.4$ |
| 162 | A3322 | I7705 | 0.133 | 0.267 | ..... | 2.15 | ..... | $13.3 \pm 2.4$ | .......... | 211.7 | ..... | $6.3 \pm 1.1$ | .......... |
| 163 | A3376 | I5167 | 0.106 | ..... | ..... | ..... | ..... | .......... | .......... | ..... | ..... | .......... | .......... |
| 164 | A3389E | I5169 | 0.093 | ..... | ..... | ..... | ..... | .......... | .......... | ..... | ..... | .......... | .......... |
| 165 | A3389W | I5169 | 0.134 | ..... | ..... | ..... | ..... | .......... | .......... | ..... | ..... | .......... | .......... |
| 166 | A3391 | I8309 | 0.138 | 0.310 | ..... | 1.66 | ..... | $12.8 \pm 0.5$ | .......... | 180.9 | ..... | $7.1 \pm 0.3$ | .......... |
| 167 | A3532 | I6173 | 0.140 | 0.335 | ..... | 2.03 | ..... | $14.4 \pm 1.6$ | .......... | 115.2 | ..... | $12.5 \pm 1.4$ | .......... |
| 168 | A3562W | I5730 | 0.131 | ..... | ..... | ..... | ..... | .......... | .......... | ..... | ..... | .......... | .......... |
| 169 | A3581 | I9980 | 0.077 | ..... | ..... | ..... | ..... | .......... | .......... | ..... | ..... | .......... | .......... |
| 170 | A3602 | I5252 | 0.106 | 0.252 | ..... | 2.09 | ..... | $13.8 \pm 2.2$ | .......... | 182.7 | ..... | $7.6 \pm 1.2$ | .......... |
| 171 | A3654 | I3289 | ..... | 0.290 | ..... | 1.43 | ..... | $12.4 \pm 1.2$ | .......... | 115.8 | ..... | $10.8 \pm 1.0$ | .......... |
| 172 | A3667 | I5735 | 0.156 | 0.331 | 0.536 | 6.58 | 14.53 | $24.6 \pm 2.1$ | $90.6 \pm 7.2$ | 135.1 | 509.2 | $18.2 \pm 1.6$ | $17.7 \pm 1.4$ |
| 173 | A370 | I245 | ..... | 0.236 | 0.617 | 17.44 | 43.48 | $32.1 \pm 5.5$ | $154.5 \pm 20.5$ | 421.9 | 752.2 | $7.2 \pm 1.2$ | $20.2 \pm 2.6$ |
| 174 | A3744 | I3044 | 0.115 | ..... | ..... | ..... | ..... | .......... | .......... | ..... | ..... | .......... | .......... |
| 175 | A376 | I1773 | 0.093 | ..... | ..... | ..... | ..... | .......... | .......... | ..... | ..... | .......... | .......... |
| 176 | A3888 | I1872 | ..... | 0.321 | 0.529 | 11.48 | 24.79 | $30.4 \pm 3.7$ | $112.3 \pm 10.8$ | 277.1 | 800.9 | $10.8 \pm 1.4$ | $14.0 \pm 1.3$ |
| 177 | A389 | I6128 | 0.100 | 0.266 | ..... | 2.14 | ..... | $14.5 \pm 0.7$ | .......... | 187.4 | ..... | $7.8 \pm 0.4$ | .......... |
| 178 | A3998 | I6385 | 0.100 | 0.264 | ..... | 1.68 | ..... | $11.6 \pm 1.3$ | .......... | 195.3 | ..... | $5.8 \pm 0.7$ | .......... |
| 179 | A4067S | I5745 | 0.103 | 0.326 | 0.488 | 1.19 | 2.29 | $11.4 \pm 1.3$ | $39.7 \pm 5.5$ | 119.4 | 296.2 | $9.3 \pm 1.1$ | $13.4 \pm 1.9$ |
| 180 | AWM4 | I10543 | 0.132 | 0.208 | 0.323 | 0.41 | 0.67 | $5.4 \pm 0.5$ | $18.5 \pm 1.9$ | 156.5 | 394.2 | $3.5 \pm 0.3$ | $4.7 \pm 0.5$ |
| 181 | AWM5 | I3302 | 0.125 | ..... | ..... | ..... | ..... | .......... | .......... | ..... | ..... | .......... | .......... |
| 182 | AWM7 | I6698 | 0.124 | 0.239 | ..... | 2.49 | ..... | $14.7 \pm 1.1$ | .......... | 180.3 | ..... | $8.2 \pm 0.6$ | .......... |
| | | H6638 | ..... | ..... | ..... | ..... | ..... | .......... | .......... | ..... | ..... | .......... | .......... |
| 183 | CENTAURUS | I298 | 0.094 | 0.209 | ..... | 1.44 | ..... | $10.3 \pm 0.8$ | .......... | 129.4 | ..... | $8.0 \pm 0.6$ | .......... |
| | | H4341 | ..... | ..... | ..... | ..... | ..... | .......... | .......... | ..... | ..... | .......... | .......... |
| 184 | CL0016+16 | H7755 | 0.142 | ..... | ..... | ..... | ..... | .......... | .......... | ..... | ..... | .......... | .......... |
| 185 | CYGNUS-A | I1807 | 0.112 | 0.193 | 0.278 | 8.51 | 11.77 | $23.7 \pm 1.9$ | $70.9 \pm 6.1$ | 216.9 | 418.3 | $10.8 \pm 0.9$ | $16.9 \pm 1.4$ |
| | | H10760 | 0.075 | ..... | ..... | ..... | ..... | .......... | .......... | ..... | ..... | .......... | .......... |
| 186 | FORNAX-A | H1885 | ..... | ..... | ..... | ..... | ..... | .......... | .......... | ..... | ..... | .......... | .......... |
| 187 | HERCULES-A | I10533 | ..... | 0.265 | ..... | 5.33 | ..... | $22.3 \pm 1.4$ | .......... | 259.3 | ..... | $8.6 \pm 0.5$ | .......... |
| 188 | HYDRA-A | I1894 | 0.113 | 0.202 | 0.255 | 6.27 | 7.79 | $21.3 \pm 1.3$ | $50.9 \pm 3.2$ | 187.0 | 411.4 | $11.4 \pm 0.7$ | $12.4 \pm 0.8$ |
| 189 | M87 | H282 | ..... | ..... | ..... | ..... | ..... | .......... | .......... | ..... | ..... | .......... | .......... |
| | | I10362 | 0.086 | ..... | ..... | ..... | ..... | .......... | .......... | ..... | ..... | .......... | .......... |

Table 4: Deprojection Results Summed To Different Radii

| # | Name | Sequence Number | $R[L_X/2]$ ($10^{44}$ erg s$^{-1}$) ($R<0.2$ Mpc) | ($R<0.5$ Mpc) | ($R<1.0$ Mpc) | $L_X$ ($10^{44}$ erg s$^{-1}$) ($R<0.5$ Mpc) | ($R<1.0$ Mpc) | $M_{gas}$ ($10^{12}$ M$_\odot$) ($R<0.5$ Mpc) | ($R<1.0$ Mpc) | $M_{grav}$ ($10^{12}$ M$_\odot$) ($R<0.5$ Mpc) | ($R<1.0$ Mpc) | $M_{gas}/M_{grav}$ (%) ($R<0.5$ Mpc) | ($R<1.0$ Mpc) |
|---|---|---|---|---|---|---|---|---|---|---|---|---|---|
| (i) | (ii) | (iii) | (iv) | (v) | (vi) | (vii) | (viii) | (ix) | (x) | (xi) | (xii) | | |
| 190 | MKW3S | I2604 | 0.119 | 0.194 | ..... | 3.08 | ..... | $14.5 \pm 1.2$ | .......... | 156.6 | ..... | $9.3 \pm 0.8$ | .......... |
| | | H4359 | 0.111 | ..... | ..... | ..... | ..... | .......... | .......... | ..... | ..... | .......... | .......... |
| 191 | MKW4 | I2601 | 0.080 | ..... | ..... | ..... | ..... | .......... | .......... | ..... | ..... | .......... | .......... |
| 192 | OPHIUCHUS | H6553 | 0.129 | ..... | ..... | ..... | ..... | .......... | .......... | ..... | ..... | .......... | .......... |
| 193 | PKS0745-19 | H6541 | 0.113 | ..... | ..... | ..... | ..... | .......... | .......... | ..... | ..... | .......... | .......... |
| 194 | SC1842-63 | I6105 | 0.117 | ..... | ..... | ..... | ..... | .......... | .......... | ..... | ..... | .......... | .......... |
| 195 | SC2316-3632 | I7569 | ..... | 0.319 | ..... | 2.51 | ..... | $16.6 \pm 2.2$ | .......... | 167.6 | ..... | $9.7 \pm 1.3$ | .......... |
| 196 | SERSIC159-03 | I1874 | 0.055 | ..... | ..... | ..... | ..... | .......... | .......... | ..... | ..... | .......... | .......... |
| 197 | ZW0258+43 | I4611 | 0.120 | 0.372 | ..... | 0.26 | ..... | $5.3 \pm 1.1$ | .......... | 74.4 | ..... | $7.0 \pm 1.5$ | .......... |
| 198 | ZW0628+25 | I4613 | 0.119 | 0.256 | ..... | 2.37 | ..... | $13.2 \pm 1.6$ | .......... | 253.3 | ..... | $5.2 \pm 0.6$ | .......... |
| 199 | ZW0712+53 | I4620 | 0.130 | ..... | ..... | ..... | ..... | .......... | .......... | ..... | ..... | .......... | .......... |
| 200 | ZW1615+35 | I322 | 0.109 | ..... | ..... | ..... | ..... | .......... | .......... | ..... | ..... | .......... | .......... |
| 201 | 3C129 | I1832 | 0.142 | 0.292 | ..... | 0.69 | ..... | $7.4 \pm 0.6$ | .......... | 201.4 | ..... | $3.7 \pm 0.3$ | .......... |
| 202 | 3C130 | I3924 | 0.104 | 0.352 | ..... | 2.23 | ..... | $14.9 \pm 2.7$ | .......... | 130.9 | ..... | $10.9 \pm 2.1$ | .......... |
| 203 | 3C295 | H4290 | 0.092 | ..... | ..... | ..... | ..... | .......... | .......... | ..... | ..... | .......... | .......... |
| 204 | 3C330 | I272 | 0.148 | 0.311 | ..... | 0.73 | ..... | $9.0 \pm 1.1$ | .......... | 131.5 | ..... | $6.7 \pm 0.8$ | .......... |
| 205 | 3C370 | I1907 | 0.120 | 0.225 | ..... | 0.31 | ..... | $5.3 \pm 0.7$ | .......... | 98.6 | ..... | $5.4 \pm 0.7$ | .......... |
| 206 | 3C411 | I6833 | 0.122 | ..... | ..... | ..... | ..... | .......... | .......... | ..... | ..... | .......... | .......... |
| 207 | 3C449 | I3916 | ..... | ..... | ..... | ..... | ..... | .......... | .......... | ..... | ..... | .......... | .......... |

Table 5: Catalogue Of Cooling Flows Detected In The Einstein Observatory Deprojections

| Rank (i) | # (ii) | Name (iii) | Sequence (iv) | $r_{\rm cool}$ ( kpc) (v) | $\dot{M}$ ( $M_{\odot}$ yr$^{-1}$ ) (vi) |
|---|---|---|---|---|---|
| 1 | 31 | A478 | I303 | $240^{+158}_{-161}$ | $736^{+114}_{-434}$ |
| 2 | 193 | PKS0745-19 | H6541 | $177^{+31}_{-44}$ | $579^{+399}_{-215}$ |
| 3 | 104 | A2029 | I138 | $192^{+160}_{-121}$ | $431^{+236}_{-288}$ |
| 4 | 203 | 3C295 | H4290 | $181^{+60}_{-78}$ | $410^{+125}_{-130}$ |
| 5 | 118 | A2142 | H1800 | $172^{+52}_{-38}$ | $369^{+130}_{-87}$ |
| 6 | 91 | A1795 | I293 | $181^{+112}_{-122}$ | $321^{+166}_{-213}$ |
| 7 | 8 | A115N | I209 | $153^{+97}_{-103}$ | $313^{+369}_{-186}$ |
| $1^b$ | 31 | A478 | H4198 | $144^{+143}_{-33}$ | $299^{+498}_{-117}$ |
| $3^b$ | 104 | A2029 | H7882 | $144^{+27}_{-22}$ | $298^{+85}_{-98}$ |
| 8 | 9 | A115S | I209 | $> 234^{+0}_{->134}$ | $> 294^{+0}_{->182}$ |
| 9 | 29 | A426 | H285 | $123^{+>14}_{-7}$ | $291^{+>-7}_{-58}$ |
| 10 | 196 | SERSIC159-03 | I1874 | $> 229^{+0}_{->141}$ | $> 288^{+0}_{->130}$ |
| $9^b$ | 29 | A426 | I283 | $145^{+25}_{-24}$ | $283^{+14}_{-12}$ |
| 11 | 185 | CYGNUS-A | I1807 | $167^{+104}_{-113}$ | $242^{+69}_{-142}$ |
| $6^b$ | 91 | A1795 | H7881 | $155^{+121}_{-41}$ | $238^{+344}_{-84}$ |
| 12 | 188 | HYDRA-A | I1894 | $170^{+80}_{-120}$ | $222^{+98}_{-132}$ |
| 13 | 127 | A2244 | H10190 | $182^{+51}_{-164}$ | $205^{+160}_{-202}$ |
| 14 | 120 | A2151 | I1801 | $146^{+36}_{-37}$ | $166^{+51}_{-41}$ |
| $13^b$ | 127 | A2244 | | $116^{+143}_{-30}$ | $155^{+121}_{-31}$ |
| $11^b$ | 185 | CYGNUS-A | H10760 | $113^{+23}_{-37}$ | $150^{+42}_{-22}$ |
| 15 | 41 | A586 | I211 | $118^{+185}_{-118}$ | $138^{+347}_{-138}$ |
| 16 | 44 | A644 | I5728 | $111^{+86}_{-111}$ | $136^{+161}_{-136}$ |
| 17 | 32 | A496 | I2348 | $138^{+79}_{-95}$ | $134^{+58}_{-85}$ |
| 18 | 190 | MKW3S | I2604 | $158^{+53}_{-31}$ | $132^{+68}_{-45}$ |
| 19 | 81 | A1650 | I6034 | $110^{+119}_{-110}$ | $122^{+168}_{-122}$ |
| 20 | 74 | A1546 | I6868 | $144^{+123}_{-144}$ | $121^{+141}_{-121}$ |
| $18^b$ | 190 | MKW3S | H4359 | $148^{+>75}_{-89}$ | $121^{+>54}_{-79}$ |
| 21 | 95 | A1877 | I6883 | $156^{+>91}_{-50}$ | $117^{+>11}_{-58}$ |
| 22 | 12 | A133 | I2333 | $148^{+137}_{-91}$ | $110^{+71}_{-67}$ |
| 23 | 5 | A85 | H6013 | $131^{+21}_{-62}$ | $108^{+58}_{-61}$ |
| $5^b$ | 118 | A2142 | I1798 | $77^{+125}_{-77}$ | $106^{+248}_{-106}$ |
| 24 | 157 | A2715 | I4517 | $109^{+198}_{-65}$ | $100^{+126}_{-64}$ |
| 25 | 124 | A2199 | H4597 | $124^{+37}_{-37}$ | $97^{+9}_{-31}$ |
| 26 | 148 | A2580 | H5751 | $91^{+73}_{-36}$ | $95^{+71}_{-34}$ |
| 27 | 107 | A2052 | I1853 | $140^{+100}_{-38}$ | $94^{+84}_{-37}$ |
| $25^b$ | 124 | A2199 | I4193 | $119^{+30}_{-30}$ | $94^{+45}_{-35}$ |
| 28 | 184 | CL0016+16 | H7755 | $106^{+226}_{-106}$ | $88^{+497}_{-88}$ |
| 29 | 146 | A2556 | I336 | $101^{+134}_{-101}$ | $81^{+105}_{-81}$ |
| $23^b$ | 5 | A85 | I292 | $105^{+145}_{-55}$ | $81^{+103}_{-54}$ |
| 30 | 125 | A2218 | H3160 | $85^{+84}_{-29}$ | $66^{+76}_{-30}$ |
| $17^b$ | 32 | A496 | H10401 | $101^{+69}_{-22}$ | $56^{+76}_{-13}$ |
| $27^b$ | 107 | A2052 | H5728 | $94^{+11}_{-8}$ | $54^{+17}_{-16}$ |
| 31 | 151 | A2626 | I201 | $114^{+50}_{-59}$ | $53^{+36}_{-30}$ |
| 32 | 182 | AWM7 | H6638 | $> 114^{+0}_{->89}$ | $> 48^{+0}_{->47}$ |
| $32^b$ | 182 | AWM7 | I6698 | $111^{+57}_{-87}$ | $45^{+50}_{-37}$ |
| 33 | 153 | A2657 | I290 | $101^{+61}_{-47}$ | $44^{+36}_{-24}$ |
| 34 | 175 | A376 | I1773 | $84^{+125}_{-42}$ | $42^{+42}_{-14}$ |
| 35 | 155 | A2670 | I314 | $80^{+129}_{-80}$ | $41^{+71}_{-41}$ |
| 36 | 192 | OPHIUCHUS | H6553 | $68^{+95}_{-45}$ | $41^{+110}_{-28}$ |
| 37 | 101 | A1991 | I6039 | $73^{+94}_{-17}$ | $37^{+36}_{-11}$ |

Table 5: Catalogue Of Cooling Flows Detected In The Einstein Observatory Deprojections

| Rank (i) | # (ii) | Name (iii) | Sequence (iv) | $r_{\rm cool}$ ( kpc) (v) | $\dot{M}$ ( $M_{\odot}$ yr$^{-1}$ ) (vi) |
|---|---|---|---|---|---|
| 38 | 133 | A2312 | I6269 | $91^{+93}_{-91}$ | $37^{+48}_{-37}$ |
| 39 | 108 | A2055 | I137 | $87^{+65}_{-37}$ | $36^{+17}_{-12}$ |
| 40 | 109 | A2063 | I162 | $87^{+18}_{-52}$ | $35^{+10}_{-19}$ |
| 41 | 20 | A278 | I7698 | $81^{+99}_{-45}$ | $33^{+44}_{-14}$ |
| 42 | 142 | A2415 | I130 | $81^{+88}_{-81}$ | $32^{+34}_{-32}$ |
| 43 | 189 | M87 | I10362 | $106^{+16}_{-10}$ | $32^{+1}_{-2}$ |
| 44 | 183 | CENTAURUS | I298 | $86^{+35}_{-53}$ | $26^{+16}_{-13}$ |
| 45 | 164 | A3389E | I5169 | $87^{+46}_{-60}$ | $22^{+21}_{-10}$ |
| 46 | 57 | A970 | I7791 | $64^{+105}_{-64}$ | $20^{+32}_{-20}$ |
| 47 | 121 | A2152 | I1855 | $118^{+66}_{-118}$ | $20^{+13}_{-20}$ |
| 48 | 150 | A2625 | I156 | $79^{+93}_{-79}$ | $19^{+18}_{-19}$ |
| 49 | 169 | A3581 | I9980 | $79^{+38}_{-9}$ | $18^{+4}_{-3}$ |
| 50 | 40 | A576 | I3455 | $69^{+117}_{-69}$ | $17^{+47}_{-17}$ |
| $40^b$ | 109 | A2063 | H4595 | $65^{+>22}_{-17}$ | $16^{+>3}_{-11}$ |
| $44^b$ | 183 | CENTAURUS | H4341 | $72^{+>29}_{-32}$ | $15^{+>14}_{-10}$ |
| 51 | 128 | A2250 | I3090 | $65^{+119}_{-65}$ | $14^{+15}_{-14}$ |
| 52 | 93 | A1837 | I141 | $74^{+110}_{-74}$ | $12^{+29}_{-12}$ |
| 53 | 80 | A1644 | I7654 | $49^{+89}_{-49}$ | $12^{+19}_{-12}$ |
| 54 | 191 | MKW4 | I2601 | $72^{+31}_{-10}$ | $10^{+4}_{-3}$ |
| 55 | 37 | A548S | I7860 | $69^{+134}_{-69}$ | $10^{+32}_{-10}$ |
| 56 | 54 | A838 | I6097 | $93.0^{+53.0}_{-45.0}$ | $9.8^{+7.3}_{-4.2}$ |
| 57 | 159 | A3158 | H5753 | $53.0^{+113.0}_{-53.0}$ | $9.6^{+67.8}_{-9.6}$ |
| 58 | 19 | A262 | I295 | $67.0^{+82.0}_{-17.0}$ | $9.4^{+21.2}_{-4.4}$ |
| 59 | 2 | A74 | I8989 | $85.0^{+166.0}_{-85.0}$ | $9.3^{+1.0}_{-9.3}$ |
| 60 | 60 | A1060 | I6114 | $68.0^{+46.0}_{-68.0}$ | $8.0^{+14.9}_{-8.0}$ |
| 61 | 165 | A3389W | I5169 | $57.0^{+182.0}_{-57.0}$ | $7.9^{+63.0}_{-7.9}$ |
| 62 | 21 | A347 | I302 | $75.0^{+21.0}_{-18.0}$ | $7.8^{+3.5}_{-2.7}$ |
| 63 | 115 | A2107 | I134 | $42.0^{+81.0}_{-42.0}$ | $7.1^{+17.2}_{-7.1}$ |
| $14^b$ | 120 | A2151 | H9264 | $50.0^{+154.0}_{-27.0}$ | $6.3^{+26.3}_{-3.2}$ |
| 64 | 163 | A3376 | I5167 | $53.0^{+88.0}_{-53.0}$ | $6.3^{+9.3}_{-6.3}$ |
| $43^b$ | 189 | M87 | H282 | $> 14.0^{+0.0}_{->2.0}$ | $> 6.2^{+0.0}_{->0.3}$ |
| 65 | 96 | A1890 | I165 | $58.0^{+107.0}_{-58.0}$ | $6.2^{+12.6}_{-6.2}$ |
| 66 | 100 | A1983 | I4190 | $47.0^{+82.0}_{-47.0}$ | $6.0^{+10.8}_{-6.0}$ |
| 67 | 39 | A569 | I1836 | $> 91.0^{+0.0}_{->78.0}$ | $> 5.2^{+0.0}_{->4.2}$ |
| 68 | 28 | A419 | I8993 | $55.0^{+143.0}_{-55.0}$ | $5.2^{+21.2}_{-5.2}$ |
| 69 | 27 | A407 | I1825 | $46.0^{+90.0}_{-46.0}$ | $4.6^{+11.8}_{-4.6}$ |
| 70 | 130 | A2256 | H10189 | $36.0^{+40.0}_{-36.0}$ | $3.3^{+7.8}_{-3.3}$ |
| 71 | 52 | A779 | I1841 | $39.0^{+31.0}_{-16.0}$ | $3.1^{+1.1}_{-1.1}$ |
| 72 | 207 | 3C449 | I3916 | $58.0^{+63.0}_{-58.0}$ | $2.5^{+6.0}_{-2.5}$ |
| 73 | 123 | A2197 | I1857 | $51.0^{+40.0}_{-51.0}$ | $2.4^{+3.0}_{-2.4}$ |
| 74 | 71 | A1367 | I296 | $40.0^{+68.0}_{-40.0}$ | $2.3^{+6.8}_{-2.3}$ |
| 75 | 35 | A539 | I2353 | $34.0^{+52.0}_{-34.0}$ | $2.1^{+6.8}_{-2.1}$ |
| 76 | 194 | SC1842-63 | I6105 | $48.0^{+24.0}_{-4.0}$ | $1.9^{+0.7}_{-0.7}$ |
| 77 | 186 | FORNAX-A | H1885 | $> 19.0^{+0.0}_{->6.0}$ | $> 1.9^{+0.0}_{->2.0}$ |
| 78 | 58 | A979 | I6098 | $54.0^{+138.0}_{-54.0}$ | $1.4^{+11.3}_{-1.4}$ |
| 79 | 59 | A999 | I7700 | $36.0^{+59.0}_{-36.0}$ | $1.1^{+0.8}_{-1.1}$ |
| 80 | 23 | A397 | I7699 | $35.0^{+72.0}_{-35.0}$ | $0.8^{+3.1}_{-0.8}$ |

**Table 6** Supplementary Reference Data From The Literature

| Cluster | $L_{\alpha}$ ($10^{40}$ erg s$^{-1}$ ) | $t_{\rm cool}$ ( $H_0^{-1}$) | $\dot{M}$ ( M$_{\odot}$ yr$^{-1}$ ) |
|---|---|---|---|
| (i) | (ii) | (iii) | (iv) |
| A85 | 17.84 | $0.291^{+0.046}_{-0.046}$ | $236^{+14}_{-18}$ |
| A119 | .......... | $1.120^{+0.713}_{-0.713}$ | $23^{+13}_{-20}$ |
| A133 | 84.05 | .......... | .......... |
| A262 | 21.94 | $0.067^{+0.031}_{-0.031}$ | $47^{+17}_{-18}$ |
| A399 | .......... | $1.948^{+0.383}_{-0.383}$ | .......... |
| A401 | .......... | $1.542^{+0.453}_{-0.453}$ | $12^{+3}_{-10}$ |
| A426 | 4502.28 | $0.037^{+0.002}_{-0.002}$ | $183^{+17}_{-19}$ |
| A478 | .......... | $0.176^{+0.038}_{-0.038}$ | $570^{+359}_{-307}$ |
| A496 | 36.44 | $0.161^{+0.023}_{-0.023}$ | $112^{+20}_{-16}$ |
| A576 | .......... | $1.158^{+0.253}_{-0.253}$ | $6^{+2}_{-2}$ |
| A644 | .......... | $0.667^{+0.084}_{-0.084}$ | $326^{+48}_{-181}$ |
| A754 | .......... | $0.729^{+0.959}_{-0.959}$ | $24^{+24}_{-18}$ |
| A1060 | .......... | $0.169^{+0.253}_{-0.253}$ | $9^{+9}_{-6}$ |
| A1126 | 547.76 | .......... | .......... |
| A1367 | .......... | $1.649^{+0.284}_{-0.284}$ | .......... |
| A1644 | .......... | $0.314^{+0.077}_{-0.077}$ | $19^{+3}_{-3}$ |
| A1650 | .......... | $4.142^{+7.900}_{-7.900}$ | .......... |
| A1656 | .......... | $7.286^{+18.331}_{-18.331}$ | .......... |
| A1689 | .......... | $1.058^{+0.176}_{-0.176}$ | $164^{+60}_{-40}$ |
| A1736 | .......... | $2.523^{+0.575}_{-0.575}$ | .......... |
| A1795 | 964.62 | $0.192^{+0.031}_{-0.031}$ | $478^{+38}_{-56}$ |
| A1991 | 28.55 | .......... | .......... |
| A2029 | .......... | $0.291^{+0.054}_{-0.054}$ | $402^{+40}_{-50}$ |
| A2052 | 38.46 | $0.084^{+0.015}_{-0.015}$ | $90^{+20}_{-19}$ |
| A2063 | .......... | $0.314^{+0.077}_{-0.077}$ | $45^{+9}_{-10}$ |
| A2065 | .......... | $2.554^{+0.629}_{-0.629}$ | .......... |
| A2142 | .......... | $0.230^{+0.061}_{-0.061}$ | $188^{+96}_{-71}$ |
| A2147 | .......... | $1.166^{+1.879}_{-1.879}$ | $54^{+29}_{-46}$ |
| A2199 | 23.54 | $0.184^{+0.138}_{-0.138}$ | $150^{+46}_{-40}$ |
| A2244 | .......... | $0.936^{+0.153}_{-0.153}$ | $82^{+109}_{-38}$ |
| A2255 | .......... | $10.968^{+20.555}_{-20.555}$ | .......... |
| A2256 | .......... | $13.422^{+2.915}_{-2.915}$ | .......... |
| A2319 | .......... | $1.005^{+0.169}_{-0.169}$ | $66^{+26}_{-33}$ |
| A2597 | 2995.82 | $0.851^{+0.153}_{-0.153}$ | $480^{+148}_{-117}$ |
| A2626 | 32.97 | .......... | .......... |
| A2670 | .......... | $0.161^{+0.038}_{-0.038}$ | .......... |
| A3112 | .......... | $0.161^{+0.077}_{-0.077}$ | $430^{+115}_{-81}$ |
| A3158 | .......... | $1.779^{+0.291}_{-0.291}$ | .......... |
| A3266 | .......... | $1.649^{+1.327}_{-1.327}$ | $10^{+15}_{-9}$ |
| A3391 | .......... | $2.439^{+0.437}_{-0.437}$ | .......... |
| A3532 | .......... | $3.382^{+1.143}_{-1.143}$ | .......... |
| A3562 | .......... | $0.345^{+0.368}_{-0.368}$ | $45^{+17}_{-28}$ |
| A3571 | .......... | $0.598^{+0.989}_{-0.989}$ | $79^{+61}_{-61}$ |
| A3581 | 40.81 | .......... | .......... |
| A3667 | .......... | $12.655^{+18.101}_{-18.101}$ | .......... |
| A4059 | .......... | $0.268^{+0.652}_{-0.652}$ | $124^{+42}_{-38}$ |
| AWM7 | .......... | $0.299^{+0.092}_{-0.092}$ | $42^{+12}_{-16}$ |
| CENTAURUS | 6.56 | $0.038^{+0.008}_{-0.008}$ | $18^{+6}_{-7}$ |
| CYGNUS-A | 6374.20 | $0.322^{+0.038}_{-0.038}$ | $187^{+20}_{-21}$ |
| HYDRA-A | 142.51 | $0.138^{+0.054}_{-0.054}$ | $315^{+149}_{-70}$ |
| MKW3 | 30.98 | $0.307^{+0.069}_{-0.069}$ | $151^{+32}_{-25}$ |
| OPHIUCHUS | .......... | $0.238^{+0.568}_{-0.568}$ | $75^{+72}_{-50}$ |
| PKS 0745-19 | 2780.39 | $0.161^{+0.038}_{-0.038}$ | $702^{+271}_{-232}$ |
| SERSIC 159-03 | 71.86 | .......... | .......... |
| TRIANGULUM AUSTRALIS | .......... | $4.985^{+19.482}_{-19.482}$ | .......... |
| M87 | 9.27 | $0.002^{+0.001}_{-0.001}$ | $10^{+3}_{-3}$ |
| 2A0336+09 | 1111.29 | $0.069^{+0.008}_{-0.008}$ | $142^{+25}_{-32}$ |
| 3C129 | .......... | $0.391^{+0.345}_{-0.345}$ | $61^{+29}_{-28}$ |